\documentclass[nofootinbib,11pt]{revtex4-1}
\usepackage{amsmath,amssymb,pgf,pgfarrows,pgfnodes,float,appendix, hyperref,slashed,breakurl,graphicx}
\definecolor{Color}{rgb}{0.28, 0.24, 0.55}
\definecolor{Orange}{rgb}{1,0.38,0.11}
\hypersetup{
    colorlinks = true,
    citecolor = Orange,
    linkcolor = Color,
    urlcolor  = Orange,
}
\usepackage{graphicx}
\usepackage{subfigure}
\usepackage[margin=0.9in]{geometry}

\newcommand{\NLDBD}{$0 \nu \beta \beta$}

\usepackage{color, colortbl}
\definecolor{Gray}{gray}{0.8}
\definecolor{GrayLight}{gray}{0.4}
\definecolor{Darkgreen}{RGB}{30,120,30}
\definecolor{granate}{rgb}{0.8039,0.2,0.2}
\newcommand{\beq}{\begin{equation}}
\newcommand{\eeq}{\end{equation}}
\newcommand{\bea}{\begin{eqnarray}}
\newcommand{\eea}{\end{eqnarray}}

\usepackage{tikz}
\usetikzlibrary{"arrows", "automata", "backgrounds", "calendar", "chains", "matrix", "mindmap", "patterns", "petri", "shadows", "shapes.geometric", "shapes.misc", "spy", "trees"}
\usetikzlibrary{arrows,shapes}
\usetikzlibrary{trees}
\usetikzlibrary{matrix,arrows} 				
\usetikzlibrary{positioning}				
\usetikzlibrary{calc,through}				
\usetikzlibrary{decorations.pathreplacing}  
\usepackage{pgffor}							

\usetikzlibrary{decorations.pathmorphing}	
\usetikzlibrary{decorations.markings}
\tikzset{
    vector/.style={decorate, decoration={snake}, draw},
	provector/.style={decorate, decoration={snake,amplitude=2.5pt}, draw},
	antivector/.style={decorate, decoration={snake,amplitude=-2.5pt}, draw},
    fermion/.style={draw=black, postaction={decorate},
        decoration={markings,mark=at position .55 with {\arrow[draw=black]{>}}}},
    fermioncyan/.style={draw=black, postaction={decorate},
        decoration={markings,mark=at position .55 with {\arrow[draw=cyan]{<}}}},
    fermiondif/.style={draw=black, postaction={decorate},
        decoration={markings,mark=at position .7 with {\arrow[draw=black]{>}}}},
            fermiondif2/.style={draw=black, postaction={decorate},
        decoration={markings,mark=at position .7 with {\arrow[draw=black]{<}}}},
    fermionend/.style={draw=black, postaction={decorate},
        decoration={markings,mark=at position 1 with {\arrow[draw=black]{>}}}},
    fermionuchannel2/.style={draw=black, postaction={decorate},
        decoration={markings,mark=at position .4 with {\arrow[draw=black]{>}}}},
    scalardif/.style={dashed,draw=black, postaction={decorate},
        decoration={markings,mark=at position .7 with {\arrow[draw=black]{>}}}},
    scalarend/.style={dashed,draw=black, postaction={decorate},
        decoration={markings,mark=at position 1 with {\arrow[draw=black]{>}}}},
    fermionbar/.style={draw=black, postaction={decorate},
        decoration={markings,mark=at position .55 with {\arrow[draw=black]{<}}}},
    fermionnoarrow/.style={draw=black},
    gluon/.style={decorate, draw=black,
        decoration={coil,amplitude=4pt, segment length=5pt}},
    scalar/.style={dashed,draw=black, postaction={decorate},
        decoration={markings,mark=at position .55 with {\arrow[draw=black]{>}}}},
    scalarcyan/.style={dashed,draw=black, postaction={decorate},
        decoration={markings,mark=at position .55 with {\arrow[draw=cyan]{>}}}},
    scalaruchannel1/.style={dashed,draw=black, postaction={decorate},
        decoration={markings,mark=at position .7 with {\arrow[draw=black]{>}}}},
                  scalaruchannel2/.style={dashed,draw=black, postaction={decorate},
        decoration={markings,mark=at position .4 with {\arrow[draw=black]{>}}}},
    scalarbar/.style={dashed,draw=black, postaction={decorate},
        decoration={markings,mark=at position .55 with {\arrow[draw=black]{<}}}},
    scalarnoarrow/.style={dashed,draw=black},
    electron/.style={draw=black, postaction={decorate},
        decoration={markings,mark=at position .55 with {\arrow[draw=black]{>}}}},
	bigvector/.style={decorate, decoration={snake,amplitude=4pt}, draw},
}

\tikzstyle{block} = [draw, rectangle, 
    minimum height=3em, minimum width=6em]

\usepackage{tikz}
\usetikzlibrary{fit}
\tikzset{%
  highlight/.style={rectangle,rounded corners,color=granate,draw,text opacity =1,
    fill opacity=0.5,thick,inner sep=0pt}
}

\tikzset{
    cross/.pic = {
    \draw[rotate = 45] (-#1,0) -- (#1,0);
    \draw[rotate = 45] (0,-#1) -- (0, #1);
    }
}

\tikzset{
    square/.style={%
        draw=none,
        circle,
        append after command={%
            \pgfextra \draw[#1] (\tikzlastnode.north-|\tikzlastnode.west) rectangle 
                (\tikzlastnode.south-|\tikzlastnode.east);\endpgfextra}
    },
    square/.default=black
}

\tikzstyle{block} = [draw, rectangle, 
    minimum height=3em, minimum width=6em]

\usepackage{xparse}
\NewDocumentCommand\semiloop{O{black}mmmO{}O{above}}
{%
\draw[#1] let \p1 = ($(#3)-(#2)$) in (#3) arc (#4:({#4+180}):({0.5*veclen(\x1,\y1)})node[midway, #6] {#5};)
}

\begin{document}

\title{
\Large{On Baryon and Lepton Number Violation}}

\author{Pavel Fileviez P\'erez (Editor)} \email{pxf112@case.edu}
\affiliation{Physics Department and Center for Education and Research in Cosmology and Astrophysics, 
Case Western Reserve University, Cleveland, OH 44106, USA}

\author{Andrea Pocar (Editor)} \email{pocar@umass.edu}
\affiliation{Amherst Center for Fundamental Interactions and Department of Physics, University of Massachusetts, Amherst, MA 01003-9337, USA}

\author{K. S. Babu}
\affiliation{Department of Physics, Oklahoma State University, Stillwater, OK 74078, USA}

\author{Leah J. Broussard}
\affiliation{Oak Ridge National Laboratory, Oak Ridge, TN 37831, USA}

\author{Vincenzo Cirigliano}
\affiliation{Institute for Nuclear Theory, University of Washington, Seattle WA 98195-1550, USA}

\author{Susan Gardner}
\affiliation{Department of Physics and Astronomy, University of Kentucky, Lexington, Kentucky 40506-0055 USA}

\author{Julian Heeck}
\affiliation{Department of Physics, University of Virginia, Charlottesville, Virginia 22904-4714, USA}

\author{Ed Kearns}
\affiliation{Department of Physics, Boston University, Boston, MA 02215, USA}

\author{Andrew J. Long}
\affiliation{Department of Physics and Astronomy, Rice University, Houston, TX 77005, USA \\}

\author{Stuart Raby}
\affiliation{Department of Physics, Ohio State University, Columbus, Ohio 43210, USA}

\author{Richard Ruiz}
\affiliation{Institute of Nuclear Physics, Polish Academy of Sciences, ul. Radzikowskiego, Kracow 31-342, Poland}

\author{Evelyn Thomson}
\affiliation{Department of Physics and Astronomy, University of Pennsylvania Philadelphia, PA 19104–6396, USA}

\author{Carlos E. M. Wagner}
\affiliation{HEP Division, Argonne National Laboratory, 9700 Cass Ave., Argonne, IL 60439, USA and Enrico Fermi Institute and Kavli Institute for Cosmological Physics, Department of Physics, University of Chicago, Chicago, IL 60637, USA}

\author{Mark B. Wise}
\affiliation{Walter Burke Institute for Theoretical Physics, California Institute of Technology, Pasadena, CA 91125, USA}

\newpage
\begin{center}
SNOWMASS 2021 REPORT: \\

RF4: BARYON AND LEPTON NUMBER VIOLATING PROCESSES \\

RARE PROCESSES AND PRECISION MEASUREMENTS FRONTIER
\end{center}

%

\vspace{1.5cm}
\begin{abstract}

\end{abstract}

\maketitle

\clearpage
\newpage

\small
\tableofcontents
\normalsize

\newpage
\section{Introduction}
The Standard Model (SM) of Particle Physics describes with high precision the properties of quarks and leptons, and how they interact through the electromagnetic, weak and strong forces. There are many reasons to believe that the Standard Model has to be modified. It is well-known that the SM does not provide a mechanism for neutrino masses and it cannot explain the origin of the matter-antimatter asymmetry in the Universe. Understanding the origin of baryon ($B$) and lepton ($L$) number violation is crucial to explain the origin of neutrino masses, the mechanisms for the matter-antimatter asymmetry in the Universe, and possible baryon and/or lepton number violating processes. As a result discovery of these phenomena would have a profound impact and would guide us
towards specific effective field theory descriptions of the Standard Model.

\begin{figure}[t]
\includegraphics[width=0.75\textwidth]{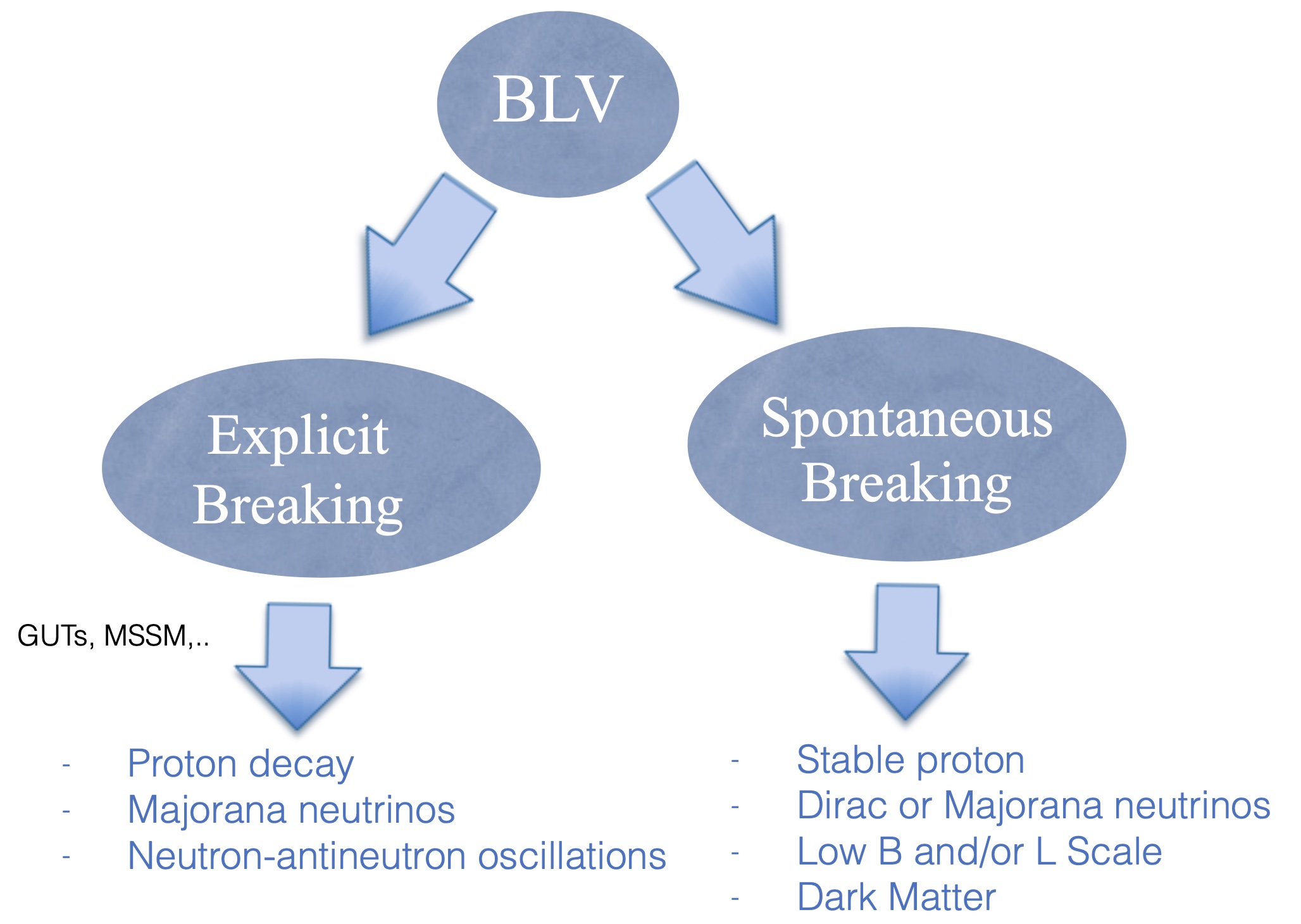}
\caption{Different ways to understand the origin of Baryon and/or Lepton number violation in physics beyond the Standard Model and the possible predictions in the different class of theories. Here BLV stands for Baryon and Lepton Number Violation. For more details see the discussion in section II.}
\label{Btheories}
\end{figure}

In this report we discuss the main ideas about the origin of baryon and lepton number violation in theories for physics beyond the Standard Model. We discuss theories such as grand unified theories and supersymmetric theories where $B$ and $L$ are explicitly broken, and theories where one has the spontaneous breaking of these symmetries at the low scale. In Fig.~\ref{Btheories} we show the two ways to understand the origin of $B$ and $L$ violation, in the case of explicit breaking one typically predicts proton decay and Majorana neutrinos, while in the simple theories for spontaneous breaking the proton is stable and the neutrinos could be Majorana or Dirac fermions.
As is well-known, in any theory with Majorana neutrinos one 
can make predictions for neutrinoless double beta and generically one could have neutron-antineutron oscillations in theories predicting baryon number violation by two units. 

The predictions for extremely rare processes such as neutrinoless double beta decay, proton decay and neutron-antineutron oscillations are discussed in detail. The discovery of any of these processes can tell us about the nature of $B$ and/or $L$ violation, and the relevant scale related to the breaking of these symmetries. We also discuss other more exotic processes where $B$ and $L$ are broken in more than two units. 
The possibility to observe $B$ and/or $L$ signatures at colliders is discussed, emphasizing the need to look for exotic signatures
associated to the different seesaw mechanisms for neutrino masses, signatures in theories with leptoquarks and in supersymmetric theories with $R$-parity violation. The simplest mechanisms to explain the matter-antimatter asymmetry are discussed. 

The discovery of any Baryon and/or Lepton violating signatures could provide an unique window to physics beyond the Standard Model. The main goal of this report is to provide a complete discussion of the main ideas in physics beyond the Standard Model where we can make predictions for $B$ and $L$ violating processes. 

This report is organized as follows: In section~\ref{sect:Theory} we discuss the theories for baryon and lepton number violation, in section~\ref{sect:NLDBD} we discuss the theoretical predictions and experiments for neutrinoless double beta decay, while in section~\ref{sect:Collider} 
we discuss the signatures of different theories for neutrino masses, $R$-parity violating signatures and other exotics at collider experiments. We discuss the predictions for proton decay and 
the experimental reach in section~\ref{sect:Proton}, while in section~\ref{sect:nnbar} we discuss the motivations and experimental efforts to search for neutron-antineutron oscillations. In section~\ref{sect:Exotic} we discuss more exotic $B$ and/or $L$ violating processes, and in section~\ref{sect:Cosmo} we provide a detailed discussion of the different mechanisms for baryogenesis. Finally, we summarise our main ideas and recommendations for the future of this field.          

\section{Theories for baryon and lepton number violation}
\begin{center}
 Pavel Fileviez P\'erez (Case Western Reserve University), Mark B. Wise (Caltech)
\end{center}
\label{sect:Theory}
In the Standard Model baryon $(B)$ and total lepton $(L)$ number are accidental global symmetries broken at the quantum level by $SU(2)$ instantons. 
$L$ and $B$ violation first occur via dimension five and six operators, respectively,  if one treats the Standard Model as an effective field theory~\cite{Weinberg:1979sa}:
\begin{equation*}
{\mathcal L} \supset \frac{1}{\Lambda_L} \ell \ell H H + \frac{1}{\Lambda_B^2} q q q \ell + \ldots.
\end{equation*}
Here $H$ is the Standard Model Higgs, $\ell$ and $q$ are the lepton and quark doublets, respectively.
Naively, the experimental bounds on neutrino masses demand $\Lambda_L \lesssim 10^{14}$ GeV, while proton decay experiments tells us that $\Lambda_B \gtrsim 10^{15}$ GeV~\cite{Nath:2006ut}. 
It is important to emphasize that we do not know the scales for $B$ and $L$ violation because these symmetries could be broken in different units. 
For example, the proton could be stable if the effective $B$ violation is $\Delta B \neq \pm 1$ allowing $\Lambda_B$ to be close to the electroweak scale.
In cosmology $B$ violation is a key ingredient needed to explain the matter-antimatter asymmetry in the Universe~\cite{Sakharov:1967dj}.
From neutrino oscillation experiments we know that the $L_{e}, L_\mu, L_\tau$ symmetries are broken in nature.  
$L$ could be conserved and neutrinos can be Dirac fermions. Neutrinos can be Majorana fermions if the effective $L$ violation is $\Delta L=\pm 2$.

The Standard Model does not provide a mechanism to explain the origin of neutrino masses. In the case of Majorana neutrinos, $\Delta L = \pm 2$, 
and one has several seesaw mechanisms 
that can explain the smallness of their masses. In Type I (or canonical) seesaw~\cite{Minkowski:1977sc,Mohapatra:1979ia,Glashow:1979nm,Gell-Mann:1979vob,Yanagida:1979as} 
the neutrino masses are suppressed by the masses of the right-handed neutrinos $M_R$, i.e. $m_\nu^I \sim v_W^2 / M_R$ with $v_W \sim 10^2$ GeV being the electroweak scale.
In the case of Type II seesaw the neutrino masses are proportional to the small vacuum expectation value of a scalar $SU(2)$ triplet, $m_\nu^{II} \sim v_\Delta$~\cite{Konetschny:1977bn,Cheng:1980qt,Mohapatra:1980yp,Lazarides:1980nt,Schechter:1980gr}.
Type III seesaw is similar to Type I, but in this case neutrino masses are suppressed by the masses of the fermionic $SU(2)$ triplets $M_\rho$, i.e. $m_\nu^{III} \sim v_W^2 / M_\rho$~\cite{Foot:1988aq}.
In a similar spirit one can consider different simplified models for $\Delta B=\pm 2$, see for example Ref.~\cite{Arnold:2012sd}.

There are several theories that predict $B$ and $L$ violation. In grand unified theories (GUTs), $B$ and $L$ 
are explicitly broken when quarks and leptons are unified in the same multiplet. The theories based on $SU(5)$~\cite{Georgi:1974sy}, and 
$SO(10)$~\cite{Georgi:1974my,Fritzsch:1974nn} are attractive extensions of the Standard Model. These theories could describe physics at 
the high-scale, $\Lambda_{GUT} \sim 10^{15-16}$ GeV, and one predicts generically the decay of the proton. $SO(10)$ GUTs predict the existence of Majorana neutrinos, $\Delta L= \pm 2$, and proton decay, $\Delta B = \pm 1$. Unfortunately, 
the minimal theory based on $SU(5)$ is ruled out because one cannot explain the values of the gauge couplings at the electroweak scale.
For simple realistic non-supersymmetric GUTs based on $SU(5)$ and $SO(10)$, see for example the studies in Refs.~\cite{Dorsner:2005fq,Bajc:2006ia,Preda:2022izo}.

The Minimal Supersymmetric Standard Model (MSSM) is an extension of the Standard Model that could describe physics around the TeV scale.
The MSSM predicts new interactions that explicitly break $B$ and $L$. The $R$-parity violating interactions could give 
rise to fast proton decay and provide a mechanism for Majorana neutrino masses. These interactions can change
the way we could discover Supersymmetry (SUSY) at colliders. For a review about R-parity violation see Ref.~\cite{Barbier:2004ez}. SUSY provides a theoretical framework 
to understand the unification of gauge interactions~\cite{Dimopoulos:1981yj,Ibanez:1981yh,Einhorn:1981sx,Marciano:1981un} since one achieves unification with good precision thanks 
to the new fields present in the MSSM. SUSY GUTs~\cite{Dimopoulos:1981zb,Sakai:1981gr} could describe physics at a high scale, $\Lambda \sim 10^{16-17}$ GeV, but generally predict fast proton decay mediated by colored Higgs(ino) fields if one does not assume heavy squarks and sleptons~\cite{Nath:2006ut}. It is important to mention that in
supersymmetric theories one has the $B-L$ conserving dimension five operator, $\hat{q}\hat{q}\hat{q} \hat{\ell} / M_{Pl}$, 
suppressed by the Planck scale that generically could give rise to fast proton decay.

$B$ and $L$ violation at a low scale can occur in theories where these symmetries are defined as local gauge symmetries. 
The idea of a local gauge theory for $B$ was first mentioned in Ref.~\cite{Pais:1973mi}. In these theories one can have the spontaneous 
$B$ and $L$  breaking close to the electroweak scale in agreement with all the experimental constraints.
Since $B$ and $L$ are not anomaly-free symmetries of the Standard Model, one needs to add new fermions~\cite{Rajpoot:1989jb,Foot:1989ts,Carone:1995pu,FileviezPerez:2010gw} to define an anomaly-free theory based on $U(1)_B$ and/or $U(1)_L$. 
The simplest realistic anomaly-free theories with these features contain only four~\cite{FileviezPerez:2014lnj} or six~\cite{Duerr:2013dza} extra fermionic representations. 
These theories~\cite{FileviezPerez:2014lnj,Duerr:2013dza} predict the stability of the proton because $B$ must be broken in 3 units and then one can have the spontaneous breaking of $U(1)_B$ at the low scale.
A fermionic dark matter candidate is also predicted in these models as a natural consequence of anomaly cancellation. The cosmological constraints 
on the dark matter density imply an upper bound on the $B$ breaking scale about $\mathcal{O}(10)$ TeV~\cite{FileviezPerez:2019jju}.  

$B$ and $L$ violation could play an important role in physics beyond the Standard Model. 
Experiments looking for $B$ and $L$ violation could be crucial to establish a new theory for physics beyond the Standard Model 
that can explain the origin of neutrino masses and explain the matter-antimatter asymmetry in the Universe.

\section{Neutrinoless double beta decay}
\begin{center}
Vincenzo Cirigliano (INT, University of Washington), Andrea Pocar (UMass Amherst)
\end{center}
\label{sect:NLDBD}

Neutrinoless double beta decay (\NLDBD) is the process where two neutrons inside an atomic nucleus are transmuted into two protons and two electrons without the emission of neutrinos. An observation of this process would indicate that lepton number (L) is not a good symmetry of nature ($\Delta L=2$)  and that the neutrino mass  has a Majorana component, implying that the mass eigenfields are self-conjugate~\cite{Schechter:1981bd}. 
Observation of \NLDBD\ would thus have profound implications on understanding  the mechanism of neutrino mass 
generation~\cite{Minkowski:1977sc,Mohapatra:1979ia,Gell-Mann:1979vob}, 
and would also give insight into leptogenesis scenarios for the generation of the 
matter-antimatter asymmetry in the universe~\cite{Davidson:2008bu}. 

Current experimental searches pose  very stringent limits  \cite{KamLAND-Zen:2016pfg,EXO-200:2019rkq,Majorana:2019nbd,CUPID:2019gpc,GERDA:2020xhi,CUORE:2021mvw,KamLAND-Zen:2022tow,Arnquist:2022zrp}, e.g. $T^{0\nu}_{1/2}>2.3\times 10^{26}$ yr for ${}^{136}$Xe
\cite{KamLAND-Zen:2022tow,KamLAND-Zen:2016pfg},
with `ton-scale' experiments aiming for one to two orders of magnitude sensitivity improvement in $T^{0\nu}_{1/2}$~\cite{nEXO:2018ylp,nEXO:2021ujk,LEGEND:2021bnm,armstrong2019cupid,CUPID:2022wpt,SNO:2021xpa,Nakamura:2020szx,adams2020sensitivity}.
When interpreted in terms of light Majorana neutrino exchange, this sensitivity corresponds roughly to an effective Majorana mass  $m_{\beta \beta} 
\equiv | \sum_{i} U_{ei}^2 m_i| \sim 10$~meV, 
modulo theoretical uncertainties  in the nuclear matrix elements discussed below. 
This corresponds to covering the region in which \NLDBD\ occurs if neutrinos are Majorana particles and the spectrum follows the so-called inverted ordering (often called inverted hierarchy), with the heaviest of the three light neutrino having the largest electron-flavor component. 
While  the simplest interpretation of $0 \nu \beta \beta$ experiments assumes that lepton-number violation (LNV) is due to the exchange of light Majorana neutrinos, 
this only covers the scenarios in which LNV originates  at very high energy (high-scale see-saw). 
In many beyond-the-SM (BSM) constructions,  other lower-scale sources of LNV exist that can induce $0\nu\beta\beta$. For example, in left-right symmetric models, apart from the exchange of a light Majorana neutrino, there appear contributions  from the exchange of heavy neutrinos and charged scalars. 
In other scenarios there may be light right-handed (sterile) neutrinos with masses much lower than the electroweak scale. 
Given the breadth of mechanisms and scales associated with LNV sources 
(for a review see \cite{Rodejohann:2011mu}), 
ton-scale searches for \NLDBD\  have a significant discovery potential that goes beyond the `inverted hierarchy' high-scale seesaw.

The current Snowmass planning exercise focuses on R\&D for `beyond ton scale' experiments.  
By the end of the ton-scale program, we will be either still in a discovery mode or in a post-discovery phase. 
In both scenarios  further  vigorous  R\&D  will be needed: 

\begin{itemize}

\item[(i)] In one case,  the emphasis will be on reaching the next discovery benchmark. 
A very well-motivated benchmark is provided in the framework  of light Majorana exchange 
by $m_{\beta \beta} \sim 1$~meV, which covers the normal-ordering spectrum modulo a small 
finely tuned region in which the Majorana phases conspire to suppress $m_{\beta \beta}$ well below 
its natural scale. Reaching this sensitivity would provide the ultimate test of the 
high-scale seesaw paradigm. 
The corresponding factor of $\sim$100 in half-life reach would also push the 
sensitivity to multi-TeV scale LNV well beyond the reach of high-energy colliders, thus 
probing uncharted territory. 

\item[(ii)] In the other case, to capitalize on a discovered non-zero \NLDBD\ rate, 
the emphasis will be on uncovering the mechanism behind LNV and neutrino mass. 
LNV sources  can be disentangled by  performing 
sensitive studies of single-electron spectra, 
electron angular distribution, and dependence of the \NLDBD\ rate on the isotope, 
in conjunction with other probes such as  $pp \to ee + 2 \  {\rm jets}$ at colliders~\cite{Keung:1983uu,Peng:2015haa}, 
neutrino oscillations, and input from Cosmology. 
\end{itemize}
This program requires developments in theory and experiment, and next we 
discuss the prospects for both:

On the theory side,  a full assessment of the impact of \NLDBD\   
searches for particle physics necessarily requires bridging widely separated energy scales, 
from the scale where LNV originates all the way down to the nuclear scale. 
Effective Field Theory (EFT) provides the framework for doing this, 
by describing in a systematically improvable expansion in ratios of energy scales  the 
LNV dynamics both at high energy and  at hadronic / nuclear scales. 
The EFT  approach to \NLDBD\ requires non-perturbative  input from lattice QCD 
and culminates in a set of LNV-mechanism-dependent  $nn \to pp$ transition operators 
to be used in nuclear many-body calculations. 
The combination of EFT, lattice QCD, and first-principle nuclear many-body methods 
provides the  only path towards 
reducing the current $O(1)$ uncertainties in \NLDBD\ 
matrix elements~\cite{Engel:2016xgb}, 
which currently plague the interpretation of \NLDBD\ in terms of LNV parameters (e.g. $m_{\beta \beta}$) 
and preclude a credible diagnosing of LNV mechanisms even in a scenario of multiple observations (e.g. the matrix elements 
uncertainties are larger than the variation of central values with isotope). 

A detailed account of the theoretical status and prospects is presented in Ref.~\cite{Cirigliano:2022oqy}. 
Encouraging progress has been made in the last few years on several fronts. 
The EFT framework for \NLDBD\ has been   developed~\cite{Prezeau:2003xn,Cirigliano:2017tvr,Cirigliano:2018hja,Cirigliano:2017djv,Cirigliano:2018yza,Cirigliano:2019vdj,Dekens:2020ttz}, 
generalizing previous approaches~\cite{Doi:1985dx,Pas:1999fc,Pas:2000vn}. 
Progress has been made in LQCD 
for the  $\pi^- \pi^- \to e e$ process~\cite{Nicholson:2018mwc,Feng:2018pdq,Tuo:2019bue,Detmold:2020jqv} 
and towards two-nucleon amplitudes~\cite{Briceno:2015tza,Feng:2020nqj,Davoudi:2020xdv}, 
and first-principles nuclear many-body methods have been 
applied to  light nuclei for benchmarking purposes~\cite{Pastore:2017ofx,Basili:2019gvn} 
and to  isotopes of experimental interest~\cite{Yao:2019rck,Belley:2020ejd,Wirth:2021pij}. 
However,  several challenges remain 
and will require  concerted efforts in order to achieve reduced theoretical uncertainties in the next decade. 
These include: 
(i) An assessment of the uncertainties in the rate and single-electron spectra for various isotopes of experimental interest. 
On the hadronic side, 
the matching step from quark-level operators to  hadronic EFTs involves  non-perturbative parameters, the 
so-called low-energy constants (LECs). 
Notably, new LECs appear at leading order in  \NLDBD\ for both light Majorana neutrino exchange~\cite{Cirigliano:2018hja}
(for which dispersive~\cite{Cirigliano:2020dmx,Cirigliano:2021qko} and large-$N_C$~\cite{Richardson:2021xiu} estimates recently appeared) 
and TeV scale LNV~\cite{Cirigliano:2018yza} (completely unknown). 
Improvements on these key quantities will be possible in the future through the analysis of suitable $\Delta I =2$ observables, 
as well as direct calculations using  lattice QCD  methods 
(for reviews and prospects see~\cite{Cirigliano:2020yhp,Davoudi:2020ngi}). 
On the nuclear structure side, a major future thrust will involve the analysis of nuclear matrix elements with ab-initio methods 
and a validation of the EFT expansion. 
(ii)  On the phenomenology side, 
interesting future challenges  include a systematic 
 connection between $0\nu\beta\beta$ and LNV observables at present (ATLAS, CMS, FASER)
and future  (EIC, ShIP, MATHUSLA) collider experiments; 
as well as a comprehensive  study of  $0\nu\beta\beta$ constraints on realistic models 
with sterile neutrinos in the keV-GeV mass  range (e.g.\ 3+3 models), 
which are relevant to low-scale leptogenesis scenarios. 

On the experimental side, the global $0\nu\beta\beta$ decay effort is growing at a fast pace, with half-life sensitivity well in excess of $10^{25}$ years firmly established with a veriety of techniques for the isotopes $^{136}$Xe, $^{76}$Ge, and $^{130}$Te (with $^{82}$Se and $^{100}$Mo following suit). An up-to-date review of the field is provided in Ref.~\cite{Dolinski:2019nrj}.
A tonne-scale program was launched under the stewardship, in the US, of the DoE Office of Nuclear Physics to reach $T_{1/2}\sim10^{28}$ years and entirely cover the inverted neutrino mass ordering parameter space for a $0\nu\beta\beta$ process mediated by the virtual exchange of light Majorana neutrinos. In a meeting held in the Fall of 2021\footnote{https://agenda.infn.it/event/27143/timetable/\#20210929.detailed}, European and North American stakeholders reiterated the importance of aggressively searching for $0\nu\beta\beta$ decay, a process "capable of reshaping our current understanding of nature" and concluded that "the best chance for success is an international campaign with more than one large ton-scale experiment implemented in the next decade with one in Europe and the other in North America."
In addition, the international community in $0\nu\beta\beta$ decay is interested in exploring “whether a more formal structure for international coordination of this research would be beneficial not only for experiments of the next decade but also for future multi-ton and/or multi-site experiments.”

The formal ton-scale program includes three projects: i) CUPID ($^{100}$Mo)~\cite{armstrong2019cupid,CUPID:2022wpt}, an array of 1596 lithium molybdenate (LMO) scintillating bolometers instrumented with 1710 light detectors and housed in the cryostat currently hosting the CUORE experiment ($^{130}$Te) at Gran Sasso; ii) LEGEND-1000 ($^{76}$Ge)~\cite{LEGEND:2021bnm}, a scaled-up successor of the LEGEND-200 experiment under construction at Gran Sasso (itself a joint successor of the GERDA and Majorana-Demonstrator experiments), using a ton array of high-purity p-type, ICPC Ge semiconductor detectors immersed in scintillating liquid argon; and iii) nEXO ($^{136}$Xe)~\cite{nEXO:2018ylp,nEXO:2021ujk}, a single-volume TPC using 5 tonnes of enriched liquid xenon and designed based on the predecessor EXO-200 experiment. Other programs with significant US involvement using technologies with good scalability prospects include NEXT ($^{136}$Xe)~\cite{adams2020sensitivity}, a high-pressure gaseous Xe TPC, KamLAND-Zen 800 ($^{136}$Xe)~\cite{KamLAND-Zen:2022tow}, running 745 kg of enriched xenon in liquid scintillator and currently setting the tightest lower bound on $0\nu\beta\beta$ decay at $T_{1/2}>2.3\times 10^{26}$ years, and SNO+ ($^{130}$Te)~\cite{SNO:2021xpa}, planning to dissolve large amounts of natural tellurium in liquid scintillator. 

Exploring $0\nu\beta\beta$ decay beyond the tonne-scale will require giant detectors with up to hundred of tons of decaying isotope, very low background and technologies that allow to preserve signal identification and background rejection capabilities through this significant scale-up. Conceptual ideas for detectors with $0\nu\beta\beta$ decay half-life sensitivity of $10^{29}-10^{30}$ years have recently been put forth. These revolve around very large, monolithic detectors that detect $0\nu\beta\beta$ decays with $\sim$100\% efficiency, build on current technologies that have demonstrated the potential for scale-up, and implement novel technologies and instrumentation for enhanced background rejection. Key challenges include the procurement of double-beta emitting isotope at the 100-tonne scale, the management of the solar neutrino background, the mitigation of radon in and around the sensitive detector volume, and efficient signal collection at MeV energy to enable signal-to-background discrimination solutions.
Two main threads are identified:
\begin{itemize}
    \item Very large time projection chambers (TPCs). 
    \item Very large, multi-purpose liquid scintillator detectors.
\end{itemize}

The plausibility of scaling pure xenon TPCs to the ton-scale was studied in Ref.~\cite{avasthi2021kiloton}. The study explores the potential of both liquid and gaseous TPC configurations, and focuses on one of the largest challenges, {\it i.e.}, the procurement of the xenon. The possibility to dissolve significant amounts of xenon in large liquid argon TPCs was also proposed in the context of the DUNE and DarkSide collaborations. 
Non-xenon TPCs have also been proposed for $0\nu\beta\beta$ decay, specifically using $^{82}$SeF$_6$~\cite{nygren2018neutrinoless} operated as an ion TPC.
Very large, loaded scintillators are also proposed to include an ambitious $0\nu\beta\beta$ decay program, such as the \textsc{Theia} concept~\cite{Theia:2019non} with $^{136}$Xe and $^{nat}$Te as leading isotopes under consideration. These large detectors would implement novel liquid scintillators with fast light readout able to separate the Cherenkov and scintillation components for single electron background discrimination.

We refer to a document prepared by the Neutrino Frontier (NF05) titled "Snowmass NF05 Report: Neutrino Properties" for a detailed account of the efforts emerged during the Snowmass 2022 process towards the definition of an experimental $0\nu\beta\beta$ decay framework to cover the normal neutrino mass ordering. The report includes a broad overview of the current R\&D activities that are vital for the success of a beyond-the-tonne-scale $0\nu\beta\beta$ decay program.

\section{Baryon and Lepton number violation at colliders}
\begin{center}
Richard Ruiz (INP Krak\'ow), Evelyn Thomson (University of Pennsylvania)
\end{center}
\label{sect:Collider}


The scales for baryon and lepton number violation are unknown and one can study the possibilities to test the origin of $B$ and $L$ violation at colliders. We know that the SM does not provide a mechanism for neutrino masses and we could look for exotic signatures at colliders that are unique to these mechanisms. There are several theories for physics beyond the Standard Model that predict $B$ and $L$ violation, and one can hope to test these ideas by looking for exotic collider signatures.

Due to their far-reaching consequences, searches for baryon number violation (BNV), lepton number violation (LNV), 
lepton flavor violation (LFV), and $R$-parity violation (RPV) are among the highest priorities for the particle, nuclear, and astroparticle communities~\cite{EuropeanStrategyforParticlePhysicsPreparatoryGroup:2019qin,EuropeanStrategyGroup:2020pow}.
Needless to say, there is tremendous sensitivity to such new phenomena at low- and high-energy experiments, and particularly at the LHC~\cite{Gluza:2002vs,Han:2006ip,
Nath:2006ut,Abada:2007ux,delAguila:2008cj,Atre:2009rg,Tello:2010am,Deppisch:2015qwa,Cai:2017jrq,Cai:2017mow,Pascoli:2018heg,Han:2022qgg,Mandal:2022zmy,Abdullahi:2022jlv,Chauhan:2022gkz}. 
In light of the upcoming high-luminosity program, we briefly summarize 
the theoretical and experimental outlook for tests of BNV, LNV, LFV, and RPV 
at the HL-LHC for a subset of highly popular models. Reviews and up-to-date summaries can be found in Refs.~\cite{Nath:2006ut,Abada:2007ux,FileviezPerez:2009ud,Deppisch:2015qwa,Cai:2017jrq,Cai:2017mow,Pascoli:2018heg,Han:2022qgg,Mandal:2022zmy,Abdullahi:2022jlv}.


Several mechanisms exist that can generate Majorana neutrino masses:
The Type I seesaw extends the SM by at least two right-handed neutrinos, $\nu_R$. After mixing with the SM's left-handed neutrinos, $\nu_L$, one has the light mass eigenstates $\nu_m$ and a collection of heavy mass eigenstates $N_{m'}$~\cite{Minkowski:1977sc, Yanagida:1979as, Gell-Mann:1979vob, Glashow:1979nm, Mohapatra:1979ia, Shrock:1980ct,Schechter:1980gr}.  
\textit{A priori}, the mass and mixing of $N_{m'}$ are unknown, and different assumptions can lead to Dirac-like (pseudo-Dirac) or Majorana neutrinos~\cite{Wolfenstein:1981kw,Petcov:1982ya}. With additional assumptions, connections can be made to dark matter, new gauge theories, supersymmetry,
and grand unification.
This is pertinent for effective field theories that are extended by right-handed neutrinos, e.g., $\nu$SMEFT~\cite{delAguila:2008ir,Bhattacharya:2015vja,Liao:2016qyd,Datta:2020ocb,Chala:2020pbn}. Light $N_{m'}$ can be produced at the LHC in the decays of mesons, baryons, $\tau$ leptons, EW bosons, and top quarks, while heavier mass states can be produced in a variety of mechanisms at the LHC~\cite{Datta:1993nm,Atre:2009rg,Degrande:2016aje,Cai:2017mow}.
Dedicated Monte Carlo tools are also available for these channels~\cite{Alva:2014gxa,Degrande:2016aje,Pascoli:2018heg}.
$N_{m'}$ can be short- or long-lived~\cite{Alimena:2019zri,Abdullahi:2022jlv}, and mediate LNV and/or LFV. 
Flagship signatures for $N_{m'}$ include: searches for same-sign lepton pairs $(\ell^\pm_i \ell^\pm_j)$ with jets~\cite{Keung:1983uu,Han:2006ip}, i.e., $pp\to \ell^\pm_i \ell^\pm_j + nj+X$, which manifestly exhibits LNV and LFV;
searches for three charged leptons and missing transverse energy (MET), i.e., $pp\to \ell^\pm_i \ell^\pm_j \ell^\mp_k +$MET, which can exhibit LFV; 
and searches for opposite-sign, different-flavor lepton pairs, i.e., 
$pp\to \ell^\pm_i \ell^\mp_j + nj+X$, which manifestly exhibits LFV for $i\neq j$.

At $\sqrt{s}=13$ TeV, LHC experiments set constraints on active-sterile mixing for a range of masses, from $m_N\approx1$ GeV-10 TeV, that far exceed LEP limits~\cite{LHCb:2014osd,ATLAS:2019kpx,LHCb:2020wxx,CMS:2022fut,CMS-PAS-EXO-21-003}.
For Dirac and Majorana $N$, constraints on active-sterile mixing reach as small as $\vert V_{\ell N}\vert^2\sim 10^{-7}~(0.5)$ for $m_N\sim10$ GeV (10 TeV). HL-LHC projections show~\cite{Cottin:2018kmq,Cottin:2018nms,Pascoli:2018heg,Alimena:2019zri,Drewes:2019fou,R:2020odv,Fuks:2020att,CMS:2022msk} that sensitivity can be pushed still by orders of magnitude for masses up to $\mathcal{O}(10)$ TeV. 
Beyond this, the HL-LHC program can be enhanced by further efforts to improve sensitivity at/with:
high-rapidity  experiments; intermediate masses in the range $m_{N_{m'}} = 20-150$ GeV; final states involving $\tau$ leptons; tracking and timing related to displaced vertices;
indirect searches for $N_{m'}$; novel production and decay modes and their simulations; boosted topologies; 
as well as connections to new LHC experiments.

In the Type II seesaw one has an extra scalar triplet, ${\Delta}\sim (1,3,1)$~\cite{Konetschny:1977bn, Cheng:1980qt,Lazarides:1980nt,Schechter:1980gr, Mohapatra:1980yp}. After EW symmetry breaking the vacuum expectation value of ${\Delta}$ generates light, left-handed Majorana neutrino masses. 
After mixing with the SM Higgs boson, the model predicts doubly and singly charged scalars, $\Delta^{\pm\pm}$ and $\Delta^\pm$, as well as neutral CP-even and CP-odd scalars, $\Delta^0$ and $A^0$. The scalars of the Type II seesaw can be produced through a variety of gauge interactions and Higgs mixing~\cite{Han:2007bk,Nemevsek:2016enw,Cai:2017mow,Fuks:2019clu}. Predictions (and tools) for many of these channels are known to NLO in QCD with parton shower-matching~\cite{Muhlleitner:2003me,Fuks:2019clu}. 
Importantly, the decays of Type II scalars are governed by the PMNS matrix and neutrino masses, and reveal a complementarity between oscillation experiments and the LHC~\cite{Han:2007bk,FileviezPerez:2008jbu,FileviezPerez:2008wbg,Han:2022qgg}. 
Flagship signatures include: 
searches for pair and associated production of $\Delta^{\pm\pm}$ and $\Delta^\pm$ into final states with multiple charged leptons, e.g., 
$pp \to \Delta^{\pm\pm}\Delta^{\mp(\mp)} \to \ell^\pm_i\ell^\pm_j\ell^\mp_k(\ell^\mp_l)+X$, which can manifestly exhibit LFV;
searches for same-sign lepton and $W$ pairs, e.g.,
$pp \to \Delta^{\pm\pm}\Delta^{\mp\mp} \to \ell^\pm_i\ell^\pm_j W^\mp W^\mp  \to \ell^\pm_i\ell^\pm_j + nj +X$, which manifestly exhibits LNV and LFV;
as well as searches for single production of $\Delta^{\pm\pm}$ from same-sign $W^\pm W^\pm$ scattering, i.e., $pp \to \Delta^{\pm\pm}jj \to \ell^\pm_i \ell^\pm_j jj +X$, 
which manifestly exhibits LNV and LFV.

Present searches for Type II scalars at the LHC constrain masses to be above $m_H = 230-870$ GeV, depending on the precise benchmark and signal category~\cite{CMS:2017pet,CMS:2017fhs,ATLAS:2017xqs,ATLAS:2018ceg,ATLAS:2021jol}.
Prospects for the HL-LHC appear promising
and can push this to at least 2 TeV~\cite{delAguila:2008cj,FileviezPerez:2008jbu,FileviezPerez:2008wbg,Fuks:2019clu,Ashanujjaman:2021txz,Mandal:2022zmy,CMS:2022msk}. 
At the same time, the HL-LHC program can benefit from further theoretical and experimental progress. This includes dedicated searches for and studies on: the states $\Delta^0$ and $A^0$;
non-degenerate mass spectra; boosted topologies; production from vector boson fusion and Higgs portals; final-states with $\tau$ leptons; indirect signatures; novel production and decay modes; and connections to new gauge sectors.

In the Type III seesaw the SM is extended by at least two fermionic triplets ${\Sigma}\sim (1,3,0)$~\cite{Foot:1988aq}. 
Here light neutrino masses are generated as in the Type I seesaw.  The field content also leads to the existence of heavy charged $(E^\pm)$ \textit{and} neutral $(N)$ lepton mass eigenstates. Like the Type II scenario, these particles carry gauge charges and can be produced at the LHC through various mechanisms~\cite{Cai:2017mow}. Predictions (and tools) for a number of these channels are known up to NLO in QCD~\cite{Biggio:2011ja,Ruiz:2015zca,Cai:2017mow}. 
Depending on initial assumptions, triplet leptons can decay into multi- and many-lepton final states that exhibit LNV and/or LFV.
Flagship signatures for the Type III seesaw include:
searches for $E^+E^-$ pairs that decay through the $Z$ to final states with  \textit{six} charged leptons, i.e., $pp \to E^+E^- \to \ell^\pm_i \ell^\mp_j ZZ \to \ell^\pm_i \ell^\mp_j \ell^+_k\ell^-_k \ell^+_l\ell^-_l+X$, which manifestly exhibits LFV; 
searches for $E^\pm N$ pairs that decay through the $W$ and $Z$ to same-sign lepton and jets, i.e., $pp \to E^\pm N \to \ell^\pm_i \ell^\pm_j Z W^\mp \to \ell^\pm_i \ell^\pm_j +nj +X$, which manifestly exhibits LNV and LFV;
and searches for final states with \textit{five} charged lepton, e.g.,
$pp \to E^\pm N \to \ell^\pm_i \ell^\pm_j Z W^\mp \to \ell^\pm_i \ell^\pm_j 
\ell^+_k\ell^-_k \ell^\mp_l+$MET, which can manifestly exhibit LFV.
Present searches for Type III leptons at the LHC constrain masses to be above $m_E = 790-1065$ GeV, depending on the benchmark and signal category~\cite{ATLAS:2020wop,CMS:2019lwf,CMS:2022nty,ATLAS:2022yhd}.
Prospects for the HL-LHC show that this can be pushed to at least 2 TeV or more~\cite{delAguila:2008cj,Franceschini:2008pz,Arhrib:2009mz,Ruiz:2015zca,Cai:2017mow}.
However, the HL-LHC program can benefit from further searches and studies into:
non-degenerate spectra; boosted topologies; 
production from vector boson fusion; final-states with $\tau$ leptons; indirect searches; novel production and decay modes and their simulations.

As in grand unified theories (GUTs)~\cite{Nath:2006ut,Croon:2019kpe}, popular radiative Seesaws hypothesize the existence of leptoquarks~\cite{AristizabalSierra:2007nf,Cai:2014kra,Pas:2015hca,Dorsner:2017wwn,Cai:2017jrq}.  
Notable signatures include:
searches for pair production of the scalar leptoquark $S_{\rm LQ}^{\pm\frac{1}{3}}$ to dilepton and jet pairs, i.e., $pp \to S_{\rm LQ}^{+\frac{1}{3}}S_{\rm LQ}^{-\frac{1}{3}} \to \ell^+_i\ell^-_j jj + X$, which manifestly exhibits LFV for $i\neq j$;
searches for single production of the scalar leptoquark $S_{\rm LQ}^{\pm\frac{2}{3}}$ to same-sign dilepton and jet pairs, i.e., $pp \to 
S_{\rm LQ}^{\pm\frac{2}{3}} \to 
S_{\rm LQ}^{\pm\frac{1}{3}}S_{\rm LQ}^{\pm\frac{1}{3}} \to \ell^\pm_i\ell^\pm_j jj + X$, which manifestly exhibits LNV and LFV;
and searches for the analogous processes with ``final-state'' top and bottom quarks, e.g.,
$S_{\rm LQ}^{-\frac{1}{3}} \to \ell^- t$
and
$S_{\rm LQ}^{-\frac{2}{3}} \to \ell^- b$, which also manifestly exhibit BNV when heavy flavor-tagging is used. Present LHC limits on leptoquarks vary on the precise signature and category but generally exclude leptoquark masses up to 2 TeV~\cite{CMS:2018txo,CMS:2018lab,CMS:2018iye,CMS:2018ncu,ATLAS:2020dsf,ATLAS:2020dsk,ATLAS:2020xov,ATLAS:2021oiz,ATLAS:2021yij,ATLAS:2021mla,CMS:2018yiq,CMS:2020wzx,ATLAS:2021jyv,CMS:2021far}. 
While some studies indicate a potentially rich phenomenology at the LHC~\cite{Porod:2000hv,Nath:2006ut,AristizabalSierra:2007nf,Kohda:2012sr,Cai:2014kra,Pas:2015hca,Dorsner:2017wwn,Cai:2017jrq,Babu:2019vff,Babu:2019mfe}, the high-luminosity program can benefit from systematic surveys that include connections to flavor anomalies and neutrino masses.

There are also models without leptoquarks that radiatively generate left-handed Majorana neutrino masses, and hence also feature LNV~\cite{Zee:1980ai,Zee:1985rj,Zee:1985id,Babu:1988ki,Babu:1988wk,Babu:1988ig,FileviezPerez:2009ud,FileviezPerez:2010ch,Cai:2014kra,Cai:2017jrq,Cai:2017mow}. Among these is the Zee-Babu model~\cite{Zee:1985rj,Zee:1985id,Babu:1988ki}, wherein neutrino masses are generated at two loops via singly $(h^\pm)$ and doubly charged $(k^{\pm\pm})$ scalars that also couple to the SM Higgs.
Both scalars carry hypercharge but neither carries SU$(3)_c$ or SU$(2)_L$ gauge charges; this leads to inherently smaller production cross sections at colliders~\cite{Gunion:1996pq,Alcaide:2017dcx}. The leading production channel of Zee-Babu pair production via the Drell-Yan mechanism, but other channels also exist. Predictions (and tools) for these of these channels are known up to NLO in QCD with parton shower-matching~\cite{Ruiz:2022sct}. Decay rates of Zee-Babu scalars are strongly correlated with neutrino oscillation parameters~\cite{Nebot:2007bc,Ohlsson:2009vk,Schmidt:2014zoa,Long:2014fja,Herrero-Garcia:2014hfa,Ruiz:2022sct}.
Notable LHC signatures include: 
$pp \to k^{++}k^{--} \to \ell^+_i\ell^+_j\ell^-_k\ell^-_l+ X$, which manifestly exhibits LFV for $i,j\neq k,l$;
and
$pp \to h^{+}h^{-} \to \ell^+_i\ell^-_j$+MET.
Establishing LNV at the LHC requires observing several different processes~\cite{Ruiz:2022sct}. With the full Run II data set at $\sqrt{s}=13$ TeV, $k$ masses as high as $m_k=890$ GeV and decay rates as small as BR$(k^{\pm\pm}\to \ell^\pm_i \ell^\pm_j)=16\%$ for $\ell\in\{e,\mu\}$ have been excluded at the LHC~\cite{ATLAS:2022yzd,Ruiz:2022sct}. This is projected to reach about $m_k = 1110$ GeV and BR$(k^{\pm\pm}\to\ell_i^\pm \ell_j^\pm)=8\%$ with about $\mathcal{L}=3$ ab$^{-1}$ at $13$ TeV~\cite{Ruiz:2022sct}.
Furthermore, with updated neutrino oscillation data, predictions for flavor-violating $h^\pm \to \ell^\pm \nu_{\ell'}$ decays in the Zee-Babu are now sufficiently precise to have discriminating power.
HL-LHC searches can benefit from further studies on searches for $h^+h^-$ and associated $k^{\pm\pm}h^{\mp}h^{\mp}$ production, as well as further investigations into the connection between oscillation experiments and the LHC.

In extensions of the SM with new gauge symmetries one can have many different collider signatures. 
In Left-Right Symmetric models, for example, characteristic signatures include:
the production of heavy Majorana neutrinos $N$ through resonant $W_R/Z_R$ bosons that decay into same-sign leptons and jets~\cite{Senjanovic:1975rk}, i.e., $pp \to W_R^\pm \to \ell^\pm_i N \to \ell^\pm_i \ell^\pm_j +nj+X$ and  $pp \to Z_R^\pm \to N N \to \ell^\pm_i \ell^\pm_j +nj+X$, which manifestly exhibit LNV and LFV; the same process through non-resonant $W_R/Z_R$ bosons~\cite{Ruiz:2017nip,Nemevsek:2018bbt}, which again manifestly exhibit LNV and LFV; boosted configurations where $N$ are clustered into a single jet $(J)$~\cite{Ferrari:2000sp,Mitra:2016kov,Mattelaer:2016ynf}, i.e., $pp \to W_R^\pm \to \ell^\pm_i N \to \ell^\pm_i J+X$ and  $pp \to Z_R^\pm \to N N \to J J+X$; and searches for long-lived $NN$ pair production through Higgs-portal couplings~\cite{Nemevsek:2016enw,Nemevsek:2018bbt}. Searches for LNV and LFV in $U(1)$ scenarios exhibit similar signatures but with potentially light $Z'$ masses and gauge couplings $g'\ll1$. Present searches for these states set various limits on masses and couplings beyond~\cite{CMS:2018agk,ATLAS:2019isd,ATLAS:2019erb,CMS:2019buh,CMS:2021dzb,CMS:2022yjm}.
Despite these constraints, the outlook at the HL-LHC is promising~\cite{Nath:2006ut,Das:2012ii,Deppisch:2015qwa,Nemevsek:2016enw,Alioli:2017ces,Cai:2017mow,Nemevsek:2016enw,Mitra:2016kov,Ruiz:2017nip,Nemevsek:2018bbt,Deppisch:2019kvs,Deppisch:2019ldi,Chiang:2019ajm,Cottin:2021lzz,Beltran:2021hpq,Buarque:2021dji,Dekens:2021bro,Han:2022qgg,Padhan:2022fak}.
The HL-LHC program can be enhanced by further searches and studies into:boosted topologies; production from vector boson fusion and Higgs portals; final-states with $\tau$ leptons; the use of machine learning techniques; indirect searches; 
as well as novel production and decay modes (and their simulations).
For LNV signatures in theories based on local $B-L$ gauge symmetry see, for example, Refs.~\cite{Basso:2008iv,FileviezPerez:2009hdc,Basso:2010pe,Accomando:2016sge,Accomando:2017fmb, Accomando:2017qcs,Deppisch:2019ldi,Deppisch:2019kvs,FileviezPerez:2020cgn,Padhan:2022fak}.

In supersymmetric models with $R$-parity Violation it may be possible that BNV and LNV also occur at low scales. Such situations arise naturally, for example, when a right-handed sneutrino acquires a vacuum expectation value~\cite{FileviezPerez:2013fsv,Barger:2008wn,FileviezPerez:2012mj}, thereby generating tiny neutrino masses. As a consequence, sparticles can be produced through many non-traditional mechanisms and mediate LNV, LFV, as well as  RPV~\cite{Barger:1989rk,Barger:2008wn,FileviezPerez:2013fsv,FileviezPerez:2012mj}.  Further the LSP is no longer stable and can decay to standard model particles, which opens the door to novel signatures for SUSY without significant missing transverse energy.  If RPV is sufficiently small, then the LSP may also be long-lived and that opens even more doors to long-lived particle signatures that are a topic of much recent development in terms of reconstruction, trigger, and analysis by the LHC experiments~\cite{Alimena:2021mdu,Alimena:2019zri,CMS:2021kdm,CMS:2021tkn,CMS:2020iwv,ATLAS:2020wjh,ATLAS:2019fwx}.  Present searches for RPV supersymmetry set various limits on sparticle masses and cross sections, with gluinos being excluded below 2.5 TeV, top squarks below 1.4 TeV, bottom squarks below 750 GeV, winos below 1.6 TeV, and sleptons below 1.2 TeV~\cite{CMS:2022yjm,CMS:2021knz,CMS:2020cpy,CMS:2018ikp,CMS:2018skt,CMS:2018ncu,CMS:2018lab,CMS:2018mts,CMS:2018pdq,CMS:2018hnz,CMS:2017szl,ATLAS:2021fbt,ATLAS:2021tar,ATLAS:2021yyr,ATLAS:2020uer,ATLAS:2020wgq,ATLAS:2019fag,ATLAS:2018umm,ATLAS:2017jnp,ATLAS:2017jvy}.   However, it is critically important to realize that the precise bound depends on the specific underlying assumptions, which can include simplified models with decoupled squarks, one specific RPV coupling from the many possible options, and the assumption of large mass differences between the particles in the SUSY spectrum.  

The outlook at the HL-LHC is very promising for searches for signatures that are rarer and/or  at higher mass. For example, the recent search~\cite{ATLAS:2021fbt} for Higgsinos decaying to quarks gained the first sensitivity to such a signature since LEP, while the first search for chargino decays to a three lepton resonance gave sensitivity up to 1 TeV~\cite{ATLAS:2020uer}. For long-lived particles, the upgrades to the detector and the trigger will expand the reach of searches for displaced leptons and jets.  For example, a recent search for displaced vertices gave new sensitivity to long-lived particles with hadronic final states~\cite{CMS:2021kdm}. The HL-LHC program can be enhanced by further searches and studies of the different exotic signatures coming from baryon and/or lepton number violation.


\section{Proton decay}
\begin{center} 
Ed Kearns (Boston University), Stuart Raby (Ohio State University)
\end{center}
\label{sect:Proton}
Theories which unify the strong and electroweak forces guarantee that nucleons should decay. The rate is determined by the grand unification group and the GUT scale, $M_G$. Effective operator analysis consistent with the Standard Model allows for dimension six operators suppressed by a new scale of physics,  $\Lambda^{-2} \sim g^2/M_G^2$. Supersymmetric extensions of the Standard Model allow for dimension four and
five baryon and lepton number violating operators which must be suppressed in order to avoid rapid proton decay. Gravity does not respect global symmetries,  therefore even Planck scale physics would be expected to violate both baryon and lepton numbers. Theories with large extra dimensions must invent new symmetries to prevent rapid proton decay. Therefore, one might say that nucleon decay is an ubiquitous consequence of any theory beyond the Standard Model~\cite{Nath:2006ut}. 

In SUSY GUTs, the GUT scale is of order $2 \times 10^{16}$ GeV with $\alpha_G^{-1} \sim 24$. In this case the proton lifetime mediated by the dimension six operators for the canonical decay mode is $\tau(p \rightarrow e^+ \pi^0) \sim 2 \times 10^{36}$ years.  This is outside the discovery range of the next generation of proton decay experiments. However, in SUSY GUTs there are additional dimension 5 operators contributing to nucleon decay.  This contribution to proton decay is suppressed by an effective color triplet Higgsino mass, $M_T$, a loop factor $LF$ which converts a dimension 5 operator, with 2 fermions and 2 scalars, into a dimension 6 four fermion operator with $LF \sim \tilde m_W/\tilde m^2$ and a product of effective Yukawa couplings, $CC$. Here $\tilde{m}_W$ and $\tilde{m}$ are the gaugino and sfermion masses, respectively. We then
have $\Lambda^2 \sim CC \times LF/M_T$. This can be sufficiently suppressed in models with $M_T \sim 10^{17}$ GeV, heavy squarks and sleptons with mass of order several TeV and light gauginos.   With gaugino masses light enough to be discoverable at the LHC and scalar masses less than 30 TeV, K. S. Babu {\it et al.}\cite{Babu:2020ncc}, find $\tau(p \rightarrow K^+ \bar \nu) \leq 1.1 \times 10^{35}$ years. For the proton decay predictions in different grand unified theories see the discussion in Ref.~\cite{Dev:2022jbf}.   

The experimental search for proton decay takes place in large detectors used in multipurpose neutrino experiments. The first generation experiments such as IMB, Kamiokande, and Soudan-II, were of order 1 kiloton in sensitive mass and completed their exposures by 1990. The first generation experiments set informative limits that excluded the initial idea that proton decay could be readily observed, in particular as predicted by the simplest SU(5) GUT. There is only one second generation nucleon decay experiment to speak of, Super-Kamiokande (Super-K), a water Cherenkov detector with 50 kilotons of total mass. Super-K began operation in 1996 and is expected to continue until at least 2027. The SK experiment has not detected any signs of nucleon decay. For example, SK has set a lifetime limit for $p \rightarrow e^+\pi^0$ at $2.4\times10^{34}$~y (90\% CL) based on a recent analysis of 450 kt-yr worth of data~\cite{Super-Kamiokande:2020wjk}. This result uses several improvements over older publications including neutron tagging and expansion of the fiducial volume from 22.5 to 27.2 ktons.
The most recent published limit for the SUSY-favored decay mode $p \rightarrow {\bar{\nu}} K^+$ is $5.9\times 10^{33}$ years based on an exposure of 260 kt-yr~\cite{Super-Kamiokande:2014otb}. Overall, the Super-K experiment has published leading limits on 30 baryon number violating processes including less conventional decay modes that are not necessarily motivated by GUTs.

It is anticipated that the 20-kt liquid-scintillator reactor experiment JUNO will begin operation in 2022. Although not sensitive to a wide range of proton decay channels, JUNO can do well on the particular mode $p \rightarrow \nu K^+$, using excellent energy resolution and the unique 12-ns timing signature of $K^+$ decay at rest. With a 10 year exposure, JUNO should reach sensitivity to this decay mode of $8 \times 10^{34}$ years\cite{JUNO:2022hxd}. Using a similar approach, the speculative THEIA experiment\cite{Theia:2019non} should perform similarly but using water-based rather than oil-based scintillator.

In the late 2020s, a third generation of proton decay experiments will take place based on new massive neutrino detectors. Around 2027, it is expected that Hyper-Kamiokande (Hyper-K), with 186 fiducial kilotons of water will begin operation and dominate the search for proton decay. Hyper-K will naturally be able to search in the same channels as Super-K and benefit from development of techniques. As seen in the experience with Super-K, extremely long exposures of such detectors are anticipated. A twenty year exposure of Hyper-K is sensitive to proton decay to $e^+ \pi^0$ lifetimes of $10^{35}$ years, and $p \rightarrow \nu K^+$ lifetimes of $3 \times 10^{34}$ years\cite{Hyper-Kamiokande:2022smq}.

The DUNE experiment is also expected to begin in the late 2020s with two 10-kt modules of liquid argon using the time projection chamber technique (LArTPC). For most nucleon decay modes, only 20 kilotons of fiducial mass is not competitive with the long exposure of Super-K or the larger mass of Hyper-K. However, the detailed imaging capabilities, particularly the ability to measure $dE/dx$ along the track of a charge particle allow DUNE to be competitive for decay modes that include charged kaons or displaced vertices. It is also considered that DUNE will be expanded with up to two more 10-kton LArTPC modules. Given an eventual exposure of 400 kton-y, the DUNE LArTPC detectors should be sensitive to $p \rightarrow \nu K^+$ lifetimes somewhat greater than $10^{34}$ years\cite{DUNE:2020fgq}. 

Somewhat more detail on the above experiments are given in a Snowmass whitepaper on Baryon Number Violation in Neutrino Experiments\cite{Dev:2022jbf} and references within. For practical reasons, the detectors that can searches for proton decay multi-purpose and generally optimized for neutrino studies. Unless or until candidate signatures are seen, it is difficult to envision funding much larger detector masses at the megaton scale. The high price of excavating rock limits the detector mass.

\section{Neutron-antineutron Oscillations}
\begin{center}
K. S. Babu (Oklahoma State University), Leah Broussard (Oak Ridge National Laboratory)
\end{center}
\label{sect:nnbar}
A robust search for the mechanism behind baryogenesis should proceed along all open experimental avenues: processes where only $\Delta L\neq 0$, where $\Delta L\neq 0$ and $\Delta B\neq 0$ simultaneously, and where only $\Delta B\neq 0$.  
Therefore, along with neutrinoless double beta decay and proton decay, which are being explored vigorously, searches for the $\Delta B\neq 0$ and $\Delta (B-L)\neq 0$ process of neutron-antineutron oscillations ($n\rightarrow\bar{n}$) are a necessary component of a multifaceted effort to experimentally observe BNV. A BSM $(B-L)$ violating process, e.g.\ $n\rightarrow\bar{n}$ or the heavy lepton decays within classic leptogenesis, is needed for the baryon asymmetry in the universe to develop and survive a ``washing-out'' by sphalerons~\cite{tHooft:1976rip}.   $n\rightarrow\bar{n}$ oscillations additionally have the attractive advantage of providing a definitive, observable, ``on-shell'' test of the mechanism behind baryogenesis~\cite{Proceedings:2020nzz,Barrow:2022gsu}. 
$n\rightarrow\bar{n}$ oscillations were originally introduced in~\cite{Kuzmin:1970nx,Glashow:1979nm,Mohapatra:1980qe} and the phenomenology has been studied in detail~\cite{Mohapatra:1980de,Kuo:1980ew,Cowsik:1980np,Chetyrkin:1980ta,Barbieri:1981yr,Caswell:1982qs,Rao:1982gt,Rao:1983sd}. 
New physics related to $n\rightarrow\bar{n}$ mixing and $\Delta B =2$ more generally has also been considered~\cite{Berezhiani:2005hv,Dutta:2005af,Babu:2006xc,Babu:2008rq,Babu:2013yca,Dev:2015uca,McKeen:2015cuz,Calibbi:2016ukt,Allahverdi:2017edd,Grojean:2018fus,Bringmann:2018sbs} with implications for LHC searches (e.g. limits on heavy scalar diquarks~\citep{CMS:2019gwf}) and primordial baryogenesis; with regards to theories with extra dimensions~\cite{Dvali:1999gf,Nussinov:2001rb,Girmohanta:2019fsx,Girmohanta:2020qfd}; as relates to discrete  $C$, $P$, $T$ symmetries~\cite{Berezhiani:2018xsx,Berezhiani:2018pcp,Tureanu:2018phm,Fujikawa:2020gsa}; as well as very low scale spontaneous $B-L$ violation by two units~\cite{Berezhiani:2015afa,Addazi:2016rgo,Babu:2016rwa}. 
Overall, the $\Delta B =2$ process of  $n\rightarrow\bar{n}$ and the possibly related $\Delta B =1$ process of $n\rightarrow n'$ (neutron to sterile neutron oscillations)~\cite{Berezhiani:2005hv,Berezhiani:2006je} are comparatively unexplored experimentally, and current and future facilities will provide for rich opportunities for discovery. 

Direct experimental limits are currently available from a free neutron search at the Institut Laue-Langevin (ILL) some decades ago with a lower limit of $\tau_{n\bar{n}}\sim10^8\,$s~\cite{BaldoCeolin:1994jz} and somewhat stronger indirect limits are taken from 
matter instability caused by $n-n'$ inside nuclear matter at Super-K (intranuclear searches)~\cite{Super-Kamiokande:2020bov}. The two techniques are experimentally as well as theoretically complementary, and both are needed to identify the source of BSM physics~\cite{Young:2019pzq}. Intranuclear searches require very large mass detectors~\cite{Dev:2022jbf} to overcome the suppression factor arising from the $\{n,\bar{n}\}$ energy splitting inside the nucleus~\citep{Barrow:2019viz,Oosterhof:2019dlo}, and are ultimately limited by irreducible atmospheric $\nu$ backgrounds. Final state interactions (intranuclear rescattering) of the annihilation products are another significant source of uncertainty~\citep{Barrow:2019viz,Super-Kamiokande:2011idx,Super-Kamiokande:2020bov}. Comparable sensitivities could be available from the NO$\nu$A Far detector~\cite{NOvA:2007rmc} if backgrounds from cosmic rays could be suppressed to the level of atmospheric $\nu$'s. Higher sensitivities will be available from Hyper-K~\citep{Hyper-Kamiokande:2018ofw} although significant improvement hinges on the ability of the improved detector to reduce backgrounds. DUNE~\citep{DUNE:2020ypp} looks most promising thanks to a substantial increase in mass as well as improved capabilities for reconstruction from the bubble-chamber-like images.  Significant efforts are ongoing to assess the impact of nuclear model configurations on improved modeling of the annihilation, and to leverage modern techniques such as deep learning and boosted decision trees in discriminating signal from background. Intranuclear searches can reach a sensitivity of $\tau_{n\bar{n}}\gtrsim10^{8-9}\,$s~\cite{Dev:2022jbf} following the traditional intranuclear suppression factor formalism~\citep{Dover:1985hk,Friedman:2008es,Barrow:2019viz}. While not competitive, the capabilities of the LArTPC can be demonstrated in a proof-of-principle search using the MicroBooNE detector~\citep{microboone1113}.

Free neutron searches are theoretically and experimentally cleaner than intranuclear searches, and offer the tantalizing possibility of an incontrovertible discovery. The ILL search detected zero background events~\cite{BaldoCeolin:1994jz}, and future higher sensitivity searches are similarly expected to be background-free~\cite{Addazi:2020nlz}. One concept is based on ultracold neutrons, neutrons which can be stored in material traps for long times~\citep{Serebrov:2016rvi,Fomin:2019oyj}.  This approach utilizes a much more compact geometry than the 50~m beamline required for the ILL experiment, with an improved sensitivity of about 10-40$\times$ the ILL result, or $\tau_{n\bar{n}}\sim10^{8-9}\,$s, depending on how $n$ reflections are modeled~\citep{Fomin:2017aiz,Fomin:2017lej,Fomin:2018qrq,Fomin:2019oje,Fomin:2019oyj}. 
The most promising proposal, the NNBAR experiment at the European Spallation Source (ESS)~\cite{Peggs:2013sgv}, will implement a cold neutron approach like the ILL but capitalizes on the technological developments in detection and especially advances in neutron optics in the decades since that experiment. 
NNBAR will improve the sensitivity to $n\rightarrow\bar{n}$ by 1000$\times$ the ILL result~\citep{Frost:2016qzt,Klinkby_2016,Santoro:2020nke,Phillips:2014fgb,Addazi:2020nlz}, reaching a limit of $\tau_{n\bar{n}}\sim10^{9-10}\,$s \citep{Nesvizhevsky:2020vwx,Addazi:2020nlz,Gudkov:2019gro}. The ESS includes critical provisions to achieve this impressive increase--the Large Beam Port earmarked for NNBAR is constructed, there is provision for an up to 300~m beamline, and the fundamental physics team is leading the effort to design a high intensity liquid deuterium lower moderator to be installed $> 2030$ and optimize it for NNBAR and other fundamental physics efforts as part of the HighNESS project~\citep{Addazi:2020nlz,Santoro:2020nke,Santoro:2022tvi}. 

While the NNBAR experiment is expected to be an O(\$100M) project, a staged program addressing complementary physics questions on a more economical scale is being developed for the 2020s~\citep{Addazi:2020nlz}. Searches for neutron conversions into sterile ``mirror'' neutrons $n\rightarrow n'$, $n\rightarrow n' \rightarrow n$~\cite{Berezhiani:2005hv,Berezhiani:2018qqw} and $n\rightarrow n' \rightarrow \bar{n}$~\cite{Berezhiani:2020vbe} probe questions of dark matter and cobaryogenesis~\cite{Kobzarev:1966qya,Blinnikov:1982eh,Foot:1991bp,Hodges:1993yb,Berezhiani:1995yi, Berezhiani:1995am,Okun:2006eb,Berezhiani:2000gw,Ignatiev:2003js,Berezhiani:2003xm,Berezhiani:2003wj,Berezhiani:2005ek,Foot:2014mia,Berezhiani:2020vbe, Babu:2021mjg}. A program at Oak Ridge National Laboratory using existing $n$ scattering instruments is already underway~\cite{Berezhiani:2017azg,Broussard:2017yev,Broussard:2019tgw}. 
A recent search at the Spallation Neutron Source~\cite{Broussard:2021eyr} has ruled out the phenomenon of $n\rightarrow n'$ oscillations as an explanation for the long-standing neutron lifetime discrepancy~\cite{Berezhiani:2018eds}, the disagreement between cold and ultracold neutron lifetime measurements. Improved sensitivity for $n\rightarrow n'$ searches generally can be obtained with an optimized experimental setup at the ESS, as part of the HIBEAM program \citep{Addazi:2020nlz}, in the late 2020s.  In particular, searches for $n\rightarrow n' \rightarrow \bar{n}$, which is experimentally unexplored, offer attractive opportunities for early R\&D for a future high sensitivity search for $n\rightarrow \bar{n}$.

\section{More exotic L and B violating processes}
\begin{center}
Susan Gardner (University of Kentucky), Julian Heeck (University of Virginia)
\end{center}
\label{sect:Exotic}
Baryon and lepton number violation need not be restricted to the familiar channels 
$p\to e^+\pi^0$ ($\Delta B = \Delta L =1$), $n\to \bar{n}$ ($\Delta B = 2$), and $0\nu\beta\beta$ ($\Delta L = 2$)
discussed in previous sections. The striking experimental signatures associated
with $\Delta B/\Delta L$ processes allow for 
a wide variety of possible channels, 
only some of which have been explored so far. 
In addition, models involving neutron decays into light new particles have received considerable attention in recent years as possible solutions~\cite{Fornal:2018eol,Fornal:2020gto} 
to the neutron-lifetime anomaly~\cite{Wietfeldt2011RvMP...83.1173W}: the new particles
may include a dark matter candidate and can carry baryon number, leading to processes with \emph{apparent} baryon number violation.
Although neutron star observations preclude certain models~\cite{McKeen:2018xwc,Baym2018PhRvL.121f1801B,Motta:2018rxp}, 
simple solutions are nevertheless still possible~\cite{Grinstein:2018ptl,Strumia:2021ybk,Berryman:2022zic}.
Thinking broadly, models with 
dark fermions that mix with the neutron can furthermore lead to a neutron-shining-through-a-wall effect~\cite{Hostert:2022ntu}, or mediate exotic H-atom decays~\cite{Berezhiani:2018udo,McKeen:2020vpf}, or new modes for $n\to{\bar n}$~\cite{Berezhiani:2020vbe},
with additional astrophysical probes possible~\cite{McKeen:2020oyr,Berryman:2022zic}.
Finally, dark matter (DM) scattering can also mediate baryon number violation, as in $\text{DM}\,p\to n e^+$~\cite{Kile:2009nn},
$\text{DM}\,p \to \text{DM}'\,K^+$~\cite{Davoudiasl:2010am,Davoudiasl:2011fj,Blinov:2012hq},
$\text{DM}\,p \to \text{DM}'\,e^+$~\cite{Huang:2013xfa}, 
$\text{DM}\,n\to \text{DM}'\,\pi^0$~\cite{Fornal:2020poq}, and 
$\text{DM}\,n\to \text{mesons}$~\cite{Jin:2018moh,Keung:2019wpw}, leading to distinct  
final-state kinematics than those scrutinized thus far.

\begin{figure}[t]
\includegraphics[width=0.75\textwidth]{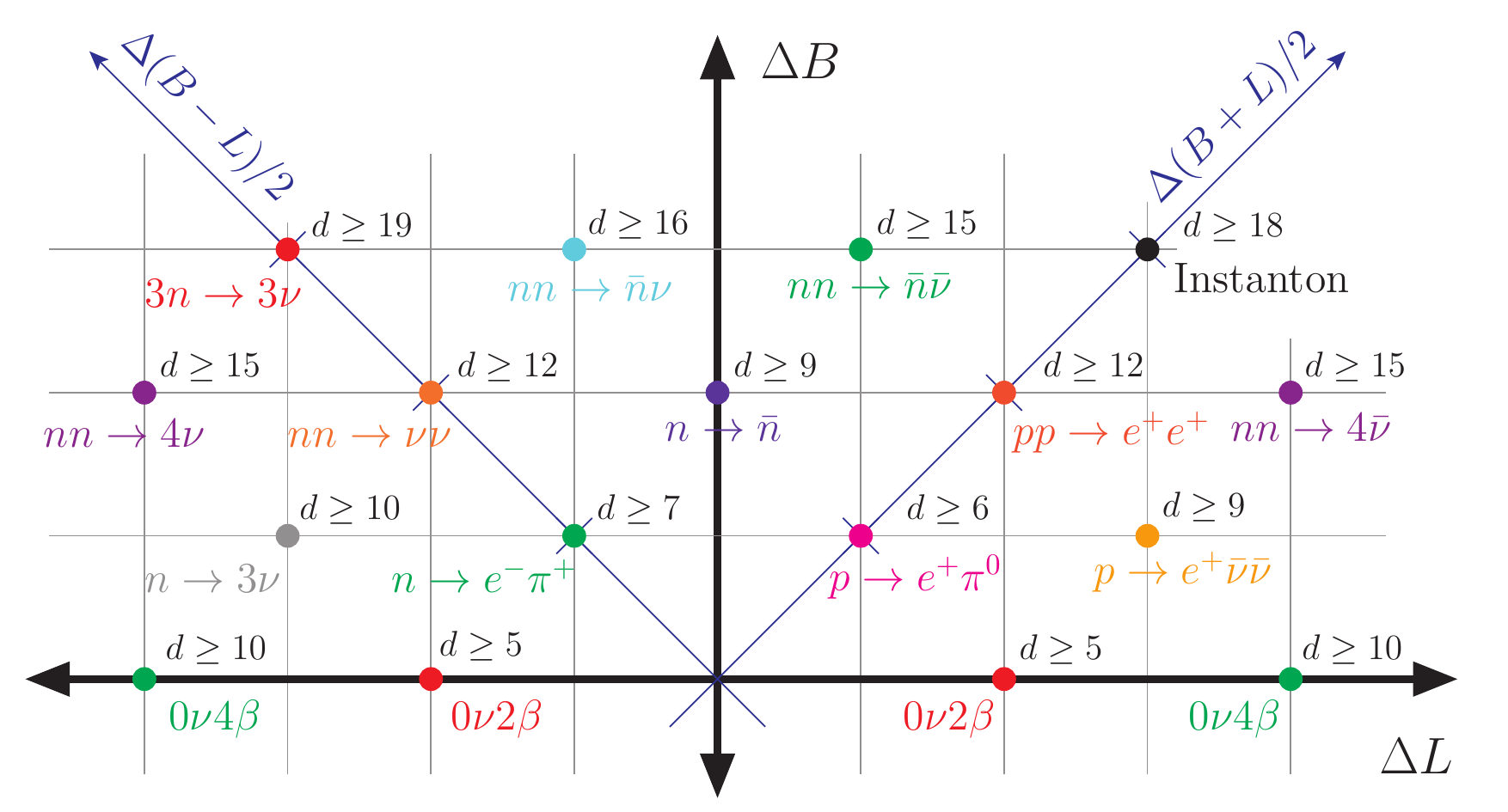}
\caption{
Landscape of baryon and lepton number violation including some representative processes and the minimal mass dimension $d$ of the underlying EFT operator~\cite{Helset:2019eyc}. From Ref.~\cite{Heeck:2019kgr}.
\label{fig:BL_grid}}
\end{figure}

As first noted by Weinberg~\cite{Weinberg:1979sa}, true $\Delta B$ and $\Delta L$ processes arise in the SM EFT through effective operators with mass dimension $d\geq 5$. If the SM's global symmetry $U(1)_B\times U(1)_L$ is only broken in a particular direction, selection rules emerge that lead to the conservation of a linear combination $\alpha\Delta B + \beta \Delta L$, automatically eliminating many forms of baryon or lepton number violation~\cite{Weinberg:1980bf,Heeck:2019kgr}. Non-perturbative instanton processes in the SM provide a useful example, as they break $U(1)_B\times U(1)_L\to \mathbb{Z}_3^{(B+L)/2}\times U(1)_{B-L}$ and thus only allow for (unobservably-small~\cite{tHooft:1976rip}) $\Delta B = \Delta L = 3$ processes. A visual guide to allowed $\Delta B$ and $\Delta L$ is given in Fig.~\ref{fig:BL_grid}. Breaking $ B$ or $ L$ by higher units requires operators of higher mass dimension $d$, yielding lower rates 
for a fixed new-physics scale; however, even large $d \gg 6$ can be testable in clean channels~\cite{Heeck:2019kgr}. We note some recent $|\Delta B| >1$ and $|\Delta L|>1$ work 
from theoretical~\cite{Heeck:2013rpa,Fonseca:2018ehk,Fonseca:2018aav,Nussinov:2020wri,He:2021sbl,He:2021mrt,Helset:2021plg} and experimental \cite{Super-Kamiokande:2014hie,Super-Kamiokande:2015pys,Super-Kamiokande:2015jbb,NEMO-3:2017gmi,Super-Kamiokande:2018apg,SNO:2018ydj,Kidd:2018fbb,EXO-200:2017hwz,Majorana:2018pdo,Barabash:2019enn}
perspectives. 
Inclusive searches such as $p\to \mu^++\text{anything}$ offer an efficient way to constrain large classes of $\Delta B$ operators~\cite{Heeck:2019kgr}.
Observations of \emph{several} distinct $B$ and/or $L$ violating processes imply the existence of other $\Delta B/\Delta L$ violations, even if not observed directly~\cite{Babu:2014tra,Gardner:2018azu,Berryman:2022zic};
this amounts to vector addition in Fig.~\ref{fig:BL_grid}, even if the particular suggested channels are different.

In addition, baryon-number-violating 
operators involving heavy quarks or leptons also give rise to nuclear decays: loops can be closed to convert heavy quarks into lighter ones and taus 
into tau neutrinos, which gives a suppression that is typically more than compensated by the nucleon-lifetime-scale sensitivity~\cite{Marciano:1994bg,Hou:2005iu,Durieux:2012gj,Alonso:2014zka,Heeck:2019kgr,Dorsner:2022twk}. Fine-tuned scenarios in which nuclear decays are negligible might exist, however, and warrant direct searches for $\Delta B$ top~\cite{Dong:2011rh,CMS:2013zol}, bottom~\cite{BaBar:2011yks}, charm~\cite{BESIII:2021hyt}, hyperon~\cite{McCracken:2015coa}, and tau 
decays~\cite{Belle:2020lfn}. Nucleon decays involving higher units of lepton flavor, e.g.~$p\to e^-\mu^+\mu^+$, can dominate in flavor-symmetric models~\cite{Hambye:2017qix,Heeck:2019kgr} and give clean signatures~\cite{Super-Kamiokande:2020tor}. Nuclear processes can also give rise to combined
$L$ and lepton-flavor violation, e.g.~in $\mu^- +{\text N}(A,Z) \to e^+ +\text{N}(A,Z-2)$~\cite{Lee:2021hnx}.

As common, our discussion has 
employed EFT language 
to identify promising $\Delta B$ and $\Delta L$ signatures,
implicitly assuming that all new physics is \emph{heavy}, say beyond the electroweak scale. 
Novel $\Delta B$ and $\Delta L$ signatures arise when \emph{light} new particles $\chi$ are introduced,
which can then appear as 
$n\to \pi^0 \chi$~\cite{Davoudiasl:2013pda,Davoudiasl:2014gfa},
$p\to \pi^+\chi$~\cite{Davoudiasl:2013pda,Davoudiasl:2014gfa,Helo:2018bgb},
$p\to e^+\chi$~\cite{McKeen:2020zni}, or in
hydrogen decays~\cite{McKeen:2020zni,McKeen:2020vpf}. The light particle could even be the graviton $G$, emitted via $p\to e^+ \pi^0 G$~\cite{Haisch:2021nos}.
Particles with somewhat larger masses in the GeV range with $\Delta B$ couplings to \emph{heavier} quarks can also be searched for in charm or bottom factories  through decays such as $B^+\to p\chi$, which offers an avalanche of new signatures that have yet to be explored experimentally~\cite{Heeck:2020nbq,Fajfer:2020tqf,Alonso-Alvarez:2021oaj,Goudzovski:2022vbt}. 
Radiative dark decays of strange baryons~\cite{Fajfer:2020tqf} can also impact the cooling properties of heavy neutron stars~\cite{Berryman:2022zic}.
 
Searches for baryon and lepton number violation are uniquely sensitive to exotic models/operators largely impossible to probe in other ways. This includes operators of very high mass dimension, breaking $B$ or $L$ by high units or through heavy flavors, and models including light new particles.
Broad searches for such exotica are encouraged given the absence of signals in more conventional channels.

\section{Connections to Cosmology}
\begin{center}
Andrew J. Long (Rice University), Carlos Wagner (U. of Chicago and Argonne National Laboratory)
\end{center}
\label{sect:Cosmo}



Cosmic wonder:
The observable universe is made of matter, rather than antimatter.  
Observations of the cosmic microwave background~\cite{Aghanim:2018eyx} yield a measurement of the cosmological matter-antimatter asymmetry, corresponding to the excess of protons and nuclei over anti-protons and anti-nuclei.  
In the language of particle physics, we say that the observable universe has a net baryon number $\textsf{B}$~\cite{Steigman:1976ev,Cohen:1997ac}.  
The origin of this cosmological baryon asymmetry is a mystery.  

It is generally assumed that there was some very early epoch in our cosmic history at which the universe had a vanishing baryon asymmetry (equal amounts of baryon and antibaryons).  
For example, cosmological inflation evacuated the universe of matter, which had to be regenerated after inflation during the period of reheating~\cite{Kofman:1994rk,Kofman:1997yn}.  
The event in our cosmic history that led to the generation of a cosmological baryon asymmetry is called baryogenesis.  

We know very little about the particles and forces that may have played a role during baryogenesis, nor do we know at what epoch in the cosmic history this event took place.  
In trying to understand how baryogenesis may occur in theories of elementary particle physics, we are guided by the three Sakharov criteria~\cite{Sakharov:1967dj}: violation of baryon number $\textsf{B}$, violation of $\textsf{C}$ and $\textsf{CP}$, and a departure from thermal equilibrium.  
These three requirements express the necessary conditions for the generation of a cosmological baryon asymmetry under the assumption of $\textsf{CPT}$ conservation.  

A plethora of different models have been proposed, studied, and tested.  
Each model finds its own way to implement the Sakharov criteria using different particle and forces, different connections to the Standard Model, different energy scales, and different epochs in our cosmic history.  
Due to this wide variety in model building efforts, experimental probes of baryogenesis must also be versatile and variegated.  
In the remainder of this short section, we highlight a narrow selection of actively explored theories of baryogenesis that offer a high degree of testability.  

Electroweak baryogenesis:
Electroweak baryogenesis~\cite{Cohen:1990py,Cohen:1990it,Cohen:1991iu,Cohen:1993nk} is the idea that the baryon asymmetry arose as a by-product of the cosmological electroweak phase transition, during which the Higgs field developed its nonzero expectation value giving mass to the weak gauge bosons and signaling that the electroweak symmetry is spontaneously broken.
Early work~\cite{Kuzmin:1985mm} identified that successful electroweak baryogenesis would require the electroweak phase transition to be first order, proceeding through the nucleation and percolation of bubbles.  
However, the Standard Model predicts that this phase transition a continuous crossover and not first order~\cite{DOnofrio:2015mpa}.  
Thus successful electroweak baryogenesis requires new physics coupled to the Higgs boson such that the phase transition is first order and also adequate \textsf{CP} violation is available.  
This new physics provides an excellent target for new physics searches at the LHC~\cite{Ramsey-Musolf:2019lsf} and some of the most well-motivated candidate models are already being tested~\cite{McDonald:1993ey,Carena:1996wj,Davies:1996qn,Huber:2000mg,Menon:2004wv,Chung:2009cb,Carena:2011jy,Cohen:2012zza,Curtin:2012aa}.  
Future colliders~\cite{CEPCStudyGroup:2018ghi,FCC:2018byv,FCC:2018evy,FCC:2018vvp,Cepeda:2019klc}, as well as the LHC at higher luminosities, will furnish precision Higgs measurements, continue to search for di-Higgs production, and further probe the new physics that may be responsible for electroweak baryogenesis~\cite{Curtin:2014jma,Huang:2015tdv,Huang:2016cjm,Chen:2017qcz,Reichert:2017puo,Carena:2018vpt,Goncalves:2018qas,Carena:2022yvx}.  
Precision probes of \textsf{CP} violation at the Intensity Frontier offer a complementary test of Higgs physics~\cite{Ramsey-Musolf:2019lsf}, and an upper limit on the electron's electric dipole moment (EDM) obtained by the ACME collaboration~\cite{ACME:2018yjb} puts strong indirect constraints on models of electroweak baryogenesis~\cite{Blum:2010by,Cesarotti:2018huy}.  
Recent developments in model-building have explored alternative cosmic histories~\cite{Caprini:2011uz,Baldes:2016rqn,Baldes:2016gaf,Cline:2017qpe,Long:2017rdo,Carena:2018cjh,Glioti:2018roy,Ellis:2019flb,Hall:2019ank,Hall:2019rld} that allow for baryogenesis from a first order phase transition while evading strong constraints on new physics.  

Leptogenesis:
Baryogenesis from leptogenesis~\cite{Buchmuller:2004nz,Davidson:2008bu} refers to a broad class of models in which the baryon asymmetry arises in conjunction with a lepton asymmetry, and typically the new sources of \textsf{CP} violation couple to leptons.  
In models of high-scale leptogenesis~\cite{Fukugita:1986hr}, the asymmetry is generated very early in the cosmic history at the seesaw energy scale, where new particles are associated with the origin of the light neutrino masses~\cite{Minkowski:1977sc,Mohapatra:1979ia,Yanagida:1980xy,Mohapatra:1980yp,Schechter:1980gr}.  
Although the new physics is not directly accessible in the laboratory, several indirect observables could be used to probe high-scale leptogenesis; these include the Majorana nature of the light neutrinos and \textsf{CP} violation in the neutrino sector~\cite{Branco:2001pq,Branco:2002xf,Endoh:2002wm,Frampton:2002qc,Pascoli:2006ci,Branco:2006ce,Anisimov:2007mw,Molinaro:2009lud,Shimizu:2017vwi,Moffat:2018smo}.  
On the other hand, models of resonant leptogenesis~\cite{Pilaftsis:2003gt} or low-scale leptogenesis~\cite{Akhmedov:1998qx} introduce new physics at the weak or GeV scale, respectively, allowing high-energy collider and beam dump experiments~\cite{SHIP:2021tpn} to probe the new degrees of freedom directly, typically through searches for displaced vertices or long-lived particles~\cite{Gorbunov:2007ak,Atre:2009rg,Deppisch:2015qwa,Cai:2017mow,Bondarenko:2018ptm,Drewes:2019fou,Drewes:2021nqr}.  

Mesogenesis:
A recent proposal~\cite{Aitken:2017wie,Elor:2018twp,Elor:2020tkc,Elahi:2021jia} suggests that new physics at the mass scale of the Standard Model mesons could give rise to baryogenesis and dark matter via \textsf{CP} violation from the Standard Model quark sector and possibly also new physics.  
In these models, a late-decaying scalar produces quark-antiquark pairs that hadronize into Standard Model mesons, which undergo out-of-equilibrium \textsf{CP}-violating processes, such as $B$ meson oscillations and charged meson decays.  
This results in equal and opposite asymmetries carried by the Standard Model baryons and dark sector baryons, such that baryon number need not be violated.  
Mesogenesis has the appealing feature that the new physics must couple to Standard Model mesons, and therefore it is exceptionally testable~\cite{Alonso-Alvarez:2021oaj,Alonso-Alvarez:2021qfd,Goudzovski:2022vbt,Barrow:2022gsu} as we noted in the previous section.  
Searches for \textsf{B}-number-violating meson decays are already underway at Belle~\cite{Belle:2021gmc} and LHCb~\cite{Rodriguez:2021urv,Borsato:2021aum}.

\section{Summary}
In this report we have discussed the main theories to understand the origin of baryon and lepton number violation in physics beyond the Standard Model. We presented the theoretical predictions for rare processes such as neutrinoless double beta decay, proton decay, and neutron-antineutron oscillation, and overviewed the prospects to discover these rare processes in the near future. The possibility to observe baryon and lepton violating signatures at current and future colliders and through precision studies of other rare processes, and the testability of different baryogenesis mechanisms is discussed in detail. 

A healthy and broad experimental program looking for proton decay, neutrinoless double beta decay and neutron-antineutron oscillations is essential to make new discoveries in this field. These searches are carried out at various experimental facilities in the US and abroad, and use instrumentation arching across traditional HEP/NP boundaries. In addition, experiments such as those at the Large Hadron Collider could discover exotic baryon and/or lepton number violating signatures connected to low energy scale theories for neutrino masses, supersymmetric models with $R$-parity violation, new gauge theories or other mechanisms for physics beyond the Standard Model. The landscape presented in this report could be crucial to discover the underlying mechanism for neutrino masses and the matter-antimatter asymmetry in the universe.

\textit{\bf Acknowledgments}:
 A.P. was supported by the U.S. NSF under Awards PHY-1821085 and PHY-2111213. 
 K.S.B. is supported in part by the U.S. DOE grant No. DE-SC0016013.
 L.J.B. was supported by the U.S. DOE Office of Nuclear Physics Contract No. DE-AC05-00OR22725.
 V.C. was supported by the U.S. DOE under Grant No. DE-FG02-00ER41132.
 S.G. acknowledges partial support from the U.S. 
 Department of Energy under contract DE-FG02-96ER40989.
 E.K. gratefully acknowledges support by the U.S. Department of Energy Award DE-SC0015628.
 The work of S.R. is supported in part by the U.S. DOE under Award No. DE-SC0011726. 
 A.J.L was supported by the U.S. NSF under Award PHY-2114024.
 E.T. gratefully acknowledges support by the U.S. Department of Energy Award DE-SC0007901.
 The work  of C.W. at Argonne National Laboratory was supported by the U.S. DOE under Contract No. DE-AC02-06CH11357.
 The work of C.W. at the University of Chicago was supported by the U.S. DOE Grant DE-SC0013642.
 M.B.W. was supported by the U.S. DOE, Office of High Energy Physics, under Award Number DE-SC0011632 and by the Walter Burke Institute for Theoretical Physics. 
 We are grateful to J. L. Barrow, Y. Kamyshkov, D. Milstead, and V. Santoro for helpful discussions.
R.R. acknowledges the support of Narodowe Centrum Nauki under Grant No. 2019/ 34/ E/ ST2/ 00186, and also acknowledges the support of the Polska Akademia Nauk (grant agreement PAN.BFD.S.BDN. 613. 022. 2021 - PASIFIC 1, POPSICLE). This work has received funding from the European Union's Horizon 2020 research and innovation program under the Sk{\l}odowska-Curie grant agreement No.  847639 and from the Polish Ministry of Education and Science.


\bibliography{RF4}

\begin{thebibliography}{477}%
\makeatletter
\providecommand \@ifxundefined [1]{%
 \@ifx{#1\undefined}
}%
\providecommand \@ifnum [1]{%
 \ifnum #1\expandafter \@firstoftwo
 \else \expandafter \@secondoftwo
 \fi
}%
\providecommand \@ifx [1]{%
 \ifx #1\expandafter \@firstoftwo
 \else \expandafter \@secondoftwo
 \fi
}%
\providecommand \natexlab [1]{#1}%
\providecommand \enquote  [1]{``#1''}%
\providecommand \bibnamefont  [1]{#1}%
\providecommand \bibfnamefont [1]{#1}%
\providecommand \citenamefont [1]{#1}%
\providecommand \href@noop [0]{\@secondoftwo}%
\providecommand \href [0]{\begingroup \@sanitize@url \@href}%
\providecommand \@href[1]{\@@startlink{#1}\@@href}%
\providecommand \@@href[1]{\endgroup#1\@@endlink}%
\providecommand \@sanitize@url [0]{\catcode `\\12\catcode `\$12\catcode
  `\&12\catcode `\#12\catcode `\^12\catcode `\_12\catcode `\%12\relax}%
\providecommand \@@startlink[1]{}%
\providecommand \@@endlink[0]{}%
\providecommand \url  [0]{\begingroup\@sanitize@url \@url }%
\providecommand \@url [1]{\endgroup\@href {#1}{\urlprefix }}%
\providecommand \urlprefix  [0]{URL }%
\providecommand \Eprint [0]{\href }%
\providecommand \doibase [0]{http://dx.doi.org/}%
\providecommand \selectlanguage [0]{\@gobble}%
\providecommand \bibinfo  [0]{\@secondoftwo}%
\providecommand \bibfield  [0]{\@secondoftwo}%
\providecommand \translation [1]{[#1]}%
\providecommand \BibitemOpen [0]{}%
\providecommand \bibitemStop [0]{}%
\providecommand \bibitemNoStop [0]{.\EOS\space}%
\providecommand \EOS [0]{\spacefactor3000\relax}%
\providecommand \BibitemShut  [1]{\csname bibitem#1\endcsname}%
\let\auto@bib@innerbib\@empty
\bibitem [{\citenamefont {Weinberg}(1979)}]{Weinberg:1979sa}%
  \BibitemOpen
  \bibfield  {author} {\bibinfo {author} {\bibfnamefont {S.}~\bibnamefont
  {Weinberg}},\ }\href {\doibase 10.1103/PhysRevLett.43.1566} {\bibfield
  {journal} {\bibinfo  {journal} {Phys. Rev. Lett.}\ }\textbf {\bibinfo
  {volume} {43}},\ \bibinfo {pages} {1566} (\bibinfo {year}
  {1979})}\BibitemShut {NoStop}%
\bibitem [{\citenamefont {Nath}\ and\ \citenamefont
  {Fileviez~Perez}(2007)}]{Nath:2006ut}%
  \BibitemOpen
  \bibfield  {author} {\bibinfo {author} {\bibfnamefont {P.}~\bibnamefont
  {Nath}}\ and\ \bibinfo {author} {\bibfnamefont {P.}~\bibnamefont
  {Fileviez~Perez}},\ }\href {\doibase 10.1016/j.physrep.2007.02.010}
  {\bibfield  {journal} {\bibinfo  {journal} {Phys. Rept.}\ }\textbf {\bibinfo
  {volume} {441}},\ \bibinfo {pages} {191} (\bibinfo {year} {2007})},\ \Eprint
  {http://arxiv.org/abs/hep-ph/0601023} {arXiv:hep-ph/0601023} \BibitemShut
  {NoStop}%
\bibitem [{\citenamefont {Sakharov}(1967)}]{Sakharov:1967dj}%
  \BibitemOpen
  \bibfield  {author} {\bibinfo {author} {\bibfnamefont {A.~D.}\ \bibnamefont
  {Sakharov}},\ }\href {\doibase 10.1070/PU1991v034n05ABEH002497} {\bibfield
  {journal} {\bibinfo  {journal} {Pisma Zh. Eksp. Teor. Fiz.}\ }\textbf
  {\bibinfo {volume} {5}},\ \bibinfo {pages} {32} (\bibinfo {year}
  {1967})}\BibitemShut {NoStop}%
\bibitem [{\citenamefont {Minkowski}(1977)}]{Minkowski:1977sc}%
  \BibitemOpen
  \bibfield  {author} {\bibinfo {author} {\bibfnamefont {P.}~\bibnamefont
  {Minkowski}},\ }\href {\doibase 10.1016/0370-2693(77)90435-X} {\bibfield
  {journal} {\bibinfo  {journal} {Phys. Lett.}\ }\textbf {\bibinfo {volume}
  {67B}},\ \bibinfo {pages} {421} (\bibinfo {year} {1977})}\BibitemShut
  {NoStop}%
\bibitem [{\citenamefont {Mohapatra}\ and\ \citenamefont
  {Senjanovic}(1980)}]{Mohapatra:1979ia}%
  \BibitemOpen
  \bibfield  {author} {\bibinfo {author} {\bibfnamefont {R.~N.}\ \bibnamefont
  {Mohapatra}}\ and\ \bibinfo {author} {\bibfnamefont {G.}~\bibnamefont
  {Senjanovic}},\ }\href {\doibase 10.1103/PhysRevLett.44.912} {\bibfield
  {journal} {\bibinfo  {journal} {Phys. Rev. Lett.}\ }\textbf {\bibinfo
  {volume} {44}},\ \bibinfo {pages} {912} (\bibinfo {year} {1980})}\BibitemShut
  {NoStop}%
\bibitem [{\citenamefont {Glashow}(1980)}]{Glashow:1979nm}%
  \BibitemOpen
  \bibfield  {author} {\bibinfo {author} {\bibfnamefont {S.~L.}\ \bibnamefont
  {Glashow}},\ }\href {\doibase 10.1007/978-1-4684-7197-7_15} {\bibfield
  {journal} {\bibinfo  {journal} {NATO Sci. Ser. B}\ }\textbf {\bibinfo
  {volume} {61}},\ \bibinfo {pages} {687} (\bibinfo {year} {1980})}\BibitemShut
  {NoStop}%
\bibitem [{\citenamefont {Gell-Mann}\ \emph {et~al.}(1979)\citenamefont
  {Gell-Mann}, \citenamefont {Ramond},\ and\ \citenamefont
  {Slansky}}]{Gell-Mann:1979vob}%
  \BibitemOpen
  \bibfield  {author} {\bibinfo {author} {\bibfnamefont {M.}~\bibnamefont
  {Gell-Mann}}, \bibinfo {author} {\bibfnamefont {P.}~\bibnamefont {Ramond}}, \
  and\ \bibinfo {author} {\bibfnamefont {R.}~\bibnamefont {Slansky}},\
  }\bibfield  {booktitle} {\emph {\bibinfo {booktitle} {{Supergravity Workshop
  Stony Brook, New York, September 27-28, 1979}}},\ }\href@noop {} {\bibfield
  {journal} {\bibinfo  {journal} {Conf. Proc.}\ }\textbf {\bibinfo {volume}
  {C790927}},\ \bibinfo {pages} {315} (\bibinfo {year} {1979})},\ \Eprint
  {http://arxiv.org/abs/1306.4669} {arXiv:1306.4669 [hep-th]} \BibitemShut
  {NoStop}%
\bibitem [{\citenamefont {Yanagida}(1979)}]{Yanagida:1979as}%
  \BibitemOpen
  \bibfield  {author} {\bibinfo {author} {\bibfnamefont {T.}~\bibnamefont
  {Yanagida}},\ }\href@noop {} {\bibfield  {journal} {\bibinfo  {journal}
  {Conf. Proc. C}\ }\textbf {\bibinfo {volume} {7902131}},\ \bibinfo {pages}
  {95} (\bibinfo {year} {1979})}\BibitemShut {NoStop}%
\bibitem [{\citenamefont {Konetschny}\ and\ \citenamefont
  {Kummer}(1977)}]{Konetschny:1977bn}%
  \BibitemOpen
  \bibfield  {author} {\bibinfo {author} {\bibfnamefont {W.}~\bibnamefont
  {Konetschny}}\ and\ \bibinfo {author} {\bibfnamefont {W.}~\bibnamefont
  {Kummer}},\ }\href {\doibase 10.1016/0370-2693(77)90407-5} {\bibfield
  {journal} {\bibinfo  {journal} {Phys. Lett. B}\ }\textbf {\bibinfo {volume}
  {70}},\ \bibinfo {pages} {433} (\bibinfo {year} {1977})}\BibitemShut
  {NoStop}%
\bibitem [{\citenamefont {Cheng}\ and\ \citenamefont
  {Li}(1980)}]{Cheng:1980qt}%
  \BibitemOpen
  \bibfield  {author} {\bibinfo {author} {\bibfnamefont {T.~P.}\ \bibnamefont
  {Cheng}}\ and\ \bibinfo {author} {\bibfnamefont {L.-F.}\ \bibnamefont {Li}},\
  }\href {\doibase 10.1103/PhysRevD.22.2860} {\bibfield  {journal} {\bibinfo
  {journal} {Phys. Rev. D}\ }\textbf {\bibinfo {volume} {22}},\ \bibinfo
  {pages} {2860} (\bibinfo {year} {1980})}\BibitemShut {NoStop}%
\bibitem [{\citenamefont {Mohapatra}\ and\ \citenamefont
  {Senjanovic}(1981)}]{Mohapatra:1980yp}%
  \BibitemOpen
  \bibfield  {author} {\bibinfo {author} {\bibfnamefont {R.~N.}\ \bibnamefont
  {Mohapatra}}\ and\ \bibinfo {author} {\bibfnamefont {G.}~\bibnamefont
  {Senjanovic}},\ }\href {\doibase 10.1103/PhysRevD.23.165} {\bibfield
  {journal} {\bibinfo  {journal} {Phys. Rev. D}\ }\textbf {\bibinfo {volume}
  {23}},\ \bibinfo {pages} {165} (\bibinfo {year} {1981})}\BibitemShut
  {NoStop}%
\bibitem [{\citenamefont {Lazarides}\ \emph {et~al.}(1981)\citenamefont
  {Lazarides}, \citenamefont {Shafi},\ and\ \citenamefont
  {Wetterich}}]{Lazarides:1980nt}%
  \BibitemOpen
  \bibfield  {author} {\bibinfo {author} {\bibfnamefont {G.}~\bibnamefont
  {Lazarides}}, \bibinfo {author} {\bibfnamefont {Q.}~\bibnamefont {Shafi}}, \
  and\ \bibinfo {author} {\bibfnamefont {C.}~\bibnamefont {Wetterich}},\ }\href
  {\doibase 10.1016/0550-3213(81)90354-0} {\bibfield  {journal} {\bibinfo
  {journal} {Nucl. Phys. B}\ }\textbf {\bibinfo {volume} {181}},\ \bibinfo
  {pages} {287} (\bibinfo {year} {1981})}\BibitemShut {NoStop}%
\bibitem [{\citenamefont {Schechter}\ and\ \citenamefont
  {Valle}(1980)}]{Schechter:1980gr}%
  \BibitemOpen
  \bibfield  {author} {\bibinfo {author} {\bibfnamefont {J.}~\bibnamefont
  {Schechter}}\ and\ \bibinfo {author} {\bibfnamefont {J.~W.~F.}\ \bibnamefont
  {Valle}},\ }\href {\doibase 10.1103/PhysRevD.22.2227} {\bibfield  {journal}
  {\bibinfo  {journal} {Phys. Rev. D}\ }\textbf {\bibinfo {volume} {22}},\
  \bibinfo {pages} {2227} (\bibinfo {year} {1980})}\BibitemShut {NoStop}%
\bibitem [{\citenamefont {Foot}\ \emph
  {et~al.}(1989{\natexlab{a}})\citenamefont {Foot}, \citenamefont {Lew},
  \citenamefont {He},\ and\ \citenamefont {Joshi}}]{Foot:1988aq}%
  \BibitemOpen
  \bibfield  {author} {\bibinfo {author} {\bibfnamefont {R.}~\bibnamefont
  {Foot}}, \bibinfo {author} {\bibfnamefont {H.}~\bibnamefont {Lew}}, \bibinfo
  {author} {\bibfnamefont {X.~G.}\ \bibnamefont {He}}, \ and\ \bibinfo {author}
  {\bibfnamefont {G.~C.}\ \bibnamefont {Joshi}},\ }\href {\doibase
  10.1007/BF01415558} {\bibfield  {journal} {\bibinfo  {journal} {Z. Phys. C}\
  }\textbf {\bibinfo {volume} {44}},\ \bibinfo {pages} {441} (\bibinfo {year}
  {1989}{\natexlab{a}})}\BibitemShut {NoStop}%
\bibitem [{\citenamefont {Arnold}\ \emph {et~al.}(2013)\citenamefont {Arnold},
  \citenamefont {Fornal},\ and\ \citenamefont {Wise}}]{Arnold:2012sd}%
  \BibitemOpen
  \bibfield  {author} {\bibinfo {author} {\bibfnamefont {J.~M.}\ \bibnamefont
  {Arnold}}, \bibinfo {author} {\bibfnamefont {B.}~\bibnamefont {Fornal}}, \
  and\ \bibinfo {author} {\bibfnamefont {M.~B.}\ \bibnamefont {Wise}},\ }\href
  {\doibase 10.1103/PhysRevD.87.075004} {\bibfield  {journal} {\bibinfo
  {journal} {Phys. Rev. D}\ }\textbf {\bibinfo {volume} {87}},\ \bibinfo
  {pages} {075004} (\bibinfo {year} {2013})},\ \Eprint
  {http://arxiv.org/abs/1212.4556} {arXiv:1212.4556 [hep-ph]} \BibitemShut
  {NoStop}%
\bibitem [{\citenamefont {Georgi}\ and\ \citenamefont
  {Glashow}(1974)}]{Georgi:1974sy}%
  \BibitemOpen
  \bibfield  {author} {\bibinfo {author} {\bibfnamefont {H.}~\bibnamefont
  {Georgi}}\ and\ \bibinfo {author} {\bibfnamefont {S.~L.}\ \bibnamefont
  {Glashow}},\ }\href {\doibase 10.1103/PhysRevLett.32.438} {\bibfield
  {journal} {\bibinfo  {journal} {Phys. Rev. Lett.}\ }\textbf {\bibinfo
  {volume} {32}},\ \bibinfo {pages} {438} (\bibinfo {year} {1974})}\BibitemShut
  {NoStop}%
\bibitem [{\citenamefont {Georgi}(1975)}]{Georgi:1974my}%
  \BibitemOpen
  \bibfield  {author} {\bibinfo {author} {\bibfnamefont {H.}~\bibnamefont
  {Georgi}},\ }\href {\doibase 10.1063/1.2947450} {\bibfield  {journal}
  {\bibinfo  {journal} {AIP Conf. Proc.}\ }\textbf {\bibinfo {volume} {23}},\
  \bibinfo {pages} {575} (\bibinfo {year} {1975})}\BibitemShut {NoStop}%
\bibitem [{\citenamefont {Fritzsch}\ and\ \citenamefont
  {Minkowski}(1975)}]{Fritzsch:1974nn}%
  \BibitemOpen
  \bibfield  {author} {\bibinfo {author} {\bibfnamefont {H.}~\bibnamefont
  {Fritzsch}}\ and\ \bibinfo {author} {\bibfnamefont {P.}~\bibnamefont
  {Minkowski}},\ }\href {\doibase 10.1016/0003-4916(75)90211-0} {\bibfield
  {journal} {\bibinfo  {journal} {Annals Phys.}\ }\textbf {\bibinfo {volume}
  {93}},\ \bibinfo {pages} {193} (\bibinfo {year} {1975})}\BibitemShut
  {NoStop}%
\bibitem [{\citenamefont {Dorsner}\ and\ \citenamefont
  {Fileviez~Perez}(2005)}]{Dorsner:2005fq}%
  \BibitemOpen
  \bibfield  {author} {\bibinfo {author} {\bibfnamefont {I.}~\bibnamefont
  {Dorsner}}\ and\ \bibinfo {author} {\bibfnamefont {P.}~\bibnamefont
  {Fileviez~Perez}},\ }\href {\doibase 10.1016/j.nuclphysb.2005.06.016}
  {\bibfield  {journal} {\bibinfo  {journal} {Nucl. Phys. B}\ }\textbf
  {\bibinfo {volume} {723}},\ \bibinfo {pages} {53} (\bibinfo {year} {2005})},\
  \Eprint {http://arxiv.org/abs/hep-ph/0504276} {arXiv:hep-ph/0504276}
  \BibitemShut {NoStop}%
\bibitem [{\citenamefont {Bajc}\ and\ \citenamefont
  {Senjanovic}(2007)}]{Bajc:2006ia}%
  \BibitemOpen
  \bibfield  {author} {\bibinfo {author} {\bibfnamefont {B.}~\bibnamefont
  {Bajc}}\ and\ \bibinfo {author} {\bibfnamefont {G.}~\bibnamefont
  {Senjanovic}},\ }\href {\doibase 10.1088/1126-6708/2007/08/014} {\bibfield
  {journal} {\bibinfo  {journal} {JHEP}\ }\textbf {\bibinfo {volume} {08}},\
  \bibinfo {pages} {014} (\bibinfo {year} {2007})},\ \Eprint
  {http://arxiv.org/abs/hep-ph/0612029} {arXiv:hep-ph/0612029} \BibitemShut
  {NoStop}%
\bibitem [{\citenamefont {Preda}\ \emph {et~al.}(2022)\citenamefont {Preda},
  \citenamefont {Senjanovic},\ and\ \citenamefont
  {Zantedeschi}}]{Preda:2022izo}%
  \BibitemOpen
  \bibfield  {author} {\bibinfo {author} {\bibfnamefont {A.}~\bibnamefont
  {Preda}}, \bibinfo {author} {\bibfnamefont {G.}~\bibnamefont {Senjanovic}}, \
  and\ \bibinfo {author} {\bibfnamefont {M.}~\bibnamefont {Zantedeschi}},\
  }\href@noop {} {\  (\bibinfo {year} {2022})},\ \Eprint
  {http://arxiv.org/abs/2201.02785} {arXiv:2201.02785 [hep-ph]} \BibitemShut
  {NoStop}%
\bibitem [{\citenamefont {Barbier}\ \emph {et~al.}(2005)\citenamefont {Barbier}
  \emph {et~al.}}]{Barbier:2004ez}%
  \BibitemOpen
  \bibfield  {author} {\bibinfo {author} {\bibfnamefont {R.}~\bibnamefont
  {Barbier}} \emph {et~al.},\ }\href {\doibase 10.1016/j.physrep.2005.08.006}
  {\bibfield  {journal} {\bibinfo  {journal} {Phys. Rept.}\ }\textbf {\bibinfo
  {volume} {420}},\ \bibinfo {pages} {1} (\bibinfo {year} {2005})},\ \Eprint
  {http://arxiv.org/abs/hep-ph/0406039} {arXiv:hep-ph/0406039} \BibitemShut
  {NoStop}%
\bibitem [{\citenamefont {Dimopoulos}\ \emph {et~al.}(1981)\citenamefont
  {Dimopoulos}, \citenamefont {Raby},\ and\ \citenamefont
  {Wilczek}}]{Dimopoulos:1981yj}%
  \BibitemOpen
  \bibfield  {author} {\bibinfo {author} {\bibfnamefont {S.}~\bibnamefont
  {Dimopoulos}}, \bibinfo {author} {\bibfnamefont {S.}~\bibnamefont {Raby}}, \
  and\ \bibinfo {author} {\bibfnamefont {F.}~\bibnamefont {Wilczek}},\ }\href
  {\doibase 10.1103/PhysRevD.24.1681} {\bibfield  {journal} {\bibinfo
  {journal} {Phys. Rev. D}\ }\textbf {\bibinfo {volume} {24}},\ \bibinfo
  {pages} {1681} (\bibinfo {year} {1981})}\BibitemShut {NoStop}%
\bibitem [{\citenamefont {Ibanez}\ and\ \citenamefont
  {Ross}(1981)}]{Ibanez:1981yh}%
  \BibitemOpen
  \bibfield  {author} {\bibinfo {author} {\bibfnamefont {L.~E.}\ \bibnamefont
  {Ibanez}}\ and\ \bibinfo {author} {\bibfnamefont {G.~G.}\ \bibnamefont
  {Ross}},\ }\href {\doibase 10.1016/0370-2693(81)91200-4} {\bibfield
  {journal} {\bibinfo  {journal} {Phys. Lett. B}\ }\textbf {\bibinfo {volume}
  {105}},\ \bibinfo {pages} {439} (\bibinfo {year} {1981})}\BibitemShut
  {NoStop}%
\bibitem [{\citenamefont {Einhorn}\ and\ \citenamefont
  {Jones}(1982)}]{Einhorn:1981sx}%
  \BibitemOpen
  \bibfield  {author} {\bibinfo {author} {\bibfnamefont {M.~B.}\ \bibnamefont
  {Einhorn}}\ and\ \bibinfo {author} {\bibfnamefont {D.~R.~T.}\ \bibnamefont
  {Jones}},\ }\href {\doibase 10.1016/0550-3213(82)90502-8} {\bibfield
  {journal} {\bibinfo  {journal} {Nucl. Phys. B}\ }\textbf {\bibinfo {volume}
  {196}},\ \bibinfo {pages} {475} (\bibinfo {year} {1982})}\BibitemShut
  {NoStop}%
\bibitem [{\citenamefont {Marciano}\ and\ \citenamefont
  {Senjanovic}(1982)}]{Marciano:1981un}%
  \BibitemOpen
  \bibfield  {author} {\bibinfo {author} {\bibfnamefont {W.~J.}\ \bibnamefont
  {Marciano}}\ and\ \bibinfo {author} {\bibfnamefont {G.}~\bibnamefont
  {Senjanovic}},\ }\href {\doibase 10.1103/PhysRevD.25.3092} {\bibfield
  {journal} {\bibinfo  {journal} {Phys. Rev. D}\ }\textbf {\bibinfo {volume}
  {25}},\ \bibinfo {pages} {3092} (\bibinfo {year} {1982})}\BibitemShut
  {NoStop}%
\bibitem [{\citenamefont {Dimopoulos}\ and\ \citenamefont
  {Georgi}(1981)}]{Dimopoulos:1981zb}%
  \BibitemOpen
  \bibfield  {author} {\bibinfo {author} {\bibfnamefont {S.}~\bibnamefont
  {Dimopoulos}}\ and\ \bibinfo {author} {\bibfnamefont {H.}~\bibnamefont
  {Georgi}},\ }\href {\doibase 10.1016/0550-3213(81)90522-8} {\bibfield
  {journal} {\bibinfo  {journal} {Nucl. Phys. B}\ }\textbf {\bibinfo {volume}
  {193}},\ \bibinfo {pages} {150} (\bibinfo {year} {1981})}\BibitemShut
  {NoStop}%
\bibitem [{\citenamefont {Sakai}(1981)}]{Sakai:1981gr}%
  \BibitemOpen
  \bibfield  {author} {\bibinfo {author} {\bibfnamefont {N.}~\bibnamefont
  {Sakai}},\ }\href {\doibase 10.1007/BF01573998} {\bibfield  {journal}
  {\bibinfo  {journal} {Z. Phys. C}\ }\textbf {\bibinfo {volume} {11}},\
  \bibinfo {pages} {153} (\bibinfo {year} {1981})}\BibitemShut {NoStop}%
\bibitem [{\citenamefont {Pais}(1973)}]{Pais:1973mi}%
  \BibitemOpen
  \bibfield  {author} {\bibinfo {author} {\bibfnamefont {A.}~\bibnamefont
  {Pais}},\ }\href {\doibase 10.1103/PhysRevD.8.1844} {\bibfield  {journal}
  {\bibinfo  {journal} {Phys. Rev. D}\ }\textbf {\bibinfo {volume} {8}},\
  \bibinfo {pages} {1844} (\bibinfo {year} {1973})}\BibitemShut {NoStop}%
\bibitem [{\citenamefont {Rajpoot}(1989)}]{Rajpoot:1989jb}%
  \BibitemOpen
  \bibfield  {author} {\bibinfo {author} {\bibfnamefont {S.}~\bibnamefont
  {Rajpoot}},\ }\href {\doibase 10.1103/PhysRevD.40.2421} {\bibfield  {journal}
  {\bibinfo  {journal} {Phys. Rev. D}\ }\textbf {\bibinfo {volume} {40}},\
  \bibinfo {pages} {2421} (\bibinfo {year} {1989})}\BibitemShut {NoStop}%
\bibitem [{\citenamefont {Foot}\ \emph
  {et~al.}(1989{\natexlab{b}})\citenamefont {Foot}, \citenamefont {Joshi},\
  and\ \citenamefont {Lew}}]{Foot:1989ts}%
  \BibitemOpen
  \bibfield  {author} {\bibinfo {author} {\bibfnamefont {R.}~\bibnamefont
  {Foot}}, \bibinfo {author} {\bibfnamefont {G.~C.}\ \bibnamefont {Joshi}}, \
  and\ \bibinfo {author} {\bibfnamefont {H.}~\bibnamefont {Lew}},\ }\href
  {\doibase 10.1103/PhysRevD.40.2487} {\bibfield  {journal} {\bibinfo
  {journal} {Phys. Rev. D}\ }\textbf {\bibinfo {volume} {40}},\ \bibinfo
  {pages} {2487} (\bibinfo {year} {1989}{\natexlab{b}})}\BibitemShut {NoStop}%
\bibitem [{\citenamefont {Carone}\ and\ \citenamefont
  {Murayama}(1995)}]{Carone:1995pu}%
  \BibitemOpen
  \bibfield  {author} {\bibinfo {author} {\bibfnamefont {C.~D.}\ \bibnamefont
  {Carone}}\ and\ \bibinfo {author} {\bibfnamefont {H.}~\bibnamefont
  {Murayama}},\ }\href {\doibase 10.1103/PhysRevD.52.484} {\bibfield  {journal}
  {\bibinfo  {journal} {Phys. Rev. D}\ }\textbf {\bibinfo {volume} {52}},\
  \bibinfo {pages} {484} (\bibinfo {year} {1995})},\ \Eprint
  {http://arxiv.org/abs/hep-ph/9501220} {arXiv:hep-ph/9501220} \BibitemShut
  {NoStop}%
\bibitem [{\citenamefont {Fileviez~Perez}\ and\ \citenamefont
  {Wise}(2010)}]{FileviezPerez:2010gw}%
  \BibitemOpen
  \bibfield  {author} {\bibinfo {author} {\bibfnamefont {P.}~\bibnamefont
  {Fileviez~Perez}}\ and\ \bibinfo {author} {\bibfnamefont {M.~B.}\
  \bibnamefont {Wise}},\ }\href {\doibase 10.1103/PhysRevD.82.079901}
  {\bibfield  {journal} {\bibinfo  {journal} {Phys. Rev. D}\ }\textbf {\bibinfo
  {volume} {82}},\ \bibinfo {pages} {011901} (\bibinfo {year} {2010})},\
  \bibinfo {note} {[Erratum: Phys.Rev.D 82, 079901 (2010)]},\ \Eprint
  {http://arxiv.org/abs/1002.1754} {arXiv:1002.1754 [hep-ph]} \BibitemShut
  {NoStop}%
\bibitem [{\citenamefont {Fileviez~Perez}\ \emph {et~al.}(2014)\citenamefont
  {Fileviez~Perez}, \citenamefont {Ohmer},\ and\ \citenamefont
  {Patel}}]{FileviezPerez:2014lnj}%
  \BibitemOpen
  \bibfield  {author} {\bibinfo {author} {\bibfnamefont {P.}~\bibnamefont
  {Fileviez~Perez}}, \bibinfo {author} {\bibfnamefont {S.}~\bibnamefont
  {Ohmer}}, \ and\ \bibinfo {author} {\bibfnamefont {H.~H.}\ \bibnamefont
  {Patel}},\ }\href {\doibase 10.1016/j.physletb.2014.06.057} {\bibfield
  {journal} {\bibinfo  {journal} {Phys. Lett. B}\ }\textbf {\bibinfo {volume}
  {735}},\ \bibinfo {pages} {283} (\bibinfo {year} {2014})},\ \Eprint
  {http://arxiv.org/abs/1403.8029} {arXiv:1403.8029 [hep-ph]} \BibitemShut
  {NoStop}%
\bibitem [{\citenamefont {Duerr}\ \emph {et~al.}(2013)\citenamefont {Duerr},
  \citenamefont {Fileviez~Perez},\ and\ \citenamefont {Wise}}]{Duerr:2013dza}%
  \BibitemOpen
  \bibfield  {author} {\bibinfo {author} {\bibfnamefont {M.}~\bibnamefont
  {Duerr}}, \bibinfo {author} {\bibfnamefont {P.}~\bibnamefont
  {Fileviez~Perez}}, \ and\ \bibinfo {author} {\bibfnamefont {M.~B.}\
  \bibnamefont {Wise}},\ }\href {\doibase 10.1103/PhysRevLett.110.231801}
  {\bibfield  {journal} {\bibinfo  {journal} {Phys. Rev. Lett.}\ }\textbf
  {\bibinfo {volume} {110}},\ \bibinfo {pages} {231801} (\bibinfo {year}
  {2013})},\ \Eprint {http://arxiv.org/abs/1304.0576} {arXiv:1304.0576
  [hep-ph]} \BibitemShut {NoStop}%
\bibitem [{\citenamefont {Fileviez~P\'erez}\ \emph {et~al.}(2019)\citenamefont
  {Fileviez~P\'erez}, \citenamefont {Golias}, \citenamefont {Li}, \citenamefont
  {Murgui},\ and\ \citenamefont {Plascencia}}]{FileviezPerez:2019jju}%
  \BibitemOpen
  \bibfield  {author} {\bibinfo {author} {\bibfnamefont {P.}~\bibnamefont
  {Fileviez~P\'erez}}, \bibinfo {author} {\bibfnamefont {E.}~\bibnamefont
  {Golias}}, \bibinfo {author} {\bibfnamefont {R.-H.}\ \bibnamefont {Li}},
  \bibinfo {author} {\bibfnamefont {C.}~\bibnamefont {Murgui}}, \ and\ \bibinfo
  {author} {\bibfnamefont {A.~D.}\ \bibnamefont {Plascencia}},\ }\href
  {\doibase 10.1103/PhysRevD.100.015017} {\bibfield  {journal} {\bibinfo
  {journal} {Phys. Rev. D}\ }\textbf {\bibinfo {volume} {100}},\ \bibinfo
  {pages} {015017} (\bibinfo {year} {2019})},\ \Eprint
  {http://arxiv.org/abs/1904.01017} {arXiv:1904.01017 [hep-ph]} \BibitemShut
  {NoStop}%
\bibitem [{\citenamefont {Schechter}\ and\ \citenamefont
  {Valle}(1982)}]{Schechter:1981bd}%
  \BibitemOpen
  \bibfield  {author} {\bibinfo {author} {\bibfnamefont {J.}~\bibnamefont
  {Schechter}}\ and\ \bibinfo {author} {\bibfnamefont {J.~W.~F.}\ \bibnamefont
  {Valle}},\ }\href {\doibase 10.1103/PhysRevD.25.2951} {\bibfield  {journal}
  {\bibinfo  {journal} {Phys. Rev.}\ }\textbf {\bibinfo {volume} {D25}},\
  \bibinfo {pages} {2951} (\bibinfo {year} {1982})}\BibitemShut {NoStop}%
\bibitem [{\citenamefont {Davidson}\ \emph {et~al.}(2008)\citenamefont
  {Davidson}, \citenamefont {Nardi},\ and\ \citenamefont
  {Nir}}]{Davidson:2008bu}%
  \BibitemOpen
  \bibfield  {author} {\bibinfo {author} {\bibfnamefont {S.}~\bibnamefont
  {Davidson}}, \bibinfo {author} {\bibfnamefont {E.}~\bibnamefont {Nardi}}, \
  and\ \bibinfo {author} {\bibfnamefont {Y.}~\bibnamefont {Nir}},\ }\href
  {\doibase 10.1016/j.physrep.2008.06.002} {\bibfield  {journal} {\bibinfo
  {journal} {Phys. Rept.}\ }\textbf {\bibinfo {volume} {466}},\ \bibinfo
  {pages} {105} (\bibinfo {year} {2008})},\ \Eprint
  {http://arxiv.org/abs/0802.2962} {arXiv:0802.2962 [hep-ph]} \BibitemShut
  {NoStop}%
\bibitem [{\citenamefont {Gando}\ \emph {et~al.}(2016)\citenamefont {Gando}
  \emph {et~al.}}]{KamLAND-Zen:2016pfg}%
  \BibitemOpen
  \bibfield  {author} {\bibinfo {author} {\bibfnamefont {A.}~\bibnamefont
  {Gando}} \emph {et~al.} (\bibinfo {collaboration} {KamLAND-Zen}),\ }\href
  {\doibase 10.1103/PhysRevLett.117.109903, 10.1103/PhysRevLett.117.082503}
  {\bibfield  {journal} {\bibinfo  {journal} {Phys. Rev. Lett.}\ }\textbf
  {\bibinfo {volume} {117}},\ \bibinfo {pages} {082503} (\bibinfo {year}
  {2016})},\ \bibinfo {note} {[Addendum: Phys. Rev.
  Lett.117,no.10,109903(2016)]},\ \Eprint {http://arxiv.org/abs/1605.02889}
  {arXiv:1605.02889 [hep-ex]} \BibitemShut {NoStop}%
\bibitem [{\citenamefont {Anton}\ \emph {et~al.}(2019)\citenamefont {Anton}
  \emph {et~al.}}]{EXO-200:2019rkq}%
  \BibitemOpen
  \bibfield  {author} {\bibinfo {author} {\bibfnamefont {G.}~\bibnamefont
  {Anton}} \emph {et~al.} (\bibinfo {collaboration} {EXO-200}),\ }\href
  {\doibase 10.1103/PhysRevLett.123.161802} {\bibfield  {journal} {\bibinfo
  {journal} {Phys. Rev. Lett.}\ }\textbf {\bibinfo {volume} {123}},\ \bibinfo
  {pages} {161802} (\bibinfo {year} {2019})},\ \Eprint
  {http://arxiv.org/abs/1906.02723} {arXiv:1906.02723 [hep-ex]} \BibitemShut
  {NoStop}%
\bibitem [{\citenamefont {Alvis}\ \emph
  {et~al.}(2019{\natexlab{a}})\citenamefont {Alvis} \emph
  {et~al.}}]{Majorana:2019nbd}%
  \BibitemOpen
  \bibfield  {author} {\bibinfo {author} {\bibfnamefont {S.~I.}\ \bibnamefont
  {Alvis}} \emph {et~al.} (\bibinfo {collaboration} {Majorana}),\ }\href
  {\doibase 10.1103/PhysRevC.100.025501} {\bibfield  {journal} {\bibinfo
  {journal} {Phys. Rev. C}\ }\textbf {\bibinfo {volume} {100}},\ \bibinfo
  {pages} {025501} (\bibinfo {year} {2019}{\natexlab{a}})},\ \Eprint
  {http://arxiv.org/abs/1902.02299} {arXiv:1902.02299 [nucl-ex]} \BibitemShut
  {NoStop}%
\bibitem [{\citenamefont {Azzolini}\ \emph {et~al.}(2019)\citenamefont
  {Azzolini} \emph {et~al.}}]{CUPID:2019gpc}%
  \BibitemOpen
  \bibfield  {author} {\bibinfo {author} {\bibfnamefont {O.}~\bibnamefont
  {Azzolini}} \emph {et~al.} (\bibinfo {collaboration} {CUPID}),\ }\href
  {\doibase 10.1103/PhysRevLett.123.032501} {\bibfield  {journal} {\bibinfo
  {journal} {Phys. Rev. Lett.}\ }\textbf {\bibinfo {volume} {123}},\ \bibinfo
  {pages} {032501} (\bibinfo {year} {2019})},\ \Eprint
  {http://arxiv.org/abs/1906.05001} {arXiv:1906.05001 [nucl-ex]} \BibitemShut
  {NoStop}%
\bibitem [{\citenamefont {Agostini}\ \emph {et~al.}(2020)\citenamefont
  {Agostini} \emph {et~al.}}]{GERDA:2020xhi}%
  \BibitemOpen
  \bibfield  {author} {\bibinfo {author} {\bibfnamefont {M.}~\bibnamefont
  {Agostini}} \emph {et~al.} (\bibinfo {collaboration} {GERDA}),\ }\href
  {\doibase 10.1103/PhysRevLett.125.252502} {\bibfield  {journal} {\bibinfo
  {journal} {Phys. Rev. Lett.}\ }\textbf {\bibinfo {volume} {125}},\ \bibinfo
  {pages} {252502} (\bibinfo {year} {2020})},\ \Eprint
  {http://arxiv.org/abs/2009.06079} {arXiv:2009.06079 [nucl-ex]} \BibitemShut
  {NoStop}%
\bibitem [{\citenamefont {Adams}\ \emph {et~al.}(2022)\citenamefont {Adams}
  \emph {et~al.}}]{CUORE:2021mvw}%
  \BibitemOpen
  \bibfield  {author} {\bibinfo {author} {\bibfnamefont {D.~Q.}\ \bibnamefont
  {Adams}} \emph {et~al.} (\bibinfo {collaboration} {CUORE}),\ }\href {\doibase
  10.1038/s41586-022-04497-4} {\bibfield  {journal} {\bibinfo  {journal}
  {Nature}\ }\textbf {\bibinfo {volume} {604}},\ \bibinfo {pages} {53}
  (\bibinfo {year} {2022})},\ \Eprint {http://arxiv.org/abs/2104.06906}
  {arXiv:2104.06906 [nucl-ex]} \BibitemShut {NoStop}%
\bibitem [{\citenamefont {Abe}\ \emph {et~al.}(2022)\citenamefont {Abe} \emph
  {et~al.}}]{KamLAND-Zen:2022tow}%
  \BibitemOpen
  \bibfield  {author} {\bibinfo {author} {\bibfnamefont {S.}~\bibnamefont
  {Abe}} \emph {et~al.} (\bibinfo {collaboration} {KamLAND-Zen}),\ }\href@noop
  {} {\  (\bibinfo {year} {2022})},\ \Eprint {http://arxiv.org/abs/2203.02139}
  {arXiv:2203.02139 [hep-ex]} \BibitemShut {NoStop}%
\bibitem [{\citenamefont {Arnquist}\ \emph {et~al.}(2022)\citenamefont
  {Arnquist} \emph {et~al.}}]{Arnquist:2022zrp}%
  \BibitemOpen
  \bibfield  {author} {\bibinfo {author} {\bibfnamefont {I.~J.}\ \bibnamefont
  {Arnquist}} \emph {et~al.},\ }\href@noop {} {\  (\bibinfo {year} {2022})},\
  \Eprint {http://arxiv.org/abs/2207.07638} {arXiv:2207.07638 [nucl-ex]}
  \BibitemShut {NoStop}%
\bibitem [{\citenamefont {Kharusi}\ \emph {et~al.}(2018)\citenamefont {Kharusi}
  \emph {et~al.}}]{nEXO:2018ylp}%
  \BibitemOpen
  \bibfield  {author} {\bibinfo {author} {\bibfnamefont {S.~A.}\ \bibnamefont
  {Kharusi}} \emph {et~al.} (\bibinfo {collaboration} {nEXO}),\ }\href@noop {}
  {\  (\bibinfo {year} {2018})},\ \Eprint {http://arxiv.org/abs/1805.11142}
  {arXiv:1805.11142 [physics.ins-det]} \BibitemShut {NoStop}%
\bibitem [{\citenamefont {Adhikari}\ \emph {et~al.}(2022)\citenamefont
  {Adhikari} \emph {et~al.}}]{nEXO:2021ujk}%
  \BibitemOpen
  \bibfield  {author} {\bibinfo {author} {\bibfnamefont {G.}~\bibnamefont
  {Adhikari}} \emph {et~al.} (\bibinfo {collaboration} {nEXO}),\ }\href
  {\doibase 10.1088/1361-6471/ac3631} {\bibfield  {journal} {\bibinfo
  {journal} {J. Phys. G}\ }\textbf {\bibinfo {volume} {49}},\ \bibinfo {pages}
  {015104} (\bibinfo {year} {2022})},\ \Eprint
  {http://arxiv.org/abs/2106.16243} {arXiv:2106.16243 [nucl-ex]} \BibitemShut
  {NoStop}%
\bibitem [{\citenamefont {Abgrall}\ \emph {et~al.}(2021)\citenamefont {Abgrall}
  \emph {et~al.}}]{LEGEND:2021bnm}%
  \BibitemOpen
  \bibfield  {author} {\bibinfo {author} {\bibfnamefont {N.}~\bibnamefont
  {Abgrall}} \emph {et~al.} (\bibinfo {collaboration} {LEGEND}),\ }\href@noop
  {} {\  (\bibinfo {year} {2021})},\ \Eprint {http://arxiv.org/abs/2107.11462}
  {arXiv:2107.11462 [physics.ins-det]} \BibitemShut {NoStop}%
\bibitem [{\citenamefont {Armstrong}\ \emph {et~al.}(2019)\citenamefont
  {Armstrong} \emph {et~al.}}]{armstrong2019cupid}%
  \BibitemOpen
  \bibfield  {author} {\bibinfo {author} {\bibfnamefont {W.~R.}\ \bibnamefont
  {Armstrong}} \emph {et~al.} (\bibinfo {collaboration} {CUPID}),\ }\href@noop
  {} {\  (\bibinfo {year} {2019})},\ \Eprint {http://arxiv.org/abs/1907.09376}
  {arXiv:1907.09376 [physics.ins-det]} \BibitemShut {NoStop}%
\bibitem [{\citenamefont {Armatol}\ \emph {et~al.}(2022)\citenamefont {Armatol}
  \emph {et~al.}}]{CUPID:2022wpt}%
  \BibitemOpen
  \bibfield  {author} {\bibinfo {author} {\bibfnamefont {A.}~\bibnamefont
  {Armatol}} \emph {et~al.} (\bibinfo {collaboration} {CUPID}),\ }\href@noop {}
  {\  (\bibinfo {year} {2022})},\ \Eprint {http://arxiv.org/abs/2203.08386}
  {arXiv:2203.08386 [nucl-ex]} \BibitemShut {NoStop}%
\bibitem [{\citenamefont {Albanese}\ \emph {et~al.}(2021)\citenamefont
  {Albanese} \emph {et~al.}}]{SNO:2021xpa}%
  \BibitemOpen
  \bibfield  {author} {\bibinfo {author} {\bibfnamefont {V.}~\bibnamefont
  {Albanese}} \emph {et~al.} (\bibinfo {collaboration} {SNO+}),\ }\href
  {\doibase 10.1088/1748-0221/16/08/P08059} {\bibfield  {journal} {\bibinfo
  {journal} {JINST}\ }\textbf {\bibinfo {volume} {16}},\ \bibinfo {pages}
  {P08059} (\bibinfo {year} {2021})},\ \Eprint
  {http://arxiv.org/abs/2104.11687} {arXiv:2104.11687 [physics.ins-det]}
  \BibitemShut {NoStop}%
\bibitem [{\citenamefont {Nakamura}\ \emph {et~al.}(2020)\citenamefont
  {Nakamura}, \citenamefont {Sambonsugi}, \citenamefont {Shiraishi},\ and\
  \citenamefont {Wada}}]{Nakamura:2020szx}%
  \BibitemOpen
  \bibfield  {author} {\bibinfo {author} {\bibfnamefont {R.}~\bibnamefont
  {Nakamura}}, \bibinfo {author} {\bibfnamefont {H.}~\bibnamefont
  {Sambonsugi}}, \bibinfo {author} {\bibfnamefont {K.}~\bibnamefont
  {Shiraishi}}, \ and\ \bibinfo {author} {\bibfnamefont {Y.}~\bibnamefont
  {Wada}},\ }\href {\doibase 10.1088/1742-6596/1468/1/012256} {\bibfield
  {journal} {\bibinfo  {journal} {J. Phys. Conf. Ser.}\ }\textbf {\bibinfo
  {volume} {1468}},\ \bibinfo {pages} {012256} (\bibinfo {year}
  {2020})}\BibitemShut {NoStop}%
\bibitem [{\citenamefont {Adams}\ \emph {et~al.}(2021)\citenamefont {Adams}
  \emph {et~al.}}]{adams2020sensitivity}%
  \BibitemOpen
  \bibfield  {author} {\bibinfo {author} {\bibfnamefont {C.}~\bibnamefont
  {Adams}} \emph {et~al.} (\bibinfo {collaboration} {NEXT}),\ }\href {\doibase
  10.1007/JHEP08(2021)164} {\bibfield  {journal} {\bibinfo  {journal} {JHEP}\
  }\textbf {\bibinfo {volume} {2021}},\ \bibinfo {pages} {164} (\bibinfo {year}
  {2021})},\ \Eprint {http://arxiv.org/abs/2005.06467} {arXiv:2005.06467
  [physics.ins-det]} \BibitemShut {NoStop}%
\bibitem [{\citenamefont {Rodejohann}(2011)}]{Rodejohann:2011mu}%
  \BibitemOpen
  \bibfield  {author} {\bibinfo {author} {\bibfnamefont {W.}~\bibnamefont
  {Rodejohann}},\ }\href {\doibase 10.1142/S0218301311020186} {\bibfield
  {journal} {\bibinfo  {journal} {Int. J. Mod. Phys.}\ }\textbf {\bibinfo
  {volume} {E20}},\ \bibinfo {pages} {1833} (\bibinfo {year} {2011})},\ \Eprint
  {http://arxiv.org/abs/1106.1334} {arXiv:1106.1334 [hep-ph]} \BibitemShut
  {NoStop}%
\bibitem [{\citenamefont {Keung}\ and\ \citenamefont
  {Senjanovic}(1983)}]{Keung:1983uu}%
  \BibitemOpen
  \bibfield  {author} {\bibinfo {author} {\bibfnamefont {W.-Y.}\ \bibnamefont
  {Keung}}\ and\ \bibinfo {author} {\bibfnamefont {G.}~\bibnamefont
  {Senjanovic}},\ }\href {\doibase 10.1103/PhysRevLett.50.1427} {\bibfield
  {journal} {\bibinfo  {journal} {Phys. Rev. Lett.}\ }\textbf {\bibinfo
  {volume} {50}},\ \bibinfo {pages} {1427} (\bibinfo {year}
  {1983})}\BibitemShut {NoStop}%
\bibitem [{\citenamefont {Peng}\ \emph {et~al.}(2016)\citenamefont {Peng},
  \citenamefont {Ramsey-Musolf},\ and\ \citenamefont {Winslow}}]{Peng:2015haa}%
  \BibitemOpen
  \bibfield  {author} {\bibinfo {author} {\bibfnamefont {T.}~\bibnamefont
  {Peng}}, \bibinfo {author} {\bibfnamefont {M.~J.}\ \bibnamefont
  {Ramsey-Musolf}}, \ and\ \bibinfo {author} {\bibfnamefont {P.}~\bibnamefont
  {Winslow}},\ }\href {\doibase 10.1103/PhysRevD.93.093002} {\bibfield
  {journal} {\bibinfo  {journal} {Phys. Rev. D}\ }\textbf {\bibinfo {volume}
  {93}},\ \bibinfo {pages} {093002} (\bibinfo {year} {2016})},\ \Eprint
  {http://arxiv.org/abs/1508.04444} {arXiv:1508.04444 [hep-ph]} \BibitemShut
  {NoStop}%
\bibitem [{\citenamefont {Engel}\ and\ \citenamefont
  {Men\'endez}(2017)}]{Engel:2016xgb}%
  \BibitemOpen
  \bibfield  {author} {\bibinfo {author} {\bibfnamefont {J.}~\bibnamefont
  {Engel}}\ and\ \bibinfo {author} {\bibfnamefont {J.}~\bibnamefont
  {Men\'endez}},\ }\href {\doibase 10.1088/1361-6633/aa5bc5} {\bibfield
  {journal} {\bibinfo  {journal} {Rept. Prog. Phys.}\ }\textbf {\bibinfo
  {volume} {80}},\ \bibinfo {pages} {046301} (\bibinfo {year} {2017})},\
  \Eprint {http://arxiv.org/abs/1610.06548} {arXiv:1610.06548 [nucl-th]}
  \BibitemShut {NoStop}%
\bibitem [{\citenamefont {Cirigliano}\ \emph {et~al.}(2022)\citenamefont
  {Cirigliano} \emph {et~al.}}]{Cirigliano:2022oqy}%
  \BibitemOpen
  \bibfield  {author} {\bibinfo {author} {\bibfnamefont {V.}~\bibnamefont
  {Cirigliano}} \emph {et~al.}\ }(\bibinfo {year} {2022})\ \Eprint
  {http://arxiv.org/abs/2203.12169} {arXiv:2203.12169 [hep-ph]} \BibitemShut
  {NoStop}%
\bibitem [{\citenamefont {Prezeau}\ \emph {et~al.}(2003)\citenamefont
  {Prezeau}, \citenamefont {Ramsey-Musolf},\ and\ \citenamefont
  {Vogel}}]{Prezeau:2003xn}%
  \BibitemOpen
  \bibfield  {author} {\bibinfo {author} {\bibfnamefont {G.}~\bibnamefont
  {Prezeau}}, \bibinfo {author} {\bibfnamefont {M.}~\bibnamefont
  {Ramsey-Musolf}}, \ and\ \bibinfo {author} {\bibfnamefont {P.}~\bibnamefont
  {Vogel}},\ }\href {\doibase 10.1103/PhysRevD.68.034016} {\bibfield  {journal}
  {\bibinfo  {journal} {Phys. Rev.}\ }\textbf {\bibinfo {volume} {D68}},\
  \bibinfo {pages} {034016} (\bibinfo {year} {2003})},\ \Eprint
  {http://arxiv.org/abs/hep-ph/0303205} {arXiv:hep-ph/0303205 [hep-ph]}
  \BibitemShut {NoStop}%
\bibitem [{\citenamefont {Cirigliano}\ \emph
  {et~al.}(2018{\natexlab{a}})\citenamefont {Cirigliano}, \citenamefont
  {Dekens}, \citenamefont {Mereghetti},\ and\ \citenamefont
  {Walker-Loud}}]{Cirigliano:2017tvr}%
  \BibitemOpen
  \bibfield  {author} {\bibinfo {author} {\bibfnamefont {V.}~\bibnamefont
  {Cirigliano}}, \bibinfo {author} {\bibfnamefont {W.}~\bibnamefont {Dekens}},
  \bibinfo {author} {\bibfnamefont {E.}~\bibnamefont {Mereghetti}}, \ and\
  \bibinfo {author} {\bibfnamefont {A.}~\bibnamefont {Walker-Loud}},\ }\href
  {\doibase 10.1103/PhysRevC.97.065501} {\bibfield  {journal} {\bibinfo
  {journal} {Phys. Rev. C}\ }\textbf {\bibinfo {volume} {97}},\ \bibinfo
  {pages} {065501} (\bibinfo {year} {2018}{\natexlab{a}})},\ \bibinfo {note}
  {[Erratum: Phys.Rev.C 100, 019903 (2019)]},\ \Eprint
  {http://arxiv.org/abs/1710.01729} {arXiv:1710.01729 [hep-ph]} \BibitemShut
  {NoStop}%
\bibitem [{\citenamefont {Cirigliano}\ \emph
  {et~al.}(2018{\natexlab{b}})\citenamefont {Cirigliano}, \citenamefont
  {Dekens}, \citenamefont {De~Vries}, \citenamefont {Graesser}, \citenamefont
  {Mereghetti}, \citenamefont {Pastore},\ and\ \citenamefont
  {Van~Kolck}}]{Cirigliano:2018hja}%
  \BibitemOpen
  \bibfield  {author} {\bibinfo {author} {\bibfnamefont {V.}~\bibnamefont
  {Cirigliano}}, \bibinfo {author} {\bibfnamefont {W.}~\bibnamefont {Dekens}},
  \bibinfo {author} {\bibfnamefont {J.}~\bibnamefont {De~Vries}}, \bibinfo
  {author} {\bibfnamefont {M.~L.}\ \bibnamefont {Graesser}}, \bibinfo {author}
  {\bibfnamefont {E.}~\bibnamefont {Mereghetti}}, \bibinfo {author}
  {\bibfnamefont {S.}~\bibnamefont {Pastore}}, \ and\ \bibinfo {author}
  {\bibfnamefont {U.}~\bibnamefont {Van~Kolck}},\ }\href {\doibase
  10.1103/PhysRevLett.120.202001} {\bibfield  {journal} {\bibinfo  {journal}
  {Phys. Rev. Lett.}\ }\textbf {\bibinfo {volume} {120}},\ \bibinfo {pages}
  {202001} (\bibinfo {year} {2018}{\natexlab{b}})},\ \Eprint
  {http://arxiv.org/abs/1802.10097} {arXiv:1802.10097 [hep-ph]} \BibitemShut
  {NoStop}%
\bibitem [{\citenamefont {Cirigliano}\ \emph {et~al.}(2017)\citenamefont
  {Cirigliano}, \citenamefont {Dekens}, \citenamefont {de~Vries}, \citenamefont
  {Graesser},\ and\ \citenamefont {Mereghetti}}]{Cirigliano:2017djv}%
  \BibitemOpen
  \bibfield  {author} {\bibinfo {author} {\bibfnamefont {V.}~\bibnamefont
  {Cirigliano}}, \bibinfo {author} {\bibfnamefont {W.}~\bibnamefont {Dekens}},
  \bibinfo {author} {\bibfnamefont {J.}~\bibnamefont {de~Vries}}, \bibinfo
  {author} {\bibfnamefont {M.~L.}\ \bibnamefont {Graesser}}, \ and\ \bibinfo
  {author} {\bibfnamefont {E.}~\bibnamefont {Mereghetti}},\ }\href {\doibase
  10.1007/JHEP12(2017)082} {\bibfield  {journal} {\bibinfo  {journal} {JHEP}\
  }\textbf {\bibinfo {volume} {12}},\ \bibinfo {pages} {082} (\bibinfo {year}
  {2017})},\ \Eprint {http://arxiv.org/abs/1708.09390} {arXiv:1708.09390
  [hep-ph]} \BibitemShut {NoStop}%
\bibitem [{\citenamefont {Cirigliano}\ \emph
  {et~al.}(2018{\natexlab{c}})\citenamefont {Cirigliano}, \citenamefont
  {Dekens}, \citenamefont {de~Vries}, \citenamefont {Graesser},\ and\
  \citenamefont {Mereghetti}}]{Cirigliano:2018yza}%
  \BibitemOpen
  \bibfield  {author} {\bibinfo {author} {\bibfnamefont {V.}~\bibnamefont
  {Cirigliano}}, \bibinfo {author} {\bibfnamefont {W.}~\bibnamefont {Dekens}},
  \bibinfo {author} {\bibfnamefont {J.}~\bibnamefont {de~Vries}}, \bibinfo
  {author} {\bibfnamefont {M.}~\bibnamefont {Graesser}}, \ and\ \bibinfo
  {author} {\bibfnamefont {E.}~\bibnamefont {Mereghetti}},\ }\href {\doibase
  10.1007/JHEP12(2018)097} {\bibfield  {journal} {\bibinfo  {journal} {JHEP}\
  }\textbf {\bibinfo {volume} {12}},\ \bibinfo {pages} {097} (\bibinfo {year}
  {2018}{\natexlab{c}})},\ \Eprint {http://arxiv.org/abs/1806.02780}
  {arXiv:1806.02780 [hep-ph]} \BibitemShut {NoStop}%
\bibitem [{\citenamefont {Cirigliano}\ \emph {et~al.}(2019)\citenamefont
  {Cirigliano}, \citenamefont {Dekens}, \citenamefont {De~Vries}, \citenamefont
  {Graesser}, \citenamefont {Mereghetti}, \citenamefont {Pastore},
  \citenamefont {Piarulli}, \citenamefont {Van~Kolck},\ and\ \citenamefont
  {Wiringa}}]{Cirigliano:2019vdj}%
  \BibitemOpen
  \bibfield  {author} {\bibinfo {author} {\bibfnamefont {V.}~\bibnamefont
  {Cirigliano}}, \bibinfo {author} {\bibfnamefont {W.}~\bibnamefont {Dekens}},
  \bibinfo {author} {\bibfnamefont {J.}~\bibnamefont {De~Vries}}, \bibinfo
  {author} {\bibfnamefont {M.}~\bibnamefont {Graesser}}, \bibinfo {author}
  {\bibfnamefont {E.}~\bibnamefont {Mereghetti}}, \bibinfo {author}
  {\bibfnamefont {S.}~\bibnamefont {Pastore}}, \bibinfo {author} {\bibfnamefont
  {M.}~\bibnamefont {Piarulli}}, \bibinfo {author} {\bibfnamefont
  {U.}~\bibnamefont {Van~Kolck}}, \ and\ \bibinfo {author} {\bibfnamefont
  {R.}~\bibnamefont {Wiringa}},\ }\href {\doibase 10.1103/PhysRevC.100.055504}
  {\bibfield  {journal} {\bibinfo  {journal} {Phys. Rev. C}\ }\textbf {\bibinfo
  {volume} {100}},\ \bibinfo {pages} {055504} (\bibinfo {year} {2019})},\
  \Eprint {http://arxiv.org/abs/1907.11254} {arXiv:1907.11254 [nucl-th]}
  \BibitemShut {NoStop}%
\bibitem [{\citenamefont {Dekens}\ \emph {et~al.}(2020)\citenamefont {Dekens},
  \citenamefont {de~Vries}, \citenamefont {Fuyuto}, \citenamefont
  {Mereghetti},\ and\ \citenamefont {Zhou}}]{Dekens:2020ttz}%
  \BibitemOpen
  \bibfield  {author} {\bibinfo {author} {\bibfnamefont {W.}~\bibnamefont
  {Dekens}}, \bibinfo {author} {\bibfnamefont {J.}~\bibnamefont {de~Vries}},
  \bibinfo {author} {\bibfnamefont {K.}~\bibnamefont {Fuyuto}}, \bibinfo
  {author} {\bibfnamefont {E.}~\bibnamefont {Mereghetti}}, \ and\ \bibinfo
  {author} {\bibfnamefont {G.}~\bibnamefont {Zhou}},\ }\href {\doibase
  10.1007/JHEP06(2020)097} {\bibfield  {journal} {\bibinfo  {journal} {JHEP}\
  }\textbf {\bibinfo {volume} {06}},\ \bibinfo {pages} {097} (\bibinfo {year}
  {2020})},\ \Eprint {http://arxiv.org/abs/2002.07182} {arXiv:2002.07182
  [hep-ph]} \BibitemShut {NoStop}%
\bibitem [{\citenamefont {Doi}\ \emph {et~al.}(1985)\citenamefont {Doi},
  \citenamefont {Kotani},\ and\ \citenamefont {Takasugi}}]{Doi:1985dx}%
  \BibitemOpen
  \bibfield  {author} {\bibinfo {author} {\bibfnamefont {M.}~\bibnamefont
  {Doi}}, \bibinfo {author} {\bibfnamefont {T.}~\bibnamefont {Kotani}}, \ and\
  \bibinfo {author} {\bibfnamefont {E.}~\bibnamefont {Takasugi}},\ }\href
  {\doibase 10.1143/PTPS.83.1} {\bibfield  {journal} {\bibinfo  {journal}
  {Prog. Theor. Phys. Suppl.}\ }\textbf {\bibinfo {volume} {83}},\ \bibinfo
  {pages} {1} (\bibinfo {year} {1985})}\BibitemShut {NoStop}%
\bibitem [{\citenamefont {Pas}\ \emph {et~al.}(1999)\citenamefont {Pas},
  \citenamefont {Hirsch}, \citenamefont {Klapdor-Kleingrothaus},\ and\
  \citenamefont {Kovalenko}}]{Pas:1999fc}%
  \BibitemOpen
  \bibfield  {author} {\bibinfo {author} {\bibfnamefont {H.}~\bibnamefont
  {Pas}}, \bibinfo {author} {\bibfnamefont {M.}~\bibnamefont {Hirsch}},
  \bibinfo {author} {\bibfnamefont {H.~V.}\ \bibnamefont
  {Klapdor-Kleingrothaus}}, \ and\ \bibinfo {author} {\bibfnamefont {S.~G.}\
  \bibnamefont {Kovalenko}},\ }\href {\doibase 10.1016/S0370-2693(99)00330-5}
  {\bibfield  {journal} {\bibinfo  {journal} {Phys. Lett.}\ }\textbf {\bibinfo
  {volume} {B453}},\ \bibinfo {pages} {194} (\bibinfo {year} {1999})},\
  \bibinfo {note} {[,393(1999)]}\BibitemShut {NoStop}%
\bibitem [{\citenamefont {Pas}\ \emph {et~al.}(2001)\citenamefont {Pas},
  \citenamefont {Hirsch}, \citenamefont {Klapdor-Kleingrothaus},\ and\
  \citenamefont {Kovalenko}}]{Pas:2000vn}%
  \BibitemOpen
  \bibfield  {author} {\bibinfo {author} {\bibfnamefont {H.}~\bibnamefont
  {Pas}}, \bibinfo {author} {\bibfnamefont {M.}~\bibnamefont {Hirsch}},
  \bibinfo {author} {\bibfnamefont {H.~V.}\ \bibnamefont
  {Klapdor-Kleingrothaus}}, \ and\ \bibinfo {author} {\bibfnamefont {S.~G.}\
  \bibnamefont {Kovalenko}},\ }\href {\doibase 10.1016/S0370-2693(00)01359-9}
  {\bibfield  {journal} {\bibinfo  {journal} {Phys. Lett.}\ }\textbf {\bibinfo
  {volume} {B498}},\ \bibinfo {pages} {35} (\bibinfo {year} {2001})},\ \Eprint
  {http://arxiv.org/abs/hep-ph/0008182} {arXiv:hep-ph/0008182 [hep-ph]}
  \BibitemShut {NoStop}%
\bibitem [{\citenamefont {Nicholson}\ \emph {et~al.}(2018)\citenamefont
  {Nicholson} \emph {et~al.}}]{Nicholson:2018mwc}%
  \BibitemOpen
  \bibfield  {author} {\bibinfo {author} {\bibfnamefont {A.}~\bibnamefont
  {Nicholson}} \emph {et~al.},\ }\href {\doibase
  10.1103/PhysRevLett.121.172501} {\bibfield  {journal} {\bibinfo  {journal}
  {Phys. Rev. Lett.}\ }\textbf {\bibinfo {volume} {121}},\ \bibinfo {pages}
  {172501} (\bibinfo {year} {2018})},\ \Eprint
  {http://arxiv.org/abs/1805.02634} {arXiv:1805.02634 [nucl-th]} \BibitemShut
  {NoStop}%
\bibitem [{\citenamefont {Feng}\ \emph {et~al.}(2019)\citenamefont {Feng},
  \citenamefont {Jin}, \citenamefont {Tuo},\ and\ \citenamefont
  {Xia}}]{Feng:2018pdq}%
  \BibitemOpen
  \bibfield  {author} {\bibinfo {author} {\bibfnamefont {X.}~\bibnamefont
  {Feng}}, \bibinfo {author} {\bibfnamefont {L.-C.}\ \bibnamefont {Jin}},
  \bibinfo {author} {\bibfnamefont {X.-Y.}\ \bibnamefont {Tuo}}, \ and\
  \bibinfo {author} {\bibfnamefont {S.-C.}\ \bibnamefont {Xia}},\ }\href
  {\doibase 10.1103/PhysRevLett.122.022001} {\bibfield  {journal} {\bibinfo
  {journal} {Phys. Rev. Lett.}\ }\textbf {\bibinfo {volume} {122}},\ \bibinfo
  {pages} {022001} (\bibinfo {year} {2019})},\ \Eprint
  {http://arxiv.org/abs/1809.10511} {arXiv:1809.10511 [hep-lat]} \BibitemShut
  {NoStop}%
\bibitem [{\citenamefont {Tuo}\ \emph {et~al.}(2019)\citenamefont {Tuo},
  \citenamefont {Feng},\ and\ \citenamefont {Jin}}]{Tuo:2019bue}%
  \BibitemOpen
  \bibfield  {author} {\bibinfo {author} {\bibfnamefont {X.-Y.}\ \bibnamefont
  {Tuo}}, \bibinfo {author} {\bibfnamefont {X.}~\bibnamefont {Feng}}, \ and\
  \bibinfo {author} {\bibfnamefont {L.-C.}\ \bibnamefont {Jin}},\ }\href
  {\doibase 10.1103/PhysRevD.100.094511} {\bibfield  {journal} {\bibinfo
  {journal} {Phys. Rev. D}\ }\textbf {\bibinfo {volume} {100}},\ \bibinfo
  {pages} {094511} (\bibinfo {year} {2019})},\ \Eprint
  {http://arxiv.org/abs/1909.13525} {arXiv:1909.13525 [hep-lat]} \BibitemShut
  {NoStop}%
\bibitem [{\citenamefont {Detmold}\ and\ \citenamefont
  {Murphy}(2020)}]{Detmold:2020jqv}%
  \BibitemOpen
  \bibfield  {author} {\bibinfo {author} {\bibfnamefont {W.}~\bibnamefont
  {Detmold}}\ and\ \bibinfo {author} {\bibfnamefont {D.}~\bibnamefont {Murphy}}
  (\bibinfo {collaboration} {NPLQCD}),\ }\href@noop {} {\  (\bibinfo {year}
  {2020})},\ \Eprint {http://arxiv.org/abs/2004.07404} {arXiv:2004.07404
  [hep-lat]} \BibitemShut {NoStop}%
\bibitem [{\citenamefont {Briceño}\ and\ \citenamefont
  {Hansen}(2016)}]{Briceno:2015tza}%
  \BibitemOpen
  \bibfield  {author} {\bibinfo {author} {\bibfnamefont {R.~A.}\ \bibnamefont
  {Briceño}}\ and\ \bibinfo {author} {\bibfnamefont {M.~T.}\ \bibnamefont
  {Hansen}},\ }\href {\doibase 10.1103/PhysRevD.94.013008} {\bibfield
  {journal} {\bibinfo  {journal} {Phys. Rev. D}\ }\textbf {\bibinfo {volume}
  {94}},\ \bibinfo {pages} {013008} (\bibinfo {year} {2016})},\ \Eprint
  {http://arxiv.org/abs/1509.08507} {arXiv:1509.08507 [hep-lat]} \BibitemShut
  {NoStop}%
\bibitem [{\citenamefont {Feng}\ \emph {et~al.}(2021)\citenamefont {Feng},
  \citenamefont {Jin}, \citenamefont {Wang},\ and\ \citenamefont
  {Zhang}}]{Feng:2020nqj}%
  \BibitemOpen
  \bibfield  {author} {\bibinfo {author} {\bibfnamefont {X.}~\bibnamefont
  {Feng}}, \bibinfo {author} {\bibfnamefont {L.-C.}\ \bibnamefont {Jin}},
  \bibinfo {author} {\bibfnamefont {Z.-Y.}\ \bibnamefont {Wang}}, \ and\
  \bibinfo {author} {\bibfnamefont {Z.}~\bibnamefont {Zhang}},\ }\href
  {\doibase 10.1103/PhysRevD.103.034508} {\bibfield  {journal} {\bibinfo
  {journal} {Phys. Rev. D}\ }\textbf {\bibinfo {volume} {103}},\ \bibinfo
  {pages} {034508} (\bibinfo {year} {2021})},\ \Eprint
  {http://arxiv.org/abs/2005.01956} {arXiv:2005.01956 [hep-lat]} \BibitemShut
  {NoStop}%
\bibitem [{\citenamefont {Davoudi}\ and\ \citenamefont
  {Kadam}(2020)}]{Davoudi:2020xdv}%
  \BibitemOpen
  \bibfield  {author} {\bibinfo {author} {\bibfnamefont {Z.}~\bibnamefont
  {Davoudi}}\ and\ \bibinfo {author} {\bibfnamefont {S.~V.}\ \bibnamefont
  {Kadam}},\ }\href {\doibase 10.1103/PhysRevD.102.114521} {\bibfield
  {journal} {\bibinfo  {journal} {Phys. Rev. D}\ }\textbf {\bibinfo {volume}
  {102}},\ \bibinfo {pages} {114521} (\bibinfo {year} {2020})},\ \Eprint
  {http://arxiv.org/abs/2007.15542} {arXiv:2007.15542 [hep-lat]} \BibitemShut
  {NoStop}%
\bibitem [{\citenamefont {Pastore}\ \emph {et~al.}(2018)\citenamefont
  {Pastore}, \citenamefont {Carlson}, \citenamefont {Cirigliano}, \citenamefont
  {Dekens}, \citenamefont {Mereghetti},\ and\ \citenamefont
  {Wiringa}}]{Pastore:2017ofx}%
  \BibitemOpen
  \bibfield  {author} {\bibinfo {author} {\bibfnamefont {S.}~\bibnamefont
  {Pastore}}, \bibinfo {author} {\bibfnamefont {J.}~\bibnamefont {Carlson}},
  \bibinfo {author} {\bibfnamefont {V.}~\bibnamefont {Cirigliano}}, \bibinfo
  {author} {\bibfnamefont {W.}~\bibnamefont {Dekens}}, \bibinfo {author}
  {\bibfnamefont {E.}~\bibnamefont {Mereghetti}}, \ and\ \bibinfo {author}
  {\bibfnamefont {R.~B.}\ \bibnamefont {Wiringa}},\ }\href {\doibase
  10.1103/PhysRevC.97.014606} {\bibfield  {journal} {\bibinfo  {journal} {Phys.
  Rev.}\ }\textbf {\bibinfo {volume} {C97}},\ \bibinfo {pages} {014606}
  (\bibinfo {year} {2018})},\ \Eprint {http://arxiv.org/abs/1710.05026}
  {arXiv:1710.05026 [nucl-th]} \BibitemShut {NoStop}%
\bibitem [{\citenamefont {Basili}\ \emph {et~al.}(2020)\citenamefont {Basili},
  \citenamefont {Yao}, \citenamefont {Engel}, \citenamefont {Hergert},
  \citenamefont {Lockner}, \citenamefont {Maris},\ and\ \citenamefont
  {Vary}}]{Basili:2019gvn}%
  \BibitemOpen
  \bibfield  {author} {\bibinfo {author} {\bibfnamefont {R.~A.~M.}\
  \bibnamefont {Basili}}, \bibinfo {author} {\bibfnamefont {J.~M.}\
  \bibnamefont {Yao}}, \bibinfo {author} {\bibfnamefont {J.}~\bibnamefont
  {Engel}}, \bibinfo {author} {\bibfnamefont {H.}~\bibnamefont {Hergert}},
  \bibinfo {author} {\bibfnamefont {M.}~\bibnamefont {Lockner}}, \bibinfo
  {author} {\bibfnamefont {P.}~\bibnamefont {Maris}}, \ and\ \bibinfo {author}
  {\bibfnamefont {J.~P.}\ \bibnamefont {Vary}},\ }\href {\doibase
  10.1103/PhysRevC.102.014302} {\bibfield  {journal} {\bibinfo  {journal}
  {Phys. Rev. C}\ }\textbf {\bibinfo {volume} {102}},\ \bibinfo {pages}
  {014302} (\bibinfo {year} {2020})},\ \Eprint
  {http://arxiv.org/abs/1909.06501} {arXiv:1909.06501 [nucl-th]} \BibitemShut
  {NoStop}%
\bibitem [{\citenamefont {Yao}\ \emph {et~al.}(2020)\citenamefont {Yao},
  \citenamefont {Bally}, \citenamefont {Engel}, \citenamefont {Wirth},
  \citenamefont {Rodríguez},\ and\ \citenamefont {Hergert}}]{Yao:2019rck}%
  \BibitemOpen
  \bibfield  {author} {\bibinfo {author} {\bibfnamefont {J.}~\bibnamefont
  {Yao}}, \bibinfo {author} {\bibfnamefont {B.}~\bibnamefont {Bally}}, \bibinfo
  {author} {\bibfnamefont {J.}~\bibnamefont {Engel}}, \bibinfo {author}
  {\bibfnamefont {R.}~\bibnamefont {Wirth}}, \bibinfo {author} {\bibfnamefont
  {T.}~\bibnamefont {Rodríguez}}, \ and\ \bibinfo {author} {\bibfnamefont
  {H.}~\bibnamefont {Hergert}},\ }\href {\doibase
  10.1103/PhysRevLett.124.232501} {\bibfield  {journal} {\bibinfo  {journal}
  {Phys. Rev. Lett.}\ }\textbf {\bibinfo {volume} {124}},\ \bibinfo {pages}
  {232501} (\bibinfo {year} {2020})},\ \Eprint
  {http://arxiv.org/abs/1908.05424} {arXiv:1908.05424 [nucl-th]} \BibitemShut
  {NoStop}%
\bibitem [{\citenamefont {Belley}\ \emph {et~al.}(2021)\citenamefont {Belley},
  \citenamefont {Payne}, \citenamefont {Stroberg}, \citenamefont {Miyagi},\
  and\ \citenamefont {Holt}}]{Belley:2020ejd}%
  \BibitemOpen
  \bibfield  {author} {\bibinfo {author} {\bibfnamefont {A.}~\bibnamefont
  {Belley}}, \bibinfo {author} {\bibfnamefont {C.~G.}\ \bibnamefont {Payne}},
  \bibinfo {author} {\bibfnamefont {S.~R.}\ \bibnamefont {Stroberg}}, \bibinfo
  {author} {\bibfnamefont {T.}~\bibnamefont {Miyagi}}, \ and\ \bibinfo {author}
  {\bibfnamefont {J.~D.}\ \bibnamefont {Holt}},\ }\href {\doibase
  10.1103/PhysRevLett.126.042502} {\bibfield  {journal} {\bibinfo  {journal}
  {Phys. Rev. Lett.}\ }\textbf {\bibinfo {volume} {126}},\ \bibinfo {pages}
  {042502} (\bibinfo {year} {2021})},\ \Eprint
  {http://arxiv.org/abs/2008.06588} {arXiv:2008.06588 [nucl-th]} \BibitemShut
  {NoStop}%
\bibitem [{\citenamefont {Wirth}\ \emph {et~al.}(2021)\citenamefont {Wirth},
  \citenamefont {Yao},\ and\ \citenamefont {Hergert}}]{Wirth:2021pij}%
  \BibitemOpen
  \bibfield  {author} {\bibinfo {author} {\bibfnamefont {R.}~\bibnamefont
  {Wirth}}, \bibinfo {author} {\bibfnamefont {J.~M.}\ \bibnamefont {Yao}}, \
  and\ \bibinfo {author} {\bibfnamefont {H.}~\bibnamefont {Hergert}},\ }\href
  {\doibase 10.1103/PhysRevLett.127.242502} {\bibfield  {journal} {\bibinfo
  {journal} {Phys. Rev. Lett.}\ }\textbf {\bibinfo {volume} {127}},\ \bibinfo
  {pages} {242502} (\bibinfo {year} {2021})},\ \Eprint
  {http://arxiv.org/abs/2105.05415} {arXiv:2105.05415 [nucl-th]} \BibitemShut
  {NoStop}%
\bibitem [{\citenamefont {Cirigliano}\ \emph
  {et~al.}(2021{\natexlab{a}})\citenamefont {Cirigliano}, \citenamefont
  {Dekens}, \citenamefont {de~Vries}, \citenamefont {Hoferichter},\ and\
  \citenamefont {Mereghetti}}]{Cirigliano:2020dmx}%
  \BibitemOpen
  \bibfield  {author} {\bibinfo {author} {\bibfnamefont {V.}~\bibnamefont
  {Cirigliano}}, \bibinfo {author} {\bibfnamefont {W.}~\bibnamefont {Dekens}},
  \bibinfo {author} {\bibfnamefont {J.}~\bibnamefont {de~Vries}}, \bibinfo
  {author} {\bibfnamefont {M.}~\bibnamefont {Hoferichter}}, \ and\ \bibinfo
  {author} {\bibfnamefont {E.}~\bibnamefont {Mereghetti}},\ }\href {\doibase
  10.1103/PhysRevLett.126.172002} {\bibfield  {journal} {\bibinfo  {journal}
  {Phys. Rev. Lett.}\ }\textbf {\bibinfo {volume} {126}},\ \bibinfo {pages}
  {172002} (\bibinfo {year} {2021}{\natexlab{a}})},\ \Eprint
  {http://arxiv.org/abs/2012.11602} {arXiv:2012.11602 [nucl-th]} \BibitemShut
  {NoStop}%
\bibitem [{\citenamefont {Cirigliano}\ \emph
  {et~al.}(2021{\natexlab{b}})\citenamefont {Cirigliano}, \citenamefont
  {Dekens}, \citenamefont {de~Vries}, \citenamefont {Hoferichter},\ and\
  \citenamefont {Mereghetti}}]{Cirigliano:2021qko}%
  \BibitemOpen
  \bibfield  {author} {\bibinfo {author} {\bibfnamefont {V.}~\bibnamefont
  {Cirigliano}}, \bibinfo {author} {\bibfnamefont {W.}~\bibnamefont {Dekens}},
  \bibinfo {author} {\bibfnamefont {J.}~\bibnamefont {de~Vries}}, \bibinfo
  {author} {\bibfnamefont {M.}~\bibnamefont {Hoferichter}}, \ and\ \bibinfo
  {author} {\bibfnamefont {E.}~\bibnamefont {Mereghetti}},\ }\href {\doibase
  10.1007/JHEP05(2021)289} {\bibfield  {journal} {\bibinfo  {journal} {JHEP}\
  }\textbf {\bibinfo {volume} {05}},\ \bibinfo {pages} {289} (\bibinfo {year}
  {2021}{\natexlab{b}})},\ \Eprint {http://arxiv.org/abs/2102.03371}
  {arXiv:2102.03371 [nucl-th]} \BibitemShut {NoStop}%
\bibitem [{\citenamefont {Richardson}\ \emph {et~al.}(2021)\citenamefont
  {Richardson}, \citenamefont {Schindler}, \citenamefont {Pastore},\ and\
  \citenamefont {Springer}}]{Richardson:2021xiu}%
  \BibitemOpen
  \bibfield  {author} {\bibinfo {author} {\bibfnamefont {T.~R.}\ \bibnamefont
  {Richardson}}, \bibinfo {author} {\bibfnamefont {M.~R.}\ \bibnamefont
  {Schindler}}, \bibinfo {author} {\bibfnamefont {S.}~\bibnamefont {Pastore}},
  \ and\ \bibinfo {author} {\bibfnamefont {R.~P.}\ \bibnamefont {Springer}},\
  }\href {\doibase 10.1103/PhysRevC.103.055501} {\bibfield  {journal} {\bibinfo
   {journal} {Phys. Rev. C}\ }\textbf {\bibinfo {volume} {103}},\ \bibinfo
  {pages} {055501} (\bibinfo {year} {2021})},\ \Eprint
  {http://arxiv.org/abs/2102.02184} {arXiv:2102.02184 [nucl-th]} \BibitemShut
  {NoStop}%
\bibitem [{\citenamefont {Cirigliano}\ \emph {et~al.}(2020)\citenamefont
  {Cirigliano}, \citenamefont {Detmold}, \citenamefont {Nicholson},\ and\
  \citenamefont {Shanahan}}]{Cirigliano:2020yhp}%
  \BibitemOpen
  \bibfield  {author} {\bibinfo {author} {\bibfnamefont {V.}~\bibnamefont
  {Cirigliano}}, \bibinfo {author} {\bibfnamefont {W.}~\bibnamefont {Detmold}},
  \bibinfo {author} {\bibfnamefont {A.}~\bibnamefont {Nicholson}}, \ and\
  \bibinfo {author} {\bibfnamefont {P.}~\bibnamefont {Shanahan}},\ }\href
  {\doibase 10.1016/j.ppnp.2020.103771} {\  (\bibinfo {year} {2020}),\
  10.1016/j.ppnp.2020.103771},\ \Eprint {http://arxiv.org/abs/2003.08493}
  {arXiv:2003.08493 [nucl-th]} \BibitemShut {NoStop}%
\bibitem [{\citenamefont {Davoudi}\ \emph {et~al.}(2021)\citenamefont
  {Davoudi}, \citenamefont {Detmold}, \citenamefont {Orginos}, \citenamefont
  {Parre\~no}, \citenamefont {Savage}, \citenamefont {Shanahan},\ and\
  \citenamefont {Wagman}}]{Davoudi:2020ngi}%
  \BibitemOpen
  \bibfield  {author} {\bibinfo {author} {\bibfnamefont {Z.}~\bibnamefont
  {Davoudi}}, \bibinfo {author} {\bibfnamefont {W.}~\bibnamefont {Detmold}},
  \bibinfo {author} {\bibfnamefont {K.}~\bibnamefont {Orginos}}, \bibinfo
  {author} {\bibfnamefont {A.}~\bibnamefont {Parre\~no}}, \bibinfo {author}
  {\bibfnamefont {M.~J.}\ \bibnamefont {Savage}}, \bibinfo {author}
  {\bibfnamefont {P.}~\bibnamefont {Shanahan}}, \ and\ \bibinfo {author}
  {\bibfnamefont {M.~L.}\ \bibnamefont {Wagman}},\ }\href {\doibase
  10.1016/j.physrep.2020.10.004} {\bibfield  {journal} {\bibinfo  {journal}
  {Phys. Rept.}\ }\textbf {\bibinfo {volume} {900}},\ \bibinfo {pages} {1}
  (\bibinfo {year} {2021})},\ \Eprint {http://arxiv.org/abs/2008.11160}
  {arXiv:2008.11160 [hep-lat]} \BibitemShut {NoStop}%
\bibitem [{\citenamefont {Dolinski}\ \emph {et~al.}(2019)\citenamefont
  {Dolinski}, \citenamefont {Poon},\ and\ \citenamefont
  {Rodejohann}}]{Dolinski:2019nrj}%
  \BibitemOpen
  \bibfield  {author} {\bibinfo {author} {\bibfnamefont {M.~J.}\ \bibnamefont
  {Dolinski}}, \bibinfo {author} {\bibfnamefont {A.~W.~P.}\ \bibnamefont
  {Poon}}, \ and\ \bibinfo {author} {\bibfnamefont {W.}~\bibnamefont
  {Rodejohann}},\ }\href {\doibase 10.1146/annurev-nucl-101918-023407}
  {\bibfield  {journal} {\bibinfo  {journal} {Ann. Rev. Nucl. Part. Sci.}\
  }\textbf {\bibinfo {volume} {69}},\ \bibinfo {pages} {219} (\bibinfo {year}
  {2019})},\ \Eprint {http://arxiv.org/abs/1902.04097} {arXiv:1902.04097
  [nucl-ex]} \BibitemShut {NoStop}%
\bibitem [{\citenamefont {Avasthi}\ \emph {et~al.}(2021)\citenamefont {Avasthi}
  \emph {et~al.}}]{avasthi2021kiloton}%
  \BibitemOpen
  \bibfield  {author} {\bibinfo {author} {\bibfnamefont {A.}~\bibnamefont
  {Avasthi}} \emph {et~al.},\ }\href {\doibase 10.1103/PhysRevD.104.112007}
  {\bibfield  {journal} {\bibinfo  {journal} {Phys. Rev. D}\ }\textbf {\bibinfo
  {volume} {104}},\ \bibinfo {pages} {112007} (\bibinfo {year} {2021})},\
  \Eprint {http://arxiv.org/abs/2110.01537} {arXiv:2110.01537
  [physics.ins-det]} \BibitemShut {NoStop}%
\bibitem [{\citenamefont {Nygren}\ \emph {et~al.}(2018)\citenamefont {Nygren},
  \citenamefont {Jones}, \citenamefont {L{\'o}pez-March}, \citenamefont {Mei},
  \citenamefont {Psihas},\ and\ \citenamefont
  {Renner}}]{nygren2018neutrinoless}%
  \BibitemOpen
  \bibfield  {author} {\bibinfo {author} {\bibfnamefont {D.}~\bibnamefont
  {Nygren}}, \bibinfo {author} {\bibfnamefont {B.}~\bibnamefont {Jones}},
  \bibinfo {author} {\bibfnamefont {N.}~\bibnamefont {L{\'o}pez-March}},
  \bibinfo {author} {\bibfnamefont {Y.}~\bibnamefont {Mei}}, \bibinfo {author}
  {\bibfnamefont {F.}~\bibnamefont {Psihas}}, \ and\ \bibinfo {author}
  {\bibfnamefont {J.}~\bibnamefont {Renner}},\ }\href@noop {} {\bibfield
  {journal} {\bibinfo  {journal} {Journal of Instrumentation}\ }\textbf
  {\bibinfo {volume} {13}},\ \bibinfo {pages} {P03015} (\bibinfo {year}
  {2018})}\BibitemShut {NoStop}%
\bibitem [{\citenamefont {Askins}\ \emph {et~al.}(2020)\citenamefont {Askins}
  \emph {et~al.}}]{Theia:2019non}%
  \BibitemOpen
  \bibfield  {author} {\bibinfo {author} {\bibfnamefont {M.}~\bibnamefont
  {Askins}} \emph {et~al.} (\bibinfo {collaboration} {Theia}),\ }\href
  {\doibase 10.1140/epjc/s10052-020-7977-8} {\bibfield  {journal} {\bibinfo
  {journal} {Eur. Phys. J. C}\ }\textbf {\bibinfo {volume} {80}},\ \bibinfo
  {pages} {416} (\bibinfo {year} {2020})},\ \Eprint
  {http://arxiv.org/abs/1911.03501} {arXiv:1911.03501 [physics.ins-det]}
  \BibitemShut {NoStop}%
\bibitem [{\citenamefont {Ellis}\ \emph
  {et~al.}(2019{\natexlab{a}})\citenamefont {Ellis} \emph
  {et~al.}}]{EuropeanStrategyforParticlePhysicsPreparatoryGroup:2019qin}%
  \BibitemOpen
  \bibfield  {author} {\bibinfo {author} {\bibfnamefont {R.~K.}\ \bibnamefont
  {Ellis}} \emph {et~al.},\ }\href@noop {} {\  (\bibinfo {year}
  {2019}{\natexlab{a}})},\ \Eprint {http://arxiv.org/abs/1910.11775}
  {arXiv:1910.11775 [hep-ex]} \BibitemShut {NoStop}%
\bibitem [{Eur(2020)}]{EuropeanStrategyGroup:2020pow}%
  \BibitemOpen
  \href {\doibase 10.17181/ESU2020} {\emph {\bibinfo {title} {{2020 Update of
  the European Strategy for Particle Physics}}}}\ (\bibinfo  {publisher} {CERN
  Council},\ \bibinfo {address} {Geneva},\ \bibinfo {year} {2020})\BibitemShut
  {NoStop}%
\bibitem [{\citenamefont {Gluza}(2002)}]{Gluza:2002vs}%
  \BibitemOpen
  \bibfield  {author} {\bibinfo {author} {\bibfnamefont {J.}~\bibnamefont
  {Gluza}},\ }\href@noop {} {\bibfield  {journal} {\bibinfo  {journal} {Acta
  Phys. Polon. B}\ }\textbf {\bibinfo {volume} {33}},\ \bibinfo {pages} {1735}
  (\bibinfo {year} {2002})},\ \Eprint {http://arxiv.org/abs/hep-ph/0201002}
  {arXiv:hep-ph/0201002} \BibitemShut {NoStop}%
\bibitem [{\citenamefont {Han}\ and\ \citenamefont {Zhang}(2006)}]{Han:2006ip}%
  \BibitemOpen
  \bibfield  {author} {\bibinfo {author} {\bibfnamefont {T.}~\bibnamefont
  {Han}}\ and\ \bibinfo {author} {\bibfnamefont {B.}~\bibnamefont {Zhang}},\
  }\href {\doibase 10.1103/PhysRevLett.97.171804} {\bibfield  {journal}
  {\bibinfo  {journal} {Phys. Rev. Lett.}\ }\textbf {\bibinfo {volume} {97}},\
  \bibinfo {pages} {171804} (\bibinfo {year} {2006})},\ \Eprint
  {http://arxiv.org/abs/hep-ph/0604064} {arXiv:hep-ph/0604064} \BibitemShut
  {NoStop}%
\bibitem [{\citenamefont {Abada}\ \emph {et~al.}(2007)\citenamefont {Abada},
  \citenamefont {Biggio}, \citenamefont {Bonnet}, \citenamefont {Gavela},\ and\
  \citenamefont {Hambye}}]{Abada:2007ux}%
  \BibitemOpen
  \bibfield  {author} {\bibinfo {author} {\bibfnamefont {A.}~\bibnamefont
  {Abada}}, \bibinfo {author} {\bibfnamefont {C.}~\bibnamefont {Biggio}},
  \bibinfo {author} {\bibfnamefont {F.}~\bibnamefont {Bonnet}}, \bibinfo
  {author} {\bibfnamefont {M.~B.}\ \bibnamefont {Gavela}}, \ and\ \bibinfo
  {author} {\bibfnamefont {T.}~\bibnamefont {Hambye}},\ }\href {\doibase
  10.1088/1126-6708/2007/12/061} {\bibfield  {journal} {\bibinfo  {journal}
  {JHEP}\ }\textbf {\bibinfo {volume} {12}},\ \bibinfo {pages} {061} (\bibinfo
  {year} {2007})},\ \Eprint {http://arxiv.org/abs/0707.4058} {arXiv:0707.4058
  [hep-ph]} \BibitemShut {NoStop}%
\bibitem [{\citenamefont {del Aguila}\ and\ \citenamefont
  {Aguilar-Saavedra}(2009)}]{delAguila:2008cj}%
  \BibitemOpen
  \bibfield  {author} {\bibinfo {author} {\bibfnamefont {F.}~\bibnamefont {del
  Aguila}}\ and\ \bibinfo {author} {\bibfnamefont {J.~A.}\ \bibnamefont
  {Aguilar-Saavedra}},\ }\href {\doibase 10.1016/j.nuclphysb.2008.12.029}
  {\bibfield  {journal} {\bibinfo  {journal} {Nucl. Phys. B}\ }\textbf
  {\bibinfo {volume} {813}},\ \bibinfo {pages} {22} (\bibinfo {year} {2009})},\
  \Eprint {http://arxiv.org/abs/0808.2468} {arXiv:0808.2468 [hep-ph]}
  \BibitemShut {NoStop}%
\bibitem [{\citenamefont {Atre}\ \emph {et~al.}(2009)\citenamefont {Atre},
  \citenamefont {Han}, \citenamefont {Pascoli},\ and\ \citenamefont
  {Zhang}}]{Atre:2009rg}%
  \BibitemOpen
  \bibfield  {author} {\bibinfo {author} {\bibfnamefont {A.}~\bibnamefont
  {Atre}}, \bibinfo {author} {\bibfnamefont {T.}~\bibnamefont {Han}}, \bibinfo
  {author} {\bibfnamefont {S.}~\bibnamefont {Pascoli}}, \ and\ \bibinfo
  {author} {\bibfnamefont {B.}~\bibnamefont {Zhang}},\ }\href {\doibase
  10.1088/1126-6708/2009/05/030} {\bibfield  {journal} {\bibinfo  {journal}
  {JHEP}\ }\textbf {\bibinfo {volume} {05}},\ \bibinfo {pages} {030} (\bibinfo
  {year} {2009})},\ \Eprint {http://arxiv.org/abs/0901.3589} {arXiv:0901.3589
  [hep-ph]} \BibitemShut {NoStop}%
\bibitem [{\citenamefont {Tello}\ \emph {et~al.}(2011)\citenamefont {Tello},
  \citenamefont {Nemevsek}, \citenamefont {Nesti}, \citenamefont {Senjanovic},\
  and\ \citenamefont {Vissani}}]{Tello:2010am}%
  \BibitemOpen
  \bibfield  {author} {\bibinfo {author} {\bibfnamefont {V.}~\bibnamefont
  {Tello}}, \bibinfo {author} {\bibfnamefont {M.}~\bibnamefont {Nemevsek}},
  \bibinfo {author} {\bibfnamefont {F.}~\bibnamefont {Nesti}}, \bibinfo
  {author} {\bibfnamefont {G.}~\bibnamefont {Senjanovic}}, \ and\ \bibinfo
  {author} {\bibfnamefont {F.}~\bibnamefont {Vissani}},\ }\href {\doibase
  10.1103/PhysRevLett.106.151801} {\bibfield  {journal} {\bibinfo  {journal}
  {Phys. Rev. Lett.}\ }\textbf {\bibinfo {volume} {106}},\ \bibinfo {pages}
  {151801} (\bibinfo {year} {2011})},\ \Eprint {http://arxiv.org/abs/1011.3522}
  {arXiv:1011.3522 [hep-ph]} \BibitemShut {NoStop}%
\bibitem [{\citenamefont {Deppisch}\ \emph {et~al.}(2015)\citenamefont
  {Deppisch}, \citenamefont {Bhupal~Dev},\ and\ \citenamefont
  {Pilaftsis}}]{Deppisch:2015qwa}%
  \BibitemOpen
  \bibfield  {author} {\bibinfo {author} {\bibfnamefont {F.~F.}\ \bibnamefont
  {Deppisch}}, \bibinfo {author} {\bibfnamefont {P.~S.}\ \bibnamefont
  {Bhupal~Dev}}, \ and\ \bibinfo {author} {\bibfnamefont {A.}~\bibnamefont
  {Pilaftsis}},\ }\href {\doibase 10.1088/1367-2630/17/7/075019} {\bibfield
  {journal} {\bibinfo  {journal} {New J. Phys.}\ }\textbf {\bibinfo {volume}
  {17}},\ \bibinfo {pages} {075019} (\bibinfo {year} {2015})},\ \Eprint
  {http://arxiv.org/abs/1502.06541} {arXiv:1502.06541 [hep-ph]} \BibitemShut
  {NoStop}%
\bibitem [{\citenamefont {Cai}\ \emph {et~al.}(2017)\citenamefont {Cai},
  \citenamefont {Herrero-Garc\'\i{}a}, \citenamefont {Schmidt}, \citenamefont
  {Vicente},\ and\ \citenamefont {Volkas}}]{Cai:2017jrq}%
  \BibitemOpen
  \bibfield  {author} {\bibinfo {author} {\bibfnamefont {Y.}~\bibnamefont
  {Cai}}, \bibinfo {author} {\bibfnamefont {J.}~\bibnamefont
  {Herrero-Garc\'\i{}a}}, \bibinfo {author} {\bibfnamefont {M.~A.}\
  \bibnamefont {Schmidt}}, \bibinfo {author} {\bibfnamefont {A.}~\bibnamefont
  {Vicente}}, \ and\ \bibinfo {author} {\bibfnamefont {R.~R.}\ \bibnamefont
  {Volkas}},\ }\href {\doibase 10.3389/fphy.2017.00063} {\bibfield  {journal}
  {\bibinfo  {journal} {Front. in Phys.}\ }\textbf {\bibinfo {volume} {5}},\
  \bibinfo {pages} {63} (\bibinfo {year} {2017})},\ \Eprint
  {http://arxiv.org/abs/1706.08524} {arXiv:1706.08524 [hep-ph]} \BibitemShut
  {NoStop}%
\bibitem [{\citenamefont {Cai}\ \emph {et~al.}(2018)\citenamefont {Cai},
  \citenamefont {Han}, \citenamefont {Li},\ and\ \citenamefont
  {Ruiz}}]{Cai:2017mow}%
  \BibitemOpen
  \bibfield  {author} {\bibinfo {author} {\bibfnamefont {Y.}~\bibnamefont
  {Cai}}, \bibinfo {author} {\bibfnamefont {T.}~\bibnamefont {Han}}, \bibinfo
  {author} {\bibfnamefont {T.}~\bibnamefont {Li}}, \ and\ \bibinfo {author}
  {\bibfnamefont {R.}~\bibnamefont {Ruiz}},\ }\href {\doibase
  10.3389/fphy.2018.00040} {\bibfield  {journal} {\bibinfo  {journal} {Front.
  in Phys.}\ }\textbf {\bibinfo {volume} {6}},\ \bibinfo {pages} {40} (\bibinfo
  {year} {2018})},\ \Eprint {http://arxiv.org/abs/1711.02180} {arXiv:1711.02180
  [hep-ph]} \BibitemShut {NoStop}%
\bibitem [{\citenamefont {Pascoli}\ \emph {et~al.}(2019)\citenamefont
  {Pascoli}, \citenamefont {Ruiz},\ and\ \citenamefont
  {Weiland}}]{Pascoli:2018heg}%
  \BibitemOpen
  \bibfield  {author} {\bibinfo {author} {\bibfnamefont {S.}~\bibnamefont
  {Pascoli}}, \bibinfo {author} {\bibfnamefont {R.}~\bibnamefont {Ruiz}}, \
  and\ \bibinfo {author} {\bibfnamefont {C.}~\bibnamefont {Weiland}},\ }\href
  {\doibase 10.1007/JHEP06(2019)049} {\bibfield  {journal} {\bibinfo  {journal}
  {JHEP}\ }\textbf {\bibinfo {volume} {06}},\ \bibinfo {pages} {049} (\bibinfo
  {year} {2019})},\ \Eprint {http://arxiv.org/abs/1812.08750} {arXiv:1812.08750
  [hep-ph]} \BibitemShut {NoStop}%
\bibitem [{\citenamefont {Han}\ \emph {et~al.}(2022)\citenamefont {Han},
  \citenamefont {Liao}, \citenamefont {Liu}, \citenamefont {Marfatia},\ and\
  \citenamefont {Ruiz}}]{Han:2022qgg}%
  \BibitemOpen
  \bibfield  {author} {\bibinfo {author} {\bibfnamefont {T.}~\bibnamefont
  {Han}}, \bibinfo {author} {\bibfnamefont {J.}~\bibnamefont {Liao}}, \bibinfo
  {author} {\bibfnamefont {H.}~\bibnamefont {Liu}}, \bibinfo {author}
  {\bibfnamefont {D.}~\bibnamefont {Marfatia}}, \ and\ \bibinfo {author}
  {\bibfnamefont {R.}~\bibnamefont {Ruiz}},\ }in\ \href@noop {} {\emph
  {\bibinfo {booktitle} {{2022 Snowmass Summer Study}}}}\ (\bibinfo {year}
  {2022})\ \Eprint {http://arxiv.org/abs/2203.06131} {arXiv:2203.06131
  [hep-ph]} \BibitemShut {NoStop}%
\bibitem [{\citenamefont {Mandal}\ \emph {et~al.}(2022)\citenamefont {Mandal},
  \citenamefont {Miranda}, \citenamefont {Garcia}, \citenamefont {Valle},\ and\
  \citenamefont {Xu}}]{Mandal:2022zmy}%
  \BibitemOpen
  \bibfield  {author} {\bibinfo {author} {\bibfnamefont {S.}~\bibnamefont
  {Mandal}}, \bibinfo {author} {\bibfnamefont {O.~G.}\ \bibnamefont {Miranda}},
  \bibinfo {author} {\bibfnamefont {G.~S.}\ \bibnamefont {Garcia}}, \bibinfo
  {author} {\bibfnamefont {J.~W.~F.}\ \bibnamefont {Valle}}, \ and\ \bibinfo
  {author} {\bibfnamefont {X.-J.}\ \bibnamefont {Xu}},\ }\href@noop {} {\
  (\bibinfo {year} {2022})},\ \Eprint {http://arxiv.org/abs/2203.06362}
  {arXiv:2203.06362 [hep-ph]} \BibitemShut {NoStop}%
\bibitem [{\citenamefont {Abdullahi}\ \emph {et~al.}(2022)\citenamefont
  {Abdullahi} \emph {et~al.}}]{Abdullahi:2022jlv}%
  \BibitemOpen
  \bibfield  {author} {\bibinfo {author} {\bibfnamefont {A.~M.}\ \bibnamefont
  {Abdullahi}} \emph {et~al.},\ }in\ \href@noop {} {\emph {\bibinfo {booktitle}
  {{2022 Snowmass Summer Study}}}}\ (\bibinfo {year} {2022})\ \Eprint
  {http://arxiv.org/abs/2203.08039} {arXiv:2203.08039 [hep-ph]} \BibitemShut
  {NoStop}%
\bibitem [{\citenamefont {Chauhan}\ \emph {et~al.}(2022)\citenamefont
  {Chauhan}, \citenamefont {Dev}, \citenamefont {Dziewit}, \citenamefont
  {Flieger}, \citenamefont {Gluza}, \citenamefont {Grzanka}, \citenamefont
  {Karmakar}, \citenamefont {Vergeest},\ and\ \citenamefont
  {Zieba}}]{Chauhan:2022gkz}%
  \BibitemOpen
  \bibfield  {author} {\bibinfo {author} {\bibfnamefont {G.}~\bibnamefont
  {Chauhan}}, \bibinfo {author} {\bibfnamefont {P.~S.~B.}\ \bibnamefont {Dev}},
  \bibinfo {author} {\bibfnamefont {B.}~\bibnamefont {Dziewit}}, \bibinfo
  {author} {\bibfnamefont {W.}~\bibnamefont {Flieger}}, \bibinfo {author}
  {\bibfnamefont {J.}~\bibnamefont {Gluza}}, \bibinfo {author} {\bibfnamefont
  {K.}~\bibnamefont {Grzanka}}, \bibinfo {author} {\bibfnamefont
  {B.}~\bibnamefont {Karmakar}}, \bibinfo {author} {\bibfnamefont
  {J.}~\bibnamefont {Vergeest}}, \ and\ \bibinfo {author} {\bibfnamefont
  {S.}~\bibnamefont {Zieba}},\ }in\ \href@noop {} {\emph {\bibinfo {booktitle}
  {{2022 Snowmass Summer Study}}}}\ (\bibinfo {year} {2022})\ \Eprint
  {http://arxiv.org/abs/2203.08105} {arXiv:2203.08105 [hep-ph]} \BibitemShut
  {NoStop}%
\bibitem [{\citenamefont {Fileviez~Perez}\ and\ \citenamefont
  {Wise}(2009)}]{FileviezPerez:2009ud}%
  \BibitemOpen
  \bibfield  {author} {\bibinfo {author} {\bibfnamefont {P.}~\bibnamefont
  {Fileviez~Perez}}\ and\ \bibinfo {author} {\bibfnamefont {M.~B.}\
  \bibnamefont {Wise}},\ }\href {\doibase 10.1103/PhysRevD.80.053006}
  {\bibfield  {journal} {\bibinfo  {journal} {Phys. Rev. D}\ }\textbf {\bibinfo
  {volume} {80}},\ \bibinfo {pages} {053006} (\bibinfo {year} {2009})},\
  \Eprint {http://arxiv.org/abs/0906.2950} {arXiv:0906.2950 [hep-ph]}
  \BibitemShut {NoStop}%
\bibitem [{\citenamefont {Shrock}(1981)}]{Shrock:1980ct}%
  \BibitemOpen
  \bibfield  {author} {\bibinfo {author} {\bibfnamefont {R.~E.}\ \bibnamefont
  {Shrock}},\ }\href {\doibase 10.1103/PhysRevD.24.1232} {\bibfield  {journal}
  {\bibinfo  {journal} {Phys. Rev. D}\ }\textbf {\bibinfo {volume} {24}},\
  \bibinfo {pages} {1232} (\bibinfo {year} {1981})}\BibitemShut {NoStop}%
\bibitem [{\citenamefont {Wolfenstein}(1981)}]{Wolfenstein:1981kw}%
  \BibitemOpen
  \bibfield  {author} {\bibinfo {author} {\bibfnamefont {L.}~\bibnamefont
  {Wolfenstein}},\ }\href {\doibase 10.1016/0550-3213(81)90096-1} {\bibfield
  {journal} {\bibinfo  {journal} {Nucl. Phys. B}\ }\textbf {\bibinfo {volume}
  {186}},\ \bibinfo {pages} {147} (\bibinfo {year} {1981})}\BibitemShut
  {NoStop}%
\bibitem [{\citenamefont {Petcov}(1982)}]{Petcov:1982ya}%
  \BibitemOpen
  \bibfield  {author} {\bibinfo {author} {\bibfnamefont {S.~T.}\ \bibnamefont
  {Petcov}},\ }\href {\doibase 10.1016/0370-2693(82)91246-1} {\bibfield
  {journal} {\bibinfo  {journal} {Phys. Lett. B}\ }\textbf {\bibinfo {volume}
  {110}},\ \bibinfo {pages} {245} (\bibinfo {year} {1982})}\BibitemShut
  {NoStop}%
\bibitem [{\citenamefont {del Aguila}\ \emph {et~al.}(2009)\citenamefont {del
  Aguila}, \citenamefont {Bar-Shalom}, \citenamefont {Soni},\ and\
  \citenamefont {Wudka}}]{delAguila:2008ir}%
  \BibitemOpen
  \bibfield  {author} {\bibinfo {author} {\bibfnamefont {F.}~\bibnamefont {del
  Aguila}}, \bibinfo {author} {\bibfnamefont {S.}~\bibnamefont {Bar-Shalom}},
  \bibinfo {author} {\bibfnamefont {A.}~\bibnamefont {Soni}}, \ and\ \bibinfo
  {author} {\bibfnamefont {J.}~\bibnamefont {Wudka}},\ }\href {\doibase
  10.1016/j.physletb.2008.11.031} {\bibfield  {journal} {\bibinfo  {journal}
  {Phys. Lett. B}\ }\textbf {\bibinfo {volume} {670}},\ \bibinfo {pages} {399}
  (\bibinfo {year} {2009})},\ \Eprint {http://arxiv.org/abs/0806.0876}
  {arXiv:0806.0876 [hep-ph]} \BibitemShut {NoStop}%
\bibitem [{\citenamefont {Bhattacharya}\ and\ \citenamefont
  {Wudka}(2016)}]{Bhattacharya:2015vja}%
  \BibitemOpen
  \bibfield  {author} {\bibinfo {author} {\bibfnamefont {S.}~\bibnamefont
  {Bhattacharya}}\ and\ \bibinfo {author} {\bibfnamefont {J.}~\bibnamefont
  {Wudka}},\ }\href {\doibase 10.1103/PhysRevD.94.055022} {\bibfield  {journal}
  {\bibinfo  {journal} {Phys. Rev. D}\ }\textbf {\bibinfo {volume} {94}},\
  \bibinfo {pages} {055022} (\bibinfo {year} {2016})},\ \bibinfo {note}
  {[Erratum: Phys.Rev.D 95, 039904 (2017)]},\ \Eprint
  {http://arxiv.org/abs/1505.05264} {arXiv:1505.05264 [hep-ph]} \BibitemShut
  {NoStop}%
\bibitem [{\citenamefont {Liao}\ and\ \citenamefont {Ma}(2017)}]{Liao:2016qyd}%
  \BibitemOpen
  \bibfield  {author} {\bibinfo {author} {\bibfnamefont {Y.}~\bibnamefont
  {Liao}}\ and\ \bibinfo {author} {\bibfnamefont {X.-D.}\ \bibnamefont {Ma}},\
  }\href {\doibase 10.1103/PhysRevD.96.015012} {\bibfield  {journal} {\bibinfo
  {journal} {Phys. Rev. D}\ }\textbf {\bibinfo {volume} {96}},\ \bibinfo
  {pages} {015012} (\bibinfo {year} {2017})},\ \Eprint
  {http://arxiv.org/abs/1612.04527} {arXiv:1612.04527 [hep-ph]} \BibitemShut
  {NoStop}%
\bibitem [{\citenamefont {Datta}\ \emph {et~al.}(2021)\citenamefont {Datta},
  \citenamefont {Kumar}, \citenamefont {Liu},\ and\ \citenamefont
  {Marfatia}}]{Datta:2020ocb}%
  \BibitemOpen
  \bibfield  {author} {\bibinfo {author} {\bibfnamefont {A.}~\bibnamefont
  {Datta}}, \bibinfo {author} {\bibfnamefont {J.}~\bibnamefont {Kumar}},
  \bibinfo {author} {\bibfnamefont {H.}~\bibnamefont {Liu}}, \ and\ \bibinfo
  {author} {\bibfnamefont {D.}~\bibnamefont {Marfatia}},\ }\href {\doibase
  10.1007/JHEP02(2021)015} {\bibfield  {journal} {\bibinfo  {journal} {JHEP}\
  }\textbf {\bibinfo {volume} {02}},\ \bibinfo {pages} {015} (\bibinfo {year}
  {2021})},\ \Eprint {http://arxiv.org/abs/2010.12109} {arXiv:2010.12109
  [hep-ph]} \BibitemShut {NoStop}%
\bibitem [{\citenamefont {Chala}\ and\ \citenamefont
  {Titov}(2020)}]{Chala:2020pbn}%
  \BibitemOpen
  \bibfield  {author} {\bibinfo {author} {\bibfnamefont {M.}~\bibnamefont
  {Chala}}\ and\ \bibinfo {author} {\bibfnamefont {A.}~\bibnamefont {Titov}},\
  }\href {\doibase 10.1007/JHEP09(2020)188} {\bibfield  {journal} {\bibinfo
  {journal} {JHEP}\ }\textbf {\bibinfo {volume} {09}},\ \bibinfo {pages} {188}
  (\bibinfo {year} {2020})},\ \Eprint {http://arxiv.org/abs/2006.14596}
  {arXiv:2006.14596 [hep-ph]} \BibitemShut {NoStop}%
\bibitem [{\citenamefont {Datta}\ \emph {et~al.}(1994)\citenamefont {Datta},
  \citenamefont {Guchait},\ and\ \citenamefont {Pilaftsis}}]{Datta:1993nm}%
  \BibitemOpen
  \bibfield  {author} {\bibinfo {author} {\bibfnamefont {A.}~\bibnamefont
  {Datta}}, \bibinfo {author} {\bibfnamefont {M.}~\bibnamefont {Guchait}}, \
  and\ \bibinfo {author} {\bibfnamefont {A.}~\bibnamefont {Pilaftsis}},\ }\href
  {\doibase 10.1103/PhysRevD.50.3195} {\bibfield  {journal} {\bibinfo
  {journal} {Phys. Rev. D}\ }\textbf {\bibinfo {volume} {50}},\ \bibinfo
  {pages} {3195} (\bibinfo {year} {1994})},\ \Eprint
  {http://arxiv.org/abs/hep-ph/9311257} {arXiv:hep-ph/9311257} \BibitemShut
  {NoStop}%
\bibitem [{\citenamefont {Degrande}\ \emph {et~al.}(2016)\citenamefont
  {Degrande}, \citenamefont {Mattelaer}, \citenamefont {Ruiz},\ and\
  \citenamefont {Turner}}]{Degrande:2016aje}%
  \BibitemOpen
  \bibfield  {author} {\bibinfo {author} {\bibfnamefont {C.}~\bibnamefont
  {Degrande}}, \bibinfo {author} {\bibfnamefont {O.}~\bibnamefont {Mattelaer}},
  \bibinfo {author} {\bibfnamefont {R.}~\bibnamefont {Ruiz}}, \ and\ \bibinfo
  {author} {\bibfnamefont {J.}~\bibnamefont {Turner}},\ }\href {\doibase
  10.1103/PhysRevD.94.053002} {\bibfield  {journal} {\bibinfo  {journal} {Phys.
  Rev. D}\ }\textbf {\bibinfo {volume} {94}},\ \bibinfo {pages} {053002}
  (\bibinfo {year} {2016})},\ \Eprint {http://arxiv.org/abs/1602.06957}
  {arXiv:1602.06957 [hep-ph]} \BibitemShut {NoStop}%
\bibitem [{\citenamefont {Alva}\ \emph {et~al.}(2015)\citenamefont {Alva},
  \citenamefont {Han},\ and\ \citenamefont {Ruiz}}]{Alva:2014gxa}%
  \BibitemOpen
  \bibfield  {author} {\bibinfo {author} {\bibfnamefont {D.}~\bibnamefont
  {Alva}}, \bibinfo {author} {\bibfnamefont {T.}~\bibnamefont {Han}}, \ and\
  \bibinfo {author} {\bibfnamefont {R.}~\bibnamefont {Ruiz}},\ }\href {\doibase
  10.1007/JHEP02(2015)072} {\bibfield  {journal} {\bibinfo  {journal} {JHEP}\
  }\textbf {\bibinfo {volume} {02}},\ \bibinfo {pages} {072} (\bibinfo {year}
  {2015})},\ \Eprint {http://arxiv.org/abs/1411.7305} {arXiv:1411.7305
  [hep-ph]} \BibitemShut {NoStop}%
\bibitem [{\citenamefont {Alimena}\ \emph {et~al.}(2020)\citenamefont {Alimena}
  \emph {et~al.}}]{Alimena:2019zri}%
  \BibitemOpen
  \bibfield  {author} {\bibinfo {author} {\bibfnamefont {J.}~\bibnamefont
  {Alimena}} \emph {et~al.},\ }\href {\doibase 10.1088/1361-6471/ab4574}
  {\bibfield  {journal} {\bibinfo  {journal} {J. Phys. G}\ }\textbf {\bibinfo
  {volume} {47}},\ \bibinfo {pages} {090501} (\bibinfo {year} {2020})},\
  \Eprint {http://arxiv.org/abs/1903.04497} {arXiv:1903.04497 [hep-ex]}
  \BibitemShut {NoStop}%
\bibitem [{\citenamefont {Aaij}\ \emph {et~al.}(2014)\citenamefont {Aaij} \emph
  {et~al.}}]{LHCb:2014osd}%
  \BibitemOpen
  \bibfield  {author} {\bibinfo {author} {\bibfnamefont {R.}~\bibnamefont
  {Aaij}} \emph {et~al.} (\bibinfo {collaboration} {LHCb}),\ }\href {\doibase
  10.1103/PhysRevLett.112.131802} {\bibfield  {journal} {\bibinfo  {journal}
  {Phys. Rev. Lett.}\ }\textbf {\bibinfo {volume} {112}},\ \bibinfo {pages}
  {131802} (\bibinfo {year} {2014})},\ \Eprint {http://arxiv.org/abs/1401.5361}
  {arXiv:1401.5361 [hep-ex]} \BibitemShut {NoStop}%
\bibitem [{\citenamefont {Aad}\ \emph {et~al.}(2019{\natexlab{a}})\citenamefont
  {Aad} \emph {et~al.}}]{ATLAS:2019kpx}%
  \BibitemOpen
  \bibfield  {author} {\bibinfo {author} {\bibfnamefont {G.}~\bibnamefont
  {Aad}} \emph {et~al.} (\bibinfo {collaboration} {ATLAS}),\ }\href {\doibase
  10.1007/JHEP10(2019)265} {\bibfield  {journal} {\bibinfo  {journal} {JHEP}\
  }\textbf {\bibinfo {volume} {10}},\ \bibinfo {pages} {265} (\bibinfo {year}
  {2019}{\natexlab{a}})},\ \Eprint {http://arxiv.org/abs/1905.09787}
  {arXiv:1905.09787 [hep-ex]} \BibitemShut {NoStop}%
\bibitem [{\citenamefont {Aaij}\ \emph {et~al.}(2021)\citenamefont {Aaij} \emph
  {et~al.}}]{LHCb:2020wxx}%
  \BibitemOpen
  \bibfield  {author} {\bibinfo {author} {\bibfnamefont {R.}~\bibnamefont
  {Aaij}} \emph {et~al.} (\bibinfo {collaboration} {LHCb}),\ }\href {\doibase
  10.1140/epjc/s10052-021-08973-5} {\bibfield  {journal} {\bibinfo  {journal}
  {Eur. Phys. J. C}\ }\textbf {\bibinfo {volume} {81}},\ \bibinfo {pages} {248}
  (\bibinfo {year} {2021})},\ \Eprint {http://arxiv.org/abs/2011.05263}
  {arXiv:2011.05263 [hep-ex]} \BibitemShut {NoStop}%
\bibitem [{\citenamefont {Tumasyan}\ \emph
  {et~al.}(2022{\natexlab{a}})\citenamefont {Tumasyan} \emph
  {et~al.}}]{CMS:2022fut}%
  \BibitemOpen
  \bibfield  {author} {\bibinfo {author} {\bibfnamefont {A.}~\bibnamefont
  {Tumasyan}} \emph {et~al.} (\bibinfo {collaboration} {CMS}),\ }\href@noop {}
  {\  (\bibinfo {year} {2022}{\natexlab{a}})},\ \Eprint
  {http://arxiv.org/abs/2201.05578} {arXiv:2201.05578 [hep-ex]} \BibitemShut
  {NoStop}%
\bibitem [{CMS(2022)}]{CMS-PAS-EXO-21-003}%
  \BibitemOpen
  \href {https://cds.cern.ch/record/2803671} {\emph {\bibinfo {title} {{Probing
  Majorana neutrinos and the Weinberg operator in the same-charge dimuon
  channel through vector boson fusion processes in proton-proton collisions at
  $\sqrt{s}=13~\mathrm{TeV}$ }}}},\ \bibinfo {type} {Tech. Rep.}\ (\bibinfo
  {institution} {CERN},\ \bibinfo {address} {Geneva},\ \bibinfo {year}
  {2022})\BibitemShut {NoStop}%
\bibitem [{\citenamefont {Cottin}\ \emph
  {et~al.}(2018{\natexlab{a}})\citenamefont {Cottin}, \citenamefont {Helo},\
  and\ \citenamefont {Hirsch}}]{Cottin:2018kmq}%
  \BibitemOpen
  \bibfield  {author} {\bibinfo {author} {\bibfnamefont {G.}~\bibnamefont
  {Cottin}}, \bibinfo {author} {\bibfnamefont {J.~C.}\ \bibnamefont {Helo}}, \
  and\ \bibinfo {author} {\bibfnamefont {M.}~\bibnamefont {Hirsch}},\ }\href
  {\doibase 10.1103/PhysRevD.97.055025} {\bibfield  {journal} {\bibinfo
  {journal} {Phys. Rev. D}\ }\textbf {\bibinfo {volume} {97}},\ \bibinfo
  {pages} {055025} (\bibinfo {year} {2018}{\natexlab{a}})},\ \Eprint
  {http://arxiv.org/abs/1801.02734} {arXiv:1801.02734 [hep-ph]} \BibitemShut
  {NoStop}%
\bibitem [{\citenamefont {Cottin}\ \emph
  {et~al.}(2018{\natexlab{b}})\citenamefont {Cottin}, \citenamefont {Helo},\
  and\ \citenamefont {Hirsch}}]{Cottin:2018nms}%
  \BibitemOpen
  \bibfield  {author} {\bibinfo {author} {\bibfnamefont {G.}~\bibnamefont
  {Cottin}}, \bibinfo {author} {\bibfnamefont {J.~C.}\ \bibnamefont {Helo}}, \
  and\ \bibinfo {author} {\bibfnamefont {M.}~\bibnamefont {Hirsch}},\ }\href
  {\doibase 10.1103/PhysRevD.98.035012} {\bibfield  {journal} {\bibinfo
  {journal} {Phys. Rev. D}\ }\textbf {\bibinfo {volume} {98}},\ \bibinfo
  {pages} {035012} (\bibinfo {year} {2018}{\natexlab{b}})},\ \Eprint
  {http://arxiv.org/abs/1806.05191} {arXiv:1806.05191 [hep-ph]} \BibitemShut
  {NoStop}%
\bibitem [{\citenamefont {Drewes}\ and\ \citenamefont
  {Hajer}(2020)}]{Drewes:2019fou}%
  \BibitemOpen
  \bibfield  {author} {\bibinfo {author} {\bibfnamefont {M.}~\bibnamefont
  {Drewes}}\ and\ \bibinfo {author} {\bibfnamefont {J.}~\bibnamefont {Hajer}},\
  }\href {\doibase 10.1007/JHEP02(2020)070} {\bibfield  {journal} {\bibinfo
  {journal} {JHEP}\ }\textbf {\bibinfo {volume} {02}},\ \bibinfo {pages} {070}
  (\bibinfo {year} {2020})},\ \Eprint {http://arxiv.org/abs/1903.06100}
  {arXiv:1903.06100 [hep-ph]} \BibitemShut {NoStop}%
\bibitem [{\citenamefont {R}\ \emph {et~al.}(2020)\citenamefont {R},
  \citenamefont {Cottin}, \citenamefont {Helo},\ and\ \citenamefont
  {Hirsch}}]{R:2020odv}%
  \BibitemOpen
  \bibfield  {author} {\bibinfo {author} {\bibfnamefont {C.~A.}\ \bibnamefont
  {R}}, \bibinfo {author} {\bibfnamefont {G.}~\bibnamefont {Cottin}}, \bibinfo
  {author} {\bibfnamefont {J.~C.}\ \bibnamefont {Helo}}, \ and\ \bibinfo
  {author} {\bibfnamefont {M.}~\bibnamefont {Hirsch}},\ }\href {\doibase
  10.1103/PhysRevD.101.095033} {\bibfield  {journal} {\bibinfo  {journal}
  {Phys. Rev. D}\ }\textbf {\bibinfo {volume} {101}},\ \bibinfo {pages}
  {095033} (\bibinfo {year} {2020})},\ \Eprint
  {http://arxiv.org/abs/2003.11494} {arXiv:2003.11494 [hep-ph]} \BibitemShut
  {NoStop}%
\bibitem [{\citenamefont {Fuks}\ \emph {et~al.}(2021)\citenamefont {Fuks},
  \citenamefont {Neundorf}, \citenamefont {Peters}, \citenamefont {Ruiz},\ and\
  \citenamefont {Saimpert}}]{Fuks:2020att}%
  \BibitemOpen
  \bibfield  {author} {\bibinfo {author} {\bibfnamefont {B.}~\bibnamefont
  {Fuks}}, \bibinfo {author} {\bibfnamefont {J.}~\bibnamefont {Neundorf}},
  \bibinfo {author} {\bibfnamefont {K.}~\bibnamefont {Peters}}, \bibinfo
  {author} {\bibfnamefont {R.}~\bibnamefont {Ruiz}}, \ and\ \bibinfo {author}
  {\bibfnamefont {M.}~\bibnamefont {Saimpert}},\ }\href {\doibase
  10.1103/PhysRevD.103.055005} {\bibfield  {journal} {\bibinfo  {journal}
  {Phys. Rev. D}\ }\textbf {\bibinfo {volume} {103}},\ \bibinfo {pages}
  {055005} (\bibinfo {year} {2021})},\ \Eprint
  {http://arxiv.org/abs/2011.02547} {arXiv:2011.02547 [hep-ph]} \BibitemShut
  {NoStop}%
\bibitem [{\citenamefont {{CMS Collaboration}}(2022)}]{CMS:2022msk}%
  \BibitemOpen
  \bibfield  {author} {\bibinfo {author} {\bibnamefont {{CMS Collaboration}}},\
  }\href@noop {} {\  (\bibinfo {year} {2022})},\ \bibinfo {note}
  {{CMS-PAS-FTR-22-003}}\BibitemShut {NoStop}%
\bibitem [{\citenamefont {Han}\ \emph {et~al.}(2007)\citenamefont {Han},
  \citenamefont {Mukhopadhyaya}, \citenamefont {Si},\ and\ \citenamefont
  {Wang}}]{Han:2007bk}%
  \BibitemOpen
  \bibfield  {author} {\bibinfo {author} {\bibfnamefont {T.}~\bibnamefont
  {Han}}, \bibinfo {author} {\bibfnamefont {B.}~\bibnamefont {Mukhopadhyaya}},
  \bibinfo {author} {\bibfnamefont {Z.}~\bibnamefont {Si}}, \ and\ \bibinfo
  {author} {\bibfnamefont {K.}~\bibnamefont {Wang}},\ }\href {\doibase
  10.1103/PhysRevD.76.075013} {\bibfield  {journal} {\bibinfo  {journal} {Phys.
  Rev. D}\ }\textbf {\bibinfo {volume} {76}},\ \bibinfo {pages} {075013}
  (\bibinfo {year} {2007})},\ \Eprint {http://arxiv.org/abs/0706.0441}
  {arXiv:0706.0441 [hep-ph]} \BibitemShut {NoStop}%
\bibitem [{\citenamefont {Nemev\v{s}ek}\ \emph {et~al.}(2017)\citenamefont
  {Nemev\v{s}ek}, \citenamefont {Nesti},\ and\ \citenamefont
  {Vasquez}}]{Nemevsek:2016enw}%
  \BibitemOpen
  \bibfield  {author} {\bibinfo {author} {\bibfnamefont {M.}~\bibnamefont
  {Nemev\v{s}ek}}, \bibinfo {author} {\bibfnamefont {F.}~\bibnamefont {Nesti}},
  \ and\ \bibinfo {author} {\bibfnamefont {J.~C.}\ \bibnamefont {Vasquez}},\
  }\href {\doibase 10.1007/JHEP04(2017)114} {\bibfield  {journal} {\bibinfo
  {journal} {JHEP}\ }\textbf {\bibinfo {volume} {04}},\ \bibinfo {pages} {114}
  (\bibinfo {year} {2017})},\ \Eprint {http://arxiv.org/abs/1612.06840}
  {arXiv:1612.06840 [hep-ph]} \BibitemShut {NoStop}%
\bibitem [{\citenamefont {Fuks}\ \emph {et~al.}(2020)\citenamefont {Fuks},
  \citenamefont {Nemev\v{s}ek},\ and\ \citenamefont {Ruiz}}]{Fuks:2019clu}%
  \BibitemOpen
  \bibfield  {author} {\bibinfo {author} {\bibfnamefont {B.}~\bibnamefont
  {Fuks}}, \bibinfo {author} {\bibfnamefont {M.}~\bibnamefont {Nemev\v{s}ek}},
  \ and\ \bibinfo {author} {\bibfnamefont {R.}~\bibnamefont {Ruiz}},\ }\href
  {\doibase 10.1103/PhysRevD.101.075022} {\bibfield  {journal} {\bibinfo
  {journal} {Phys. Rev. D}\ }\textbf {\bibinfo {volume} {101}},\ \bibinfo
  {pages} {075022} (\bibinfo {year} {2020})},\ \Eprint
  {http://arxiv.org/abs/1912.08975} {arXiv:1912.08975 [hep-ph]} \BibitemShut
  {NoStop}%
\bibitem [{\citenamefont {Muhlleitner}\ and\ \citenamefont
  {Spira}(2003)}]{Muhlleitner:2003me}%
  \BibitemOpen
  \bibfield  {author} {\bibinfo {author} {\bibfnamefont {M.}~\bibnamefont
  {Muhlleitner}}\ and\ \bibinfo {author} {\bibfnamefont {M.}~\bibnamefont
  {Spira}},\ }\href {\doibase 10.1103/PhysRevD.68.117701} {\bibfield  {journal}
  {\bibinfo  {journal} {Phys. Rev. D}\ }\textbf {\bibinfo {volume} {68}},\
  \bibinfo {pages} {117701} (\bibinfo {year} {2003})},\ \Eprint
  {http://arxiv.org/abs/hep-ph/0305288} {arXiv:hep-ph/0305288} \BibitemShut
  {NoStop}%
\bibitem [{\citenamefont {Fileviez~Perez}\ \emph
  {et~al.}(2008{\natexlab{a}})\citenamefont {Fileviez~Perez}, \citenamefont
  {Han}, \citenamefont {Huang}, \citenamefont {Li},\ and\ \citenamefont
  {Wang}}]{FileviezPerez:2008jbu}%
  \BibitemOpen
  \bibfield  {author} {\bibinfo {author} {\bibfnamefont {P.}~\bibnamefont
  {Fileviez~Perez}}, \bibinfo {author} {\bibfnamefont {T.}~\bibnamefont {Han}},
  \bibinfo {author} {\bibfnamefont {G.-y.}\ \bibnamefont {Huang}}, \bibinfo
  {author} {\bibfnamefont {T.}~\bibnamefont {Li}}, \ and\ \bibinfo {author}
  {\bibfnamefont {K.}~\bibnamefont {Wang}},\ }\href {\doibase
  10.1103/PhysRevD.78.015018} {\bibfield  {journal} {\bibinfo  {journal} {Phys.
  Rev. D}\ }\textbf {\bibinfo {volume} {78}},\ \bibinfo {pages} {015018}
  (\bibinfo {year} {2008}{\natexlab{a}})},\ \Eprint
  {http://arxiv.org/abs/0805.3536} {arXiv:0805.3536 [hep-ph]} \BibitemShut
  {NoStop}%
\bibitem [{\citenamefont {Fileviez~Perez}\ \emph
  {et~al.}(2008{\natexlab{b}})\citenamefont {Fileviez~Perez}, \citenamefont
  {Han}, \citenamefont {Huang}, \citenamefont {Li},\ and\ \citenamefont
  {Wang}}]{FileviezPerez:2008wbg}%
  \BibitemOpen
  \bibfield  {author} {\bibinfo {author} {\bibfnamefont {P.}~\bibnamefont
  {Fileviez~Perez}}, \bibinfo {author} {\bibfnamefont {T.}~\bibnamefont {Han}},
  \bibinfo {author} {\bibfnamefont {G.-Y.}\ \bibnamefont {Huang}}, \bibinfo
  {author} {\bibfnamefont {T.}~\bibnamefont {Li}}, \ and\ \bibinfo {author}
  {\bibfnamefont {K.}~\bibnamefont {Wang}},\ }\href {\doibase
  10.1103/PhysRevD.78.071301} {\bibfield  {journal} {\bibinfo  {journal} {Phys.
  Rev. D}\ }\textbf {\bibinfo {volume} {78}},\ \bibinfo {pages} {071301}
  (\bibinfo {year} {2008}{\natexlab{b}})},\ \Eprint
  {http://arxiv.org/abs/0803.3450} {arXiv:0803.3450 [hep-ph]} \BibitemShut
  {NoStop}%
\bibitem [{\citenamefont {{CMS Collaboration}}(2017)}]{CMS:2017pet}%
  \BibitemOpen
  \bibfield  {author} {\bibinfo {author} {\bibnamefont {{CMS Collaboration}}},\
  }\href@noop {} {\  (\bibinfo {year} {2017})},\ \bibinfo {note}
  {{CMS-PAS-HIG-16-036}}\BibitemShut {NoStop}%
\bibitem [{\citenamefont {Sirunyan}\ \emph
  {et~al.}(2018{\natexlab{a}})\citenamefont {Sirunyan} \emph
  {et~al.}}]{CMS:2017fhs}%
  \BibitemOpen
  \bibfield  {author} {\bibinfo {author} {\bibfnamefont {A.~M.}\ \bibnamefont
  {Sirunyan}} \emph {et~al.} (\bibinfo {collaboration} {CMS}),\ }\href
  {\doibase 10.1103/PhysRevLett.120.081801} {\bibfield  {journal} {\bibinfo
  {journal} {Phys. Rev. Lett.}\ }\textbf {\bibinfo {volume} {120}},\ \bibinfo
  {pages} {081801} (\bibinfo {year} {2018}{\natexlab{a}})},\ \Eprint
  {http://arxiv.org/abs/1709.05822} {arXiv:1709.05822 [hep-ex]} \BibitemShut
  {NoStop}%
\bibitem [{\citenamefont {Aaboud}\ \emph
  {et~al.}(2018{\natexlab{a}})\citenamefont {Aaboud} \emph
  {et~al.}}]{ATLAS:2017xqs}%
  \BibitemOpen
  \bibfield  {author} {\bibinfo {author} {\bibfnamefont {M.}~\bibnamefont
  {Aaboud}} \emph {et~al.} (\bibinfo {collaboration} {ATLAS}),\ }\href
  {\doibase 10.1140/epjc/s10052-018-5661-z} {\bibfield  {journal} {\bibinfo
  {journal} {Eur. Phys. J. C}\ }\textbf {\bibinfo {volume} {78}},\ \bibinfo
  {pages} {199} (\bibinfo {year} {2018}{\natexlab{a}})},\ \Eprint
  {http://arxiv.org/abs/1710.09748} {arXiv:1710.09748 [hep-ex]} \BibitemShut
  {NoStop}%
\bibitem [{\citenamefont {Aaboud}\ \emph
  {et~al.}(2019{\natexlab{a}})\citenamefont {Aaboud} \emph
  {et~al.}}]{ATLAS:2018ceg}%
  \BibitemOpen
  \bibfield  {author} {\bibinfo {author} {\bibfnamefont {M.}~\bibnamefont
  {Aaboud}} \emph {et~al.} (\bibinfo {collaboration} {ATLAS}),\ }\href
  {\doibase 10.1140/epjc/s10052-018-6500-y} {\bibfield  {journal} {\bibinfo
  {journal} {Eur. Phys. J. C}\ }\textbf {\bibinfo {volume} {79}},\ \bibinfo
  {pages} {58} (\bibinfo {year} {2019}{\natexlab{a}})},\ \Eprint
  {http://arxiv.org/abs/1808.01899} {arXiv:1808.01899 [hep-ex]} \BibitemShut
  {NoStop}%
\bibitem [{\citenamefont {Aad}\ \emph {et~al.}(2021{\natexlab{a}})\citenamefont
  {Aad} \emph {et~al.}}]{ATLAS:2021jol}%
  \BibitemOpen
  \bibfield  {author} {\bibinfo {author} {\bibfnamefont {G.}~\bibnamefont
  {Aad}} \emph {et~al.} (\bibinfo {collaboration} {ATLAS}),\ }\href {\doibase
  10.1007/JHEP06(2021)146} {\bibfield  {journal} {\bibinfo  {journal} {JHEP}\
  }\textbf {\bibinfo {volume} {06}},\ \bibinfo {pages} {146} (\bibinfo {year}
  {2021}{\natexlab{a}})},\ \Eprint {http://arxiv.org/abs/2101.11961}
  {arXiv:2101.11961 [hep-ex]} \BibitemShut {NoStop}%
\bibitem [{\citenamefont {Ashanujjaman}\ and\ \citenamefont
  {Ghosh}(2021)}]{Ashanujjaman:2021txz}%
  \BibitemOpen
  \bibfield  {author} {\bibinfo {author} {\bibfnamefont {S.}~\bibnamefont
  {Ashanujjaman}}\ and\ \bibinfo {author} {\bibfnamefont {K.}~\bibnamefont
  {Ghosh}},\ }\href@noop {} {\  (\bibinfo {year} {2021})},\ \Eprint
  {http://arxiv.org/abs/2108.10952} {arXiv:2108.10952 [hep-ph]} \BibitemShut
  {NoStop}%
\bibitem [{\citenamefont {Biggio}\ and\ \citenamefont
  {Bonnet}(2012)}]{Biggio:2011ja}%
  \BibitemOpen
  \bibfield  {author} {\bibinfo {author} {\bibfnamefont {C.}~\bibnamefont
  {Biggio}}\ and\ \bibinfo {author} {\bibfnamefont {F.}~\bibnamefont
  {Bonnet}},\ }\href {\doibase 10.1140/epjc/s10052-012-1899-z} {\bibfield
  {journal} {\bibinfo  {journal} {Eur. Phys. J. C}\ }\textbf {\bibinfo {volume}
  {72}},\ \bibinfo {pages} {1899} (\bibinfo {year} {2012})},\ \Eprint
  {http://arxiv.org/abs/1107.3463} {arXiv:1107.3463 [hep-ph]} \BibitemShut
  {NoStop}%
\bibitem [{\citenamefont {Ruiz}(2015)}]{Ruiz:2015zca}%
  \BibitemOpen
  \bibfield  {author} {\bibinfo {author} {\bibfnamefont {R.}~\bibnamefont
  {Ruiz}},\ }\href {\doibase 10.1007/JHEP12(2015)165} {\bibfield  {journal}
  {\bibinfo  {journal} {JHEP}\ }\textbf {\bibinfo {volume} {12}},\ \bibinfo
  {pages} {165} (\bibinfo {year} {2015})},\ \Eprint
  {http://arxiv.org/abs/1509.05416} {arXiv:1509.05416 [hep-ph]} \BibitemShut
  {NoStop}%
\bibitem [{\citenamefont {Aad}\ \emph {et~al.}(2021{\natexlab{b}})\citenamefont
  {Aad} \emph {et~al.}}]{ATLAS:2020wop}%
  \BibitemOpen
  \bibfield  {author} {\bibinfo {author} {\bibfnamefont {G.}~\bibnamefont
  {Aad}} \emph {et~al.} (\bibinfo {collaboration} {ATLAS}),\ }\href {\doibase
  10.1140/epjc/s10052-021-08929-9} {\bibfield  {journal} {\bibinfo  {journal}
  {Eur. Phys. J. C}\ }\textbf {\bibinfo {volume} {81}},\ \bibinfo {pages} {218}
  (\bibinfo {year} {2021}{\natexlab{b}})},\ \Eprint
  {http://arxiv.org/abs/2008.07949} {arXiv:2008.07949 [hep-ex]} \BibitemShut
  {NoStop}%
\bibitem [{\citenamefont {Sirunyan}\ \emph
  {et~al.}(2020{\natexlab{a}})\citenamefont {Sirunyan} \emph
  {et~al.}}]{CMS:2019lwf}%
  \BibitemOpen
  \bibfield  {author} {\bibinfo {author} {\bibfnamefont {A.~M.}\ \bibnamefont
  {Sirunyan}} \emph {et~al.} (\bibinfo {collaboration} {CMS}),\ }\href
  {\doibase 10.1007/JHEP03(2020)051} {\bibfield  {journal} {\bibinfo  {journal}
  {JHEP}\ }\textbf {\bibinfo {volume} {03}},\ \bibinfo {pages} {051} (\bibinfo
  {year} {2020}{\natexlab{a}})},\ \Eprint {http://arxiv.org/abs/1911.04968}
  {arXiv:1911.04968 [hep-ex]} \BibitemShut {NoStop}%
\bibitem [{\citenamefont {Tumasyan}\ \emph
  {et~al.}(2022{\natexlab{b}})\citenamefont {Tumasyan} \emph
  {et~al.}}]{CMS:2022nty}%
  \BibitemOpen
  \bibfield  {author} {\bibinfo {author} {\bibfnamefont {A.}~\bibnamefont
  {Tumasyan}} \emph {et~al.} (\bibinfo {collaboration} {CMS}),\ }\href@noop {}
  {\  (\bibinfo {year} {2022}{\natexlab{b}})},\ \Eprint
  {http://arxiv.org/abs/2202.08676} {arXiv:2202.08676 [hep-ex]} \BibitemShut
  {NoStop}%
\bibitem [{\citenamefont {Aad}\ \emph {et~al.}(2022)\citenamefont {Aad} \emph
  {et~al.}}]{ATLAS:2022yhd}%
  \BibitemOpen
  \bibfield  {author} {\bibinfo {author} {\bibfnamefont {G.}~\bibnamefont
  {Aad}} \emph {et~al.} (\bibinfo {collaboration} {ATLAS}),\ }\href@noop {} {\
  (\bibinfo {year} {2022})},\ \Eprint {http://arxiv.org/abs/2202.02039}
  {arXiv:2202.02039 [hep-ex]} \BibitemShut {NoStop}%
\bibitem [{\citenamefont {Franceschini}\ \emph {et~al.}(2008)\citenamefont
  {Franceschini}, \citenamefont {Hambye},\ and\ \citenamefont
  {Strumia}}]{Franceschini:2008pz}%
  \BibitemOpen
  \bibfield  {author} {\bibinfo {author} {\bibfnamefont {R.}~\bibnamefont
  {Franceschini}}, \bibinfo {author} {\bibfnamefont {T.}~\bibnamefont
  {Hambye}}, \ and\ \bibinfo {author} {\bibfnamefont {A.}~\bibnamefont
  {Strumia}},\ }\href {\doibase 10.1103/PhysRevD.78.033002} {\bibfield
  {journal} {\bibinfo  {journal} {Phys. Rev. D}\ }\textbf {\bibinfo {volume}
  {78}},\ \bibinfo {pages} {033002} (\bibinfo {year} {2008})},\ \Eprint
  {http://arxiv.org/abs/0805.1613} {arXiv:0805.1613 [hep-ph]} \BibitemShut
  {NoStop}%
\bibitem [{\citenamefont {Arhrib}\ \emph {et~al.}(2010)\citenamefont {Arhrib},
  \citenamefont {Bajc}, \citenamefont {Ghosh}, \citenamefont {Han},
  \citenamefont {Huang}, \citenamefont {Puljak},\ and\ \citenamefont
  {Senjanovic}}]{Arhrib:2009mz}%
  \BibitemOpen
  \bibfield  {author} {\bibinfo {author} {\bibfnamefont {A.}~\bibnamefont
  {Arhrib}}, \bibinfo {author} {\bibfnamefont {B.}~\bibnamefont {Bajc}},
  \bibinfo {author} {\bibfnamefont {D.~K.}\ \bibnamefont {Ghosh}}, \bibinfo
  {author} {\bibfnamefont {T.}~\bibnamefont {Han}}, \bibinfo {author}
  {\bibfnamefont {G.-Y.}\ \bibnamefont {Huang}}, \bibinfo {author}
  {\bibfnamefont {I.}~\bibnamefont {Puljak}}, \ and\ \bibinfo {author}
  {\bibfnamefont {G.}~\bibnamefont {Senjanovic}},\ }\href {\doibase
  10.1103/PhysRevD.82.053004} {\bibfield  {journal} {\bibinfo  {journal} {Phys.
  Rev. D}\ }\textbf {\bibinfo {volume} {82}},\ \bibinfo {pages} {053004}
  (\bibinfo {year} {2010})},\ \Eprint {http://arxiv.org/abs/0904.2390}
  {arXiv:0904.2390 [hep-ph]} \BibitemShut {NoStop}%
\bibitem [{\citenamefont {Croon}\ \emph {et~al.}(2019)\citenamefont {Croon},
  \citenamefont {Gonzalo}, \citenamefont {Graf}, \citenamefont {Ko\v{s}nik},\
  and\ \citenamefont {White}}]{Croon:2019kpe}%
  \BibitemOpen
  \bibfield  {author} {\bibinfo {author} {\bibfnamefont {D.}~\bibnamefont
  {Croon}}, \bibinfo {author} {\bibfnamefont {T.~E.}\ \bibnamefont {Gonzalo}},
  \bibinfo {author} {\bibfnamefont {L.}~\bibnamefont {Graf}}, \bibinfo {author}
  {\bibfnamefont {N.}~\bibnamefont {Ko\v{s}nik}}, \ and\ \bibinfo {author}
  {\bibfnamefont {G.}~\bibnamefont {White}},\ }\href {\doibase
  10.3389/fphy.2019.00076} {\bibfield  {journal} {\bibinfo  {journal} {Front.
  in Phys.}\ }\textbf {\bibinfo {volume} {7}},\ \bibinfo {pages} {76} (\bibinfo
  {year} {2019})},\ \Eprint {http://arxiv.org/abs/1903.04977} {arXiv:1903.04977
  [hep-ph]} \BibitemShut {NoStop}%
\bibitem [{\citenamefont {Aristizabal~Sierra}\ \emph
  {et~al.}(2008)\citenamefont {Aristizabal~Sierra}, \citenamefont {Hirsch},\
  and\ \citenamefont {Kovalenko}}]{AristizabalSierra:2007nf}%
  \BibitemOpen
  \bibfield  {author} {\bibinfo {author} {\bibfnamefont {D.}~\bibnamefont
  {Aristizabal~Sierra}}, \bibinfo {author} {\bibfnamefont {M.}~\bibnamefont
  {Hirsch}}, \ and\ \bibinfo {author} {\bibfnamefont {S.~G.}\ \bibnamefont
  {Kovalenko}},\ }\href {\doibase 10.1103/PhysRevD.77.055011} {\bibfield
  {journal} {\bibinfo  {journal} {Phys. Rev. D}\ }\textbf {\bibinfo {volume}
  {77}},\ \bibinfo {pages} {055011} (\bibinfo {year} {2008})},\ \Eprint
  {http://arxiv.org/abs/0710.5699} {arXiv:0710.5699 [hep-ph]} \BibitemShut
  {NoStop}%
\bibitem [{\citenamefont {Cai}\ \emph {et~al.}(2015)\citenamefont {Cai},
  \citenamefont {Clarke}, \citenamefont {Schmidt},\ and\ \citenamefont
  {Volkas}}]{Cai:2014kra}%
  \BibitemOpen
  \bibfield  {author} {\bibinfo {author} {\bibfnamefont {Y.}~\bibnamefont
  {Cai}}, \bibinfo {author} {\bibfnamefont {J.~D.}\ \bibnamefont {Clarke}},
  \bibinfo {author} {\bibfnamefont {M.~A.}\ \bibnamefont {Schmidt}}, \ and\
  \bibinfo {author} {\bibfnamefont {R.~R.}\ \bibnamefont {Volkas}},\ }\href
  {\doibase 10.1007/JHEP02(2015)161} {\bibfield  {journal} {\bibinfo  {journal}
  {JHEP}\ }\textbf {\bibinfo {volume} {02}},\ \bibinfo {pages} {161} (\bibinfo
  {year} {2015})},\ \Eprint {http://arxiv.org/abs/1410.0689} {arXiv:1410.0689
  [hep-ph]} \BibitemShut {NoStop}%
\bibitem [{\citenamefont {P\"as}\ and\ \citenamefont
  {Schumacher}(2015)}]{Pas:2015hca}%
  \BibitemOpen
  \bibfield  {author} {\bibinfo {author} {\bibfnamefont {H.}~\bibnamefont
  {P\"as}}\ and\ \bibinfo {author} {\bibfnamefont {E.}~\bibnamefont
  {Schumacher}},\ }\href {\doibase 10.1103/PhysRevD.92.114025} {\bibfield
  {journal} {\bibinfo  {journal} {Phys. Rev. D}\ }\textbf {\bibinfo {volume}
  {92}},\ \bibinfo {pages} {114025} (\bibinfo {year} {2015})},\ \Eprint
  {http://arxiv.org/abs/1510.08757} {arXiv:1510.08757 [hep-ph]} \BibitemShut
  {NoStop}%
\bibitem [{\citenamefont {Dor\v{s}ner}\ \emph {et~al.}(2017)\citenamefont
  {Dor\v{s}ner}, \citenamefont {Fajfer},\ and\ \citenamefont
  {Ko\v{s}nik}}]{Dorsner:2017wwn}%
  \BibitemOpen
  \bibfield  {author} {\bibinfo {author} {\bibfnamefont {I.}~\bibnamefont
  {Dor\v{s}ner}}, \bibinfo {author} {\bibfnamefont {S.}~\bibnamefont {Fajfer}},
  \ and\ \bibinfo {author} {\bibfnamefont {N.}~\bibnamefont {Ko\v{s}nik}},\
  }\href {\doibase 10.1140/epjc/s10052-017-4987-2} {\bibfield  {journal}
  {\bibinfo  {journal} {Eur. Phys. J. C}\ }\textbf {\bibinfo {volume} {77}},\
  \bibinfo {pages} {417} (\bibinfo {year} {2017})},\ \Eprint
  {http://arxiv.org/abs/1701.08322} {arXiv:1701.08322 [hep-ph]} \BibitemShut
  {NoStop}%
\bibitem [{\citenamefont {Sirunyan}\ \emph
  {et~al.}(2018{\natexlab{b}})\citenamefont {Sirunyan} \emph
  {et~al.}}]{CMS:2018txo}%
  \BibitemOpen
  \bibfield  {author} {\bibinfo {author} {\bibfnamefont {A.~M.}\ \bibnamefont
  {Sirunyan}} \emph {et~al.} (\bibinfo {collaboration} {CMS}),\ }\href
  {\doibase 10.1007/JHEP07(2018)115} {\bibfield  {journal} {\bibinfo  {journal}
  {JHEP}\ }\textbf {\bibinfo {volume} {07}},\ \bibinfo {pages} {115} (\bibinfo
  {year} {2018}{\natexlab{b}})},\ \Eprint {http://arxiv.org/abs/1806.03472}
  {arXiv:1806.03472 [hep-ex]} \BibitemShut {NoStop}%
\bibitem [{\citenamefont {Sirunyan}\ \emph
  {et~al.}(2019{\natexlab{a}})\citenamefont {Sirunyan} \emph
  {et~al.}}]{CMS:2018lab}%
  \BibitemOpen
  \bibfield  {author} {\bibinfo {author} {\bibfnamefont {A.~M.}\ \bibnamefont
  {Sirunyan}} \emph {et~al.} (\bibinfo {collaboration} {CMS}),\ }\href
  {\doibase 10.1103/PhysRevD.99.032014} {\bibfield  {journal} {\bibinfo
  {journal} {Phys. Rev. D}\ }\textbf {\bibinfo {volume} {99}},\ \bibinfo
  {pages} {032014} (\bibinfo {year} {2019}{\natexlab{a}})},\ \Eprint
  {http://arxiv.org/abs/1808.05082} {arXiv:1808.05082 [hep-ex]} \BibitemShut
  {NoStop}%
\bibitem [{\citenamefont {Sirunyan}\ \emph
  {et~al.}(2019{\natexlab{b}})\citenamefont {Sirunyan} \emph
  {et~al.}}]{CMS:2018iye}%
  \BibitemOpen
  \bibfield  {author} {\bibinfo {author} {\bibfnamefont {A.~M.}\ \bibnamefont
  {Sirunyan}} \emph {et~al.} (\bibinfo {collaboration} {CMS}),\ }\href
  {\doibase 10.1007/JHEP03(2019)170} {\bibfield  {journal} {\bibinfo  {journal}
  {JHEP}\ }\textbf {\bibinfo {volume} {03}},\ \bibinfo {pages} {170} (\bibinfo
  {year} {2019}{\natexlab{b}})},\ \Eprint {http://arxiv.org/abs/1811.00806}
  {arXiv:1811.00806 [hep-ex]} \BibitemShut {NoStop}%
\bibitem [{\citenamefont {Sirunyan}\ \emph
  {et~al.}(2019{\natexlab{c}})\citenamefont {Sirunyan} \emph
  {et~al.}}]{CMS:2018ncu}%
  \BibitemOpen
  \bibfield  {author} {\bibinfo {author} {\bibfnamefont {A.~M.}\ \bibnamefont
  {Sirunyan}} \emph {et~al.} (\bibinfo {collaboration} {CMS}),\ }\href
  {\doibase 10.1103/PhysRevD.99.052002} {\bibfield  {journal} {\bibinfo
  {journal} {Phys. Rev. D}\ }\textbf {\bibinfo {volume} {99}},\ \bibinfo
  {pages} {052002} (\bibinfo {year} {2019}{\natexlab{c}})},\ \Eprint
  {http://arxiv.org/abs/1811.01197} {arXiv:1811.01197 [hep-ex]} \BibitemShut
  {NoStop}%
\bibitem [{\citenamefont {Aad}\ \emph {et~al.}(2020{\natexlab{a}})\citenamefont
  {Aad} \emph {et~al.}}]{ATLAS:2020dsf}%
  \BibitemOpen
  \bibfield  {author} {\bibinfo {author} {\bibfnamefont {G.}~\bibnamefont
  {Aad}} \emph {et~al.} (\bibinfo {collaboration} {ATLAS}),\ }\href {\doibase
  10.1140/epjc/s10052-020-8102-8} {\bibfield  {journal} {\bibinfo  {journal}
  {Eur. Phys. J. C}\ }\textbf {\bibinfo {volume} {80}},\ \bibinfo {pages} {737}
  (\bibinfo {year} {2020}{\natexlab{a}})},\ \Eprint
  {http://arxiv.org/abs/2004.14060} {arXiv:2004.14060 [hep-ex]} \BibitemShut
  {NoStop}%
\bibitem [{\citenamefont {Aad}\ \emph {et~al.}(2020{\natexlab{b}})\citenamefont
  {Aad} \emph {et~al.}}]{ATLAS:2020dsk}%
  \BibitemOpen
  \bibfield  {author} {\bibinfo {author} {\bibfnamefont {G.}~\bibnamefont
  {Aad}} \emph {et~al.} (\bibinfo {collaboration} {ATLAS}),\ }\href {\doibase
  10.1007/JHEP10(2020)112} {\bibfield  {journal} {\bibinfo  {journal} {JHEP}\
  }\textbf {\bibinfo {volume} {10}},\ \bibinfo {pages} {112} (\bibinfo {year}
  {2020}{\natexlab{b}})},\ \Eprint {http://arxiv.org/abs/2006.05872}
  {arXiv:2006.05872 [hep-ex]} \BibitemShut {NoStop}%
\bibitem [{\citenamefont {Aad}\ \emph {et~al.}(2021{\natexlab{c}})\citenamefont
  {Aad} \emph {et~al.}}]{ATLAS:2020xov}%
  \BibitemOpen
  \bibfield  {author} {\bibinfo {author} {\bibfnamefont {G.}~\bibnamefont
  {Aad}} \emph {et~al.} (\bibinfo {collaboration} {ATLAS}),\ }\href {\doibase
  10.1140/epjc/s10052-021-09009-8} {\bibfield  {journal} {\bibinfo  {journal}
  {Eur. Phys. J. C}\ }\textbf {\bibinfo {volume} {81}},\ \bibinfo {pages} {313}
  (\bibinfo {year} {2021}{\natexlab{c}})},\ \Eprint
  {http://arxiv.org/abs/2010.02098} {arXiv:2010.02098 [hep-ex]} \BibitemShut
  {NoStop}%
\bibitem [{\citenamefont {Aad}\ \emph {et~al.}(2021{\natexlab{d}})\citenamefont
  {Aad} \emph {et~al.}}]{ATLAS:2021oiz}%
  \BibitemOpen
  \bibfield  {author} {\bibinfo {author} {\bibfnamefont {G.}~\bibnamefont
  {Aad}} \emph {et~al.} (\bibinfo {collaboration} {ATLAS}),\ }\href {\doibase
  10.1007/JHEP06(2021)179} {\bibfield  {journal} {\bibinfo  {journal} {JHEP}\
  }\textbf {\bibinfo {volume} {06}},\ \bibinfo {pages} {179} (\bibinfo {year}
  {2021}{\natexlab{d}})},\ \Eprint {http://arxiv.org/abs/2101.11582}
  {arXiv:2101.11582 [hep-ex]} \BibitemShut {NoStop}%
\bibitem [{\citenamefont {Aad}\ \emph {et~al.}(2021{\natexlab{e}})\citenamefont
  {Aad} \emph {et~al.}}]{ATLAS:2021yij}%
  \BibitemOpen
  \bibfield  {author} {\bibinfo {author} {\bibfnamefont {G.}~\bibnamefont
  {Aad}} \emph {et~al.} (\bibinfo {collaboration} {ATLAS}),\ }\href {\doibase
  10.1007/JHEP05(2021)093} {\bibfield  {journal} {\bibinfo  {journal} {JHEP}\
  }\textbf {\bibinfo {volume} {05}},\ \bibinfo {pages} {093} (\bibinfo {year}
  {2021}{\natexlab{e}})},\ \Eprint {http://arxiv.org/abs/2101.12527}
  {arXiv:2101.12527 [hep-ex]} \BibitemShut {NoStop}%
\bibitem [{\citenamefont {Aad}\ \emph {et~al.}(2021{\natexlab{f}})\citenamefont
  {Aad} \emph {et~al.}}]{ATLAS:2021mla}%
  \BibitemOpen
  \bibfield  {author} {\bibinfo {author} {\bibfnamefont {G.}~\bibnamefont
  {Aad}} \emph {et~al.} (\bibinfo {collaboration} {ATLAS}),\ }\href {\doibase
  10.1103/PhysRevLett.127.141801} {\bibfield  {journal} {\bibinfo  {journal}
  {Phys. Rev. Lett.}\ }\textbf {\bibinfo {volume} {127}},\ \bibinfo {pages}
  {141801} (\bibinfo {year} {2021}{\natexlab{f}})},\ \Eprint
  {http://arxiv.org/abs/2105.13847} {arXiv:2105.13847 [hep-ex]} \BibitemShut
  {NoStop}%
\bibitem [{\citenamefont {Sirunyan}\ \emph
  {et~al.}(2019{\natexlab{d}})\citenamefont {Sirunyan} \emph
  {et~al.}}]{CMS:2018yiq}%
  \BibitemOpen
  \bibfield  {author} {\bibinfo {author} {\bibfnamefont {A.~M.}\ \bibnamefont
  {Sirunyan}} \emph {et~al.} (\bibinfo {collaboration} {CMS}),\ }\href
  {\doibase 10.1016/j.physletb.2019.05.046} {\bibfield  {journal} {\bibinfo
  {journal} {Phys. Lett. B}\ }\textbf {\bibinfo {volume} {795}},\ \bibinfo
  {pages} {76} (\bibinfo {year} {2019}{\natexlab{d}})},\ \Eprint
  {http://arxiv.org/abs/1811.10151} {arXiv:1811.10151 [hep-ex]} \BibitemShut
  {NoStop}%
\bibitem [{\citenamefont {Sirunyan}\ \emph
  {et~al.}(2021{\natexlab{a}})\citenamefont {Sirunyan} \emph
  {et~al.}}]{CMS:2020wzx}%
  \BibitemOpen
  \bibfield  {author} {\bibinfo {author} {\bibfnamefont {A.~M.}\ \bibnamefont
  {Sirunyan}} \emph {et~al.} (\bibinfo {collaboration} {CMS}),\ }\href
  {\doibase 10.1016/j.physletb.2021.136446} {\bibfield  {journal} {\bibinfo
  {journal} {Phys. Lett. B}\ }\textbf {\bibinfo {volume} {819}},\ \bibinfo
  {pages} {136446} (\bibinfo {year} {2021}{\natexlab{a}})},\ \Eprint
  {http://arxiv.org/abs/2012.04178} {arXiv:2012.04178 [hep-ex]} \BibitemShut
  {NoStop}%
\bibitem [{\citenamefont {Aad}\ \emph {et~al.}(2021{\natexlab{g}})\citenamefont
  {Aad} \emph {et~al.}}]{ATLAS:2021jyv}%
  \BibitemOpen
  \bibfield  {author} {\bibinfo {author} {\bibfnamefont {G.}~\bibnamefont
  {Aad}} \emph {et~al.} (\bibinfo {collaboration} {ATLAS}),\ }\href {\doibase
  10.1103/PhysRevD.104.112005} {\bibfield  {journal} {\bibinfo  {journal}
  {Phys. Rev. D}\ }\textbf {\bibinfo {volume} {104}},\ \bibinfo {pages}
  {112005} (\bibinfo {year} {2021}{\natexlab{g}})},\ \Eprint
  {http://arxiv.org/abs/2108.07665} {arXiv:2108.07665 [hep-ex]} \BibitemShut
  {NoStop}%
\bibitem [{\citenamefont {Tumasyan}\ \emph
  {et~al.}(2021{\natexlab{a}})\citenamefont {Tumasyan} \emph
  {et~al.}}]{CMS:2021far}%
  \BibitemOpen
  \bibfield  {author} {\bibinfo {author} {\bibfnamefont {A.}~\bibnamefont
  {Tumasyan}} \emph {et~al.} (\bibinfo {collaboration} {CMS}),\ }\href
  {\doibase 10.1007/JHEP11(2021)153} {\bibfield  {journal} {\bibinfo  {journal}
  {JHEP}\ }\textbf {\bibinfo {volume} {11}},\ \bibinfo {pages} {153} (\bibinfo
  {year} {2021}{\natexlab{a}})},\ \Eprint {http://arxiv.org/abs/2107.13021}
  {arXiv:2107.13021 [hep-ex]} \BibitemShut {NoStop}%
\bibitem [{\citenamefont {Porod}\ \emph {et~al.}(2001)\citenamefont {Porod},
  \citenamefont {Hirsch}, \citenamefont {Romao},\ and\ \citenamefont
  {Valle}}]{Porod:2000hv}%
  \BibitemOpen
  \bibfield  {author} {\bibinfo {author} {\bibfnamefont {W.}~\bibnamefont
  {Porod}}, \bibinfo {author} {\bibfnamefont {M.}~\bibnamefont {Hirsch}},
  \bibinfo {author} {\bibfnamefont {J.}~\bibnamefont {Romao}}, \ and\ \bibinfo
  {author} {\bibfnamefont {J.~W.~F.}\ \bibnamefont {Valle}},\ }\href {\doibase
  10.1103/PhysRevD.63.115004} {\bibfield  {journal} {\bibinfo  {journal} {Phys.
  Rev. D}\ }\textbf {\bibinfo {volume} {63}},\ \bibinfo {pages} {115004}
  (\bibinfo {year} {2001})},\ \Eprint {http://arxiv.org/abs/hep-ph/0011248}
  {arXiv:hep-ph/0011248} \BibitemShut {NoStop}%
\bibitem [{\citenamefont {Kohda}\ \emph {et~al.}(2013)\citenamefont {Kohda},
  \citenamefont {Sugiyama},\ and\ \citenamefont {Tsumura}}]{Kohda:2012sr}%
  \BibitemOpen
  \bibfield  {author} {\bibinfo {author} {\bibfnamefont {M.}~\bibnamefont
  {Kohda}}, \bibinfo {author} {\bibfnamefont {H.}~\bibnamefont {Sugiyama}}, \
  and\ \bibinfo {author} {\bibfnamefont {K.}~\bibnamefont {Tsumura}},\ }\href
  {\doibase 10.1016/j.physletb.2012.12.048} {\bibfield  {journal} {\bibinfo
  {journal} {Phys. Lett. B}\ }\textbf {\bibinfo {volume} {718}},\ \bibinfo
  {pages} {1436} (\bibinfo {year} {2013})},\ \Eprint
  {http://arxiv.org/abs/1210.5622} {arXiv:1210.5622 [hep-ph]} \BibitemShut
  {NoStop}%
\bibitem [{\citenamefont {Babu}\ \emph
  {et~al.}(2020{\natexlab{a}})\citenamefont {Babu}, \citenamefont {Dev},
  \citenamefont {Jana},\ and\ \citenamefont {Sui}}]{Babu:2019vff}%
  \BibitemOpen
  \bibfield  {author} {\bibinfo {author} {\bibfnamefont {K.~S.}\ \bibnamefont
  {Babu}}, \bibinfo {author} {\bibfnamefont {P.~S.}\ \bibnamefont {Dev}},
  \bibinfo {author} {\bibfnamefont {S.}~\bibnamefont {Jana}}, \ and\ \bibinfo
  {author} {\bibfnamefont {Y.}~\bibnamefont {Sui}},\ }\href {\doibase
  10.1103/PhysRevLett.124.041805} {\bibfield  {journal} {\bibinfo  {journal}
  {Phys. Rev. Lett.}\ }\textbf {\bibinfo {volume} {124}},\ \bibinfo {pages}
  {041805} (\bibinfo {year} {2020}{\natexlab{a}})},\ \Eprint
  {http://arxiv.org/abs/1908.02779} {arXiv:1908.02779 [hep-ph]} \BibitemShut
  {NoStop}%
\bibitem [{\citenamefont {Babu}\ \emph
  {et~al.}(2020{\natexlab{b}})\citenamefont {Babu}, \citenamefont {Dev},
  \citenamefont {Jana},\ and\ \citenamefont {Thapa}}]{Babu:2019mfe}%
  \BibitemOpen
  \bibfield  {author} {\bibinfo {author} {\bibfnamefont {K.~S.}\ \bibnamefont
  {Babu}}, \bibinfo {author} {\bibfnamefont {P.~S.~B.}\ \bibnamefont {Dev}},
  \bibinfo {author} {\bibfnamefont {S.}~\bibnamefont {Jana}}, \ and\ \bibinfo
  {author} {\bibfnamefont {A.}~\bibnamefont {Thapa}},\ }\href {\doibase
  10.1007/JHEP03(2020)006} {\bibfield  {journal} {\bibinfo  {journal} {JHEP}\
  }\textbf {\bibinfo {volume} {03}},\ \bibinfo {pages} {006} (\bibinfo {year}
  {2020}{\natexlab{b}})},\ \Eprint {http://arxiv.org/abs/1907.09498}
  {arXiv:1907.09498 [hep-ph]} \BibitemShut {NoStop}%
\bibitem [{\citenamefont {Zee}(1980)}]{Zee:1980ai}%
  \BibitemOpen
  \bibfield  {author} {\bibinfo {author} {\bibfnamefont {A.}~\bibnamefont
  {Zee}},\ }\href {\doibase 10.1016/0370-2693(80)90349-4,
  10.1016/0370-2693(80)90193-8} {\bibfield  {journal} {\bibinfo  {journal}
  {Phys. Lett.}\ }\textbf {\bibinfo {volume} {93B}},\ \bibinfo {pages} {389}
  (\bibinfo {year} {1980})},\ \bibinfo {note} {[Erratum: Phys.
  Lett.95B,461(1980)]}\BibitemShut {NoStop}%
\bibitem [{\citenamefont {Zee}(1985)}]{Zee:1985rj}%
  \BibitemOpen
  \bibfield  {author} {\bibinfo {author} {\bibfnamefont {A.}~\bibnamefont
  {Zee}},\ }\href {\doibase 10.1016/0370-2693(85)90625-2} {\bibfield  {journal}
  {\bibinfo  {journal} {Phys. Lett. B}\ }\textbf {\bibinfo {volume} {161}},\
  \bibinfo {pages} {141} (\bibinfo {year} {1985})}\BibitemShut {NoStop}%
\bibitem [{\citenamefont {Zee}(1986)}]{Zee:1985id}%
  \BibitemOpen
  \bibfield  {author} {\bibinfo {author} {\bibfnamefont {A.}~\bibnamefont
  {Zee}},\ }\href {\doibase 10.1016/0550-3213(86)90475-X} {\bibfield  {journal}
  {\bibinfo  {journal} {Nucl. Phys.}\ }\textbf {\bibinfo {volume} {B264}},\
  \bibinfo {pages} {99} (\bibinfo {year} {1986})}\BibitemShut {NoStop}%
\bibitem [{\citenamefont {Babu}(1988)}]{Babu:1988ki}%
  \BibitemOpen
  \bibfield  {author} {\bibinfo {author} {\bibfnamefont {K.~S.}\ \bibnamefont
  {Babu}},\ }\href {\doibase 10.1016/0370-2693(88)91584-5} {\bibfield
  {journal} {\bibinfo  {journal} {Phys. Lett.}\ }\textbf {\bibinfo {volume}
  {B203}},\ \bibinfo {pages} {132} (\bibinfo {year} {1988})}\BibitemShut
  {NoStop}%
\bibitem [{\citenamefont {Babu}\ \emph {et~al.}(1989)\citenamefont {Babu},
  \citenamefont {Ma},\ and\ \citenamefont {Pantaleone}}]{Babu:1988wk}%
  \BibitemOpen
  \bibfield  {author} {\bibinfo {author} {\bibfnamefont {K.~S.}\ \bibnamefont
  {Babu}}, \bibinfo {author} {\bibfnamefont {E.}~\bibnamefont {Ma}}, \ and\
  \bibinfo {author} {\bibfnamefont {J.~T.}\ \bibnamefont {Pantaleone}},\ }\href
  {\doibase 10.1016/0370-2693(89)91425-1} {\bibfield  {journal} {\bibinfo
  {journal} {Phys. Lett.}\ }\textbf {\bibinfo {volume} {B218}},\ \bibinfo
  {pages} {233} (\bibinfo {year} {1989})}\BibitemShut {NoStop}%
\bibitem [{\citenamefont {Babu}\ and\ \citenamefont {Ma}(1988)}]{Babu:1988ig}%
  \BibitemOpen
  \bibfield  {author} {\bibinfo {author} {\bibfnamefont {K.~S.}\ \bibnamefont
  {Babu}}\ and\ \bibinfo {author} {\bibfnamefont {E.}~\bibnamefont {Ma}},\
  }\href {\doibase 10.1103/PhysRevLett.61.674} {\bibfield  {journal} {\bibinfo
  {journal} {Phys. Rev. Lett.}\ }\textbf {\bibinfo {volume} {61}},\ \bibinfo
  {pages} {674} (\bibinfo {year} {1988})}\BibitemShut {NoStop}%
\bibitem [{\citenamefont {Fileviez~Perez}\ \emph {et~al.}(2011)\citenamefont
  {Fileviez~Perez}, \citenamefont {Han}, \citenamefont {Spinner},\ and\
  \citenamefont {Trenkel}}]{FileviezPerez:2010ch}%
  \BibitemOpen
  \bibfield  {author} {\bibinfo {author} {\bibfnamefont {P.}~\bibnamefont
  {Fileviez~Perez}}, \bibinfo {author} {\bibfnamefont {T.}~\bibnamefont {Han}},
  \bibinfo {author} {\bibfnamefont {S.}~\bibnamefont {Spinner}}, \ and\
  \bibinfo {author} {\bibfnamefont {M.~K.}\ \bibnamefont {Trenkel}},\ }\href
  {\doibase 10.1007/JHEP01(2011)046} {\bibfield  {journal} {\bibinfo  {journal}
  {JHEP}\ }\textbf {\bibinfo {volume} {01}},\ \bibinfo {pages} {046} (\bibinfo
  {year} {2011})},\ \Eprint {http://arxiv.org/abs/1010.5802} {arXiv:1010.5802
  [hep-ph]} \BibitemShut {NoStop}%
\bibitem [{\citenamefont {Gunion}\ \emph {et~al.}(1996)\citenamefont {Gunion},
  \citenamefont {Loomis},\ and\ \citenamefont {Pitts}}]{Gunion:1996pq}%
  \BibitemOpen
  \bibfield  {author} {\bibinfo {author} {\bibfnamefont {J.~F.}\ \bibnamefont
  {Gunion}}, \bibinfo {author} {\bibfnamefont {C.}~\bibnamefont {Loomis}}, \
  and\ \bibinfo {author} {\bibfnamefont {K.~T.}\ \bibnamefont {Pitts}},\
  }\href@noop {} {\bibfield  {journal} {\bibinfo  {journal} {eConf}\ }\textbf
  {\bibinfo {volume} {C960625}},\ \bibinfo {pages} {LTH096} (\bibinfo {year}
  {1996})},\ \Eprint {http://arxiv.org/abs/hep-ph/9610237}
  {arXiv:hep-ph/9610237} \BibitemShut {NoStop}%
\bibitem [{\citenamefont {Alcaide}\ \emph {et~al.}(2018)\citenamefont
  {Alcaide}, \citenamefont {Chala},\ and\ \citenamefont
  {Santamaria}}]{Alcaide:2017dcx}%
  \BibitemOpen
  \bibfield  {author} {\bibinfo {author} {\bibfnamefont {J.}~\bibnamefont
  {Alcaide}}, \bibinfo {author} {\bibfnamefont {M.}~\bibnamefont {Chala}}, \
  and\ \bibinfo {author} {\bibfnamefont {A.}~\bibnamefont {Santamaria}},\
  }\href {\doibase 10.1016/j.physletb.2018.02.001} {\bibfield  {journal}
  {\bibinfo  {journal} {Phys. Lett. B}\ }\textbf {\bibinfo {volume} {779}},\
  \bibinfo {pages} {107} (\bibinfo {year} {2018})},\ \Eprint
  {http://arxiv.org/abs/1710.05885} {arXiv:1710.05885 [hep-ph]} \BibitemShut
  {NoStop}%
\bibitem [{\citenamefont {Ruiz}(2022)}]{Ruiz:2022sct}%
  \BibitemOpen
  \bibfield  {author} {\bibinfo {author} {\bibfnamefont {R.}~\bibnamefont
  {Ruiz}},\ }\href@noop {} {\  (\bibinfo {year} {2022})},\ \Eprint
  {http://arxiv.org/abs/2206.14833} {arXiv:2206.14833 [hep-ph]} \BibitemShut
  {NoStop}%
\bibitem [{\citenamefont {Nebot}\ \emph {et~al.}(2008)\citenamefont {Nebot},
  \citenamefont {Oliver}, \citenamefont {Palao},\ and\ \citenamefont
  {Santamaria}}]{Nebot:2007bc}%
  \BibitemOpen
  \bibfield  {author} {\bibinfo {author} {\bibfnamefont {M.}~\bibnamefont
  {Nebot}}, \bibinfo {author} {\bibfnamefont {J.~F.}\ \bibnamefont {Oliver}},
  \bibinfo {author} {\bibfnamefont {D.}~\bibnamefont {Palao}}, \ and\ \bibinfo
  {author} {\bibfnamefont {A.}~\bibnamefont {Santamaria}},\ }\href {\doibase
  10.1103/PhysRevD.77.093013} {\bibfield  {journal} {\bibinfo  {journal} {Phys.
  Rev. D}\ }\textbf {\bibinfo {volume} {77}},\ \bibinfo {pages} {093013}
  (\bibinfo {year} {2008})},\ \Eprint {http://arxiv.org/abs/0711.0483}
  {arXiv:0711.0483 [hep-ph]} \BibitemShut {NoStop}%
\bibitem [{\citenamefont {Ohlsson}\ \emph {et~al.}(2009)\citenamefont
  {Ohlsson}, \citenamefont {Schwetz},\ and\ \citenamefont
  {Zhang}}]{Ohlsson:2009vk}%
  \BibitemOpen
  \bibfield  {author} {\bibinfo {author} {\bibfnamefont {T.}~\bibnamefont
  {Ohlsson}}, \bibinfo {author} {\bibfnamefont {T.}~\bibnamefont {Schwetz}}, \
  and\ \bibinfo {author} {\bibfnamefont {H.}~\bibnamefont {Zhang}},\ }\href
  {\doibase 10.1016/j.physletb.2009.10.025} {\bibfield  {journal} {\bibinfo
  {journal} {Phys. Lett. B}\ }\textbf {\bibinfo {volume} {681}},\ \bibinfo
  {pages} {269} (\bibinfo {year} {2009})},\ \Eprint
  {http://arxiv.org/abs/0909.0455} {arXiv:0909.0455 [hep-ph]} \BibitemShut
  {NoStop}%
\bibitem [{\citenamefont {Schmidt}\ \emph {et~al.}(2014)\citenamefont
  {Schmidt}, \citenamefont {Schwetz},\ and\ \citenamefont
  {Zhang}}]{Schmidt:2014zoa}%
  \BibitemOpen
  \bibfield  {author} {\bibinfo {author} {\bibfnamefont {D.}~\bibnamefont
  {Schmidt}}, \bibinfo {author} {\bibfnamefont {T.}~\bibnamefont {Schwetz}}, \
  and\ \bibinfo {author} {\bibfnamefont {H.}~\bibnamefont {Zhang}},\ }\href
  {\doibase 10.1016/j.nuclphysb.2014.05.024} {\bibfield  {journal} {\bibinfo
  {journal} {Nucl. Phys. B}\ }\textbf {\bibinfo {volume} {885}},\ \bibinfo
  {pages} {524} (\bibinfo {year} {2014})},\ \Eprint
  {http://arxiv.org/abs/1402.2251} {arXiv:1402.2251 [hep-ph]} \BibitemShut
  {NoStop}%
\bibitem [{\citenamefont {Long}\ and\ \citenamefont
  {Vien}(2014)}]{Long:2014fja}%
  \BibitemOpen
  \bibfield  {author} {\bibinfo {author} {\bibfnamefont {H.~N.}\ \bibnamefont
  {Long}}\ and\ \bibinfo {author} {\bibfnamefont {V.~V.}\ \bibnamefont
  {Vien}},\ }\href {\doibase 10.1142/S0217751X14500729} {\bibfield  {journal}
  {\bibinfo  {journal} {Int. J. Mod. Phys. A}\ }\textbf {\bibinfo {volume}
  {29}},\ \bibinfo {pages} {1450072} (\bibinfo {year} {2014})},\ \Eprint
  {http://arxiv.org/abs/1405.1622} {arXiv:1405.1622 [hep-ph]} \BibitemShut
  {NoStop}%
\bibitem [{\citenamefont {Herrero-Garcia}\ \emph {et~al.}(2014)\citenamefont
  {Herrero-Garcia}, \citenamefont {Nebot}, \citenamefont {Rius},\ and\
  \citenamefont {Santamaria}}]{Herrero-Garcia:2014hfa}%
  \BibitemOpen
  \bibfield  {author} {\bibinfo {author} {\bibfnamefont {J.}~\bibnamefont
  {Herrero-Garcia}}, \bibinfo {author} {\bibfnamefont {M.}~\bibnamefont
  {Nebot}}, \bibinfo {author} {\bibfnamefont {N.}~\bibnamefont {Rius}}, \ and\
  \bibinfo {author} {\bibfnamefont {A.}~\bibnamefont {Santamaria}},\ }\href
  {\doibase 10.1016/j.nuclphysb.2014.06.001} {\bibfield  {journal} {\bibinfo
  {journal} {Nucl. Phys. B}\ }\textbf {\bibinfo {volume} {885}},\ \bibinfo
  {pages} {542} (\bibinfo {year} {2014})},\ \Eprint
  {http://arxiv.org/abs/1402.4491} {arXiv:1402.4491 [hep-ph]} \BibitemShut
  {NoStop}%
\bibitem [{\citenamefont {{ATLAS Collaboration}}(2022)}]{ATLAS:2022yzd}%
  \BibitemOpen
  \bibfield  {author} {\bibinfo {author} {\bibnamefont {{ATLAS
  Collaboration}}},\ }\href {http://cds.cern.ch/record/2805214} {\  (\bibinfo
  {year} {2022})},\ \bibinfo {note} {{ATLAS-CONF-2022-010}}\BibitemShut
  {NoStop}%
\bibitem [{\citenamefont {Senjanovic}\ and\ \citenamefont
  {Mohapatra}(1975)}]{Senjanovic:1975rk}%
  \BibitemOpen
  \bibfield  {author} {\bibinfo {author} {\bibfnamefont {G.}~\bibnamefont
  {Senjanovic}}\ and\ \bibinfo {author} {\bibfnamefont {R.~N.}\ \bibnamefont
  {Mohapatra}},\ }\href {\doibase 10.1103/PhysRevD.12.1502} {\bibfield
  {journal} {\bibinfo  {journal} {Phys. Rev. D}\ }\textbf {\bibinfo {volume}
  {12}},\ \bibinfo {pages} {1502} (\bibinfo {year} {1975})}\BibitemShut
  {NoStop}%
\bibitem [{\citenamefont {Ruiz}(2017)}]{Ruiz:2017nip}%
  \BibitemOpen
  \bibfield  {author} {\bibinfo {author} {\bibfnamefont {R.}~\bibnamefont
  {Ruiz}},\ }\href {\doibase 10.1140/epjc/s10052-017-4950-2} {\bibfield
  {journal} {\bibinfo  {journal} {Eur. Phys. J. C}\ }\textbf {\bibinfo {volume}
  {77}},\ \bibinfo {pages} {375} (\bibinfo {year} {2017})},\ \Eprint
  {http://arxiv.org/abs/1703.04669} {arXiv:1703.04669 [hep-ph]} \BibitemShut
  {NoStop}%
\bibitem [{\citenamefont {Nemev\v{s}ek}\ \emph {et~al.}(2018)\citenamefont
  {Nemev\v{s}ek}, \citenamefont {Nesti},\ and\ \citenamefont
  {Popara}}]{Nemevsek:2018bbt}%
  \BibitemOpen
  \bibfield  {author} {\bibinfo {author} {\bibfnamefont {M.}~\bibnamefont
  {Nemev\v{s}ek}}, \bibinfo {author} {\bibfnamefont {F.}~\bibnamefont {Nesti}},
  \ and\ \bibinfo {author} {\bibfnamefont {G.}~\bibnamefont {Popara}},\ }\href
  {\doibase 10.1103/PhysRevD.97.115018} {\bibfield  {journal} {\bibinfo
  {journal} {Phys. Rev. D}\ }\textbf {\bibinfo {volume} {97}},\ \bibinfo
  {pages} {115018} (\bibinfo {year} {2018})},\ \Eprint
  {http://arxiv.org/abs/1801.05813} {arXiv:1801.05813 [hep-ph]} \BibitemShut
  {NoStop}%
\bibitem [{\citenamefont {Ferrari}\ \emph {et~al.}(2000)\citenamefont
  {Ferrari}, \citenamefont {Collot}, \citenamefont {Andrieux}, \citenamefont
  {Belhorma}, \citenamefont {de~Saintignon}, \citenamefont {Hostachy},
  \citenamefont {Martin},\ and\ \citenamefont {Wielers}}]{Ferrari:2000sp}%
  \BibitemOpen
  \bibfield  {author} {\bibinfo {author} {\bibfnamefont {A.}~\bibnamefont
  {Ferrari}}, \bibinfo {author} {\bibfnamefont {J.}~\bibnamefont {Collot}},
  \bibinfo {author} {\bibfnamefont {M.-L.}\ \bibnamefont {Andrieux}}, \bibinfo
  {author} {\bibfnamefont {B.}~\bibnamefont {Belhorma}}, \bibinfo {author}
  {\bibfnamefont {P.}~\bibnamefont {de~Saintignon}}, \bibinfo {author}
  {\bibfnamefont {J.-Y.}\ \bibnamefont {Hostachy}}, \bibinfo {author}
  {\bibfnamefont {P.}~\bibnamefont {Martin}}, \ and\ \bibinfo {author}
  {\bibfnamefont {M.}~\bibnamefont {Wielers}},\ }\href {\doibase
  10.1103/PhysRevD.62.013001} {\bibfield  {journal} {\bibinfo  {journal} {Phys.
  Rev. D}\ }\textbf {\bibinfo {volume} {62}},\ \bibinfo {pages} {013001}
  (\bibinfo {year} {2000})}\BibitemShut {NoStop}%
\bibitem [{\citenamefont {Mitra}\ \emph {et~al.}(2016)\citenamefont {Mitra},
  \citenamefont {Ruiz}, \citenamefont {Scott},\ and\ \citenamefont
  {Spannowsky}}]{Mitra:2016kov}%
  \BibitemOpen
  \bibfield  {author} {\bibinfo {author} {\bibfnamefont {M.}~\bibnamefont
  {Mitra}}, \bibinfo {author} {\bibfnamefont {R.}~\bibnamefont {Ruiz}},
  \bibinfo {author} {\bibfnamefont {D.~J.}\ \bibnamefont {Scott}}, \ and\
  \bibinfo {author} {\bibfnamefont {M.}~\bibnamefont {Spannowsky}},\ }\href
  {\doibase 10.1103/PhysRevD.94.095016} {\bibfield  {journal} {\bibinfo
  {journal} {Phys. Rev. D}\ }\textbf {\bibinfo {volume} {94}},\ \bibinfo
  {pages} {095016} (\bibinfo {year} {2016})},\ \Eprint
  {http://arxiv.org/abs/1607.03504} {arXiv:1607.03504 [hep-ph]} \BibitemShut
  {NoStop}%
\bibitem [{\citenamefont {Mattelaer}\ \emph {et~al.}(2016)\citenamefont
  {Mattelaer}, \citenamefont {Mitra},\ and\ \citenamefont
  {Ruiz}}]{Mattelaer:2016ynf}%
  \BibitemOpen
  \bibfield  {author} {\bibinfo {author} {\bibfnamefont {O.}~\bibnamefont
  {Mattelaer}}, \bibinfo {author} {\bibfnamefont {M.}~\bibnamefont {Mitra}}, \
  and\ \bibinfo {author} {\bibfnamefont {R.}~\bibnamefont {Ruiz}},\ }\href@noop
  {} {\  (\bibinfo {year} {2016})},\ \Eprint {http://arxiv.org/abs/1610.08985}
  {arXiv:1610.08985 [hep-ph]} \BibitemShut {NoStop}%
\bibitem [{\citenamefont {Sirunyan}\ \emph
  {et~al.}(2018{\natexlab{c}})\citenamefont {Sirunyan} \emph
  {et~al.}}]{CMS:2018agk}%
  \BibitemOpen
  \bibfield  {author} {\bibinfo {author} {\bibfnamefont {A.~M.}\ \bibnamefont
  {Sirunyan}} \emph {et~al.} (\bibinfo {collaboration} {CMS}),\ }\href
  {\doibase 10.1007/JHEP05(2018)148} {\bibfield  {journal} {\bibinfo  {journal}
  {JHEP}\ }\textbf {\bibinfo {volume} {05}},\ \bibinfo {pages} {148} (\bibinfo
  {year} {2018}{\natexlab{c}})},\ \Eprint {http://arxiv.org/abs/1803.11116}
  {arXiv:1803.11116 [hep-ex]} \BibitemShut {NoStop}%
\bibitem [{\citenamefont {Aaboud}\ \emph
  {et~al.}(2019{\natexlab{b}})\citenamefont {Aaboud} \emph
  {et~al.}}]{ATLAS:2019isd}%
  \BibitemOpen
  \bibfield  {author} {\bibinfo {author} {\bibfnamefont {M.}~\bibnamefont
  {Aaboud}} \emph {et~al.} (\bibinfo {collaboration} {ATLAS}),\ }\href
  {\doibase 10.1016/j.physletb.2019.134942} {\bibfield  {journal} {\bibinfo
  {journal} {Phys. Lett. B}\ }\textbf {\bibinfo {volume} {798}},\ \bibinfo
  {pages} {134942} (\bibinfo {year} {2019}{\natexlab{b}})},\ \Eprint
  {http://arxiv.org/abs/1904.12679} {arXiv:1904.12679 [hep-ex]} \BibitemShut
  {NoStop}%
\bibitem [{\citenamefont {Aad}\ \emph {et~al.}(2019{\natexlab{b}})\citenamefont
  {Aad} \emph {et~al.}}]{ATLAS:2019erb}%
  \BibitemOpen
  \bibfield  {author} {\bibinfo {author} {\bibfnamefont {G.}~\bibnamefont
  {Aad}} \emph {et~al.} (\bibinfo {collaboration} {ATLAS}),\ }\href {\doibase
  10.1016/j.physletb.2019.07.016} {\bibfield  {journal} {\bibinfo  {journal}
  {Phys. Lett. B}\ }\textbf {\bibinfo {volume} {796}},\ \bibinfo {pages} {68}
  (\bibinfo {year} {2019}{\natexlab{b}})},\ \Eprint
  {http://arxiv.org/abs/1903.06248} {arXiv:1903.06248 [hep-ex]} \BibitemShut
  {NoStop}%
\bibitem [{\citenamefont {Sirunyan}\ \emph
  {et~al.}(2020{\natexlab{b}})\citenamefont {Sirunyan} \emph
  {et~al.}}]{CMS:2019buh}%
  \BibitemOpen
  \bibfield  {author} {\bibinfo {author} {\bibfnamefont {A.~M.}\ \bibnamefont
  {Sirunyan}} \emph {et~al.} (\bibinfo {collaboration} {CMS}),\ }\href
  {\doibase 10.1103/PhysRevLett.124.131802} {\bibfield  {journal} {\bibinfo
  {journal} {Phys. Rev. Lett.}\ }\textbf {\bibinfo {volume} {124}},\ \bibinfo
  {pages} {131802} (\bibinfo {year} {2020}{\natexlab{b}})},\ \Eprint
  {http://arxiv.org/abs/1912.04776} {arXiv:1912.04776 [hep-ex]} \BibitemShut
  {NoStop}%
\bibitem [{\citenamefont {Tumasyan}\ \emph
  {et~al.}(2021{\natexlab{b}})\citenamefont {Tumasyan} \emph
  {et~al.}}]{CMS:2021dzb}%
  \BibitemOpen
  \bibfield  {author} {\bibinfo {author} {\bibfnamefont {A.}~\bibnamefont
  {Tumasyan}} \emph {et~al.} (\bibinfo {collaboration} {CMS}),\ }\href@noop {}
  {\  (\bibinfo {year} {2021}{\natexlab{b}})},\ \Eprint
  {http://arxiv.org/abs/2112.03949} {arXiv:2112.03949 [hep-ex]} \BibitemShut
  {NoStop}%
\bibitem [{\citenamefont {Tumasyan}\ \emph
  {et~al.}(2022{\natexlab{c}})\citenamefont {Tumasyan} \emph
  {et~al.}}]{CMS:2022yjm}%
  \BibitemOpen
  \bibfield  {author} {\bibinfo {author} {\bibfnamefont {A.}~\bibnamefont
  {Tumasyan}} \emph {et~al.} (\bibinfo {collaboration} {CMS}),\ }\href
  {\doibase 10.1007/JHEP07(2022)067} {\bibfield  {journal} {\bibinfo  {journal}
  {JHEP}\ }\textbf {\bibinfo {volume} {07}},\ \bibinfo {pages} {067} (\bibinfo
  {year} {2022}{\natexlab{c}})},\ \Eprint {http://arxiv.org/abs/2202.06075}
  {arXiv:2202.06075 [hep-ex]} \BibitemShut {NoStop}%
\bibitem [{\citenamefont {Das}\ \emph {et~al.}(2012)\citenamefont {Das},
  \citenamefont {Deppisch}, \citenamefont {Kittel},\ and\ \citenamefont
  {Valle}}]{Das:2012ii}%
  \BibitemOpen
  \bibfield  {author} {\bibinfo {author} {\bibfnamefont {S.~P.}\ \bibnamefont
  {Das}}, \bibinfo {author} {\bibfnamefont {F.~F.}\ \bibnamefont {Deppisch}},
  \bibinfo {author} {\bibfnamefont {O.}~\bibnamefont {Kittel}}, \ and\ \bibinfo
  {author} {\bibfnamefont {J.~W.~F.}\ \bibnamefont {Valle}},\ }\href {\doibase
  10.1103/PhysRevD.86.055006} {\bibfield  {journal} {\bibinfo  {journal} {Phys.
  Rev. D}\ }\textbf {\bibinfo {volume} {86}},\ \bibinfo {pages} {055006}
  (\bibinfo {year} {2012})},\ \Eprint {http://arxiv.org/abs/1206.0256}
  {arXiv:1206.0256 [hep-ph]} \BibitemShut {NoStop}%
\bibitem [{\citenamefont {Alioli}\ \emph {et~al.}(2017)\citenamefont {Alioli},
  \citenamefont {Cirigliano}, \citenamefont {Dekens}, \citenamefont
  {de~Vries},\ and\ \citenamefont {Mereghetti}}]{Alioli:2017ces}%
  \BibitemOpen
  \bibfield  {author} {\bibinfo {author} {\bibfnamefont {S.}~\bibnamefont
  {Alioli}}, \bibinfo {author} {\bibfnamefont {V.}~\bibnamefont {Cirigliano}},
  \bibinfo {author} {\bibfnamefont {W.}~\bibnamefont {Dekens}}, \bibinfo
  {author} {\bibfnamefont {J.}~\bibnamefont {de~Vries}}, \ and\ \bibinfo
  {author} {\bibfnamefont {E.}~\bibnamefont {Mereghetti}},\ }\href {\doibase
  10.1007/JHEP05(2017)086} {\bibfield  {journal} {\bibinfo  {journal} {JHEP}\
  }\textbf {\bibinfo {volume} {05}},\ \bibinfo {pages} {086} (\bibinfo {year}
  {2017})},\ \Eprint {http://arxiv.org/abs/1703.04751} {arXiv:1703.04751
  [hep-ph]} \BibitemShut {NoStop}%
\bibitem [{\citenamefont {Deppisch}\ \emph
  {et~al.}(2019{\natexlab{a}})\citenamefont {Deppisch}, \citenamefont
  {Kulkarni},\ and\ \citenamefont {Liu}}]{Deppisch:2019kvs}%
  \BibitemOpen
  \bibfield  {author} {\bibinfo {author} {\bibfnamefont {F.}~\bibnamefont
  {Deppisch}}, \bibinfo {author} {\bibfnamefont {S.}~\bibnamefont {Kulkarni}},
  \ and\ \bibinfo {author} {\bibfnamefont {W.}~\bibnamefont {Liu}},\ }\href
  {\doibase 10.1103/PhysRevD.100.035005} {\bibfield  {journal} {\bibinfo
  {journal} {Phys. Rev. D}\ }\textbf {\bibinfo {volume} {100}},\ \bibinfo
  {pages} {035005} (\bibinfo {year} {2019}{\natexlab{a}})},\ \Eprint
  {http://arxiv.org/abs/1905.11889} {arXiv:1905.11889 [hep-ph]} \BibitemShut
  {NoStop}%
\bibitem [{\citenamefont {Deppisch}\ \emph
  {et~al.}(2019{\natexlab{b}})\citenamefont {Deppisch}, \citenamefont
  {Kulkarni},\ and\ \citenamefont {Liu}}]{Deppisch:2019ldi}%
  \BibitemOpen
  \bibfield  {author} {\bibinfo {author} {\bibfnamefont {F.~F.}\ \bibnamefont
  {Deppisch}}, \bibinfo {author} {\bibfnamefont {S.}~\bibnamefont {Kulkarni}},
  \ and\ \bibinfo {author} {\bibfnamefont {W.}~\bibnamefont {Liu}},\ }\href
  {\doibase 10.1103/PhysRevD.100.115023} {\bibfield  {journal} {\bibinfo
  {journal} {Phys. Rev. D}\ }\textbf {\bibinfo {volume} {100}},\ \bibinfo
  {pages} {115023} (\bibinfo {year} {2019}{\natexlab{b}})},\ \Eprint
  {http://arxiv.org/abs/1908.11741} {arXiv:1908.11741 [hep-ph]} \BibitemShut
  {NoStop}%
\bibitem [{\citenamefont {Chiang}\ \emph {et~al.}(2019)\citenamefont {Chiang},
  \citenamefont {Cottin}, \citenamefont {Das},\ and\ \citenamefont
  {Mandal}}]{Chiang:2019ajm}%
  \BibitemOpen
  \bibfield  {author} {\bibinfo {author} {\bibfnamefont {C.-W.}\ \bibnamefont
  {Chiang}}, \bibinfo {author} {\bibfnamefont {G.}~\bibnamefont {Cottin}},
  \bibinfo {author} {\bibfnamefont {A.}~\bibnamefont {Das}}, \ and\ \bibinfo
  {author} {\bibfnamefont {S.}~\bibnamefont {Mandal}},\ }\href {\doibase
  10.1007/JHEP12(2019)070} {\bibfield  {journal} {\bibinfo  {journal} {JHEP}\
  }\textbf {\bibinfo {volume} {12}},\ \bibinfo {pages} {070} (\bibinfo {year}
  {2019})},\ \Eprint {http://arxiv.org/abs/1908.09838} {arXiv:1908.09838
  [hep-ph]} \BibitemShut {NoStop}%
\bibitem [{\citenamefont {Cottin}\ \emph {et~al.}(2021)\citenamefont {Cottin},
  \citenamefont {Helo}, \citenamefont {Hirsch}, \citenamefont {Titov},\ and\
  \citenamefont {Wang}}]{Cottin:2021lzz}%
  \BibitemOpen
  \bibfield  {author} {\bibinfo {author} {\bibfnamefont {G.}~\bibnamefont
  {Cottin}}, \bibinfo {author} {\bibfnamefont {J.~C.}\ \bibnamefont {Helo}},
  \bibinfo {author} {\bibfnamefont {M.}~\bibnamefont {Hirsch}}, \bibinfo
  {author} {\bibfnamefont {A.}~\bibnamefont {Titov}}, \ and\ \bibinfo {author}
  {\bibfnamefont {Z.~S.}\ \bibnamefont {Wang}},\ }\href {\doibase
  10.1007/JHEP09(2021)039} {\bibfield  {journal} {\bibinfo  {journal} {JHEP}\
  }\textbf {\bibinfo {volume} {09}},\ \bibinfo {pages} {039} (\bibinfo {year}
  {2021})},\ \Eprint {http://arxiv.org/abs/2105.13851} {arXiv:2105.13851
  [hep-ph]} \BibitemShut {NoStop}%
\bibitem [{\citenamefont {Beltr\'an}\ \emph {et~al.}(2022)\citenamefont
  {Beltr\'an}, \citenamefont {Cottin}, \citenamefont {Helo}, \citenamefont
  {Hirsch}, \citenamefont {Titov},\ and\ \citenamefont
  {Wang}}]{Beltran:2021hpq}%
  \BibitemOpen
  \bibfield  {author} {\bibinfo {author} {\bibfnamefont {R.}~\bibnamefont
  {Beltr\'an}}, \bibinfo {author} {\bibfnamefont {G.}~\bibnamefont {Cottin}},
  \bibinfo {author} {\bibfnamefont {J.~C.}\ \bibnamefont {Helo}}, \bibinfo
  {author} {\bibfnamefont {M.}~\bibnamefont {Hirsch}}, \bibinfo {author}
  {\bibfnamefont {A.}~\bibnamefont {Titov}}, \ and\ \bibinfo {author}
  {\bibfnamefont {Z.~S.}\ \bibnamefont {Wang}},\ }\href {\doibase
  10.1007/JHEP01(2022)044} {\bibfield  {journal} {\bibinfo  {journal} {JHEP}\
  }\textbf {\bibinfo {volume} {01}},\ \bibinfo {pages} {044} (\bibinfo {year}
  {2022})},\ \Eprint {http://arxiv.org/abs/2110.15096} {arXiv:2110.15096
  [hep-ph]} \BibitemShut {NoStop}%
\bibitem [{\citenamefont {Buarque~Franzosi}\ \emph {et~al.}(2022)\citenamefont
  {Buarque~Franzosi} \emph {et~al.}}]{Buarque:2021dji}%
  \BibitemOpen
  \bibfield  {author} {\bibinfo {author} {\bibfnamefont {D.}~\bibnamefont
  {Buarque~Franzosi}} \emph {et~al.},\ }\href {\doibase
  10.1016/j.revip.2022.100071} {\bibfield  {journal} {\bibinfo  {journal} {Rev.
  Phys.}\ }\textbf {\bibinfo {volume} {8}},\ \bibinfo {pages} {100071}
  (\bibinfo {year} {2022})},\ \Eprint {http://arxiv.org/abs/2106.01393}
  {arXiv:2106.01393 [hep-ph]} \BibitemShut {NoStop}%
\bibitem [{\citenamefont {Dekens}\ \emph {et~al.}(2021)\citenamefont {Dekens},
  \citenamefont {Andreoli}, \citenamefont {de~Vries}, \citenamefont
  {Mereghetti},\ and\ \citenamefont {Oosterhof}}]{Dekens:2021bro}%
  \BibitemOpen
  \bibfield  {author} {\bibinfo {author} {\bibfnamefont {W.}~\bibnamefont
  {Dekens}}, \bibinfo {author} {\bibfnamefont {L.}~\bibnamefont {Andreoli}},
  \bibinfo {author} {\bibfnamefont {J.}~\bibnamefont {de~Vries}}, \bibinfo
  {author} {\bibfnamefont {E.}~\bibnamefont {Mereghetti}}, \ and\ \bibinfo
  {author} {\bibfnamefont {F.}~\bibnamefont {Oosterhof}},\ }\href {\doibase
  10.1007/JHEP11(2021)127} {\bibfield  {journal} {\bibinfo  {journal} {JHEP}\
  }\textbf {\bibinfo {volume} {11}},\ \bibinfo {pages} {127} (\bibinfo {year}
  {2021})},\ \Eprint {http://arxiv.org/abs/2107.10852} {arXiv:2107.10852
  [hep-ph]} \BibitemShut {NoStop}%
\bibitem [{\citenamefont {Padhan}\ \emph {et~al.}(2022)\citenamefont {Padhan},
  \citenamefont {Mitra}, \citenamefont {Kulkarni},\ and\ \citenamefont
  {Deppisch}}]{Padhan:2022fak}%
  \BibitemOpen
  \bibfield  {author} {\bibinfo {author} {\bibfnamefont {R.}~\bibnamefont
  {Padhan}}, \bibinfo {author} {\bibfnamefont {M.}~\bibnamefont {Mitra}},
  \bibinfo {author} {\bibfnamefont {S.}~\bibnamefont {Kulkarni}}, \ and\
  \bibinfo {author} {\bibfnamefont {F.~F.}\ \bibnamefont {Deppisch}},\ }in\
  \href@noop {} {\emph {\bibinfo {booktitle} {{2022 Snowmass Summer Study}}}}\
  (\bibinfo {year} {2022})\ \Eprint {http://arxiv.org/abs/2203.06114}
  {arXiv:2203.06114 [hep-ph]} \BibitemShut {NoStop}%
\bibitem [{\citenamefont {Basso}\ \emph {et~al.}(2009)\citenamefont {Basso},
  \citenamefont {Belyaev}, \citenamefont {Moretti},\ and\ \citenamefont
  {Shepherd-Themistocleous}}]{Basso:2008iv}%
  \BibitemOpen
  \bibfield  {author} {\bibinfo {author} {\bibfnamefont {L.}~\bibnamefont
  {Basso}}, \bibinfo {author} {\bibfnamefont {A.}~\bibnamefont {Belyaev}},
  \bibinfo {author} {\bibfnamefont {S.}~\bibnamefont {Moretti}}, \ and\
  \bibinfo {author} {\bibfnamefont {C.~H.}\ \bibnamefont
  {Shepherd-Themistocleous}},\ }\href {\doibase 10.1103/PhysRevD.80.055030}
  {\bibfield  {journal} {\bibinfo  {journal} {Phys. Rev. D}\ }\textbf {\bibinfo
  {volume} {80}},\ \bibinfo {pages} {055030} (\bibinfo {year} {2009})},\
  \Eprint {http://arxiv.org/abs/0812.4313} {arXiv:0812.4313 [hep-ph]}
  \BibitemShut {NoStop}%
\bibitem [{\citenamefont {Fileviez~Perez}\ \emph {et~al.}(2009)\citenamefont
  {Fileviez~Perez}, \citenamefont {Han},\ and\ \citenamefont
  {Li}}]{FileviezPerez:2009hdc}%
  \BibitemOpen
  \bibfield  {author} {\bibinfo {author} {\bibfnamefont {P.}~\bibnamefont
  {Fileviez~Perez}}, \bibinfo {author} {\bibfnamefont {T.}~\bibnamefont {Han}},
  \ and\ \bibinfo {author} {\bibfnamefont {T.}~\bibnamefont {Li}},\ }\href
  {\doibase 10.1103/PhysRevD.80.073015} {\bibfield  {journal} {\bibinfo
  {journal} {Phys. Rev. D}\ }\textbf {\bibinfo {volume} {80}},\ \bibinfo
  {pages} {073015} (\bibinfo {year} {2009})},\ \Eprint
  {http://arxiv.org/abs/0907.4186} {arXiv:0907.4186 [hep-ph]} \BibitemShut
  {NoStop}%
\bibitem [{\citenamefont {Basso}\ \emph {et~al.}(2011)\citenamefont {Basso},
  \citenamefont {Belyaev}, \citenamefont {Moretti}, \citenamefont {Pruna},\
  and\ \citenamefont {Shepherd-Themistocleous}}]{Basso:2010pe}%
  \BibitemOpen
  \bibfield  {author} {\bibinfo {author} {\bibfnamefont {L.}~\bibnamefont
  {Basso}}, \bibinfo {author} {\bibfnamefont {A.}~\bibnamefont {Belyaev}},
  \bibinfo {author} {\bibfnamefont {S.}~\bibnamefont {Moretti}}, \bibinfo
  {author} {\bibfnamefont {G.~M.}\ \bibnamefont {Pruna}}, \ and\ \bibinfo
  {author} {\bibfnamefont {C.~H.}\ \bibnamefont {Shepherd-Themistocleous}},\
  }\href {\doibase 10.1140/epjc/s10052-011-1613-6} {\bibfield  {journal}
  {\bibinfo  {journal} {Eur. Phys. J. C}\ }\textbf {\bibinfo {volume} {71}},\
  \bibinfo {pages} {1613} (\bibinfo {year} {2011})},\ \Eprint
  {http://arxiv.org/abs/1002.3586} {arXiv:1002.3586 [hep-ph]} \BibitemShut
  {NoStop}%
\bibitem [{\citenamefont {Accomando}\ \emph {et~al.}(2016)\citenamefont
  {Accomando}, \citenamefont {Coriano}, \citenamefont {Delle~Rose},
  \citenamefont {Fiaschi}, \citenamefont {Marzo},\ and\ \citenamefont
  {Moretti}}]{Accomando:2016sge}%
  \BibitemOpen
  \bibfield  {author} {\bibinfo {author} {\bibfnamefont {E.}~\bibnamefont
  {Accomando}}, \bibinfo {author} {\bibfnamefont {C.}~\bibnamefont {Coriano}},
  \bibinfo {author} {\bibfnamefont {L.}~\bibnamefont {Delle~Rose}}, \bibinfo
  {author} {\bibfnamefont {J.}~\bibnamefont {Fiaschi}}, \bibinfo {author}
  {\bibfnamefont {C.}~\bibnamefont {Marzo}}, \ and\ \bibinfo {author}
  {\bibfnamefont {S.}~\bibnamefont {Moretti}},\ }\href {\doibase
  10.1007/JHEP07(2016)086} {\bibfield  {journal} {\bibinfo  {journal} {JHEP}\
  }\textbf {\bibinfo {volume} {07}},\ \bibinfo {pages} {086} (\bibinfo {year}
  {2016})},\ \Eprint {http://arxiv.org/abs/1605.02910} {arXiv:1605.02910
  [hep-ph]} \BibitemShut {NoStop}%
\bibitem [{\citenamefont {Accomando}\ \emph {et~al.}(2017)\citenamefont
  {Accomando}, \citenamefont {Fiaschi}, \citenamefont {Moretti},\ and\
  \citenamefont {Shepherd-Themistocleous}}]{Accomando:2017fmb}%
  \BibitemOpen
  \bibfield  {author} {\bibinfo {author} {\bibfnamefont {E.}~\bibnamefont
  {Accomando}}, \bibinfo {author} {\bibfnamefont {J.}~\bibnamefont {Fiaschi}},
  \bibinfo {author} {\bibfnamefont {S.}~\bibnamefont {Moretti}}, \ and\
  \bibinfo {author} {\bibfnamefont {C.~H.}\ \bibnamefont
  {Shepherd-Themistocleous}},\ }\href {\doibase 10.1103/PhysRevD.96.075019}
  {\bibfield  {journal} {\bibinfo  {journal} {Phys. Rev. D}\ }\textbf {\bibinfo
  {volume} {96}},\ \bibinfo {pages} {075019} (\bibinfo {year} {2017})},\
  \Eprint {http://arxiv.org/abs/1703.04360} {arXiv:1703.04360 [hep-ph]}
  \BibitemShut {NoStop}%
\bibitem [{\citenamefont {Accomando}\ \emph {et~al.}(2018)\citenamefont
  {Accomando}, \citenamefont {Delle~Rose}, \citenamefont {Moretti},
  \citenamefont {Olaiya},\ and\ \citenamefont
  {Shepherd-Themistocleous}}]{Accomando:2017qcs}%
  \BibitemOpen
  \bibfield  {author} {\bibinfo {author} {\bibfnamefont {E.}~\bibnamefont
  {Accomando}}, \bibinfo {author} {\bibfnamefont {L.}~\bibnamefont
  {Delle~Rose}}, \bibinfo {author} {\bibfnamefont {S.}~\bibnamefont {Moretti}},
  \bibinfo {author} {\bibfnamefont {E.}~\bibnamefont {Olaiya}}, \ and\ \bibinfo
  {author} {\bibfnamefont {C.~H.}\ \bibnamefont {Shepherd-Themistocleous}},\
  }\href {\doibase 10.1007/JHEP02(2018)109} {\bibfield  {journal} {\bibinfo
  {journal} {JHEP}\ }\textbf {\bibinfo {volume} {02}},\ \bibinfo {pages} {109}
  (\bibinfo {year} {2018})},\ \Eprint {http://arxiv.org/abs/1708.03650}
  {arXiv:1708.03650 [hep-ph]} \BibitemShut {NoStop}%
\bibitem [{\citenamefont {Fileviez~P\'erez}\ and\ \citenamefont
  {Plascencia}(2020)}]{FileviezPerez:2020cgn}%
  \BibitemOpen
  \bibfield  {author} {\bibinfo {author} {\bibfnamefont {P.}~\bibnamefont
  {Fileviez~P\'erez}}\ and\ \bibinfo {author} {\bibfnamefont {A.~D.}\
  \bibnamefont {Plascencia}},\ }\href {\doibase 10.1103/PhysRevD.102.015010}
  {\bibfield  {journal} {\bibinfo  {journal} {Phys. Rev. D}\ }\textbf {\bibinfo
  {volume} {102}},\ \bibinfo {pages} {015010} (\bibinfo {year} {2020})},\
  \Eprint {http://arxiv.org/abs/2005.04235} {arXiv:2005.04235 [hep-ph]}
  \BibitemShut {NoStop}%
\bibitem [{\citenamefont {Fileviez~Perez}\ and\ \citenamefont
  {Spinner}(2014)}]{FileviezPerez:2013fsv}%
  \BibitemOpen
  \bibfield  {author} {\bibinfo {author} {\bibfnamefont {P.}~\bibnamefont
  {Fileviez~Perez}}\ and\ \bibinfo {author} {\bibfnamefont {S.}~\bibnamefont
  {Spinner}},\ }\href {\doibase 10.1016/j.physletb.2013.12.022} {\bibfield
  {journal} {\bibinfo  {journal} {Phys. Lett. B}\ }\textbf {\bibinfo {volume}
  {728}},\ \bibinfo {pages} {489} (\bibinfo {year} {2014})},\ \Eprint
  {http://arxiv.org/abs/1308.0524} {arXiv:1308.0524 [hep-ph]} \BibitemShut
  {NoStop}%
\bibitem [{\citenamefont {Barger}\ \emph {et~al.}(2009)\citenamefont {Barger},
  \citenamefont {Fileviez~Perez},\ and\ \citenamefont
  {Spinner}}]{Barger:2008wn}%
  \BibitemOpen
  \bibfield  {author} {\bibinfo {author} {\bibfnamefont {V.}~\bibnamefont
  {Barger}}, \bibinfo {author} {\bibfnamefont {P.}~\bibnamefont
  {Fileviez~Perez}}, \ and\ \bibinfo {author} {\bibfnamefont {S.}~\bibnamefont
  {Spinner}},\ }\href {\doibase 10.1103/PhysRevLett.102.181802} {\bibfield
  {journal} {\bibinfo  {journal} {Phys. Rev. Lett.}\ }\textbf {\bibinfo
  {volume} {102}},\ \bibinfo {pages} {181802} (\bibinfo {year} {2009})},\
  \Eprint {http://arxiv.org/abs/0812.3661} {arXiv:0812.3661 [hep-ph]}
  \BibitemShut {NoStop}%
\bibitem [{\citenamefont {Fileviez~Perez}\ and\ \citenamefont
  {Spinner}(2012)}]{FileviezPerez:2012mj}%
  \BibitemOpen
  \bibfield  {author} {\bibinfo {author} {\bibfnamefont {P.}~\bibnamefont
  {Fileviez~Perez}}\ and\ \bibinfo {author} {\bibfnamefont {S.}~\bibnamefont
  {Spinner}},\ }\href {\doibase 10.1007/JHEP04(2012)118} {\bibfield  {journal}
  {\bibinfo  {journal} {JHEP}\ }\textbf {\bibinfo {volume} {04}},\ \bibinfo
  {pages} {118} (\bibinfo {year} {2012})},\ \Eprint
  {http://arxiv.org/abs/1201.5923} {arXiv:1201.5923 [hep-ph]} \BibitemShut
  {NoStop}%
\bibitem [{\citenamefont {Barger}\ \emph {et~al.}(1989)\citenamefont {Barger},
  \citenamefont {Giudice},\ and\ \citenamefont {Han}}]{Barger:1989rk}%
  \BibitemOpen
  \bibfield  {author} {\bibinfo {author} {\bibfnamefont {V.~D.}\ \bibnamefont
  {Barger}}, \bibinfo {author} {\bibfnamefont {G.~F.}\ \bibnamefont {Giudice}},
  \ and\ \bibinfo {author} {\bibfnamefont {T.}~\bibnamefont {Han}},\ }\href
  {\doibase 10.1103/PhysRevD.40.2987} {\bibfield  {journal} {\bibinfo
  {journal} {Phys. Rev. D}\ }\textbf {\bibinfo {volume} {40}},\ \bibinfo
  {pages} {2987} (\bibinfo {year} {1989})}\BibitemShut {NoStop}%
\bibitem [{\citenamefont {Acosta}\ \emph {et~al.}(2021)\citenamefont {Acosta}
  \emph {et~al.}}]{Alimena:2021mdu}%
  \BibitemOpen
  \bibfield  {author} {\bibinfo {author} {\bibfnamefont {D.}~\bibnamefont
  {Acosta}} \emph {et~al.},\ }\href@noop {} {\  (\bibinfo {year} {2021})},\
  \Eprint {http://arxiv.org/abs/2110.14675} {arXiv:2110.14675 [hep-ex]}
  \BibitemShut {NoStop}%
\bibitem [{\citenamefont {Tumasyan}\ \emph
  {et~al.}(2022{\natexlab{d}})\citenamefont {Tumasyan} \emph
  {et~al.}}]{CMS:2021kdm}%
  \BibitemOpen
  \bibfield  {author} {\bibinfo {author} {\bibfnamefont {A.}~\bibnamefont
  {Tumasyan}} \emph {et~al.} (\bibinfo {collaboration} {CMS}),\ }\href
  {\doibase 10.1140/epjc/s10052-022-10027-3} {\bibfield  {journal} {\bibinfo
  {journal} {Eur. Phys. J. C}\ }\textbf {\bibinfo {volume} {82}},\ \bibinfo
  {pages} {153} (\bibinfo {year} {2022}{\natexlab{d}})},\ \Eprint
  {http://arxiv.org/abs/2110.04809} {arXiv:2110.04809 [hep-ex]} \BibitemShut
  {NoStop}%
\bibitem [{\citenamefont {Sirunyan}\ \emph
  {et~al.}(2021{\natexlab{b}})\citenamefont {Sirunyan} \emph
  {et~al.}}]{CMS:2021tkn}%
  \BibitemOpen
  \bibfield  {author} {\bibinfo {author} {\bibfnamefont {A.~M.}\ \bibnamefont
  {Sirunyan}} \emph {et~al.} (\bibinfo {collaboration} {CMS}),\ }\href
  {\doibase 10.1103/PhysRevD.104.052011} {\bibfield  {journal} {\bibinfo
  {journal} {Phys. Rev. D}\ }\textbf {\bibinfo {volume} {104}},\ \bibinfo
  {pages} {052011} (\bibinfo {year} {2021}{\natexlab{b}})},\ \Eprint
  {http://arxiv.org/abs/2104.13474} {arXiv:2104.13474 [hep-ex]} \BibitemShut
  {NoStop}%
\bibitem [{\citenamefont {Sirunyan}\ \emph
  {et~al.}(2021{\natexlab{c}})\citenamefont {Sirunyan} \emph
  {et~al.}}]{CMS:2020iwv}%
  \BibitemOpen
  \bibfield  {author} {\bibinfo {author} {\bibfnamefont {A.~M.}\ \bibnamefont
  {Sirunyan}} \emph {et~al.} (\bibinfo {collaboration} {CMS}),\ }\href
  {\doibase 10.1103/PhysRevD.104.012015} {\bibfield  {journal} {\bibinfo
  {journal} {Phys. Rev. D}\ }\textbf {\bibinfo {volume} {104}},\ \bibinfo
  {pages} {012015} (\bibinfo {year} {2021}{\natexlab{c}})},\ \Eprint
  {http://arxiv.org/abs/2012.01581} {arXiv:2012.01581 [hep-ex]} \BibitemShut
  {NoStop}%
\bibitem [{\citenamefont {Aad}\ \emph {et~al.}(2021{\natexlab{h}})\citenamefont
  {Aad} \emph {et~al.}}]{ATLAS:2020wjh}%
  \BibitemOpen
  \bibfield  {author} {\bibinfo {author} {\bibfnamefont {G.}~\bibnamefont
  {Aad}} \emph {et~al.} (\bibinfo {collaboration} {ATLAS}),\ }\href {\doibase
  10.1103/PhysRevLett.127.051802} {\bibfield  {journal} {\bibinfo  {journal}
  {Phys. Rev. Lett.}\ }\textbf {\bibinfo {volume} {127}},\ \bibinfo {pages}
  {051802} (\bibinfo {year} {2021}{\natexlab{h}})},\ \Eprint
  {http://arxiv.org/abs/2011.07812} {arXiv:2011.07812 [hep-ex]} \BibitemShut
  {NoStop}%
\bibitem [{\citenamefont {Aad}\ \emph {et~al.}(2020{\natexlab{c}})\citenamefont
  {Aad} \emph {et~al.}}]{ATLAS:2019fwx}%
  \BibitemOpen
  \bibfield  {author} {\bibinfo {author} {\bibfnamefont {G.}~\bibnamefont
  {Aad}} \emph {et~al.} (\bibinfo {collaboration} {ATLAS}),\ }\href {\doibase
  10.1016/j.physletb.2019.135114} {\bibfield  {journal} {\bibinfo  {journal}
  {Phys. Lett. B}\ }\textbf {\bibinfo {volume} {801}},\ \bibinfo {pages}
  {135114} (\bibinfo {year} {2020}{\natexlab{c}})},\ \Eprint
  {http://arxiv.org/abs/1907.10037} {arXiv:1907.10037 [hep-ex]} \BibitemShut
  {NoStop}%
\bibitem [{\citenamefont {Sirunyan}\ \emph
  {et~al.}(2021{\natexlab{d}})\citenamefont {Sirunyan} \emph
  {et~al.}}]{CMS:2021knz}%
  \BibitemOpen
  \bibfield  {author} {\bibinfo {author} {\bibfnamefont {A.~M.}\ \bibnamefont
  {Sirunyan}} \emph {et~al.} (\bibinfo {collaboration} {CMS}),\ }\href
  {\doibase 10.1103/PhysRevD.104.032006} {\bibfield  {journal} {\bibinfo
  {journal} {Phys. Rev. D}\ }\textbf {\bibinfo {volume} {104}},\ \bibinfo
  {pages} {032006} (\bibinfo {year} {2021}{\natexlab{d}})},\ \Eprint
  {http://arxiv.org/abs/2102.06976} {arXiv:2102.06976 [hep-ex]} \BibitemShut
  {NoStop}%
\bibitem [{\citenamefont {Sirunyan}\ \emph
  {et~al.}(2020{\natexlab{c}})\citenamefont {Sirunyan} \emph
  {et~al.}}]{CMS:2020cpy}%
  \BibitemOpen
  \bibfield  {author} {\bibinfo {author} {\bibfnamefont {A.~M.}\ \bibnamefont
  {Sirunyan}} \emph {et~al.} (\bibinfo {collaboration} {CMS}),\ }\href
  {\doibase 10.1140/epjc/s10052-020-8168-3} {\bibfield  {journal} {\bibinfo
  {journal} {Eur. Phys. J. C}\ }\textbf {\bibinfo {volume} {80}},\ \bibinfo
  {pages} {752} (\bibinfo {year} {2020}{\natexlab{c}})},\ \Eprint
  {http://arxiv.org/abs/2001.10086} {arXiv:2001.10086 [hep-ex]} \BibitemShut
  {NoStop}%
\bibitem [{\citenamefont {Sirunyan}\ \emph
  {et~al.}(2019{\natexlab{e}})\citenamefont {Sirunyan} \emph
  {et~al.}}]{CMS:2018ikp}%
  \BibitemOpen
  \bibfield  {author} {\bibinfo {author} {\bibfnamefont {A.~M.}\ \bibnamefont
  {Sirunyan}} \emph {et~al.} (\bibinfo {collaboration} {CMS}),\ }\href
  {\doibase 10.1103/PhysRevD.99.012010} {\bibfield  {journal} {\bibinfo
  {journal} {Phys. Rev. D}\ }\textbf {\bibinfo {volume} {99}},\ \bibinfo
  {pages} {012010} (\bibinfo {year} {2019}{\natexlab{e}})},\ \Eprint
  {http://arxiv.org/abs/1810.10092} {arXiv:1810.10092 [hep-ex]} \BibitemShut
  {NoStop}%
\bibitem [{\citenamefont {Sirunyan}\ \emph
  {et~al.}(2019{\natexlab{f}})\citenamefont {Sirunyan} \emph
  {et~al.}}]{CMS:2018skt}%
  \BibitemOpen
  \bibfield  {author} {\bibinfo {author} {\bibfnamefont {A.~M.}\ \bibnamefont
  {Sirunyan}} \emph {et~al.} (\bibinfo {collaboration} {CMS}),\ }\href
  {\doibase 10.1140/epjc/s10052-019-6800-x} {\bibfield  {journal} {\bibinfo
  {journal} {Eur. Phys. J. C}\ }\textbf {\bibinfo {volume} {79}},\ \bibinfo
  {pages} {305} (\bibinfo {year} {2019}{\natexlab{f}})},\ \Eprint
  {http://arxiv.org/abs/1811.09760} {arXiv:1811.09760 [hep-ex]} \BibitemShut
  {NoStop}%
\bibitem [{\citenamefont {Sirunyan}\ \emph
  {et~al.}(2018{\natexlab{d}})\citenamefont {Sirunyan} \emph
  {et~al.}}]{CMS:2018mts}%
  \BibitemOpen
  \bibfield  {author} {\bibinfo {author} {\bibfnamefont {A.~M.}\ \bibnamefont
  {Sirunyan}} \emph {et~al.} (\bibinfo {collaboration} {CMS}),\ }\href
  {\doibase 10.1103/PhysRevD.98.112014} {\bibfield  {journal} {\bibinfo
  {journal} {Phys. Rev. D}\ }\textbf {\bibinfo {volume} {98}},\ \bibinfo
  {pages} {112014} (\bibinfo {year} {2018}{\natexlab{d}})},\ \Eprint
  {http://arxiv.org/abs/1808.03124} {arXiv:1808.03124 [hep-ex]} \BibitemShut
  {NoStop}%
\bibitem [{\citenamefont {Sirunyan}\ \emph
  {et~al.}(2018{\natexlab{e}})\citenamefont {Sirunyan} \emph
  {et~al.}}]{CMS:2018pdq}%
  \BibitemOpen
  \bibfield  {author} {\bibinfo {author} {\bibfnamefont {A.~M.}\ \bibnamefont
  {Sirunyan}} \emph {et~al.} (\bibinfo {collaboration} {CMS}),\ }\href
  {\doibase 10.1103/PhysRevLett.121.141802} {\bibfield  {journal} {\bibinfo
  {journal} {Phys. Rev. Lett.}\ }\textbf {\bibinfo {volume} {121}},\ \bibinfo
  {pages} {141802} (\bibinfo {year} {2018}{\natexlab{e}})},\ \Eprint
  {http://arxiv.org/abs/1806.01058} {arXiv:1806.01058 [hep-ex]} \BibitemShut
  {NoStop}%
\bibitem [{\citenamefont {Sirunyan}\ \emph
  {et~al.}(2018{\natexlab{f}})\citenamefont {Sirunyan} \emph
  {et~al.}}]{CMS:2018hnz}%
  \BibitemOpen
  \bibfield  {author} {\bibinfo {author} {\bibfnamefont {A.~M.}\ \bibnamefont
  {Sirunyan}} \emph {et~al.} (\bibinfo {collaboration} {CMS}),\ }\href
  {\doibase 10.1007/JHEP04(2018)073} {\bibfield  {journal} {\bibinfo  {journal}
  {JHEP}\ }\textbf {\bibinfo {volume} {04}},\ \bibinfo {pages} {073} (\bibinfo
  {year} {2018}{\natexlab{f}})},\ \Eprint {http://arxiv.org/abs/1802.01122}
  {arXiv:1802.01122 [hep-ex]} \BibitemShut {NoStop}%
\bibitem [{\citenamefont {Sirunyan}\ \emph
  {et~al.}(2018{\natexlab{g}})\citenamefont {Sirunyan} \emph
  {et~al.}}]{CMS:2017szl}%
  \BibitemOpen
  \bibfield  {author} {\bibinfo {author} {\bibfnamefont {A.~M.}\ \bibnamefont
  {Sirunyan}} \emph {et~al.} (\bibinfo {collaboration} {CMS}),\ }\href
  {\doibase 10.1016/j.physletb.2018.06.028} {\bibfield  {journal} {\bibinfo
  {journal} {Phys. Lett. B}\ }\textbf {\bibinfo {volume} {783}},\ \bibinfo
  {pages} {114} (\bibinfo {year} {2018}{\natexlab{g}})},\ \Eprint
  {http://arxiv.org/abs/1712.08920} {arXiv:1712.08920 [hep-ex]} \BibitemShut
  {NoStop}%
\bibitem [{\citenamefont {Aad}\ \emph {et~al.}(2021{\natexlab{i}})\citenamefont
  {Aad} \emph {et~al.}}]{ATLAS:2021fbt}%
  \BibitemOpen
  \bibfield  {author} {\bibinfo {author} {\bibfnamefont {G.}~\bibnamefont
  {Aad}} \emph {et~al.} (\bibinfo {collaboration} {ATLAS}),\ }\href {\doibase
  10.1140/epjc/s10052-021-09761-x} {\bibfield  {journal} {\bibinfo  {journal}
  {Eur. Phys. J. C}\ }\textbf {\bibinfo {volume} {81}},\ \bibinfo {pages}
  {1023} (\bibinfo {year} {2021}{\natexlab{i}})},\ \Eprint
  {http://arxiv.org/abs/2106.09609} {arXiv:2106.09609 [hep-ex]} \BibitemShut
  {NoStop}%
\bibitem [{\citenamefont {Aad}\ \emph {et~al.}(2021{\natexlab{j}})\citenamefont
  {Aad} \emph {et~al.}}]{ATLAS:2021tar}%
  \BibitemOpen
  \bibfield  {author} {\bibinfo {author} {\bibfnamefont {G.}~\bibnamefont
  {Aad}} \emph {et~al.} (\bibinfo {collaboration} {ATLAS}),\ }\href@noop {} {\
  (\bibinfo {year} {2021}{\natexlab{j}})},\ \Eprint
  {http://arxiv.org/abs/2112.08090} {arXiv:2112.08090 [hep-ex]} \BibitemShut
  {NoStop}%
\bibitem [{\citenamefont {Aad}\ \emph {et~al.}(2021{\natexlab{k}})\citenamefont
  {Aad} \emph {et~al.}}]{ATLAS:2021yyr}%
  \BibitemOpen
  \bibfield  {author} {\bibinfo {author} {\bibfnamefont {G.}~\bibnamefont
  {Aad}} \emph {et~al.} (\bibinfo {collaboration} {ATLAS}),\ }\href {\doibase
  10.1007/JHEP07(2021)167} {\bibfield  {journal} {\bibinfo  {journal} {JHEP}\
  }\textbf {\bibinfo {volume} {07}},\ \bibinfo {pages} {167} (\bibinfo {year}
  {2021}{\natexlab{k}})},\ \Eprint {http://arxiv.org/abs/2103.11684}
  {arXiv:2103.11684 [hep-ex]} \BibitemShut {NoStop}%
\bibitem [{\citenamefont {Aad}\ \emph {et~al.}(2021{\natexlab{l}})\citenamefont
  {Aad} \emph {et~al.}}]{ATLAS:2020uer}%
  \BibitemOpen
  \bibfield  {author} {\bibinfo {author} {\bibfnamefont {G.}~\bibnamefont
  {Aad}} \emph {et~al.} (\bibinfo {collaboration} {ATLAS}),\ }\href {\doibase
  10.1103/PhysRevD.103.112003} {\bibfield  {journal} {\bibinfo  {journal}
  {Phys. Rev. D}\ }\textbf {\bibinfo {volume} {103}},\ \bibinfo {pages}
  {112003} (\bibinfo {year} {2021}{\natexlab{l}})},\ \Eprint
  {http://arxiv.org/abs/2011.10543} {arXiv:2011.10543 [hep-ex]} \BibitemShut
  {NoStop}%
\bibitem [{\citenamefont {Aad}\ \emph {et~al.}(2021{\natexlab{m}})\citenamefont
  {Aad} \emph {et~al.}}]{ATLAS:2020wgq}%
  \BibitemOpen
  \bibfield  {author} {\bibinfo {author} {\bibfnamefont {G.}~\bibnamefont
  {Aad}} \emph {et~al.} (\bibinfo {collaboration} {ATLAS}),\ }\href {\doibase
  10.1140/epjc/s10052-020-08730-0} {\bibfield  {journal} {\bibinfo  {journal}
  {Eur. Phys. J. C}\ }\textbf {\bibinfo {volume} {81}},\ \bibinfo {pages} {11}
  (\bibinfo {year} {2021}{\natexlab{m}})},\ \bibinfo {note} {[Erratum:
  Eur.Phys.J.C 81, 249 (2021)]},\ \Eprint {http://arxiv.org/abs/2010.01015}
  {arXiv:2010.01015 [hep-ex]} \BibitemShut {NoStop}%
\bibitem [{\citenamefont {Aad}\ \emph {et~al.}(2020{\natexlab{d}})\citenamefont
  {Aad} \emph {et~al.}}]{ATLAS:2019fag}%
  \BibitemOpen
  \bibfield  {author} {\bibinfo {author} {\bibfnamefont {G.}~\bibnamefont
  {Aad}} \emph {et~al.} (\bibinfo {collaboration} {ATLAS}),\ }\href {\doibase
  10.1007/JHEP06(2020)046} {\bibfield  {journal} {\bibinfo  {journal} {JHEP}\
  }\textbf {\bibinfo {volume} {06}},\ \bibinfo {pages} {046} (\bibinfo {year}
  {2020}{\natexlab{d}})},\ \Eprint {http://arxiv.org/abs/1909.08457}
  {arXiv:1909.08457 [hep-ex]} \BibitemShut {NoStop}%
\bibitem [{\citenamefont {Aaboud}\ \emph
  {et~al.}(2018{\natexlab{b}})\citenamefont {Aaboud} \emph
  {et~al.}}]{ATLAS:2018umm}%
  \BibitemOpen
  \bibfield  {author} {\bibinfo {author} {\bibfnamefont {M.}~\bibnamefont
  {Aaboud}} \emph {et~al.} (\bibinfo {collaboration} {ATLAS}),\ }\href
  {\doibase 10.1016/j.physletb.2018.08.021} {\bibfield  {journal} {\bibinfo
  {journal} {Phys. Lett. B}\ }\textbf {\bibinfo {volume} {785}},\ \bibinfo
  {pages} {136} (\bibinfo {year} {2018}{\natexlab{b}})},\ \Eprint
  {http://arxiv.org/abs/1804.03568} {arXiv:1804.03568 [hep-ex]} \BibitemShut
  {NoStop}%
\bibitem [{\citenamefont {Aaboud}\ \emph
  {et~al.}(2018{\natexlab{c}})\citenamefont {Aaboud} \emph
  {et~al.}}]{ATLAS:2017jnp}%
  \BibitemOpen
  \bibfield  {author} {\bibinfo {author} {\bibfnamefont {M.}~\bibnamefont
  {Aaboud}} \emph {et~al.} (\bibinfo {collaboration} {ATLAS}),\ }\href
  {\doibase 10.1140/epjc/s10052-018-5693-4} {\bibfield  {journal} {\bibinfo
  {journal} {Eur. Phys. J. C}\ }\textbf {\bibinfo {volume} {78}},\ \bibinfo
  {pages} {250} (\bibinfo {year} {2018}{\natexlab{c}})},\ \Eprint
  {http://arxiv.org/abs/1710.07171} {arXiv:1710.07171 [hep-ex]} \BibitemShut
  {NoStop}%
\bibitem [{\citenamefont {Aaboud}\ \emph
  {et~al.}(2018{\natexlab{d}})\citenamefont {Aaboud} \emph
  {et~al.}}]{ATLAS:2017jvy}%
  \BibitemOpen
  \bibfield  {author} {\bibinfo {author} {\bibfnamefont {M.}~\bibnamefont
  {Aaboud}} \emph {et~al.} (\bibinfo {collaboration} {ATLAS}),\ }\href
  {\doibase 10.1103/PhysRevD.97.032003} {\bibfield  {journal} {\bibinfo
  {journal} {Phys. Rev. D}\ }\textbf {\bibinfo {volume} {97}},\ \bibinfo
  {pages} {032003} (\bibinfo {year} {2018}{\natexlab{d}})},\ \Eprint
  {http://arxiv.org/abs/1710.05544} {arXiv:1710.05544 [hep-ex]} \BibitemShut
  {NoStop}%
\bibitem [{\citenamefont {Babu}\ \emph {et~al.}(2022)\citenamefont {Babu},
  \citenamefont {Gogoladze},\ and\ \citenamefont {Un}}]{Babu:2020ncc}%
  \BibitemOpen
  \bibfield  {author} {\bibinfo {author} {\bibfnamefont {K.~S.}\ \bibnamefont
  {Babu}}, \bibinfo {author} {\bibfnamefont {I.}~\bibnamefont {Gogoladze}}, \
  and\ \bibinfo {author} {\bibfnamefont {C.~S.}\ \bibnamefont {Un}},\ }\href
  {\doibase 10.1007/JHEP02(2022)164} {\bibfield  {journal} {\bibinfo  {journal}
  {JHEP}\ }\textbf {\bibinfo {volume} {02}},\ \bibinfo {pages} {164} (\bibinfo
  {year} {2022})},\ \Eprint {http://arxiv.org/abs/2012.14411} {arXiv:2012.14411
  [hep-ph]} \BibitemShut {NoStop}%
\bibitem [{\citenamefont {Dev}\ \emph {et~al.}(2022)\citenamefont {Dev} \emph
  {et~al.}}]{Dev:2022jbf}%
  \BibitemOpen
  \bibfield  {author} {\bibinfo {author} {\bibfnamefont {P.~S.~B.}\
  \bibnamefont {Dev}} \emph {et~al.},\ }\href@noop {} {\  (\bibinfo {year}
  {2022})},\ \Eprint {http://arxiv.org/abs/2203.08771} {arXiv:2203.08771
  [hep-ex]} \BibitemShut {NoStop}%
\bibitem [{\citenamefont {Takenaka}\ \emph {et~al.}(2020)\citenamefont
  {Takenaka} \emph {et~al.}}]{Super-Kamiokande:2020wjk}%
  \BibitemOpen
  \bibfield  {author} {\bibinfo {author} {\bibfnamefont {A.}~\bibnamefont
  {Takenaka}} \emph {et~al.} (\bibinfo {collaboration} {Super-Kamiokande}),\
  }\href {\doibase 10.1103/PhysRevD.102.112011} {\bibfield  {journal} {\bibinfo
   {journal} {Phys. Rev. D}\ }\textbf {\bibinfo {volume} {102}},\ \bibinfo
  {pages} {112011} (\bibinfo {year} {2020})},\ \Eprint
  {http://arxiv.org/abs/2010.16098} {arXiv:2010.16098 [hep-ex]} \BibitemShut
  {NoStop}%
\bibitem [{\citenamefont {Abe}\ \emph {et~al.}(2014)\citenamefont {Abe} \emph
  {et~al.}}]{Super-Kamiokande:2014otb}%
  \BibitemOpen
  \bibfield  {author} {\bibinfo {author} {\bibfnamefont {K.}~\bibnamefont
  {Abe}} \emph {et~al.} (\bibinfo {collaboration} {Super-Kamiokande}),\ }\href
  {\doibase 10.1103/PhysRevD.90.072005} {\bibfield  {journal} {\bibinfo
  {journal} {Phys. Rev. D}\ }\textbf {\bibinfo {volume} {90}},\ \bibinfo
  {pages} {072005} (\bibinfo {year} {2014})},\ \Eprint
  {http://arxiv.org/abs/1408.1195} {arXiv:1408.1195 [hep-ex]} \BibitemShut
  {NoStop}%
\bibitem [{JUN(2022)}]{JUNO:2022hxd}%
  \BibitemOpen
  \href {\doibase 10.1016/j.ppnp.2021.103927} {\bibfield  {journal} {\bibinfo
  {journal} {Prog. Part. Nucl. Phys.}\ }\textbf {\bibinfo {volume} {123}},\
  \bibinfo {pages} {103927} (\bibinfo {year} {2022})}\BibitemShut {NoStop}%
\bibitem [{\citenamefont {Bian}\ \emph {et~al.}(2022)\citenamefont {Bian} \emph
  {et~al.}}]{Hyper-Kamiokande:2022smq}%
  \BibitemOpen
  \bibfield  {author} {\bibinfo {author} {\bibfnamefont {J.}~\bibnamefont
  {Bian}} \emph {et~al.} (\bibinfo {collaboration} {Hyper-Kamiokande}),\ }in\
  \href@noop {} {\emph {\bibinfo {booktitle} {{2022 Snowmass Summer Study}}}}\
  (\bibinfo {year} {2022})\ \Eprint {http://arxiv.org/abs/2203.02029}
  {arXiv:2203.02029 [hep-ex]} \BibitemShut {NoStop}%
\bibitem [{\citenamefont {Abi}\ \emph {et~al.}(2021)\citenamefont {Abi} \emph
  {et~al.}}]{DUNE:2020fgq}%
  \BibitemOpen
  \bibfield  {author} {\bibinfo {author} {\bibfnamefont {B.}~\bibnamefont
  {Abi}} \emph {et~al.} (\bibinfo {collaboration} {DUNE}),\ }\href {\doibase
  10.1140/epjc/s10052-021-09007-w} {\bibfield  {journal} {\bibinfo  {journal}
  {Eur. Phys. J. C}\ }\textbf {\bibinfo {volume} {81}},\ \bibinfo {pages} {322}
  (\bibinfo {year} {2021})},\ \Eprint {http://arxiv.org/abs/2008.12769}
  {arXiv:2008.12769 [hep-ex]} \BibitemShut {NoStop}%
\bibitem [{\citenamefont {'t~Hooft}(1976)}]{tHooft:1976rip}%
  \BibitemOpen
  \bibfield  {author} {\bibinfo {author} {\bibfnamefont {G.}~\bibnamefont
  {'t~Hooft}},\ }\href {\doibase 10.1103/PhysRevLett.37.8} {\bibfield
  {journal} {\bibinfo  {journal} {Phys. Rev. Lett.}\ }\textbf {\bibinfo
  {volume} {37}},\ \bibinfo {pages} {8} (\bibinfo {year} {1976})}\BibitemShut
  {NoStop}%
\bibitem [{Pro(2020)}]{Proceedings:2020nzz}%
  \BibitemOpen
  \href@noop {} {\emph {\bibinfo {title} {{$|\Delta \mathcal{B}| =2$: A State
  of the Field, and Looking Forward--A brief status report of theoretical and
  experimental physics opportunities}}}}\ (\bibinfo {year} {2020})\ \Eprint
  {http://arxiv.org/abs/2010.02299} {arXiv:2010.02299 [hep-ph]} \BibitemShut
  {NoStop}%
\bibitem [{\citenamefont {Barrow}\ \emph {et~al.}(2022)\citenamefont {Barrow}
  \emph {et~al.}}]{Barrow:2022gsu}%
  \BibitemOpen
  \bibfield  {author} {\bibinfo {author} {\bibfnamefont {J.~L.}\ \bibnamefont
  {Barrow}} \emph {et~al.},\ }\href@noop {} {\  (\bibinfo {year} {2022})},\
  \Eprint {http://arxiv.org/abs/2203.07059} {arXiv:2203.07059 [hep-ph]}
  \BibitemShut {NoStop}%
\bibitem [{\citenamefont {Kuzmin}(1970)}]{Kuzmin:1970nx}%
  \BibitemOpen
  \bibfield  {author} {\bibinfo {author} {\bibfnamefont {V.~A.}\ \bibnamefont
  {Kuzmin}},\ }\href@noop {} {\bibfield  {journal} {\bibinfo  {journal} {Pisma
  Zh. Eksp. Teor. Fiz.}\ }\textbf {\bibinfo {volume} {12}},\ \bibinfo {pages}
  {335} (\bibinfo {year} {1970})}\BibitemShut {NoStop}%
\bibitem [{\citenamefont {Mohapatra}\ and\ \citenamefont
  {Marshak}(1980{\natexlab{a}})}]{Mohapatra:1980qe}%
  \BibitemOpen
  \bibfield  {author} {\bibinfo {author} {\bibfnamefont {R.~N.}\ \bibnamefont
  {Mohapatra}}\ and\ \bibinfo {author} {\bibfnamefont {R.~E.}\ \bibnamefont
  {Marshak}},\ }\href {\doibase 10.1103/PhysRevLett.44.1316} {\bibfield
  {journal} {\bibinfo  {journal} {Phys. Rev. Lett.}\ }\textbf {\bibinfo
  {volume} {44}},\ \bibinfo {pages} {1316} (\bibinfo {year}
  {1980}{\natexlab{a}})},\ \bibinfo {note} {[Erratum: Phys.Rev.Lett. 44, 1643
  (1980)]}\BibitemShut {NoStop}%
\bibitem [{\citenamefont {Mohapatra}\ and\ \citenamefont
  {Marshak}(1980{\natexlab{b}})}]{Mohapatra:1980de}%
  \BibitemOpen
  \bibfield  {author} {\bibinfo {author} {\bibfnamefont {R.~N.}\ \bibnamefont
  {Mohapatra}}\ and\ \bibinfo {author} {\bibfnamefont {R.~E.}\ \bibnamefont
  {Marshak}},\ }\href {\doibase 10.1016/0370-2693(80)90853-9} {\bibfield
  {journal} {\bibinfo  {journal} {Phys. Lett. B}\ }\textbf {\bibinfo {volume}
  {94}},\ \bibinfo {pages} {183} (\bibinfo {year} {1980}{\natexlab{b}})},\
  \bibinfo {note} {[Erratum: Phys.Lett.B 96, 444--444 (1980)]}\BibitemShut
  {NoStop}%
\bibitem [{\citenamefont {Kuo}\ and\ \citenamefont {Love}(1980)}]{Kuo:1980ew}%
  \BibitemOpen
  \bibfield  {author} {\bibinfo {author} {\bibfnamefont {T.-K.}\ \bibnamefont
  {Kuo}}\ and\ \bibinfo {author} {\bibfnamefont {S.~T.}\ \bibnamefont {Love}},\
  }\href {\doibase 10.1103/PhysRevLett.45.93} {\bibfield  {journal} {\bibinfo
  {journal} {Phys. Rev. Lett.}\ }\textbf {\bibinfo {volume} {45}},\ \bibinfo
  {pages} {93} (\bibinfo {year} {1980})}\BibitemShut {NoStop}%
\bibitem [{\citenamefont {Cowsik}\ and\ \citenamefont
  {Nussinov}(1981)}]{Cowsik:1980np}%
  \BibitemOpen
  \bibfield  {author} {\bibinfo {author} {\bibfnamefont {R.}~\bibnamefont
  {Cowsik}}\ and\ \bibinfo {author} {\bibfnamefont {S.}~\bibnamefont
  {Nussinov}},\ }\href {\doibase 10.1016/0370-2693(81)90302-6} {\bibfield
  {journal} {\bibinfo  {journal} {Phys. Lett. B}\ }\textbf {\bibinfo {volume}
  {101}},\ \bibinfo {pages} {237} (\bibinfo {year} {1981})}\BibitemShut
  {NoStop}%
\bibitem [{\citenamefont {Chetyrkin}\ \emph {et~al.}(1981)\citenamefont
  {Chetyrkin}, \citenamefont {Kazarnovsky}, \citenamefont {Kuzmin},\ and\
  \citenamefont {Shaposhnikov}}]{Chetyrkin:1980ta}%
  \BibitemOpen
  \bibfield  {author} {\bibinfo {author} {\bibfnamefont {K.~G.}\ \bibnamefont
  {Chetyrkin}}, \bibinfo {author} {\bibfnamefont {M.~V.}\ \bibnamefont
  {Kazarnovsky}}, \bibinfo {author} {\bibfnamefont {V.~A.}\ \bibnamefont
  {Kuzmin}}, \ and\ \bibinfo {author} {\bibfnamefont {M.~E.}\ \bibnamefont
  {Shaposhnikov}},\ }\href {\doibase 10.1016/0370-2693(81)90117-9} {\bibfield
  {journal} {\bibinfo  {journal} {Phys. Lett. B}\ }\textbf {\bibinfo {volume}
  {99}},\ \bibinfo {pages} {358} (\bibinfo {year} {1981})}\BibitemShut
  {NoStop}%
\bibitem [{\citenamefont {Barbieri}\ and\ \citenamefont
  {Mohapatra}(1981)}]{Barbieri:1981yr}%
  \BibitemOpen
  \bibfield  {author} {\bibinfo {author} {\bibfnamefont {R.}~\bibnamefont
  {Barbieri}}\ and\ \bibinfo {author} {\bibfnamefont {R.~N.}\ \bibnamefont
  {Mohapatra}},\ }\href {\doibase 10.1007/BF01574002} {\bibfield  {journal}
  {\bibinfo  {journal} {Z. Phys. C}\ }\textbf {\bibinfo {volume} {11}},\
  \bibinfo {pages} {175} (\bibinfo {year} {1981})}\BibitemShut {NoStop}%
\bibitem [{\citenamefont {Caswell}\ \emph {et~al.}(1983)\citenamefont
  {Caswell}, \citenamefont {Milutinovic},\ and\ \citenamefont
  {Senjanovic}}]{Caswell:1982qs}%
  \BibitemOpen
  \bibfield  {author} {\bibinfo {author} {\bibfnamefont {W.~E.}\ \bibnamefont
  {Caswell}}, \bibinfo {author} {\bibfnamefont {J.}~\bibnamefont
  {Milutinovic}}, \ and\ \bibinfo {author} {\bibfnamefont {G.}~\bibnamefont
  {Senjanovic}},\ }\href {\doibase 10.1016/0370-2693(83)91585-X} {\bibfield
  {journal} {\bibinfo  {journal} {Phys. Lett. B}\ }\textbf {\bibinfo {volume}
  {122}},\ \bibinfo {pages} {373} (\bibinfo {year} {1983})}\BibitemShut
  {NoStop}%
\bibitem [{\citenamefont {Rao}\ and\ \citenamefont
  {Shrock}(1982)}]{Rao:1982gt}%
  \BibitemOpen
  \bibfield  {author} {\bibinfo {author} {\bibfnamefont {S.}~\bibnamefont
  {Rao}}\ and\ \bibinfo {author} {\bibfnamefont {R.}~\bibnamefont {Shrock}},\
  }\href {\doibase 10.1016/0370-2693(82)90333-1} {\bibfield  {journal}
  {\bibinfo  {journal} {Phys. Lett. B}\ }\textbf {\bibinfo {volume} {116}},\
  \bibinfo {pages} {238} (\bibinfo {year} {1982})}\BibitemShut {NoStop}%
\bibitem [{\citenamefont {Rao}\ and\ \citenamefont
  {Shrock}(1984)}]{Rao:1983sd}%
  \BibitemOpen
  \bibfield  {author} {\bibinfo {author} {\bibfnamefont {S.}~\bibnamefont
  {Rao}}\ and\ \bibinfo {author} {\bibfnamefont {R.~E.}\ \bibnamefont
  {Shrock}},\ }\href {\doibase 10.1016/0550-3213(84)90365-1} {\bibfield
  {journal} {\bibinfo  {journal} {Nucl. Phys. B}\ }\textbf {\bibinfo {volume}
  {232}},\ \bibinfo {pages} {143} (\bibinfo {year} {1984})}\BibitemShut
  {NoStop}%
\bibitem [{\citenamefont {Berezhiani}\ and\ \citenamefont
  {Bento}(2006{\natexlab{a}})}]{Berezhiani:2005hv}%
  \BibitemOpen
  \bibfield  {author} {\bibinfo {author} {\bibfnamefont {Z.}~\bibnamefont
  {Berezhiani}}\ and\ \bibinfo {author} {\bibfnamefont {L.}~\bibnamefont
  {Bento}},\ }\href {\doibase 10.1103/PhysRevLett.96.081801} {\bibfield
  {journal} {\bibinfo  {journal} {Phys. Rev. Lett.}\ }\textbf {\bibinfo
  {volume} {96}},\ \bibinfo {pages} {081801} (\bibinfo {year}
  {2006}{\natexlab{a}})},\ \Eprint {http://arxiv.org/abs/hep-ph/0507031}
  {arXiv:hep-ph/0507031} \BibitemShut {NoStop}%
\bibitem [{\citenamefont {Dutta}\ \emph {et~al.}(2006)\citenamefont {Dutta},
  \citenamefont {Mimura},\ and\ \citenamefont {Mohapatra}}]{Dutta:2005af}%
  \BibitemOpen
  \bibfield  {author} {\bibinfo {author} {\bibfnamefont {B.}~\bibnamefont
  {Dutta}}, \bibinfo {author} {\bibfnamefont {Y.}~\bibnamefont {Mimura}}, \
  and\ \bibinfo {author} {\bibfnamefont {R.~N.}\ \bibnamefont {Mohapatra}},\
  }\href {\doibase 10.1103/PhysRevLett.96.061801} {\bibfield  {journal}
  {\bibinfo  {journal} {Phys. Rev. Lett.}\ }\textbf {\bibinfo {volume} {96}},\
  \bibinfo {pages} {061801} (\bibinfo {year} {2006})},\ \Eprint
  {http://arxiv.org/abs/hep-ph/0510291} {arXiv:hep-ph/0510291} \BibitemShut
  {NoStop}%
\bibitem [{\citenamefont {Babu}\ \emph {et~al.}(2006)\citenamefont {Babu},
  \citenamefont {Mohapatra},\ and\ \citenamefont {Nasri}}]{Babu:2006xc}%
  \BibitemOpen
  \bibfield  {author} {\bibinfo {author} {\bibfnamefont {K.~S.}\ \bibnamefont
  {Babu}}, \bibinfo {author} {\bibfnamefont {R.~N.}\ \bibnamefont {Mohapatra}},
  \ and\ \bibinfo {author} {\bibfnamefont {S.}~\bibnamefont {Nasri}},\ }\href
  {\doibase 10.1103/PhysRevLett.97.131301} {\bibfield  {journal} {\bibinfo
  {journal} {Phys. Rev. Lett.}\ }\textbf {\bibinfo {volume} {97}},\ \bibinfo
  {pages} {131301} (\bibinfo {year} {2006})},\ \Eprint
  {http://arxiv.org/abs/hep-ph/0606144} {arXiv:hep-ph/0606144} \BibitemShut
  {NoStop}%
\bibitem [{\citenamefont {Babu}\ \emph {et~al.}(2009)\citenamefont {Babu},
  \citenamefont {Bhupal~Dev},\ and\ \citenamefont {Mohapatra}}]{Babu:2008rq}%
  \BibitemOpen
  \bibfield  {author} {\bibinfo {author} {\bibfnamefont {K.~S.}\ \bibnamefont
  {Babu}}, \bibinfo {author} {\bibfnamefont {P.~S.}\ \bibnamefont
  {Bhupal~Dev}}, \ and\ \bibinfo {author} {\bibfnamefont {R.~N.}\ \bibnamefont
  {Mohapatra}},\ }\href {\doibase 10.1103/PhysRevD.79.015017} {\bibfield
  {journal} {\bibinfo  {journal} {Phys. Rev. D}\ }\textbf {\bibinfo {volume}
  {79}},\ \bibinfo {pages} {015017} (\bibinfo {year} {2009})},\ \Eprint
  {http://arxiv.org/abs/0811.3411} {arXiv:0811.3411 [hep-ph]} \BibitemShut
  {NoStop}%
\bibitem [{\citenamefont {Babu}\ \emph {et~al.}(2013)\citenamefont {Babu},
  \citenamefont {Bhupal~Dev}, \citenamefont {Fortes},\ and\ \citenamefont
  {Mohapatra}}]{Babu:2013yca}%
  \BibitemOpen
  \bibfield  {author} {\bibinfo {author} {\bibfnamefont {K.~S.}\ \bibnamefont
  {Babu}}, \bibinfo {author} {\bibfnamefont {P.~S.}\ \bibnamefont
  {Bhupal~Dev}}, \bibinfo {author} {\bibfnamefont {E.~C. F.~S.}\ \bibnamefont
  {Fortes}}, \ and\ \bibinfo {author} {\bibfnamefont {R.~N.}\ \bibnamefont
  {Mohapatra}},\ }\href {\doibase 10.1103/PhysRevD.87.115019} {\bibfield
  {journal} {\bibinfo  {journal} {Phys. Rev. D}\ }\textbf {\bibinfo {volume}
  {87}},\ \bibinfo {pages} {115019} (\bibinfo {year} {2013})},\ \Eprint
  {http://arxiv.org/abs/1303.6918} {arXiv:1303.6918 [hep-ph]} \BibitemShut
  {NoStop}%
\bibitem [{\citenamefont {Dev}\ and\ \citenamefont
  {Mohapatra}(2015)}]{Dev:2015uca}%
  \BibitemOpen
  \bibfield  {author} {\bibinfo {author} {\bibfnamefont {P.~S.~B.}\
  \bibnamefont {Dev}}\ and\ \bibinfo {author} {\bibfnamefont {R.~N.}\
  \bibnamefont {Mohapatra}},\ }\href {\doibase 10.1103/PhysRevD.92.016007}
  {\bibfield  {journal} {\bibinfo  {journal} {Phys. Rev. D}\ }\textbf {\bibinfo
  {volume} {92}},\ \bibinfo {pages} {016007} (\bibinfo {year} {2015})},\
  \Eprint {http://arxiv.org/abs/1504.07196} {arXiv:1504.07196 [hep-ph]}
  \BibitemShut {NoStop}%
\bibitem [{\citenamefont {McKeen}\ and\ \citenamefont
  {Nelson}(2016)}]{McKeen:2015cuz}%
  \BibitemOpen
  \bibfield  {author} {\bibinfo {author} {\bibfnamefont {D.}~\bibnamefont
  {McKeen}}\ and\ \bibinfo {author} {\bibfnamefont {A.~E.}\ \bibnamefont
  {Nelson}},\ }\href {\doibase 10.1103/PhysRevD.94.076002} {\bibfield
  {journal} {\bibinfo  {journal} {Phys. Rev. D}\ }\textbf {\bibinfo {volume}
  {94}},\ \bibinfo {pages} {076002} (\bibinfo {year} {2016})},\ \Eprint
  {http://arxiv.org/abs/1512.05359} {arXiv:1512.05359 [hep-ph]} \BibitemShut
  {NoStop}%
\bibitem [{\citenamefont {Calibbi}\ \emph {et~al.}(2016)\citenamefont
  {Calibbi}, \citenamefont {Ferretti}, \citenamefont {Milstead}, \citenamefont
  {Petersson},\ and\ \citenamefont {P\"ottgen}}]{Calibbi:2016ukt}%
  \BibitemOpen
  \bibfield  {author} {\bibinfo {author} {\bibfnamefont {L.}~\bibnamefont
  {Calibbi}}, \bibinfo {author} {\bibfnamefont {G.}~\bibnamefont {Ferretti}},
  \bibinfo {author} {\bibfnamefont {D.}~\bibnamefont {Milstead}}, \bibinfo
  {author} {\bibfnamefont {C.}~\bibnamefont {Petersson}}, \ and\ \bibinfo
  {author} {\bibfnamefont {R.}~\bibnamefont {P\"ottgen}},\ }\href {\doibase
  10.1007/JHEP05(2016)144} {\bibfield  {journal} {\bibinfo  {journal} {JHEP}\
  }\textbf {\bibinfo {volume} {05}},\ \bibinfo {pages} {144} (\bibinfo {year}
  {2016})},\ \bibinfo {note} {[Erratum: JHEP 10, 195 (2017)]},\ \Eprint
  {http://arxiv.org/abs/1602.04821} {arXiv:1602.04821 [hep-ph]} \BibitemShut
  {NoStop}%
\bibitem [{\citenamefont {Allahverdi}\ \emph {et~al.}(2018)\citenamefont
  {Allahverdi}, \citenamefont {Dev},\ and\ \citenamefont
  {Dutta}}]{Allahverdi:2017edd}%
  \BibitemOpen
  \bibfield  {author} {\bibinfo {author} {\bibfnamefont {R.}~\bibnamefont
  {Allahverdi}}, \bibinfo {author} {\bibfnamefont {P.~S.~B.}\ \bibnamefont
  {Dev}}, \ and\ \bibinfo {author} {\bibfnamefont {B.}~\bibnamefont {Dutta}},\
  }\href {\doibase 10.1016/j.physletb.2018.02.019} {\bibfield  {journal}
  {\bibinfo  {journal} {Phys. Lett. B}\ }\textbf {\bibinfo {volume} {779}},\
  \bibinfo {pages} {262} (\bibinfo {year} {2018})},\ \Eprint
  {http://arxiv.org/abs/1712.02713} {arXiv:1712.02713 [hep-ph]} \BibitemShut
  {NoStop}%
\bibitem [{\citenamefont {Grojean}\ \emph {et~al.}(2018)\citenamefont
  {Grojean}, \citenamefont {Shakya}, \citenamefont {Wells},\ and\ \citenamefont
  {Zhang}}]{Grojean:2018fus}%
  \BibitemOpen
  \bibfield  {author} {\bibinfo {author} {\bibfnamefont {C.}~\bibnamefont
  {Grojean}}, \bibinfo {author} {\bibfnamefont {B.}~\bibnamefont {Shakya}},
  \bibinfo {author} {\bibfnamefont {J.~D.}\ \bibnamefont {Wells}}, \ and\
  \bibinfo {author} {\bibfnamefont {Z.}~\bibnamefont {Zhang}},\ }\href
  {\doibase 10.1103/PhysRevLett.121.171801} {\bibfield  {journal} {\bibinfo
  {journal} {Phys. Rev. Lett.}\ }\textbf {\bibinfo {volume} {121}},\ \bibinfo
  {pages} {171801} (\bibinfo {year} {2018})},\ \Eprint
  {http://arxiv.org/abs/1806.00011} {arXiv:1806.00011 [hep-ph]} \BibitemShut
  {NoStop}%
\bibitem [{\citenamefont {Bringmann}\ \emph {et~al.}(2019)\citenamefont
  {Bringmann}, \citenamefont {Cline},\ and\ \citenamefont
  {Cornell}}]{Bringmann:2018sbs}%
  \BibitemOpen
  \bibfield  {author} {\bibinfo {author} {\bibfnamefont {T.}~\bibnamefont
  {Bringmann}}, \bibinfo {author} {\bibfnamefont {J.~M.}\ \bibnamefont
  {Cline}}, \ and\ \bibinfo {author} {\bibfnamefont {J.~M.}\ \bibnamefont
  {Cornell}},\ }\href {\doibase 10.1103/PhysRevD.99.035024} {\bibfield
  {journal} {\bibinfo  {journal} {Phys. Rev. D}\ }\textbf {\bibinfo {volume}
  {99}},\ \bibinfo {pages} {035024} (\bibinfo {year} {2019})},\ \Eprint
  {http://arxiv.org/abs/1810.08215} {arXiv:1810.08215 [hep-ph]} \BibitemShut
  {NoStop}%
\bibitem [{\citenamefont {Sirunyan}\ \emph
  {et~al.}(2020{\natexlab{d}})\citenamefont {Sirunyan} \emph
  {et~al.}}]{CMS:2019gwf}%
  \BibitemOpen
  \bibfield  {author} {\bibinfo {author} {\bibfnamefont {A.}~\bibnamefont
  {Sirunyan}} \emph {et~al.} (\bibinfo {collaboration} {CMS}),\ }\href
  {\doibase 10.1007/JHEP05(2020)033} {\bibfield  {journal} {\bibinfo  {journal}
  {JHEP}\ }\textbf {\bibinfo {volume} {05}},\ \bibinfo {pages} {033} (\bibinfo
  {year} {2020}{\natexlab{d}})},\ \Eprint {http://arxiv.org/abs/1911.03947}
  {arXiv:1911.03947 [hep-ex]} \BibitemShut {NoStop}%
\bibitem [{\citenamefont {Dvali}\ and\ \citenamefont
  {Gabadadze}(1999)}]{Dvali:1999gf}%
  \BibitemOpen
  \bibfield  {author} {\bibinfo {author} {\bibfnamefont {G.~R.}\ \bibnamefont
  {Dvali}}\ and\ \bibinfo {author} {\bibfnamefont {G.}~\bibnamefont
  {Gabadadze}},\ }\href {\doibase 10.1016/S0370-2693(99)00766-2} {\bibfield
  {journal} {\bibinfo  {journal} {Phys. Lett. B}\ }\textbf {\bibinfo {volume}
  {460}},\ \bibinfo {pages} {47} (\bibinfo {year} {1999})},\ \Eprint
  {http://arxiv.org/abs/hep-ph/9904221} {arXiv:hep-ph/9904221} \BibitemShut
  {NoStop}%
\bibitem [{\citenamefont {Nussinov}\ and\ \citenamefont
  {Shrock}(2002)}]{Nussinov:2001rb}%
  \BibitemOpen
  \bibfield  {author} {\bibinfo {author} {\bibfnamefont {S.}~\bibnamefont
  {Nussinov}}\ and\ \bibinfo {author} {\bibfnamefont {R.}~\bibnamefont
  {Shrock}},\ }\href {\doibase 10.1103/PhysRevLett.88.171601} {\bibfield
  {journal} {\bibinfo  {journal} {Phys. Rev. Lett.}\ }\textbf {\bibinfo
  {volume} {88}},\ \bibinfo {pages} {171601} (\bibinfo {year} {2002})},\
  \Eprint {http://arxiv.org/abs/hep-ph/0112337} {arXiv:hep-ph/0112337}
  \BibitemShut {NoStop}%
\bibitem [{\citenamefont {Girmohanta}\ and\ \citenamefont
  {Shrock}(2020{\natexlab{a}})}]{Girmohanta:2019fsx}%
  \BibitemOpen
  \bibfield  {author} {\bibinfo {author} {\bibfnamefont {S.}~\bibnamefont
  {Girmohanta}}\ and\ \bibinfo {author} {\bibfnamefont {R.}~\bibnamefont
  {Shrock}},\ }\href {\doibase 10.1103/PhysRevD.101.015017} {\bibfield
  {journal} {\bibinfo  {journal} {Phys. Rev. D}\ }\textbf {\bibinfo {volume}
  {101}},\ \bibinfo {pages} {015017} (\bibinfo {year} {2020}{\natexlab{a}})},\
  \Eprint {http://arxiv.org/abs/1911.05102} {arXiv:1911.05102 [hep-ph]}
  \BibitemShut {NoStop}%
\bibitem [{\citenamefont {Girmohanta}\ and\ \citenamefont
  {Shrock}(2020{\natexlab{b}})}]{Girmohanta:2020qfd}%
  \BibitemOpen
  \bibfield  {author} {\bibinfo {author} {\bibfnamefont {S.}~\bibnamefont
  {Girmohanta}}\ and\ \bibinfo {author} {\bibfnamefont {R.}~\bibnamefont
  {Shrock}},\ }\href {\doibase 10.1103/PhysRevD.101.095012} {\bibfield
  {journal} {\bibinfo  {journal} {Phys. Rev. D}\ }\textbf {\bibinfo {volume}
  {101}},\ \bibinfo {pages} {095012} (\bibinfo {year} {2020}{\natexlab{b}})},\
  \Eprint {http://arxiv.org/abs/2003.14185} {arXiv:2003.14185 [hep-ph]}
  \BibitemShut {NoStop}%
\bibitem [{\citenamefont {Berezhiani}\ and\ \citenamefont
  {Vainshtein}(2019)}]{Berezhiani:2018xsx}%
  \BibitemOpen
  \bibfield  {author} {\bibinfo {author} {\bibfnamefont {Z.}~\bibnamefont
  {Berezhiani}}\ and\ \bibinfo {author} {\bibfnamefont {A.}~\bibnamefont
  {Vainshtein}},\ }\href {\doibase 10.1016/j.physletb.2018.11.014} {\bibfield
  {journal} {\bibinfo  {journal} {Phys. Lett. B}\ }\textbf {\bibinfo {volume}
  {788}},\ \bibinfo {pages} {58} (\bibinfo {year} {2019})},\ \Eprint
  {http://arxiv.org/abs/1809.00997} {arXiv:1809.00997 [hep-ph]} \BibitemShut
  {NoStop}%
\bibitem [{\citenamefont {Berezhiani}\ and\ \citenamefont
  {Vainshtein}(2018)}]{Berezhiani:2018pcp}%
  \BibitemOpen
  \bibfield  {author} {\bibinfo {author} {\bibfnamefont {Z.}~\bibnamefont
  {Berezhiani}}\ and\ \bibinfo {author} {\bibfnamefont {A.}~\bibnamefont
  {Vainshtein}},\ }\href {\doibase 10.1142/S0217751X18440165} {\bibfield
  {journal} {\bibinfo  {journal} {Int. J. Mod. Phys. A}\ }\textbf {\bibinfo
  {volume} {33}},\ \bibinfo {pages} {1844016} (\bibinfo {year}
  {2018})}\BibitemShut {NoStop}%
\bibitem [{\citenamefont {Tureanu}(2018)}]{Tureanu:2018phm}%
  \BibitemOpen
  \bibfield  {author} {\bibinfo {author} {\bibfnamefont {A.}~\bibnamefont
  {Tureanu}},\ }\href {\doibase 10.1103/PhysRevD.98.015019} {\bibfield
  {journal} {\bibinfo  {journal} {Phys. Rev. D}\ }\textbf {\bibinfo {volume}
  {98}},\ \bibinfo {pages} {015019} (\bibinfo {year} {2018})},\ \Eprint
  {http://arxiv.org/abs/1804.06433} {arXiv:1804.06433 [hep-ph]} \BibitemShut
  {NoStop}%
\bibitem [{\citenamefont {Fujikawa}\ and\ \citenamefont
  {Tureanu}(2021)}]{Fujikawa:2020gsa}%
  \BibitemOpen
  \bibfield  {author} {\bibinfo {author} {\bibfnamefont {K.}~\bibnamefont
  {Fujikawa}}\ and\ \bibinfo {author} {\bibfnamefont {A.}~\bibnamefont
  {Tureanu}},\ }\href {\doibase 10.1103/PhysRevD.103.065017} {\bibfield
  {journal} {\bibinfo  {journal} {Phys. Rev. D}\ }\textbf {\bibinfo {volume}
  {103}},\ \bibinfo {pages} {065017} (\bibinfo {year} {2021})},\ \Eprint
  {http://arxiv.org/abs/2009.12843} {arXiv:2009.12843 [hep-ph]} \BibitemShut
  {NoStop}%
\bibitem [{\citenamefont {Berezhiani}(2016)}]{Berezhiani:2015afa}%
  \BibitemOpen
  \bibfield  {author} {\bibinfo {author} {\bibfnamefont {Z.}~\bibnamefont
  {Berezhiani}},\ }\href {\doibase 10.1140/epjc/s10052-016-4564-0} {\bibfield
  {journal} {\bibinfo  {journal} {Eur. Phys. J. C}\ }\textbf {\bibinfo {volume}
  {76}},\ \bibinfo {pages} {705} (\bibinfo {year} {2016})},\ \Eprint
  {http://arxiv.org/abs/1507.05478} {arXiv:1507.05478 [hep-ph]} \BibitemShut
  {NoStop}%
\bibitem [{\citenamefont {Addazi}\ \emph {et~al.}(2017)\citenamefont {Addazi},
  \citenamefont {Berezhiani},\ and\ \citenamefont
  {Kamyshkov}}]{Addazi:2016rgo}%
  \BibitemOpen
  \bibfield  {author} {\bibinfo {author} {\bibfnamefont {A.}~\bibnamefont
  {Addazi}}, \bibinfo {author} {\bibfnamefont {Z.}~\bibnamefont {Berezhiani}},
  \ and\ \bibinfo {author} {\bibfnamefont {Y.}~\bibnamefont {Kamyshkov}},\
  }\href {\doibase 10.1140/epjc/s10052-017-4870-1} {\bibfield  {journal}
  {\bibinfo  {journal} {Eur. Phys. J. C}\ }\textbf {\bibinfo {volume} {77}},\
  \bibinfo {pages} {301} (\bibinfo {year} {2017})},\ \Eprint
  {http://arxiv.org/abs/1607.00348} {arXiv:1607.00348 [hep-ph]} \BibitemShut
  {NoStop}%
\bibitem [{\citenamefont {Babu}\ and\ \citenamefont
  {Mohapatra}(2016)}]{Babu:2016rwa}%
  \BibitemOpen
  \bibfield  {author} {\bibinfo {author} {\bibfnamefont {K.~S.}\ \bibnamefont
  {Babu}}\ and\ \bibinfo {author} {\bibfnamefont {R.~N.}\ \bibnamefont
  {Mohapatra}},\ }\href {\doibase 10.1103/PhysRevD.94.054034} {\bibfield
  {journal} {\bibinfo  {journal} {Phys. Rev. D}\ }\textbf {\bibinfo {volume}
  {94}},\ \bibinfo {pages} {054034} (\bibinfo {year} {2016})},\ \Eprint
  {http://arxiv.org/abs/1606.08374} {arXiv:1606.08374 [hep-ph]} \BibitemShut
  {NoStop}%
\bibitem [{\citenamefont {Berezhiani}\ and\ \citenamefont
  {Bento}(2006{\natexlab{b}})}]{Berezhiani:2006je}%
  \BibitemOpen
  \bibfield  {author} {\bibinfo {author} {\bibfnamefont {Z.}~\bibnamefont
  {Berezhiani}}\ and\ \bibinfo {author} {\bibfnamefont {L.}~\bibnamefont
  {Bento}},\ }\href {\doibase 10.1016/j.physletb.2006.03.008} {\bibfield
  {journal} {\bibinfo  {journal} {Phys. Lett. B}\ }\textbf {\bibinfo {volume}
  {635}},\ \bibinfo {pages} {253} (\bibinfo {year} {2006}{\natexlab{b}})},\
  \Eprint {http://arxiv.org/abs/hep-ph/0602227} {arXiv:hep-ph/0602227}
  \BibitemShut {NoStop}%
\bibitem [{\citenamefont {Baldo-Ceolin}\ \emph {et~al.}(1994)\citenamefont
  {Baldo-Ceolin} \emph {et~al.}}]{BaldoCeolin:1994jz}%
  \BibitemOpen
  \bibfield  {author} {\bibinfo {author} {\bibfnamefont {M.}~\bibnamefont
  {Baldo-Ceolin}} \emph {et~al.},\ }\href {\doibase 10.1007/BF01580321}
  {\bibfield  {journal} {\bibinfo  {journal} {Z. Phys. C}\ }\textbf {\bibinfo
  {volume} {63}},\ \bibinfo {pages} {409} (\bibinfo {year} {1994})}\BibitemShut
  {NoStop}%
\bibitem [{\citenamefont {Abe}\ \emph {et~al.}(2021)\citenamefont {Abe} \emph
  {et~al.}}]{Super-Kamiokande:2020bov}%
  \BibitemOpen
  \bibfield  {author} {\bibinfo {author} {\bibfnamefont {K.}~\bibnamefont
  {Abe}} \emph {et~al.} (\bibinfo {collaboration} {Super-Kamiokande}),\ }\href
  {\doibase 10.1103/PhysRevD.103.012008} {\bibfield  {journal} {\bibinfo
  {journal} {Phys. Rev. D}\ }\textbf {\bibinfo {volume} {103}},\ \bibinfo
  {pages} {012008} (\bibinfo {year} {2021})},\ \Eprint
  {http://arxiv.org/abs/2012.02607} {arXiv:2012.02607 [hep-ex]} \BibitemShut
  {NoStop}%
\bibitem [{\citenamefont {Young}\ and\ \citenamefont
  {Barrow}(2019)}]{Young:2019pzq}%
  \BibitemOpen
  \bibfield  {author} {\bibinfo {author} {\bibfnamefont {A.}~\bibnamefont
  {Young}}\ and\ \bibinfo {author} {\bibfnamefont {J.}~\bibnamefont {Barrow}},\
  }\href {\doibase 10.1051/epjconf/201921907005} {\bibfield  {journal}
  {\bibinfo  {journal} {EPJ Web Conf.}\ }\textbf {\bibinfo {volume} {219}},\
  \bibinfo {pages} {07005} (\bibinfo {year} {2019})}\BibitemShut {NoStop}%
\bibitem [{\citenamefont {Barrow}\ \emph {et~al.}(2020)\citenamefont {Barrow},
  \citenamefont {Golubeva}, \citenamefont {Paryev},\ and\ \citenamefont
  {Richard}}]{Barrow:2019viz}%
  \BibitemOpen
  \bibfield  {author} {\bibinfo {author} {\bibfnamefont {J.~L.}\ \bibnamefont
  {Barrow}}, \bibinfo {author} {\bibfnamefont {E.~S.}\ \bibnamefont
  {Golubeva}}, \bibinfo {author} {\bibfnamefont {E.}~\bibnamefont {Paryev}}, \
  and\ \bibinfo {author} {\bibfnamefont {J.-M.}\ \bibnamefont {Richard}},\
  }\href {\doibase 10.1103/PhysRevD.101.036008} {\bibfield  {journal} {\bibinfo
   {journal} {Phys. Rev. D}\ }\textbf {\bibinfo {volume} {101}},\ \bibinfo
  {pages} {036008} (\bibinfo {year} {2020})},\ \Eprint
  {http://arxiv.org/abs/1906.02833} {arXiv:1906.02833 [hep-ex]} \BibitemShut
  {NoStop}%
\bibitem [{\citenamefont {Oosterhof}\ \emph {et~al.}(2019)\citenamefont
  {Oosterhof}, \citenamefont {Long}, \citenamefont {de~Vries}, \citenamefont
  {Timmermans},\ and\ \citenamefont {van Kolck}}]{Oosterhof:2019dlo}%
  \BibitemOpen
  \bibfield  {author} {\bibinfo {author} {\bibfnamefont {F.}~\bibnamefont
  {Oosterhof}}, \bibinfo {author} {\bibfnamefont {B.}~\bibnamefont {Long}},
  \bibinfo {author} {\bibfnamefont {J.}~\bibnamefont {de~Vries}}, \bibinfo
  {author} {\bibfnamefont {R.~G.~E.}\ \bibnamefont {Timmermans}}, \ and\
  \bibinfo {author} {\bibfnamefont {U.}~\bibnamefont {van Kolck}},\ }\href
  {\doibase 10.1103/PhysRevLett.122.172501} {\bibfield  {journal} {\bibinfo
  {journal} {Phys. Rev. Lett.}\ }\textbf {\bibinfo {volume} {122}},\ \bibinfo
  {pages} {172501} (\bibinfo {year} {2019})},\ \Eprint
  {http://arxiv.org/abs/1902.05342} {arXiv:1902.05342 [hep-ph]} \BibitemShut
  {NoStop}%
\bibitem [{\citenamefont {Abe}\ \emph {et~al.}(2015)\citenamefont {Abe} \emph
  {et~al.}}]{Super-Kamiokande:2011idx}%
  \BibitemOpen
  \bibfield  {author} {\bibinfo {author} {\bibfnamefont {K.}~\bibnamefont
  {Abe}} \emph {et~al.} (\bibinfo {collaboration} {Super-Kamiokande}),\ }\href
  {\doibase 10.1103/PhysRevD.91.072006} {\bibfield  {journal} {\bibinfo
  {journal} {Phys. Rev. D}\ }\textbf {\bibinfo {volume} {91}},\ \bibinfo
  {pages} {072006} (\bibinfo {year} {2015})},\ \Eprint
  {http://arxiv.org/abs/1109.4227} {arXiv:1109.4227 [hep-ex]} \BibitemShut
  {NoStop}%
\bibitem [{\citenamefont {Ayres}\ \emph {et~al.}(2007)\citenamefont {Ayres}
  \emph {et~al.}}]{NOvA:2007rmc}%
  \BibitemOpen
  \bibfield  {author} {\bibinfo {author} {\bibfnamefont {D.~S.}\ \bibnamefont
  {Ayres}} \emph {et~al.} (\bibinfo {collaboration} {NOvA}),\ }\href {\doibase
  10.2172/935497} {\  (\bibinfo {year} {2007}),\ 10.2172/935497}\BibitemShut
  {NoStop}%
\bibitem [{\citenamefont {Abe}\ \emph {et~al.}(2018)\citenamefont {Abe} \emph
  {et~al.}}]{Hyper-Kamiokande:2018ofw}%
  \BibitemOpen
  \bibfield  {author} {\bibinfo {author} {\bibfnamefont {K.}~\bibnamefont
  {Abe}} \emph {et~al.} (\bibinfo {collaboration} {Hyper-Kamiokande}),\
  }\href@noop {} {\  (\bibinfo {year} {2018})},\ \Eprint
  {http://arxiv.org/abs/1805.04163} {arXiv:1805.04163 [physics.ins-det]}
  \BibitemShut {NoStop}%
\bibitem [{\citenamefont {Abi}\ \emph {et~al.}(2020)\citenamefont {Abi} \emph
  {et~al.}}]{DUNE:2020ypp}%
  \BibitemOpen
  \bibfield  {author} {\bibinfo {author} {\bibfnamefont {B.}~\bibnamefont
  {Abi}} \emph {et~al.} (\bibinfo {collaboration} {DUNE}),\ }\href@noop {} {\
  (\bibinfo {year} {2020})},\ \Eprint {http://arxiv.org/abs/2002.03005}
  {arXiv:2002.03005 [hep-ex]} \BibitemShut {NoStop}%
\bibitem [{\citenamefont {Dover}\ \emph {et~al.}(1985)\citenamefont {Dover},
  \citenamefont {Gal},\ and\ \citenamefont {Richard}}]{Dover:1985hk}%
  \BibitemOpen
  \bibfield  {author} {\bibinfo {author} {\bibfnamefont {C.~B.}\ \bibnamefont
  {Dover}}, \bibinfo {author} {\bibfnamefont {A.}~\bibnamefont {Gal}}, \ and\
  \bibinfo {author} {\bibfnamefont {J.~M.}\ \bibnamefont {Richard}},\ }\href
  {\doibase 10.1103/PhysRevC.31.1423} {\bibfield  {journal} {\bibinfo
  {journal} {Phys. Rev. C}\ }\textbf {\bibinfo {volume} {31}},\ \bibinfo
  {pages} {1423} (\bibinfo {year} {1985})}\BibitemShut {NoStop}%
\bibitem [{\citenamefont {Friedman}\ and\ \citenamefont
  {Gal}(2008)}]{Friedman:2008es}%
  \BibitemOpen
  \bibfield  {author} {\bibinfo {author} {\bibfnamefont {E.}~\bibnamefont
  {Friedman}}\ and\ \bibinfo {author} {\bibfnamefont {A.}~\bibnamefont {Gal}},\
  }\href {\doibase 10.1103/PhysRevD.78.016002} {\bibfield  {journal} {\bibinfo
  {journal} {Phys. Rev. D}\ }\textbf {\bibinfo {volume} {78}},\ \bibinfo
  {pages} {016002} (\bibinfo {year} {2008})},\ \Eprint
  {http://arxiv.org/abs/0803.3696} {arXiv:0803.3696 [hep-ph]} \BibitemShut
  {NoStop}%
\bibitem [{\citenamefont {Abratenko}\ \emph {et~al.}(2022)\citenamefont
  {Abratenko} \emph {et~al.}}]{microboone1113}%
  \BibitemOpen
  \bibfield  {author} {\bibinfo {author} {\bibfnamefont {P.}~\bibnamefont
  {Abratenko}} \emph {et~al.} (\bibinfo {collaboration} {MicroBooNE}),\
  }\href@noop {} {}\bibinfo {howpublished}
  {\href{https://microboone.fnal.gov/document/microboone-note-1113-pub/}{MICROBOONE-NOTE-1113-PUB}}
  (\bibinfo {year} {2022})\BibitemShut {NoStop}%
\bibitem [{\citenamefont {Addazi}\ \emph {et~al.}(2021)\citenamefont {Addazi}
  \emph {et~al.}}]{Addazi:2020nlz}%
  \BibitemOpen
  \bibfield  {author} {\bibinfo {author} {\bibfnamefont {A.}~\bibnamefont
  {Addazi}} \emph {et~al.},\ }\href {\doibase 10.1088/1361-6471/abf429}
  {\bibfield  {journal} {\bibinfo  {journal} {J. Phys. G}\ }\textbf {\bibinfo
  {volume} {48}},\ \bibinfo {pages} {070501} (\bibinfo {year} {2021})},\
  \Eprint {http://arxiv.org/abs/2006.04907} {arXiv:2006.04907
  [physics.ins-det]} \BibitemShut {NoStop}%
\bibitem [{\citenamefont {Serebrov}\ \emph {et~al.}(2016)\citenamefont
  {Serebrov}, \citenamefont {Fomin},\ and\ \citenamefont
  {Kamyshkov}}]{Serebrov:2016rvi}%
  \BibitemOpen
  \bibfield  {author} {\bibinfo {author} {\bibfnamefont {A.~P.}\ \bibnamefont
  {Serebrov}}, \bibinfo {author} {\bibfnamefont {A.~K.}\ \bibnamefont {Fomin}},
  \ and\ \bibinfo {author} {\bibfnamefont {Y.~A.}\ \bibnamefont {Kamyshkov}},\
  }\href {\doibase 10.1134/S1063785016010314} {\bibfield  {journal} {\bibinfo
  {journal} {Tech. Phys. Lett.}\ }\textbf {\bibinfo {volume} {42}},\ \bibinfo
  {pages} {99} (\bibinfo {year} {2016})}\BibitemShut {NoStop}%
\bibitem [{\citenamefont {Fomin}\ \emph
  {et~al.}(2019{\natexlab{a}})\citenamefont {Fomin}, \citenamefont {Serebrov},
  \citenamefont {Chaikovskii}, \citenamefont {Zherebtsov}, \citenamefont
  {Murashkin},\ and\ \citenamefont {Golubeva}}]{Fomin:2019oyj}%
  \BibitemOpen
  \bibfield  {author} {\bibinfo {author} {\bibfnamefont {A.}~\bibnamefont
  {Fomin}}, \bibinfo {author} {\bibfnamefont {A.}~\bibnamefont {Serebrov}},
  \bibinfo {author} {\bibfnamefont {M.}~\bibnamefont {Chaikovskii}}, \bibinfo
  {author} {\bibfnamefont {O.}~\bibnamefont {Zherebtsov}}, \bibinfo {author}
  {\bibfnamefont {A.}~\bibnamefont {Murashkin}}, \ and\ \bibinfo {author}
  {\bibfnamefont {E.}~\bibnamefont {Golubeva}},\ }\href {\doibase
  10.1051/epjconf/201921907003} {\bibfield  {journal} {\bibinfo  {journal} {EPJ
  Web Conf.}\ }\textbf {\bibinfo {volume} {219}},\ \bibinfo {pages} {07003}
  (\bibinfo {year} {2019}{\natexlab{a}})}\BibitemShut {NoStop}%
\bibitem [{\citenamefont {Fomin}\ \emph {et~al.}(2017)\citenamefont {Fomin},
  \citenamefont {Serebrov}, \citenamefont {Zherebtsov}, \citenamefont
  {Leonova},\ and\ \citenamefont {Chaikovskii}}]{Fomin:2017aiz}%
  \BibitemOpen
  \bibfield  {author} {\bibinfo {author} {\bibfnamefont {A.~K.}\ \bibnamefont
  {Fomin}}, \bibinfo {author} {\bibfnamefont {A.~P.}\ \bibnamefont {Serebrov}},
  \bibinfo {author} {\bibfnamefont {O.~M.}\ \bibnamefont {Zherebtsov}},
  \bibinfo {author} {\bibfnamefont {E.~N.}\ \bibnamefont {Leonova}}, \ and\
  \bibinfo {author} {\bibfnamefont {M.~E.}\ \bibnamefont {Chaikovskii}},\
  }\href {\doibase 10.1088/1742-6596/798/1/012115} {\bibfield  {journal}
  {\bibinfo  {journal} {J. Phys. Conf. Ser.}\ }\textbf {\bibinfo {volume}
  {798}},\ \bibinfo {pages} {012115} (\bibinfo {year} {2017})}\BibitemShut
  {NoStop}%
\bibitem [{\citenamefont {Fomin}(2017)}]{Fomin:2017lej}%
  \BibitemOpen
  \bibfield  {author} {\bibinfo {author} {\bibfnamefont {A.}~\bibnamefont
  {Fomin}},\ }\href {\doibase 10.22323/1.281.0189} {\bibfield  {journal}
  {\bibinfo  {journal} {PoS}\ }\textbf {\bibinfo {volume} {INPC2016}},\
  \bibinfo {pages} {189} (\bibinfo {year} {2017})}\BibitemShut {NoStop}%
\bibitem [{\citenamefont {Fomin}\ \emph {et~al.}(2018)\citenamefont {Fomin}
  \emph {et~al.}}]{Fomin:2018qrq}%
  \BibitemOpen
  \bibfield  {author} {\bibinfo {author} {\bibfnamefont {A.~K.}\ \bibnamefont
  {Fomin}} \emph {et~al.},\ }\href {\doibase 10.18502/ken.v3i1.1731} {\bibfield
   {journal} {\bibinfo  {journal} {KnE Energ. Phys.}\ }\textbf {\bibinfo
  {volume} {3}},\ \bibinfo {pages} {109} (\bibinfo {year} {2018})}\BibitemShut
  {NoStop}%
\bibitem [{\citenamefont {Fomin}\ \emph
  {et~al.}(2019{\natexlab{b}})\citenamefont {Fomin}, \citenamefont {Serebrov},
  \citenamefont {Chaikovskii}, \citenamefont {Zherebtsov}, \citenamefont
  {Murashkin},\ and\ \citenamefont {Golubeva}}]{Fomin:2019oje}%
  \BibitemOpen
  \bibfield  {author} {\bibinfo {author} {\bibfnamefont {A.}~\bibnamefont
  {Fomin}}, \bibinfo {author} {\bibfnamefont {A.}~\bibnamefont {Serebrov}},
  \bibinfo {author} {\bibfnamefont {M.}~\bibnamefont {Chaikovskii}}, \bibinfo
  {author} {\bibfnamefont {O.}~\bibnamefont {Zherebtsov}}, \bibinfo {author}
  {\bibfnamefont {A.}~\bibnamefont {Murashkin}}, \ and\ \bibinfo {author}
  {\bibfnamefont {E.}~\bibnamefont {Golubeva}},\ }\href {\doibase
  10.1088/1742-6596/1390/1/012133} {\bibfield  {journal} {\bibinfo  {journal}
  {J. Phys. Conf. Ser.}\ }\textbf {\bibinfo {volume} {1390}},\ \bibinfo {pages}
  {012133} (\bibinfo {year} {2019}{\natexlab{b}})}\BibitemShut {NoStop}%
\bibitem [{Peg(2013)}]{Peggs:2013sgv}%
  \BibitemOpen
  \href@noop {} {\  (\bibinfo {year} {2013})}\BibitemShut {NoStop}%
\bibitem [{\citenamefont {Frost}(2017)}]{Frost:2016qzt}%
  \BibitemOpen
  \bibfield  {author} {\bibinfo {author} {\bibfnamefont {M.~J.}\ \bibnamefont
  {Frost}} (\bibinfo {collaboration} {NNbar}),\ }in\ \href {\doibase
  10.1142/9789813148505_0070} {\emph {\bibinfo {booktitle} {{7th Meeting on CPT
  and Lorentz Symmetry}}}}\ (\bibinfo {year} {2017})\ pp.\ \bibinfo {pages}
  {265--267},\ \Eprint {http://arxiv.org/abs/1607.07271} {arXiv:1607.07271
  [hep-ph]} \BibitemShut {NoStop}%
\bibitem [{\citenamefont {Klinkby}\ \emph {et~al.}(2016)\citenamefont
  {Klinkby}, \citenamefont {Soldner} \emph {et~al.}}]{Klinkby_2016}%
  \BibitemOpen
  \bibfield  {author} {\bibinfo {author} {\bibfnamefont {E.}~\bibnamefont
  {Klinkby}}, \bibinfo {author} {\bibfnamefont {T.}~\bibnamefont {Soldner}},
  \emph {et~al.},\ }\href {\doibase 10.1088/1742-6596/746/1/012051} {\bibfield
  {journal} {\bibinfo  {journal} {Journal of Physics: Conference Series}\
  }\textbf {\bibinfo {volume} {746}},\ \bibinfo {pages} {012051} (\bibinfo
  {year} {2016})}\BibitemShut {NoStop}%
\bibitem [{\citenamefont {Santoro}\ \emph {et~al.}(2020)\citenamefont {Santoro}
  \emph {et~al.}}]{Santoro:2020nke}%
  \BibitemOpen
  \bibfield  {author} {\bibinfo {author} {\bibfnamefont {V.}~\bibnamefont
  {Santoro}} \emph {et~al.},\ }in\ \href@noop {} {\emph {\bibinfo {booktitle}
  {{23rd meeting of the International Collaboration on Advanced Neutron Sources
  (ICANS XXIII)}}}}\ (\bibinfo {year} {2020})\ \Eprint
  {http://arxiv.org/abs/2002.03883} {arXiv:2002.03883 [physics.ins-det]}
  \BibitemShut {NoStop}%
\bibitem [{\citenamefont {Phillips}\ \emph {et~al.}(2016)\citenamefont
  {Phillips} \emph {et~al.}}]{Phillips:2014fgb}%
  \BibitemOpen
  \bibfield  {author} {\bibinfo {author} {\bibfnamefont {D.~G.}\ \bibnamefont
  {Phillips}, \bibfnamefont {II}} \emph {et~al.},\ }\href {\doibase
  10.1016/j.physrep.2015.11.001} {\bibfield  {journal} {\bibinfo  {journal}
  {Phys. Rept.}\ }\textbf {\bibinfo {volume} {612}},\ \bibinfo {pages} {1}
  (\bibinfo {year} {2016})},\ \Eprint {http://arxiv.org/abs/1410.1100}
  {arXiv:1410.1100 [hep-ex]} \BibitemShut {NoStop}%
\bibitem [{\citenamefont {Nesvizhevsky}\ \emph {et~al.}(2020)\citenamefont
  {Nesvizhevsky}, \citenamefont {Gudkov}, \citenamefont {Protasov},
  \citenamefont {Snow},\ and\ \citenamefont {Voronin}}]{Nesvizhevsky:2020vwx}%
  \BibitemOpen
  \bibfield  {author} {\bibinfo {author} {\bibfnamefont {V.~V.}\ \bibnamefont
  {Nesvizhevsky}}, \bibinfo {author} {\bibfnamefont {V.}~\bibnamefont
  {Gudkov}}, \bibinfo {author} {\bibfnamefont {K.~V.}\ \bibnamefont
  {Protasov}}, \bibinfo {author} {\bibfnamefont {W.~M.}\ \bibnamefont {Snow}},
  \ and\ \bibinfo {author} {\bibfnamefont {A.~Y.}\ \bibnamefont {Voronin}},\
  }\href {\doibase 10.1016/j.physletb.2020.135357} {\bibfield  {journal}
  {\bibinfo  {journal} {Phys. Lett. B}\ }\textbf {\bibinfo {volume} {803}},\
  \bibinfo {pages} {135357} (\bibinfo {year} {2020})}\BibitemShut {NoStop}%
\bibitem [{\citenamefont {Gudkov}\ \emph {et~al.}(2020)\citenamefont {Gudkov},
  \citenamefont {Nesvizhevsky}, \citenamefont {Protasov}, \citenamefont
  {Snow},\ and\ \citenamefont {Voronin}}]{Gudkov:2019gro}%
  \BibitemOpen
  \bibfield  {author} {\bibinfo {author} {\bibfnamefont {V.}~\bibnamefont
  {Gudkov}}, \bibinfo {author} {\bibfnamefont {V.~V.}\ \bibnamefont
  {Nesvizhevsky}}, \bibinfo {author} {\bibfnamefont {K.~V.}\ \bibnamefont
  {Protasov}}, \bibinfo {author} {\bibfnamefont {W.~M.}\ \bibnamefont {Snow}},
  \ and\ \bibinfo {author} {\bibfnamefont {A.~Y.}\ \bibnamefont {Voronin}},\
  }\href {\doibase 10.1016/j.physletb.2020.135636} {\bibfield  {journal}
  {\bibinfo  {journal} {Phys. Lett. B}\ }\textbf {\bibinfo {volume} {808}},\
  \bibinfo {pages} {135636} (\bibinfo {year} {2020})},\ \Eprint
  {http://arxiv.org/abs/1912.06730} {arXiv:1912.06730 [hep-ph]} \BibitemShut
  {NoStop}%
\bibitem [{\citenamefont {Santoro}\ \emph {et~al.}(2022)\citenamefont {Santoro}
  \emph {et~al.}}]{Santoro:2022tvi}%
  \BibitemOpen
  \bibfield  {author} {\bibinfo {author} {\bibfnamefont {V.}~\bibnamefont
  {Santoro}} \emph {et~al.},\ }in\ \href@noop {} {\emph {\bibinfo {booktitle}
  {{14th International Topical Meeting on Nuclear Applications of
  Accelerators}}}}\ (\bibinfo {year} {2022})\ \Eprint
  {http://arxiv.org/abs/2204.04051} {arXiv:2204.04051 [physics.ins-det]}
  \BibitemShut {NoStop}%
\bibitem [{\citenamefont {Berezhiani}\ \emph {et~al.}(2019)\citenamefont
  {Berezhiani}, \citenamefont {Biondi}, \citenamefont {Kamyshkov},\ and\
  \citenamefont {Varriano}}]{Berezhiani:2018qqw}%
  \BibitemOpen
  \bibfield  {author} {\bibinfo {author} {\bibfnamefont {Z.}~\bibnamefont
  {Berezhiani}}, \bibinfo {author} {\bibfnamefont {R.}~\bibnamefont {Biondi}},
  \bibinfo {author} {\bibfnamefont {Y.}~\bibnamefont {Kamyshkov}}, \ and\
  \bibinfo {author} {\bibfnamefont {L.}~\bibnamefont {Varriano}},\ }\href
  {\doibase 10.3390/physics1020021} {\bibfield  {journal} {\bibinfo  {journal}
  {MDPI Physics}\ }\textbf {\bibinfo {volume} {1}},\ \bibinfo {pages} {271}
  (\bibinfo {year} {2019})},\ \Eprint {http://arxiv.org/abs/1812.11141}
  {arXiv:1812.11141 [nucl-th]} \BibitemShut {NoStop}%
\bibitem [{\citenamefont {Berezhiani}(2021)}]{Berezhiani:2020vbe}%
  \BibitemOpen
  \bibfield  {author} {\bibinfo {author} {\bibfnamefont {Z.}~\bibnamefont
  {Berezhiani}},\ }\href {\doibase 10.1140/epjc/s10052-020-08824-9} {\bibfield
  {journal} {\bibinfo  {journal} {Eur. Phys. J. C}\ }\textbf {\bibinfo {volume}
  {81}},\ \bibinfo {pages} {33} (\bibinfo {year} {2021})},\ \Eprint
  {http://arxiv.org/abs/2002.05609} {arXiv:2002.05609 [hep-ph]} \BibitemShut
  {NoStop}%
\bibitem [{\citenamefont {Kobzarev}\ \emph {et~al.}(1966)\citenamefont
  {Kobzarev}, \citenamefont {Okun},\ and\ \citenamefont
  {Pomeranchuk}}]{Kobzarev:1966qya}%
  \BibitemOpen
  \bibfield  {author} {\bibinfo {author} {\bibfnamefont {I.~{\relax Yu}.}\
  \bibnamefont {Kobzarev}}, \bibinfo {author} {\bibfnamefont {L.~B.}\
  \bibnamefont {Okun}}, \ and\ \bibinfo {author} {\bibfnamefont {I.~{\relax
  Ya}.}\ \bibnamefont {Pomeranchuk}},\ }\href@noop {} {\bibfield  {journal}
  {\bibinfo  {journal} {Sov. J. Nucl. Phys.}\ }\textbf {\bibinfo {volume}
  {3}},\ \bibinfo {pages} {837} (\bibinfo {year} {1966})},\ \bibinfo {note}
  {[Yad. Fiz.3,1154(1966)]}\BibitemShut {NoStop}%
\bibitem [{\citenamefont {Blinnikov}\ and\ \citenamefont
  {Khlopov}(1982)}]{Blinnikov:1982eh}%
  \BibitemOpen
  \bibfield  {author} {\bibinfo {author} {\bibfnamefont {S.~I.}\ \bibnamefont
  {Blinnikov}}\ and\ \bibinfo {author} {\bibfnamefont {M.~{\relax Yu}.}\
  \bibnamefont {Khlopov}},\ }\href@noop {} {\bibfield  {journal} {\bibinfo
  {journal} {Sov. J. Nucl. Phys.}\ }\textbf {\bibinfo {volume} {36}},\ \bibinfo
  {pages} {472} (\bibinfo {year} {1982})},\ \bibinfo {note} {[Yad.
  Fiz.36,809(1982)]}\BibitemShut {NoStop}%
\bibitem [{\citenamefont {Foot}\ \emph {et~al.}(1991)\citenamefont {Foot},
  \citenamefont {Lew},\ and\ \citenamefont {Volkas}}]{Foot:1991bp}%
  \BibitemOpen
  \bibfield  {author} {\bibinfo {author} {\bibfnamefont {R.}~\bibnamefont
  {Foot}}, \bibinfo {author} {\bibfnamefont {H.}~\bibnamefont {Lew}}, \ and\
  \bibinfo {author} {\bibfnamefont {R.~R.}\ \bibnamefont {Volkas}},\ }\href
  {\doibase 10.1016/0370-2693(91)91013-L} {\bibfield  {journal} {\bibinfo
  {journal} {Phys. Lett.}\ }\textbf {\bibinfo {volume} {B272}},\ \bibinfo
  {pages} {67} (\bibinfo {year} {1991})}\BibitemShut {NoStop}%
\bibitem [{\citenamefont {Hodges}(1993)}]{Hodges:1993yb}%
  \BibitemOpen
  \bibfield  {author} {\bibinfo {author} {\bibfnamefont {H.~M.}\ \bibnamefont
  {Hodges}},\ }\href {\doibase 10.1103/PhysRevD.47.456} {\bibfield  {journal}
  {\bibinfo  {journal} {Phys. Rev.}\ }\textbf {\bibinfo {volume} {D47}},\
  \bibinfo {pages} {456} (\bibinfo {year} {1993})}\BibitemShut {NoStop}%
\bibitem [{\citenamefont {Berezhiani}\ and\ \citenamefont
  {Mohapatra}(1995)}]{Berezhiani:1995yi}%
  \BibitemOpen
  \bibfield  {author} {\bibinfo {author} {\bibfnamefont {Z.~G.}\ \bibnamefont
  {Berezhiani}}\ and\ \bibinfo {author} {\bibfnamefont {R.~N.}\ \bibnamefont
  {Mohapatra}},\ }\href {\doibase 10.1103/PhysRevD.52.6607} {\bibfield
  {journal} {\bibinfo  {journal} {Phys. Rev. D}\ }\textbf {\bibinfo {volume}
  {52}},\ \bibinfo {pages} {6607} (\bibinfo {year} {1995})},\ \Eprint
  {http://arxiv.org/abs/hep-ph/9505385} {arXiv:hep-ph/9505385} \BibitemShut
  {NoStop}%
\bibitem [{\citenamefont {Berezhiani}\ \emph {et~al.}(1996)\citenamefont
  {Berezhiani}, \citenamefont {Dolgov},\ and\ \citenamefont
  {Mohapatra}}]{Berezhiani:1995am}%
  \BibitemOpen
  \bibfield  {author} {\bibinfo {author} {\bibfnamefont {Z.~G.}\ \bibnamefont
  {Berezhiani}}, \bibinfo {author} {\bibfnamefont {A.~D.}\ \bibnamefont
  {Dolgov}}, \ and\ \bibinfo {author} {\bibfnamefont {R.~N.}\ \bibnamefont
  {Mohapatra}},\ }\href {\doibase 10.1016/0370-2693(96)00219-5} {\bibfield
  {journal} {\bibinfo  {journal} {Phys. Lett. B}\ }\textbf {\bibinfo {volume}
  {375}},\ \bibinfo {pages} {26} (\bibinfo {year} {1996})},\ \Eprint
  {http://arxiv.org/abs/hep-ph/9511221} {arXiv:hep-ph/9511221} \BibitemShut
  {NoStop}%
\bibitem [{\citenamefont {Okun}(2007)}]{Okun:2006eb}%
  \BibitemOpen
  \bibfield  {author} {\bibinfo {author} {\bibfnamefont {L.~B.}\ \bibnamefont
  {Okun}},\ }\bibfield  {booktitle} {\emph {\bibinfo {booktitle} {{ITEP Meeting
  on the Future of Heavy Flavor Physics Moscow, Russia, July 24-25, 2006}}},\
  }\href {\doibase 10.1070/PU2007v050n04ABEH006227} {\bibfield  {journal}
  {\bibinfo  {journal} {Phys. Usp.}\ }\textbf {\bibinfo {volume} {50}},\
  \bibinfo {pages} {380} (\bibinfo {year} {2007})},\ \Eprint
  {http://arxiv.org/abs/hep-ph/0606202} {arXiv:hep-ph/0606202 [hep-ph]}
  \BibitemShut {NoStop}%
\bibitem [{\citenamefont {Berezhiani}\ \emph {et~al.}(2001)\citenamefont
  {Berezhiani}, \citenamefont {Comelli},\ and\ \citenamefont
  {Villante}}]{Berezhiani:2000gw}%
  \BibitemOpen
  \bibfield  {author} {\bibinfo {author} {\bibfnamefont {Z.}~\bibnamefont
  {Berezhiani}}, \bibinfo {author} {\bibfnamefont {D.}~\bibnamefont {Comelli}},
  \ and\ \bibinfo {author} {\bibfnamefont {F.~L.}\ \bibnamefont {Villante}},\
  }\href {\doibase 10.1016/S0370-2693(01)00217-9} {\bibfield  {journal}
  {\bibinfo  {journal} {Phys. Lett. B}\ }\textbf {\bibinfo {volume} {503}},\
  \bibinfo {pages} {362} (\bibinfo {year} {2001})},\ \Eprint
  {http://arxiv.org/abs/hep-ph/0008105} {arXiv:hep-ph/0008105} \BibitemShut
  {NoStop}%
\bibitem [{\citenamefont {Ignatiev}\ and\ \citenamefont
  {Volkas}(2003)}]{Ignatiev:2003js}%
  \BibitemOpen
  \bibfield  {author} {\bibinfo {author} {\bibfnamefont {A.~Y.}\ \bibnamefont
  {Ignatiev}}\ and\ \bibinfo {author} {\bibfnamefont {R.~R.}\ \bibnamefont
  {Volkas}},\ }\href {\doibase 10.1103/PhysRevD.68.023518} {\bibfield
  {journal} {\bibinfo  {journal} {Phys. Rev. D}\ }\textbf {\bibinfo {volume}
  {68}},\ \bibinfo {pages} {023518} (\bibinfo {year} {2003})},\ \Eprint
  {http://arxiv.org/abs/hep-ph/0304260} {arXiv:hep-ph/0304260} \BibitemShut
  {NoStop}%
\bibitem [{\citenamefont {Berezhiani}(2004)}]{Berezhiani:2003xm}%
  \BibitemOpen
  \bibfield  {author} {\bibinfo {author} {\bibfnamefont {Z.}~\bibnamefont
  {Berezhiani}},\ }\href {\doibase 10.1142/S0217751X04020075} {\bibfield
  {journal} {\bibinfo  {journal} {Int. J. Mod. Phys. A}\ }\textbf {\bibinfo
  {volume} {19}},\ \bibinfo {pages} {3775} (\bibinfo {year} {2004})},\ \Eprint
  {http://arxiv.org/abs/hep-ph/0312335} {arXiv:hep-ph/0312335} \BibitemShut
  {NoStop}%
\bibitem [{\citenamefont {Berezhiani}\ \emph {et~al.}(2005)\citenamefont
  {Berezhiani}, \citenamefont {Ciarcelluti}, \citenamefont {Comelli},\ and\
  \citenamefont {Villante}}]{Berezhiani:2003wj}%
  \BibitemOpen
  \bibfield  {author} {\bibinfo {author} {\bibfnamefont {Z.}~\bibnamefont
  {Berezhiani}}, \bibinfo {author} {\bibfnamefont {P.}~\bibnamefont
  {Ciarcelluti}}, \bibinfo {author} {\bibfnamefont {D.}~\bibnamefont
  {Comelli}}, \ and\ \bibinfo {author} {\bibfnamefont {F.~L.}\ \bibnamefont
  {Villante}},\ }\href {\doibase 10.1142/S0218271805005165} {\bibfield
  {journal} {\bibinfo  {journal} {Int. J. Mod. Phys. D}\ }\textbf {\bibinfo
  {volume} {14}},\ \bibinfo {pages} {107} (\bibinfo {year} {2005})},\ \Eprint
  {http://arxiv.org/abs/astro-ph/0312605} {arXiv:astro-ph/0312605} \BibitemShut
  {NoStop}%
\bibitem [{\citenamefont {Berezhiani}(2005)}]{Berezhiani:2005ek}%
  \BibitemOpen
  \bibfield  {author} {\bibinfo {author} {\bibfnamefont {Z.}~\bibnamefont
  {Berezhiani}},\ }in\ \href {\doibase 10.1142/9789812775344_0055} {\emph
  {\bibinfo {booktitle} {{From Fields to Strings: Circumnavigating Theoretical
  Physics: A Conference in Tribute to Ian Kogan}}}}\ (\bibinfo {year} {2005})\
  pp.\ \bibinfo {pages} {2147--2195},\ \Eprint
  {http://arxiv.org/abs/hep-ph/0508233} {arXiv:hep-ph/0508233} \BibitemShut
  {NoStop}%
\bibitem [{\citenamefont {Foot}(2014)}]{Foot:2014mia}%
  \BibitemOpen
  \bibfield  {author} {\bibinfo {author} {\bibfnamefont {R.}~\bibnamefont
  {Foot}},\ }\href {\doibase 10.1142/S0217751X14300130} {\bibfield  {journal}
  {\bibinfo  {journal} {Int. J. Mod. Phys.}\ }\textbf {\bibinfo {volume}
  {A29}},\ \bibinfo {pages} {1430013} (\bibinfo {year} {2014})},\ \Eprint
  {http://arxiv.org/abs/1401.3965} {arXiv:1401.3965 [astro-ph.CO]} \BibitemShut
  {NoStop}%
\bibitem [{\citenamefont {Babu}\ and\ \citenamefont
  {Mohapatra}(2022)}]{Babu:2021mjg}%
  \BibitemOpen
  \bibfield  {author} {\bibinfo {author} {\bibfnamefont {K.~S.}\ \bibnamefont
  {Babu}}\ and\ \bibinfo {author} {\bibfnamefont {R.~N.}\ \bibnamefont
  {Mohapatra}},\ }\href {\doibase 10.3390/sym14040731} {\bibfield  {journal}
  {\bibinfo  {journal} {Symmetry}\ }\textbf {\bibinfo {volume} {14}},\ \bibinfo
  {pages} {731} (\bibinfo {year} {2022})},\ \Eprint
  {http://arxiv.org/abs/2112.11443} {arXiv:2112.11443 [hep-ph]} \BibitemShut
  {NoStop}%
\bibitem [{\citenamefont {Berezhiani}\ \emph {et~al.}(2017)\citenamefont
  {Berezhiani}, \citenamefont {Frost}, \citenamefont {Kamyshkov}, \citenamefont
  {Rybolt},\ and\ \citenamefont {Varriano}}]{Berezhiani:2017azg}%
  \BibitemOpen
  \bibfield  {author} {\bibinfo {author} {\bibfnamefont {Z.}~\bibnamefont
  {Berezhiani}}, \bibinfo {author} {\bibfnamefont {M.}~\bibnamefont {Frost}},
  \bibinfo {author} {\bibfnamefont {Y.}~\bibnamefont {Kamyshkov}}, \bibinfo
  {author} {\bibfnamefont {B.}~\bibnamefont {Rybolt}}, \ and\ \bibinfo {author}
  {\bibfnamefont {L.}~\bibnamefont {Varriano}},\ }\href {\doibase
  10.1103/PhysRevD.96.035039} {\bibfield  {journal} {\bibinfo  {journal} {Phys.
  Rev. D}\ }\textbf {\bibinfo {volume} {96}},\ \bibinfo {pages} {035039}
  (\bibinfo {year} {2017})},\ \Eprint {http://arxiv.org/abs/1703.06735}
  {arXiv:1703.06735 [hep-ex]} \BibitemShut {NoStop}%
\bibitem [{\citenamefont {Broussard}\ \emph {et~al.}(2017)\citenamefont
  {Broussard} \emph {et~al.}}]{Broussard:2017yev}%
  \BibitemOpen
  \bibfield  {author} {\bibinfo {author} {\bibfnamefont {L.~J.}\ \bibnamefont
  {Broussard}} \emph {et~al.},\ }in\ \href@noop {} {\emph {\bibinfo {booktitle}
  {{Meeting of the APS Division of Particles and Fields}}}}\ (\bibinfo {year}
  {2017})\ \Eprint {http://arxiv.org/abs/1710.00767} {arXiv:1710.00767
  [hep-ex]} \BibitemShut {NoStop}%
\bibitem [{\citenamefont {Broussard}\ \emph {et~al.}(2019)\citenamefont
  {Broussard} \emph {et~al.}}]{Broussard:2019tgw}%
  \BibitemOpen
  \bibfield  {author} {\bibinfo {author} {\bibfnamefont {L.~J.}\ \bibnamefont
  {Broussard}} \emph {et~al.},\ }\href {\doibase 10.1051/epjconf/201921907002}
  {\bibfield  {journal} {\bibinfo  {journal} {EPJ Web Conf.}\ }\textbf
  {\bibinfo {volume} {219}},\ \bibinfo {pages} {07002} (\bibinfo {year}
  {2019})},\ \Eprint {http://arxiv.org/abs/1912.08264} {arXiv:1912.08264
  [physics.ins-det]} \BibitemShut {NoStop}%
\bibitem [{\citenamefont {Broussard}\ \emph {et~al.}(2022)\citenamefont
  {Broussard} \emph {et~al.}}]{Broussard:2021eyr}%
  \BibitemOpen
  \bibfield  {author} {\bibinfo {author} {\bibfnamefont {L.~J.}\ \bibnamefont
  {Broussard}} \emph {et~al.},\ }\href {\doibase
  10.1103/PhysRevLett.128.212503} {\bibfield  {journal} {\bibinfo  {journal}
  {Phys. Rev. Lett.}\ }\textbf {\bibinfo {volume} {128}},\ \bibinfo {pages}
  {212503} (\bibinfo {year} {2022})},\ \Eprint
  {http://arxiv.org/abs/2111.05543} {arXiv:2111.05543 [nucl-ex]} \BibitemShut
  {NoStop}%
\bibitem [{\citenamefont
  {Berezhiani}(2019{\natexlab{a}})}]{Berezhiani:2018eds}%
  \BibitemOpen
  \bibfield  {author} {\bibinfo {author} {\bibfnamefont {Z.}~\bibnamefont
  {Berezhiani}},\ }\href {\doibase 10.1140/epjc/s10052-019-6995-x} {\bibfield
  {journal} {\bibinfo  {journal} {Eur. Phys. J. C}\ }\textbf {\bibinfo {volume}
  {79}},\ \bibinfo {pages} {484} (\bibinfo {year} {2019}{\natexlab{a}})},\
  \Eprint {http://arxiv.org/abs/1807.07906} {arXiv:1807.07906 [hep-ph]}
  \BibitemShut {NoStop}%
\bibitem [{\citenamefont {Fornal}\ and\ \citenamefont
  {Grinstein}(2018)}]{Fornal:2018eol}%
  \BibitemOpen
  \bibfield  {author} {\bibinfo {author} {\bibfnamefont {B.}~\bibnamefont
  {Fornal}}\ and\ \bibinfo {author} {\bibfnamefont {B.}~\bibnamefont
  {Grinstein}},\ }\href {\doibase 10.1103/PhysRevLett.120.191801} {\bibfield
  {journal} {\bibinfo  {journal} {Phys. Rev. Lett.}\ }\textbf {\bibinfo
  {volume} {120}},\ \bibinfo {pages} {191801} (\bibinfo {year} {2018})},\
  \bibinfo {note} {[Erratum: Phys. Rev. Lett. 124, 219901 (2020)]},\ \Eprint
  {http://arxiv.org/abs/1801.01124} {arXiv:1801.01124 [hep-ph]} \BibitemShut
  {NoStop}%
\bibitem [{\citenamefont {Fornal}\ and\ \citenamefont
  {Grinstein}(2020)}]{Fornal:2020gto}%
  \BibitemOpen
  \bibfield  {author} {\bibinfo {author} {\bibfnamefont {B.}~\bibnamefont
  {Fornal}}\ and\ \bibinfo {author} {\bibfnamefont {B.}~\bibnamefont
  {Grinstein}},\ }\href {\doibase 10.1142/S0217732320300190} {\bibfield
  {journal} {\bibinfo  {journal} {Mod. Phys. Lett. A}\ }\textbf {\bibinfo
  {volume} {35}},\ \bibinfo {pages} {2030019} (\bibinfo {year} {2020})},\
  \Eprint {http://arxiv.org/abs/2007.13931} {arXiv:2007.13931 [hep-ph]}
  \BibitemShut {NoStop}%
\bibitem [{\citenamefont {{Wietfeldt}}\ and\ \citenamefont
  {{Greene}}(2011)}]{Wietfeldt2011RvMP...83.1173W}%
  \BibitemOpen
  \bibfield  {author} {\bibinfo {author} {\bibfnamefont {F.~E.}\ \bibnamefont
  {{Wietfeldt}}}\ and\ \bibinfo {author} {\bibfnamefont {G.~L.}\ \bibnamefont
  {{Greene}}},\ }\href {\doibase 10.1103/RevModPhys.83.1173} {\bibfield
  {journal} {\bibinfo  {journal} {Reviews of Modern Physics}\ }\textbf
  {\bibinfo {volume} {83}},\ \bibinfo {pages} {1173} (\bibinfo {year}
  {2011})}\BibitemShut {NoStop}%
\bibitem [{\citenamefont {McKeen}\ \emph {et~al.}(2018)\citenamefont {McKeen},
  \citenamefont {Nelson}, \citenamefont {Reddy},\ and\ \citenamefont
  {Zhou}}]{McKeen:2018xwc}%
  \BibitemOpen
  \bibfield  {author} {\bibinfo {author} {\bibfnamefont {D.}~\bibnamefont
  {McKeen}}, \bibinfo {author} {\bibfnamefont {A.~E.}\ \bibnamefont {Nelson}},
  \bibinfo {author} {\bibfnamefont {S.}~\bibnamefont {Reddy}}, \ and\ \bibinfo
  {author} {\bibfnamefont {D.}~\bibnamefont {Zhou}},\ }\href {\doibase
  10.1103/PhysRevLett.121.061802} {\bibfield  {journal} {\bibinfo  {journal}
  {Phys. Rev. Lett.}\ }\textbf {\bibinfo {volume} {121}},\ \bibinfo {pages}
  {061802} (\bibinfo {year} {2018})},\ \Eprint
  {http://arxiv.org/abs/1802.08244} {arXiv:1802.08244 [hep-ph]} \BibitemShut
  {NoStop}%
\bibitem [{\citenamefont {{Baym}}\ \emph {et~al.}(2018)\citenamefont {{Baym}},
  \citenamefont {{Beck}}, \citenamefont {{Geltenbort}},\ and\ \citenamefont
  {{Shelton}}}]{Baym2018PhRvL.121f1801B}%
  \BibitemOpen
  \bibfield  {author} {\bibinfo {author} {\bibfnamefont {G.}~\bibnamefont
  {{Baym}}}, \bibinfo {author} {\bibfnamefont {D.~H.}\ \bibnamefont {{Beck}}},
  \bibinfo {author} {\bibfnamefont {P.}~\bibnamefont {{Geltenbort}}}, \ and\
  \bibinfo {author} {\bibfnamefont {J.}~\bibnamefont {{Shelton}}},\ }\href
  {\doibase 10.1103/PhysRevLett.121.061801} {\bibfield  {journal} {\bibinfo
  {journal} {Phys. Rev. Lett.}\ }\textbf {\bibinfo {volume} {121}},\ \bibinfo
  {eid} {061801} (\bibinfo {year} {2018})},\ \Eprint
  {http://arxiv.org/abs/1802.08282} {arXiv:1802.08282 [hep-ph]} \BibitemShut
  {NoStop}%
\bibitem [{\citenamefont {Motta}\ \emph {et~al.}(2018)\citenamefont {Motta},
  \citenamefont {Guichon},\ and\ \citenamefont {Thomas}}]{Motta:2018rxp}%
  \BibitemOpen
  \bibfield  {author} {\bibinfo {author} {\bibfnamefont {T.~F.}\ \bibnamefont
  {Motta}}, \bibinfo {author} {\bibfnamefont {P.~A.~M.}\ \bibnamefont
  {Guichon}}, \ and\ \bibinfo {author} {\bibfnamefont {A.~W.}\ \bibnamefont
  {Thomas}},\ }\href {\doibase 10.1088/1361-6471/aab689} {\bibfield  {journal}
  {\bibinfo  {journal} {J. Phys. G}\ }\textbf {\bibinfo {volume} {45}},\
  \bibinfo {pages} {05LT01} (\bibinfo {year} {2018})},\ \Eprint
  {http://arxiv.org/abs/1802.08427} {arXiv:1802.08427 [nucl-th]} \BibitemShut
  {NoStop}%
\bibitem [{\citenamefont {Grinstein}\ \emph {et~al.}(2019)\citenamefont
  {Grinstein}, \citenamefont {Kouvaris},\ and\ \citenamefont
  {Nielsen}}]{Grinstein:2018ptl}%
  \BibitemOpen
  \bibfield  {author} {\bibinfo {author} {\bibfnamefont {B.}~\bibnamefont
  {Grinstein}}, \bibinfo {author} {\bibfnamefont {C.}~\bibnamefont {Kouvaris}},
  \ and\ \bibinfo {author} {\bibfnamefont {N.~G.}\ \bibnamefont {Nielsen}},\
  }\href {\doibase 10.1103/PhysRevLett.123.091601} {\bibfield  {journal}
  {\bibinfo  {journal} {Phys. Rev. Lett.}\ }\textbf {\bibinfo {volume} {123}},\
  \bibinfo {pages} {091601} (\bibinfo {year} {2019})},\ \Eprint
  {http://arxiv.org/abs/1811.06546} {arXiv:1811.06546 [hep-ph]} \BibitemShut
  {NoStop}%
\bibitem [{\citenamefont {Strumia}(2022)}]{Strumia:2021ybk}%
  \BibitemOpen
  \bibfield  {author} {\bibinfo {author} {\bibfnamefont {A.}~\bibnamefont
  {Strumia}},\ }\href {\doibase 10.1007/JHEP02(2022)067} {\bibfield  {journal}
  {\bibinfo  {journal} {JHEP}\ }\textbf {\bibinfo {volume} {02}},\ \bibinfo
  {pages} {067} (\bibinfo {year} {2022})},\ \Eprint
  {http://arxiv.org/abs/2112.09111} {arXiv:2112.09111 [hep-ph]} \BibitemShut
  {NoStop}%
\bibitem [{\citenamefont {Berryman}\ \emph {et~al.}(2022)\citenamefont
  {Berryman}, \citenamefont {Gardner},\ and\ \citenamefont
  {Zakeri}}]{Berryman:2022zic}%
  \BibitemOpen
  \bibfield  {author} {\bibinfo {author} {\bibfnamefont {J.~M.}\ \bibnamefont
  {Berryman}}, \bibinfo {author} {\bibfnamefont {S.}~\bibnamefont {Gardner}}, \
  and\ \bibinfo {author} {\bibfnamefont {M.}~\bibnamefont {Zakeri}},\ }\href
  {\doibase 10.3390/sym14030518} {\bibfield  {journal} {\bibinfo  {journal}
  {Symmetry}\ }\textbf {\bibinfo {volume} {14}},\ \bibinfo {pages} {518}
  (\bibinfo {year} {2022})},\ \Eprint {http://arxiv.org/abs/2201.02637}
  {arXiv:2201.02637 [hep-ph]} \BibitemShut {NoStop}%
\bibitem [{\citenamefont {Hostert}\ \emph {et~al.}(2022)\citenamefont
  {Hostert}, \citenamefont {McKeen}, \citenamefont {Pospelov},\ and\
  \citenamefont {Raj}}]{Hostert:2022ntu}%
  \BibitemOpen
  \bibfield  {author} {\bibinfo {author} {\bibfnamefont {M.}~\bibnamefont
  {Hostert}}, \bibinfo {author} {\bibfnamefont {D.}~\bibnamefont {McKeen}},
  \bibinfo {author} {\bibfnamefont {M.}~\bibnamefont {Pospelov}}, \ and\
  \bibinfo {author} {\bibfnamefont {N.}~\bibnamefont {Raj}},\ }\href@noop {} {\
   (\bibinfo {year} {2022})},\ \Eprint {http://arxiv.org/abs/2201.02603}
  {arXiv:2201.02603 [hep-ph]} \BibitemShut {NoStop}%
\bibitem [{\citenamefont
  {Berezhiani}(2019{\natexlab{b}})}]{Berezhiani:2018udo}%
  \BibitemOpen
  \bibfield  {author} {\bibinfo {author} {\bibfnamefont {Z.}~\bibnamefont
  {Berezhiani}},\ }\href {\doibase 10.31526/lhep.1.2019.118} {\bibfield
  {journal} {\bibinfo  {journal} {LHEP}\ }\textbf {\bibinfo {volume} {2}},\
  \bibinfo {pages} {118} (\bibinfo {year} {2019}{\natexlab{b}})},\ \Eprint
  {http://arxiv.org/abs/1812.11089} {arXiv:1812.11089 [hep-ph]} \BibitemShut
  {NoStop}%
\bibitem [{\citenamefont {McKeen}\ \emph {et~al.}(2020)\citenamefont {McKeen},
  \citenamefont {Pospelov},\ and\ \citenamefont {Raj}}]{McKeen:2020vpf}%
  \BibitemOpen
  \bibfield  {author} {\bibinfo {author} {\bibfnamefont {D.}~\bibnamefont
  {McKeen}}, \bibinfo {author} {\bibfnamefont {M.}~\bibnamefont {Pospelov}}, \
  and\ \bibinfo {author} {\bibfnamefont {N.}~\bibnamefont {Raj}},\ }\href
  {\doibase 10.1103/PhysRevLett.125.231803} {\bibfield  {journal} {\bibinfo
  {journal} {Phys. Rev. Lett.}\ }\textbf {\bibinfo {volume} {125}},\ \bibinfo
  {pages} {231803} (\bibinfo {year} {2020})},\ \Eprint
  {http://arxiv.org/abs/2006.15140} {arXiv:2006.15140 [hep-ph]} \BibitemShut
  {NoStop}%
\bibitem [{\citenamefont {McKeen}\ \emph {et~al.}(2021)\citenamefont {McKeen},
  \citenamefont {Pospelov},\ and\ \citenamefont {Raj}}]{McKeen:2020oyr}%
  \BibitemOpen
  \bibfield  {author} {\bibinfo {author} {\bibfnamefont {D.}~\bibnamefont
  {McKeen}}, \bibinfo {author} {\bibfnamefont {M.}~\bibnamefont {Pospelov}}, \
  and\ \bibinfo {author} {\bibfnamefont {N.}~\bibnamefont {Raj}},\ }\href
  {\doibase 10.1103/PhysRevD.103.115002} {\bibfield  {journal} {\bibinfo
  {journal} {Phys. Rev. D}\ }\textbf {\bibinfo {volume} {103}},\ \bibinfo
  {pages} {115002} (\bibinfo {year} {2021})},\ \Eprint
  {http://arxiv.org/abs/2012.09865} {arXiv:2012.09865 [hep-ph]} \BibitemShut
  {NoStop}%
\bibitem [{\citenamefont {Kile}\ and\ \citenamefont
  {Soni}(2009)}]{Kile:2009nn}%
  \BibitemOpen
  \bibfield  {author} {\bibinfo {author} {\bibfnamefont {J.}~\bibnamefont
  {Kile}}\ and\ \bibinfo {author} {\bibfnamefont {A.}~\bibnamefont {Soni}},\
  }\href {\doibase 10.1103/PhysRevD.80.115017} {\bibfield  {journal} {\bibinfo
  {journal} {Phys. Rev. D}\ }\textbf {\bibinfo {volume} {80}},\ \bibinfo
  {pages} {115017} (\bibinfo {year} {2009})},\ \Eprint
  {http://arxiv.org/abs/0908.3892} {arXiv:0908.3892 [hep-ph]} \BibitemShut
  {NoStop}%
\bibitem [{\citenamefont {Davoudiasl}\ \emph {et~al.}(2010)\citenamefont
  {Davoudiasl}, \citenamefont {Morrissey}, \citenamefont {Sigurdson},\ and\
  \citenamefont {Tulin}}]{Davoudiasl:2010am}%
  \BibitemOpen
  \bibfield  {author} {\bibinfo {author} {\bibfnamefont {H.}~\bibnamefont
  {Davoudiasl}}, \bibinfo {author} {\bibfnamefont {D.~E.}\ \bibnamefont
  {Morrissey}}, \bibinfo {author} {\bibfnamefont {K.}~\bibnamefont
  {Sigurdson}}, \ and\ \bibinfo {author} {\bibfnamefont {S.}~\bibnamefont
  {Tulin}},\ }\href {\doibase 10.1103/PhysRevLett.105.211304} {\bibfield
  {journal} {\bibinfo  {journal} {Phys. Rev. Lett.}\ }\textbf {\bibinfo
  {volume} {105}},\ \bibinfo {pages} {211304} (\bibinfo {year} {2010})},\
  \Eprint {http://arxiv.org/abs/1008.2399} {arXiv:1008.2399 [hep-ph]}
  \BibitemShut {NoStop}%
\bibitem [{\citenamefont {Davoudiasl}\ \emph {et~al.}(2011)\citenamefont
  {Davoudiasl}, \citenamefont {Morrissey}, \citenamefont {Sigurdson},\ and\
  \citenamefont {Tulin}}]{Davoudiasl:2011fj}%
  \BibitemOpen
  \bibfield  {author} {\bibinfo {author} {\bibfnamefont {H.}~\bibnamefont
  {Davoudiasl}}, \bibinfo {author} {\bibfnamefont {D.~E.}\ \bibnamefont
  {Morrissey}}, \bibinfo {author} {\bibfnamefont {K.}~\bibnamefont
  {Sigurdson}}, \ and\ \bibinfo {author} {\bibfnamefont {S.}~\bibnamefont
  {Tulin}},\ }\href {\doibase 10.1103/PhysRevD.84.096008} {\bibfield  {journal}
  {\bibinfo  {journal} {Phys. Rev. D}\ }\textbf {\bibinfo {volume} {84}},\
  \bibinfo {pages} {096008} (\bibinfo {year} {2011})},\ \Eprint
  {http://arxiv.org/abs/1106.4320} {arXiv:1106.4320 [hep-ph]} \BibitemShut
  {NoStop}%
\bibitem [{\citenamefont {Blinov}\ \emph {et~al.}(2012)\citenamefont {Blinov},
  \citenamefont {Morrissey}, \citenamefont {Sigurdson},\ and\ \citenamefont
  {Tulin}}]{Blinov:2012hq}%
  \BibitemOpen
  \bibfield  {author} {\bibinfo {author} {\bibfnamefont {N.}~\bibnamefont
  {Blinov}}, \bibinfo {author} {\bibfnamefont {D.~E.}\ \bibnamefont
  {Morrissey}}, \bibinfo {author} {\bibfnamefont {K.}~\bibnamefont
  {Sigurdson}}, \ and\ \bibinfo {author} {\bibfnamefont {S.}~\bibnamefont
  {Tulin}},\ }\href {\doibase 10.1103/PhysRevD.86.095021} {\bibfield  {journal}
  {\bibinfo  {journal} {Phys. Rev. D}\ }\textbf {\bibinfo {volume} {86}},\
  \bibinfo {pages} {095021} (\bibinfo {year} {2012})},\ \Eprint
  {http://arxiv.org/abs/1206.3304} {arXiv:1206.3304 [hep-ph]} \BibitemShut
  {NoStop}%
\bibitem [{\citenamefont {Huang}\ and\ \citenamefont
  {Zhao}(2014)}]{Huang:2013xfa}%
  \BibitemOpen
  \bibfield  {author} {\bibinfo {author} {\bibfnamefont {J.}~\bibnamefont
  {Huang}}\ and\ \bibinfo {author} {\bibfnamefont {Y.}~\bibnamefont {Zhao}},\
  }\href {\doibase 10.1007/JHEP02(2014)077} {\bibfield  {journal} {\bibinfo
  {journal} {JHEP}\ }\textbf {\bibinfo {volume} {02}},\ \bibinfo {pages} {077}
  (\bibinfo {year} {2014})},\ \Eprint {http://arxiv.org/abs/1312.0011}
  {arXiv:1312.0011 [hep-ph]} \BibitemShut {NoStop}%
\bibitem [{\citenamefont {Fornal}\ \emph {et~al.}(2021)\citenamefont {Fornal},
  \citenamefont {Hewitt},\ and\ \citenamefont {Zhao}}]{Fornal:2020poq}%
  \BibitemOpen
  \bibfield  {author} {\bibinfo {author} {\bibfnamefont {B.}~\bibnamefont
  {Fornal}}, \bibinfo {author} {\bibfnamefont {A.}~\bibnamefont {Hewitt}}, \
  and\ \bibinfo {author} {\bibfnamefont {Y.}~\bibnamefont {Zhao}},\ }\href
  {\doibase 10.1016/j.physletb.2021.136151} {\bibfield  {journal} {\bibinfo
  {journal} {Phys. Lett. B}\ }\textbf {\bibinfo {volume} {815}},\ \bibinfo
  {pages} {136151} (\bibinfo {year} {2021})},\ \Eprint
  {http://arxiv.org/abs/2011.09014} {arXiv:2011.09014 [hep-ph]} \BibitemShut
  {NoStop}%
\bibitem [{\citenamefont {Jin}\ and\ \citenamefont {Gao}(2018)}]{Jin:2018moh}%
  \BibitemOpen
  \bibfield  {author} {\bibinfo {author} {\bibfnamefont {M.}~\bibnamefont
  {Jin}}\ and\ \bibinfo {author} {\bibfnamefont {Y.}~\bibnamefont {Gao}},\
  }\href {\doibase 10.1103/PhysRevD.98.075026} {\bibfield  {journal} {\bibinfo
  {journal} {Phys. Rev. D}\ }\textbf {\bibinfo {volume} {98}},\ \bibinfo
  {pages} {075026} (\bibinfo {year} {2018})},\ \Eprint
  {http://arxiv.org/abs/1808.10644} {arXiv:1808.10644 [hep-ph]} \BibitemShut
  {NoStop}%
\bibitem [{\citenamefont {Keung}\ \emph {et~al.}(2019)\citenamefont {Keung},
  \citenamefont {Marfatia},\ and\ \citenamefont {Tseng}}]{Keung:2019wpw}%
  \BibitemOpen
  \bibfield  {author} {\bibinfo {author} {\bibfnamefont {W.-Y.}\ \bibnamefont
  {Keung}}, \bibinfo {author} {\bibfnamefont {D.}~\bibnamefont {Marfatia}}, \
  and\ \bibinfo {author} {\bibfnamefont {P.-Y.}\ \bibnamefont {Tseng}},\ }\href
  {\doibase 10.1007/JHEP09(2019)053} {\bibfield  {journal} {\bibinfo  {journal}
  {JHEP}\ }\textbf {\bibinfo {volume} {09}},\ \bibinfo {pages} {053} (\bibinfo
  {year} {2019})},\ \Eprint {http://arxiv.org/abs/1905.03401} {arXiv:1905.03401
  [hep-ph]} \BibitemShut {NoStop}%
\bibitem [{\citenamefont {Weinberg}(1980)}]{Weinberg:1980bf}%
  \BibitemOpen
  \bibfield  {author} {\bibinfo {author} {\bibfnamefont {S.}~\bibnamefont
  {Weinberg}},\ }\href {\doibase 10.1103/PhysRevD.22.1694} {\bibfield
  {journal} {\bibinfo  {journal} {Phys. Rev. D}\ }\textbf {\bibinfo {volume}
  {22}},\ \bibinfo {pages} {1694} (\bibinfo {year} {1980})}\BibitemShut
  {NoStop}%
\bibitem [{\citenamefont {Heeck}\ and\ \citenamefont
  {Takhistov}(2020)}]{Heeck:2019kgr}%
  \BibitemOpen
  \bibfield  {author} {\bibinfo {author} {\bibfnamefont {J.}~\bibnamefont
  {Heeck}}\ and\ \bibinfo {author} {\bibfnamefont {V.}~\bibnamefont
  {Takhistov}},\ }\href {\doibase 10.1103/PhysRevD.101.015005} {\bibfield
  {journal} {\bibinfo  {journal} {Phys. Rev. D}\ }\textbf {\bibinfo {volume}
  {101}},\ \bibinfo {pages} {015005} (\bibinfo {year} {2020})},\ \Eprint
  {http://arxiv.org/abs/1910.07647} {arXiv:1910.07647 [hep-ph]} \BibitemShut
  {NoStop}%
\bibitem [{\citenamefont {Heeck}\ and\ \citenamefont
  {Rodejohann}(2013)}]{Heeck:2013rpa}%
  \BibitemOpen
  \bibfield  {author} {\bibinfo {author} {\bibfnamefont {J.}~\bibnamefont
  {Heeck}}\ and\ \bibinfo {author} {\bibfnamefont {W.}~\bibnamefont
  {Rodejohann}},\ }\href {\doibase 10.1209/0295-5075/103/32001} {\bibfield
  {journal} {\bibinfo  {journal} {EPL}\ }\textbf {\bibinfo {volume} {103}},\
  \bibinfo {pages} {32001} (\bibinfo {year} {2013})},\ \Eprint
  {http://arxiv.org/abs/1306.0580} {arXiv:1306.0580 [hep-ph]} \BibitemShut
  {NoStop}%
\bibitem [{\citenamefont {Fonseca}\ \emph {et~al.}(2018)\citenamefont
  {Fonseca}, \citenamefont {Hirsch},\ and\ \citenamefont
  {Srivastava}}]{Fonseca:2018ehk}%
  \BibitemOpen
  \bibfield  {author} {\bibinfo {author} {\bibfnamefont {R.~M.}\ \bibnamefont
  {Fonseca}}, \bibinfo {author} {\bibfnamefont {M.}~\bibnamefont {Hirsch}}, \
  and\ \bibinfo {author} {\bibfnamefont {R.}~\bibnamefont {Srivastava}},\
  }\href {\doibase 10.1103/PhysRevD.97.075026} {\bibfield  {journal} {\bibinfo
  {journal} {Phys. Rev. D}\ }\textbf {\bibinfo {volume} {97}},\ \bibinfo
  {pages} {075026} (\bibinfo {year} {2018})},\ \Eprint
  {http://arxiv.org/abs/1802.04814} {arXiv:1802.04814 [hep-ph]} \BibitemShut
  {NoStop}%
\bibitem [{\citenamefont {Fonseca}\ and\ \citenamefont
  {Hirsch}(2018)}]{Fonseca:2018aav}%
  \BibitemOpen
  \bibfield  {author} {\bibinfo {author} {\bibfnamefont {R.~M.}\ \bibnamefont
  {Fonseca}}\ and\ \bibinfo {author} {\bibfnamefont {M.}~\bibnamefont
  {Hirsch}},\ }\href {\doibase 10.1103/PhysRevD.98.015035} {\bibfield
  {journal} {\bibinfo  {journal} {Phys. Rev. D}\ }\textbf {\bibinfo {volume}
  {98}},\ \bibinfo {pages} {015035} (\bibinfo {year} {2018})},\ \Eprint
  {http://arxiv.org/abs/1804.10545} {arXiv:1804.10545 [hep-ph]} \BibitemShut
  {NoStop}%
\bibitem [{\citenamefont {Nussinov}\ and\ \citenamefont
  {Shrock}(2020)}]{Nussinov:2020wri}%
  \BibitemOpen
  \bibfield  {author} {\bibinfo {author} {\bibfnamefont {S.}~\bibnamefont
  {Nussinov}}\ and\ \bibinfo {author} {\bibfnamefont {R.}~\bibnamefont
  {Shrock}},\ }\href {\doibase 10.1103/PhysRevD.102.035003} {\bibfield
  {journal} {\bibinfo  {journal} {Phys. Rev. D}\ }\textbf {\bibinfo {volume}
  {102}},\ \bibinfo {pages} {035003} (\bibinfo {year} {2020})},\ \Eprint
  {http://arxiv.org/abs/2005.12493} {arXiv:2005.12493 [hep-ph]} \BibitemShut
  {NoStop}%
\bibitem [{\citenamefont {He}\ and\ \citenamefont
  {Ma}(2021{\natexlab{a}})}]{He:2021sbl}%
  \BibitemOpen
  \bibfield  {author} {\bibinfo {author} {\bibfnamefont {X.-G.}\ \bibnamefont
  {He}}\ and\ \bibinfo {author} {\bibfnamefont {X.-D.}\ \bibnamefont {Ma}},\
  }\href {\doibase 10.1016/j.physletb.2021.136298} {\bibfield  {journal}
  {\bibinfo  {journal} {Phys. Lett. B}\ }\textbf {\bibinfo {volume} {817}},\
  \bibinfo {pages} {136298} (\bibinfo {year} {2021}{\natexlab{a}})},\ \Eprint
  {http://arxiv.org/abs/2101.01405} {arXiv:2101.01405 [hep-ph]} \BibitemShut
  {NoStop}%
\bibitem [{\citenamefont {He}\ and\ \citenamefont
  {Ma}(2021{\natexlab{b}})}]{He:2021mrt}%
  \BibitemOpen
  \bibfield  {author} {\bibinfo {author} {\bibfnamefont {X.-G.}\ \bibnamefont
  {He}}\ and\ \bibinfo {author} {\bibfnamefont {X.-D.}\ \bibnamefont {Ma}},\
  }\href {\doibase 10.1007/JHEP06(2021)047} {\bibfield  {journal} {\bibinfo
  {journal} {JHEP}\ }\textbf {\bibinfo {volume} {06}},\ \bibinfo {pages} {047}
  (\bibinfo {year} {2021}{\natexlab{b}})},\ \Eprint
  {http://arxiv.org/abs/2102.02562} {arXiv:2102.02562 [hep-ph]} \BibitemShut
  {NoStop}%
\bibitem [{\citenamefont {Helset}\ \emph {et~al.}(2021)\citenamefont {Helset},
  \citenamefont {Murgui},\ and\ \citenamefont {Wise}}]{Helset:2021plg}%
  \BibitemOpen
  \bibfield  {author} {\bibinfo {author} {\bibfnamefont {A.}~\bibnamefont
  {Helset}}, \bibinfo {author} {\bibfnamefont {C.}~\bibnamefont {Murgui}}, \
  and\ \bibinfo {author} {\bibfnamefont {M.~B.}\ \bibnamefont {Wise}},\ }\href
  {\doibase 10.1103/PhysRevD.104.015029} {\bibfield  {journal} {\bibinfo
  {journal} {Phys. Rev. D}\ }\textbf {\bibinfo {volume} {104}},\ \bibinfo
  {pages} {015029} (\bibinfo {year} {2021})},\ \Eprint
  {http://arxiv.org/abs/2104.03316} {arXiv:2104.03316 [hep-ph]} \BibitemShut
  {NoStop}%
\bibitem [{\citenamefont {Litos}\ \emph {et~al.}(2014)\citenamefont {Litos}
  \emph {et~al.}}]{Super-Kamiokande:2014hie}%
  \BibitemOpen
  \bibfield  {author} {\bibinfo {author} {\bibfnamefont {M.}~\bibnamefont
  {Litos}} \emph {et~al.} (\bibinfo {collaboration} {Super-Kamiokande}),\
  }\href {\doibase 10.1103/PhysRevLett.112.131803} {\bibfield  {journal}
  {\bibinfo  {journal} {Phys. Rev. Lett.}\ }\textbf {\bibinfo {volume} {112}},\
  \bibinfo {pages} {131803} (\bibinfo {year} {2014})}\BibitemShut {NoStop}%
\bibitem [{\citenamefont {Takhistov}\ \emph {et~al.}(2015)\citenamefont
  {Takhistov} \emph {et~al.}}]{Super-Kamiokande:2015pys}%
  \BibitemOpen
  \bibfield  {author} {\bibinfo {author} {\bibfnamefont {V.}~\bibnamefont
  {Takhistov}} \emph {et~al.} (\bibinfo {collaboration} {Super-Kamiokande}),\
  }\href {\doibase 10.1103/PhysRevLett.115.121803} {\bibfield  {journal}
  {\bibinfo  {journal} {Phys. Rev. Lett.}\ }\textbf {\bibinfo {volume} {115}},\
  \bibinfo {pages} {121803} (\bibinfo {year} {2015})},\ \Eprint
  {http://arxiv.org/abs/1508.05530} {arXiv:1508.05530 [hep-ex]} \BibitemShut
  {NoStop}%
\bibitem [{\citenamefont {Gustafson}\ \emph {et~al.}(2015)\citenamefont
  {Gustafson} \emph {et~al.}}]{Super-Kamiokande:2015jbb}%
  \BibitemOpen
  \bibfield  {author} {\bibinfo {author} {\bibfnamefont {J.}~\bibnamefont
  {Gustafson}} \emph {et~al.} (\bibinfo {collaboration} {Super-Kamiokande}),\
  }\href {\doibase 10.1103/PhysRevD.91.072009} {\bibfield  {journal} {\bibinfo
  {journal} {Phys. Rev. D}\ }\textbf {\bibinfo {volume} {91}},\ \bibinfo
  {pages} {072009} (\bibinfo {year} {2015})},\ \Eprint
  {http://arxiv.org/abs/1504.01041} {arXiv:1504.01041 [hep-ex]} \BibitemShut
  {NoStop}%
\bibitem [{\citenamefont {Arnold}\ \emph {et~al.}(2017)\citenamefont {Arnold}
  \emph {et~al.}}]{NEMO-3:2017gmi}%
  \BibitemOpen
  \bibfield  {author} {\bibinfo {author} {\bibfnamefont {R.}~\bibnamefont
  {Arnold}} \emph {et~al.} (\bibinfo {collaboration} {NEMO-3}),\ }\href
  {\doibase 10.1103/PhysRevLett.119.041801} {\bibfield  {journal} {\bibinfo
  {journal} {Phys. Rev. Lett.}\ }\textbf {\bibinfo {volume} {119}},\ \bibinfo
  {pages} {041801} (\bibinfo {year} {2017})},\ \Eprint
  {http://arxiv.org/abs/1705.08847} {arXiv:1705.08847 [hep-ex]} \BibitemShut
  {NoStop}%
\bibitem [{\citenamefont {Sussman}\ \emph {et~al.}(2018)\citenamefont {Sussman}
  \emph {et~al.}}]{Super-Kamiokande:2018apg}%
  \BibitemOpen
  \bibfield  {author} {\bibinfo {author} {\bibfnamefont {S.}~\bibnamefont
  {Sussman}} \emph {et~al.} (\bibinfo {collaboration} {Super-Kamiokande}),\
  }\href@noop {} {\  (\bibinfo {year} {2018})},\ \Eprint
  {http://arxiv.org/abs/1811.12430} {arXiv:1811.12430 [hep-ex]} \BibitemShut
  {NoStop}%
\bibitem [{\citenamefont {Anderson}\ \emph {et~al.}(2019)\citenamefont
  {Anderson} \emph {et~al.}}]{SNO:2018ydj}%
  \BibitemOpen
  \bibfield  {author} {\bibinfo {author} {\bibfnamefont {M.}~\bibnamefont
  {Anderson}} \emph {et~al.} (\bibinfo {collaboration} {SNO+}),\ }\href
  {\doibase 10.1103/PhysRevD.99.032008} {\bibfield  {journal} {\bibinfo
  {journal} {Phys. Rev. D}\ }\textbf {\bibinfo {volume} {99}},\ \bibinfo
  {pages} {032008} (\bibinfo {year} {2019})},\ \Eprint
  {http://arxiv.org/abs/1812.05552} {arXiv:1812.05552 [hep-ex]} \BibitemShut
  {NoStop}%
\bibitem [{\citenamefont {Kidd}\ and\ \citenamefont
  {Tornow}(2018)}]{Kidd:2018fbb}%
  \BibitemOpen
  \bibfield  {author} {\bibinfo {author} {\bibfnamefont {M.~F.}\ \bibnamefont
  {Kidd}}\ and\ \bibinfo {author} {\bibfnamefont {W.}~\bibnamefont {Tornow}},\
  }\href {\doibase 10.1103/PhysRevC.98.055501} {\bibfield  {journal} {\bibinfo
  {journal} {Phys. Rev. C}\ }\textbf {\bibinfo {volume} {98}},\ \bibinfo
  {pages} {055501} (\bibinfo {year} {2018})}\BibitemShut {NoStop}%
\bibitem [{\citenamefont {Albert}\ \emph {et~al.}(2018)\citenamefont {Albert}
  \emph {et~al.}}]{EXO-200:2017hwz}%
  \BibitemOpen
  \bibfield  {author} {\bibinfo {author} {\bibfnamefont {J.~B.}\ \bibnamefont
  {Albert}} \emph {et~al.} (\bibinfo {collaboration} {EXO-200}),\ }\href
  {\doibase 10.1103/PhysRevD.97.072007} {\bibfield  {journal} {\bibinfo
  {journal} {Phys. Rev. D}\ }\textbf {\bibinfo {volume} {97}},\ \bibinfo
  {pages} {072007} (\bibinfo {year} {2018})},\ \Eprint
  {http://arxiv.org/abs/1710.07670} {arXiv:1710.07670 [hep-ex]} \BibitemShut
  {NoStop}%
\bibitem [{\citenamefont {Alvis}\ \emph
  {et~al.}(2019{\natexlab{b}})\citenamefont {Alvis} \emph
  {et~al.}}]{Majorana:2018pdo}%
  \BibitemOpen
  \bibfield  {author} {\bibinfo {author} {\bibfnamefont {S.~I.}\ \bibnamefont
  {Alvis}} \emph {et~al.} (\bibinfo {collaboration} {Majorana}),\ }\href
  {\doibase 10.1103/PhysRevD.99.072004} {\bibfield  {journal} {\bibinfo
  {journal} {Phys. Rev. D}\ }\textbf {\bibinfo {volume} {99}},\ \bibinfo
  {pages} {072004} (\bibinfo {year} {2019}{\natexlab{b}})},\ \Eprint
  {http://arxiv.org/abs/1812.01090} {arXiv:1812.01090 [hep-ex]} \BibitemShut
  {NoStop}%
\bibitem [{\citenamefont {Barabash}\ \emph {et~al.}(2019)\citenamefont
  {Barabash}, \citenamefont {Hubert}, \citenamefont {Nachab},\ and\
  \citenamefont {Umatov}}]{Barabash:2019enn}%
  \BibitemOpen
  \bibfield  {author} {\bibinfo {author} {\bibfnamefont {A.~S.}\ \bibnamefont
  {Barabash}}, \bibinfo {author} {\bibfnamefont {P.}~\bibnamefont {Hubert}},
  \bibinfo {author} {\bibfnamefont {A.}~\bibnamefont {Nachab}}, \ and\ \bibinfo
  {author} {\bibfnamefont {V.~I.}\ \bibnamefont {Umatov}},\ }\href {\doibase
  10.1103/PhysRevC.100.045502} {\bibfield  {journal} {\bibinfo  {journal}
  {Phys. Rev. C}\ }\textbf {\bibinfo {volume} {100}},\ \bibinfo {pages}
  {045502} (\bibinfo {year} {2019})},\ \Eprint
  {http://arxiv.org/abs/1906.07180} {arXiv:1906.07180 [nucl-ex]} \BibitemShut
  {NoStop}%
\bibitem [{\citenamefont {Babu}\ and\ \citenamefont
  {Mohapatra}(2015)}]{Babu:2014tra}%
  \BibitemOpen
  \bibfield  {author} {\bibinfo {author} {\bibfnamefont {K.~S.}\ \bibnamefont
  {Babu}}\ and\ \bibinfo {author} {\bibfnamefont {R.~N.}\ \bibnamefont
  {Mohapatra}},\ }\href {\doibase 10.1103/PhysRevD.91.013008} {\bibfield
  {journal} {\bibinfo  {journal} {Phys. Rev. D}\ }\textbf {\bibinfo {volume}
  {91}},\ \bibinfo {pages} {013008} (\bibinfo {year} {2015})},\ \Eprint
  {http://arxiv.org/abs/1408.0803} {arXiv:1408.0803 [hep-ph]} \BibitemShut
  {NoStop}%
\bibitem [{\citenamefont {Gardner}\ and\ \citenamefont
  {Yan}(2019)}]{Gardner:2018azu}%
  \BibitemOpen
  \bibfield  {author} {\bibinfo {author} {\bibfnamefont {S.}~\bibnamefont
  {Gardner}}\ and\ \bibinfo {author} {\bibfnamefont {X.}~\bibnamefont {Yan}},\
  }\href {\doibase 10.1016/j.physletb.2019.01.054} {\bibfield  {journal}
  {\bibinfo  {journal} {Phys. Lett. B}\ }\textbf {\bibinfo {volume} {790}},\
  \bibinfo {pages} {421} (\bibinfo {year} {2019})},\ \Eprint
  {http://arxiv.org/abs/1808.05288} {arXiv:1808.05288 [hep-ph]} \BibitemShut
  {NoStop}%
\bibitem [{\citenamefont {Helset}\ and\ \citenamefont
  {Kobach}(2020)}]{Helset:2019eyc}%
  \BibitemOpen
  \bibfield  {author} {\bibinfo {author} {\bibfnamefont {A.}~\bibnamefont
  {Helset}}\ and\ \bibinfo {author} {\bibfnamefont {A.}~\bibnamefont
  {Kobach}},\ }\href {\doibase 10.1016/j.physletb.2019.135132} {\bibfield
  {journal} {\bibinfo  {journal} {Phys. Lett. B}\ }\textbf {\bibinfo {volume}
  {800}},\ \bibinfo {pages} {135132} (\bibinfo {year} {2020})},\ \Eprint
  {http://arxiv.org/abs/1909.05853} {arXiv:1909.05853 [hep-ph]} \BibitemShut
  {NoStop}%
\bibitem [{\citenamefont {Marciano}(1995)}]{Marciano:1994bg}%
  \BibitemOpen
  \bibfield  {author} {\bibinfo {author} {\bibfnamefont {W.~J.}\ \bibnamefont
  {Marciano}},\ }\href {\doibase 10.1016/0920-5632(95)00126-T} {\bibfield
  {journal} {\bibinfo  {journal} {Nucl. Phys. B Proc. Suppl.}\ }\textbf
  {\bibinfo {volume} {40}},\ \bibinfo {pages} {3} (\bibinfo {year}
  {1995})}\BibitemShut {NoStop}%
\bibitem [{\citenamefont {Hou}\ \emph {et~al.}(2005)\citenamefont {Hou},
  \citenamefont {Nagashima},\ and\ \citenamefont {Soddu}}]{Hou:2005iu}%
  \BibitemOpen
  \bibfield  {author} {\bibinfo {author} {\bibfnamefont {W.-S.}\ \bibnamefont
  {Hou}}, \bibinfo {author} {\bibfnamefont {M.}~\bibnamefont {Nagashima}}, \
  and\ \bibinfo {author} {\bibfnamefont {A.}~\bibnamefont {Soddu}},\ }\href
  {\doibase 10.1103/PhysRevD.72.095001} {\bibfield  {journal} {\bibinfo
  {journal} {Phys. Rev. D}\ }\textbf {\bibinfo {volume} {72}},\ \bibinfo
  {pages} {095001} (\bibinfo {year} {2005})},\ \Eprint
  {http://arxiv.org/abs/hep-ph/0509006} {arXiv:hep-ph/0509006} \BibitemShut
  {NoStop}%
\bibitem [{\citenamefont {Durieux}\ \emph {et~al.}(2013)\citenamefont
  {Durieux}, \citenamefont {Gerard}, \citenamefont {Maltoni},\ and\
  \citenamefont {Smith}}]{Durieux:2012gj}%
  \BibitemOpen
  \bibfield  {author} {\bibinfo {author} {\bibfnamefont {G.}~\bibnamefont
  {Durieux}}, \bibinfo {author} {\bibfnamefont {J.-M.}\ \bibnamefont {Gerard}},
  \bibinfo {author} {\bibfnamefont {F.}~\bibnamefont {Maltoni}}, \ and\
  \bibinfo {author} {\bibfnamefont {C.}~\bibnamefont {Smith}},\ }\href
  {\doibase 10.1016/j.physletb.2013.02.052} {\bibfield  {journal} {\bibinfo
  {journal} {Phys. Lett. B}\ }\textbf {\bibinfo {volume} {721}},\ \bibinfo
  {pages} {82} (\bibinfo {year} {2013})},\ \Eprint
  {http://arxiv.org/abs/1210.6598} {arXiv:1210.6598 [hep-ph]} \BibitemShut
  {NoStop}%
\bibitem [{\citenamefont {Alonso}\ \emph {et~al.}(2014)\citenamefont {Alonso},
  \citenamefont {Chang}, \citenamefont {Jenkins}, \citenamefont {Manohar},\
  and\ \citenamefont {Shotwell}}]{Alonso:2014zka}%
  \BibitemOpen
  \bibfield  {author} {\bibinfo {author} {\bibfnamefont {R.}~\bibnamefont
  {Alonso}}, \bibinfo {author} {\bibfnamefont {H.-M.}\ \bibnamefont {Chang}},
  \bibinfo {author} {\bibfnamefont {E.~E.}\ \bibnamefont {Jenkins}}, \bibinfo
  {author} {\bibfnamefont {A.~V.}\ \bibnamefont {Manohar}}, \ and\ \bibinfo
  {author} {\bibfnamefont {B.}~\bibnamefont {Shotwell}},\ }\href {\doibase
  10.1016/j.physletb.2014.05.065} {\bibfield  {journal} {\bibinfo  {journal}
  {Phys. Lett. B}\ }\textbf {\bibinfo {volume} {734}},\ \bibinfo {pages} {302}
  (\bibinfo {year} {2014})},\ \Eprint {http://arxiv.org/abs/1405.0486}
  {arXiv:1405.0486 [hep-ph]} \BibitemShut {NoStop}%
\bibitem [{\citenamefont {Dor\v{s}ner}\ \emph {et~al.}(2022)\citenamefont
  {Dor\v{s}ner}, \citenamefont {Fajfer},\ and\ \citenamefont
  {Sumensari}}]{Dorsner:2022twk}%
  \BibitemOpen
  \bibfield  {author} {\bibinfo {author} {\bibfnamefont {I.}~\bibnamefont
  {Dor\v{s}ner}}, \bibinfo {author} {\bibfnamefont {S.}~\bibnamefont {Fajfer}},
  \ and\ \bibinfo {author} {\bibfnamefont {O.}~\bibnamefont {Sumensari}},\
  }\href {\doibase 10.1007/JHEP05(2022)183} {\bibfield  {journal} {\bibinfo
  {journal} {JHEP}\ }\textbf {\bibinfo {volume} {05}},\ \bibinfo {pages} {183}
  (\bibinfo {year} {2022})},\ \Eprint {http://arxiv.org/abs/2202.08287}
  {arXiv:2202.08287 [hep-ph]} \BibitemShut {NoStop}%
\bibitem [{\citenamefont {Dong}\ \emph {et~al.}(2012)\citenamefont {Dong},
  \citenamefont {Durieux}, \citenamefont {Gerard}, \citenamefont {Han},\ and\
  \citenamefont {Maltoni}}]{Dong:2011rh}%
  \BibitemOpen
  \bibfield  {author} {\bibinfo {author} {\bibfnamefont {Z.}~\bibnamefont
  {Dong}}, \bibinfo {author} {\bibfnamefont {G.}~\bibnamefont {Durieux}},
  \bibinfo {author} {\bibfnamefont {J.-M.}\ \bibnamefont {Gerard}}, \bibinfo
  {author} {\bibfnamefont {T.}~\bibnamefont {Han}}, \ and\ \bibinfo {author}
  {\bibfnamefont {F.}~\bibnamefont {Maltoni}},\ }\href {\doibase
  10.1103/PhysRevD.85.016006} {\bibfield  {journal} {\bibinfo  {journal} {Phys.
  Rev. D}\ }\textbf {\bibinfo {volume} {85}},\ \bibinfo {pages} {016006}
  (\bibinfo {year} {2012})},\ \Eprint {http://arxiv.org/abs/1107.3805}
  {arXiv:1107.3805 [hep-ph]} \BibitemShut {NoStop}%
\bibitem [{\citenamefont {Chatrchyan}\ \emph {et~al.}(2014)\citenamefont
  {Chatrchyan} \emph {et~al.}}]{CMS:2013zol}%
  \BibitemOpen
  \bibfield  {author} {\bibinfo {author} {\bibfnamefont {S.}~\bibnamefont
  {Chatrchyan}} \emph {et~al.} (\bibinfo {collaboration} {CMS}),\ }\href
  {\doibase 10.1016/j.physletb.2014.02.033} {\bibfield  {journal} {\bibinfo
  {journal} {Phys. Lett. B}\ }\textbf {\bibinfo {volume} {731}},\ \bibinfo
  {pages} {173} (\bibinfo {year} {2014})},\ \Eprint
  {http://arxiv.org/abs/1310.1618} {arXiv:1310.1618 [hep-ex]} \BibitemShut
  {NoStop}%
\bibitem [{\citenamefont {del Amo~Sanchez}\ \emph {et~al.}(2011)\citenamefont
  {del Amo~Sanchez} \emph {et~al.}}]{BaBar:2011yks}%
  \BibitemOpen
  \bibfield  {author} {\bibinfo {author} {\bibfnamefont {P.}~\bibnamefont {del
  Amo~Sanchez}} \emph {et~al.} (\bibinfo {collaboration} {BaBar}),\ }\href
  {\doibase 10.1103/PhysRevD.83.091101} {\bibfield  {journal} {\bibinfo
  {journal} {Phys. Rev. D}\ }\textbf {\bibinfo {volume} {83}},\ \bibinfo
  {pages} {091101} (\bibinfo {year} {2011})},\ \Eprint
  {http://arxiv.org/abs/1101.3830} {arXiv:1101.3830 [hep-ex]} \BibitemShut
  {NoStop}%
\bibitem [{\citenamefont {Ablikim}\ \emph {et~al.}(2022)\citenamefont {Ablikim}
  \emph {et~al.}}]{BESIII:2021hyt}%
  \BibitemOpen
  \bibfield  {author} {\bibinfo {author} {\bibfnamefont {M.}~\bibnamefont
  {Ablikim}} \emph {et~al.} (\bibinfo {collaboration} {BESIII}),\ }\href
  {\doibase 10.1103/PhysRevD.105.032006} {\bibfield  {journal} {\bibinfo
  {journal} {Phys. Rev. D}\ }\textbf {\bibinfo {volume} {105}},\ \bibinfo
  {pages} {032006} (\bibinfo {year} {2022})},\ \Eprint
  {http://arxiv.org/abs/2112.10972} {arXiv:2112.10972 [hep-ex]} \BibitemShut
  {NoStop}%
\bibitem [{\citenamefont {McCracken}\ \emph {et~al.}(2015)\citenamefont
  {McCracken} \emph {et~al.}}]{McCracken:2015coa}%
  \BibitemOpen
  \bibfield  {author} {\bibinfo {author} {\bibfnamefont {M.~E.}\ \bibnamefont
  {McCracken}} \emph {et~al.},\ }\href {\doibase 10.1103/PhysRevD.92.072002}
  {\bibfield  {journal} {\bibinfo  {journal} {Phys. Rev. D}\ }\textbf {\bibinfo
  {volume} {92}},\ \bibinfo {pages} {072002} (\bibinfo {year} {2015})},\
  \Eprint {http://arxiv.org/abs/1507.03859} {arXiv:1507.03859 [hep-ex]}
  \BibitemShut {NoStop}%
\bibitem [{\citenamefont {Sahoo}\ \emph {et~al.}(2020)\citenamefont {Sahoo}
  \emph {et~al.}}]{Belle:2020lfn}%
  \BibitemOpen
  \bibfield  {author} {\bibinfo {author} {\bibfnamefont {D.}~\bibnamefont
  {Sahoo}} \emph {et~al.} (\bibinfo {collaboration} {Belle}),\ }\href {\doibase
  10.1103/PhysRevD.102.111101} {\bibfield  {journal} {\bibinfo  {journal}
  {Phys. Rev. D}\ }\textbf {\bibinfo {volume} {102}},\ \bibinfo {pages}
  {111101} (\bibinfo {year} {2020})},\ \Eprint
  {http://arxiv.org/abs/2010.15361} {arXiv:2010.15361 [hep-ex]} \BibitemShut
  {NoStop}%
\bibitem [{\citenamefont {Hambye}\ and\ \citenamefont
  {Heeck}(2018)}]{Hambye:2017qix}%
  \BibitemOpen
  \bibfield  {author} {\bibinfo {author} {\bibfnamefont {T.}~\bibnamefont
  {Hambye}}\ and\ \bibinfo {author} {\bibfnamefont {J.}~\bibnamefont {Heeck}},\
  }\href {\doibase 10.1103/PhysRevLett.120.171801} {\bibfield  {journal}
  {\bibinfo  {journal} {Phys. Rev. Lett.}\ }\textbf {\bibinfo {volume} {120}},\
  \bibinfo {pages} {171801} (\bibinfo {year} {2018})},\ \Eprint
  {http://arxiv.org/abs/1712.04871} {arXiv:1712.04871 [hep-ph]} \BibitemShut
  {NoStop}%
\bibitem [{\citenamefont {Tanaka}\ \emph {et~al.}(2020)\citenamefont {Tanaka}
  \emph {et~al.}}]{Super-Kamiokande:2020tor}%
  \BibitemOpen
  \bibfield  {author} {\bibinfo {author} {\bibfnamefont {M.}~\bibnamefont
  {Tanaka}} \emph {et~al.} (\bibinfo {collaboration} {Super-Kamiokande}),\
  }\href {\doibase 10.1103/PhysRevD.101.052011} {\bibfield  {journal} {\bibinfo
   {journal} {Phys. Rev. D}\ }\textbf {\bibinfo {volume} {101}},\ \bibinfo
  {pages} {052011} (\bibinfo {year} {2020})},\ \Eprint
  {http://arxiv.org/abs/2001.08011} {arXiv:2001.08011 [hep-ex]} \BibitemShut
  {NoStop}%
\bibitem [{\citenamefont {Lee}\ and\ \citenamefont
  {MacKenzie}(2022)}]{Lee:2021hnx}%
  \BibitemOpen
  \bibfield  {author} {\bibinfo {author} {\bibfnamefont {M.}~\bibnamefont
  {Lee}}\ and\ \bibinfo {author} {\bibfnamefont {M.}~\bibnamefont
  {MacKenzie}},\ }\href {\doibase 10.3390/universe8040227} {\bibfield
  {journal} {\bibinfo  {journal} {Universe}\ }\textbf {\bibinfo {volume} {8}},\
  \bibinfo {pages} {227} (\bibinfo {year} {2022})},\ \Eprint
  {http://arxiv.org/abs/2110.07093} {arXiv:2110.07093 [hep-ex]} \BibitemShut
  {NoStop}%
\bibitem [{\citenamefont {Davoudiasl}(2013)}]{Davoudiasl:2013pda}%
  \BibitemOpen
  \bibfield  {author} {\bibinfo {author} {\bibfnamefont {H.}~\bibnamefont
  {Davoudiasl}},\ }\href {\doibase 10.1103/PhysRevD.88.095004} {\bibfield
  {journal} {\bibinfo  {journal} {Phys. Rev. D}\ }\textbf {\bibinfo {volume}
  {88}},\ \bibinfo {pages} {095004} (\bibinfo {year} {2013})},\ \Eprint
  {http://arxiv.org/abs/1308.3473} {arXiv:1308.3473 [hep-ph]} \BibitemShut
  {NoStop}%
\bibitem [{\citenamefont {Davoudiasl}(2015)}]{Davoudiasl:2014gfa}%
  \BibitemOpen
  \bibfield  {author} {\bibinfo {author} {\bibfnamefont {H.}~\bibnamefont
  {Davoudiasl}},\ }\href {\doibase 10.1103/PhysRevLett.114.051802} {\bibfield
  {journal} {\bibinfo  {journal} {Phys. Rev. Lett.}\ }\textbf {\bibinfo
  {volume} {114}},\ \bibinfo {pages} {051802} (\bibinfo {year} {2015})},\
  \Eprint {http://arxiv.org/abs/1409.4823} {arXiv:1409.4823 [hep-ph]}
  \BibitemShut {NoStop}%
\bibitem [{\citenamefont {Helo}\ \emph {et~al.}(2018)\citenamefont {Helo},
  \citenamefont {Hirsch},\ and\ \citenamefont {Ota}}]{Helo:2018bgb}%
  \BibitemOpen
  \bibfield  {author} {\bibinfo {author} {\bibfnamefont {J.~C.}\ \bibnamefont
  {Helo}}, \bibinfo {author} {\bibfnamefont {M.}~\bibnamefont {Hirsch}}, \ and\
  \bibinfo {author} {\bibfnamefont {T.}~\bibnamefont {Ota}},\ }\href {\doibase
  10.1007/JHEP06(2018)047} {\bibfield  {journal} {\bibinfo  {journal} {JHEP}\
  }\textbf {\bibinfo {volume} {06}},\ \bibinfo {pages} {047} (\bibinfo {year}
  {2018})},\ \Eprint {http://arxiv.org/abs/1803.00035} {arXiv:1803.00035
  [hep-ph]} \BibitemShut {NoStop}%
\bibitem [{\citenamefont {McKeen}\ and\ \citenamefont
  {Pospelov}(2020)}]{McKeen:2020zni}%
  \BibitemOpen
  \bibfield  {author} {\bibinfo {author} {\bibfnamefont {D.}~\bibnamefont
  {McKeen}}\ and\ \bibinfo {author} {\bibfnamefont {M.}~\bibnamefont
  {Pospelov}},\ }\href@noop {} {\  (\bibinfo {year} {2020})},\ \Eprint
  {http://arxiv.org/abs/2003.02270} {arXiv:2003.02270 [hep-ph]} \BibitemShut
  {NoStop}%
\bibitem [{\citenamefont {Haisch}\ and\ \citenamefont
  {Hala}(2021)}]{Haisch:2021nos}%
  \BibitemOpen
  \bibfield  {author} {\bibinfo {author} {\bibfnamefont {U.}~\bibnamefont
  {Haisch}}\ and\ \bibinfo {author} {\bibfnamefont {A.}~\bibnamefont {Hala}},\
  }\href {\doibase 10.1007/JHEP11(2021)144} {\bibfield  {journal} {\bibinfo
  {journal} {JHEP}\ }\textbf {\bibinfo {volume} {11}},\ \bibinfo {pages} {144}
  (\bibinfo {year} {2021})},\ \Eprint {http://arxiv.org/abs/2108.06111}
  {arXiv:2108.06111 [hep-ph]} \BibitemShut {NoStop}%
\bibitem [{\citenamefont {Heeck}(2021)}]{Heeck:2020nbq}%
  \BibitemOpen
  \bibfield  {author} {\bibinfo {author} {\bibfnamefont {J.}~\bibnamefont
  {Heeck}},\ }\href {\doibase 10.1016/j.physletb.2020.136043} {\bibfield
  {journal} {\bibinfo  {journal} {Phys. Lett. B}\ }\textbf {\bibinfo {volume}
  {813}},\ \bibinfo {pages} {136043} (\bibinfo {year} {2021})},\ \Eprint
  {http://arxiv.org/abs/2009.01256} {arXiv:2009.01256 [hep-ph]} \BibitemShut
  {NoStop}%
\bibitem [{\citenamefont {Fajfer}\ and\ \citenamefont
  {Susi\v{c}}(2021)}]{Fajfer:2020tqf}%
  \BibitemOpen
  \bibfield  {author} {\bibinfo {author} {\bibfnamefont {S.}~\bibnamefont
  {Fajfer}}\ and\ \bibinfo {author} {\bibfnamefont {D.}~\bibnamefont
  {Susi\v{c}}},\ }\href {\doibase 10.1103/PhysRevD.103.055012} {\bibfield
  {journal} {\bibinfo  {journal} {Phys. Rev. D}\ }\textbf {\bibinfo {volume}
  {103}},\ \bibinfo {pages} {055012} (\bibinfo {year} {2021})},\ \Eprint
  {http://arxiv.org/abs/2010.08367} {arXiv:2010.08367 [hep-ph]} \BibitemShut
  {NoStop}%
\bibitem [{\citenamefont {Alonso-\'Alvarez}\ \emph {et~al.}(2022)\citenamefont
  {Alonso-\'Alvarez}, \citenamefont {Elor}, \citenamefont {Escudero},
  \citenamefont {Fornal}, \citenamefont {Grinstein},\ and\ \citenamefont
  {Martin~Camalich}}]{Alonso-Alvarez:2021oaj}%
  \BibitemOpen
  \bibfield  {author} {\bibinfo {author} {\bibfnamefont {G.}~\bibnamefont
  {Alonso-\'Alvarez}}, \bibinfo {author} {\bibfnamefont {G.}~\bibnamefont
  {Elor}}, \bibinfo {author} {\bibfnamefont {M.}~\bibnamefont {Escudero}},
  \bibinfo {author} {\bibfnamefont {B.}~\bibnamefont {Fornal}}, \bibinfo
  {author} {\bibfnamefont {B.}~\bibnamefont {Grinstein}}, \ and\ \bibinfo
  {author} {\bibfnamefont {J.}~\bibnamefont {Martin~Camalich}},\ }\href
  {\doibase 10.1103/PhysRevD.105.115005} {\bibfield  {journal} {\bibinfo
  {journal} {Phys. Rev. D}\ }\textbf {\bibinfo {volume} {105}},\ \bibinfo
  {pages} {115005} (\bibinfo {year} {2022})},\ \Eprint
  {http://arxiv.org/abs/2111.12712} {arXiv:2111.12712 [hep-ph]} \BibitemShut
  {NoStop}%
\bibitem [{\citenamefont {Goudzovski}\ \emph {et~al.}(2022)\citenamefont
  {Goudzovski} \emph {et~al.}}]{Goudzovski:2022vbt}%
  \BibitemOpen
  \bibfield  {author} {\bibinfo {author} {\bibfnamefont {E.}~\bibnamefont
  {Goudzovski}} \emph {et~al.},\ }\href@noop {} {\  (\bibinfo {year} {2022})},\
  \Eprint {http://arxiv.org/abs/2201.07805} {arXiv:2201.07805 [hep-ph]}
  \BibitemShut {NoStop}%
\bibitem [{\citenamefont {Aghanim}\ \emph {et~al.}(2020)\citenamefont {Aghanim}
  \emph {et~al.}}]{Aghanim:2018eyx}%
  \BibitemOpen
  \bibfield  {author} {\bibinfo {author} {\bibfnamefont {N.}~\bibnamefont
  {Aghanim}} \emph {et~al.} (\bibinfo {collaboration} {Planck}),\ }\href
  {\doibase 10.1051/0004-6361/201833910} {\bibfield  {journal} {\bibinfo
  {journal} {Astron. Astrophys.}\ }\textbf {\bibinfo {volume} {641}},\ \bibinfo
  {pages} {A6} (\bibinfo {year} {2020})},\ \bibinfo {note} {[Erratum:
  Astron.Astrophys. 652, C4 (2021)]},\ \Eprint
  {http://arxiv.org/abs/1807.06209} {arXiv:1807.06209 [astro-ph.CO]}
  \BibitemShut {NoStop}%
\bibitem [{\citenamefont {Steigman}(1976)}]{Steigman:1976ev}%
  \BibitemOpen
  \bibfield  {author} {\bibinfo {author} {\bibfnamefont {G.}~\bibnamefont
  {Steigman}},\ }\href@noop {} {\bibfield  {journal} {\bibinfo  {journal}
  {Ann.Rev.Astron.Astrophys.}\ }\textbf {\bibinfo {volume} {14}},\ \bibinfo
  {pages} {339} (\bibinfo {year} {1976})}\BibitemShut {NoStop}%
\bibitem [{\citenamefont {Cohen}\ \emph {et~al.}(1998)\citenamefont {Cohen},
  \citenamefont {De~Rujula},\ and\ \citenamefont {Glashow}}]{Cohen:1997ac}%
  \BibitemOpen
  \bibfield  {author} {\bibinfo {author} {\bibfnamefont {A.~G.}\ \bibnamefont
  {Cohen}}, \bibinfo {author} {\bibfnamefont {A.}~\bibnamefont {De~Rujula}}, \
  and\ \bibinfo {author} {\bibfnamefont {S.}~\bibnamefont {Glashow}},\ }\href
  {\doibase 10.1086/305328} {\bibfield  {journal} {\bibinfo  {journal}
  {Astrophys.J.}\ }\textbf {\bibinfo {volume} {495}},\ \bibinfo {pages} {539}
  (\bibinfo {year} {1998})},\ \Eprint {http://arxiv.org/abs/astro-ph/9707087}
  {arXiv:astro-ph/9707087 [astro-ph]} \BibitemShut {NoStop}%
\bibitem [{\citenamefont {Kofman}\ \emph {et~al.}(1994)\citenamefont {Kofman},
  \citenamefont {Linde},\ and\ \citenamefont {Starobinsky}}]{Kofman:1994rk}%
  \BibitemOpen
  \bibfield  {author} {\bibinfo {author} {\bibfnamefont {L.}~\bibnamefont
  {Kofman}}, \bibinfo {author} {\bibfnamefont {A.~D.}\ \bibnamefont {Linde}}, \
  and\ \bibinfo {author} {\bibfnamefont {A.~A.}\ \bibnamefont {Starobinsky}},\
  }\href {\doibase 10.1103/PhysRevLett.73.3195} {\bibfield  {journal} {\bibinfo
   {journal} {Phys.Rev.Lett.}\ }\textbf {\bibinfo {volume} {73}},\ \bibinfo
  {pages} {3195} (\bibinfo {year} {1994})},\ \Eprint
  {http://arxiv.org/abs/hep-th/9405187} {arXiv:hep-th/9405187 [hep-th]}
  \BibitemShut {NoStop}%
\bibitem [{\citenamefont {Kofman}\ \emph {et~al.}(1997)\citenamefont {Kofman},
  \citenamefont {Linde},\ and\ \citenamefont {Starobinsky}}]{Kofman:1997yn}%
  \BibitemOpen
  \bibfield  {author} {\bibinfo {author} {\bibfnamefont {L.}~\bibnamefont
  {Kofman}}, \bibinfo {author} {\bibfnamefont {A.~D.}\ \bibnamefont {Linde}}, \
  and\ \bibinfo {author} {\bibfnamefont {A.~A.}\ \bibnamefont {Starobinsky}},\
  }\href {\doibase 10.1103/PhysRevD.56.3258} {\bibfield  {journal} {\bibinfo
  {journal} {Phys.Rev.}\ }\textbf {\bibinfo {volume} {D56}},\ \bibinfo {pages}
  {3258} (\bibinfo {year} {1997})},\ \Eprint
  {http://arxiv.org/abs/hep-ph/9704452} {arXiv:hep-ph/9704452 [hep-ph]}
  \BibitemShut {NoStop}%
\bibitem [{\citenamefont {Cohen}\ \emph {et~al.}(1990)\citenamefont {Cohen},
  \citenamefont {Kaplan},\ and\ \citenamefont {Nelson}}]{Cohen:1990py}%
  \BibitemOpen
  \bibfield  {author} {\bibinfo {author} {\bibfnamefont {A.~G.}\ \bibnamefont
  {Cohen}}, \bibinfo {author} {\bibfnamefont {D.~B.}\ \bibnamefont {Kaplan}}, \
  and\ \bibinfo {author} {\bibfnamefont {A.~E.}\ \bibnamefont {Nelson}},\
  }\href {\doibase 10.1016/0370-2693(90)90690-8} {\bibfield  {journal}
  {\bibinfo  {journal} {Phys.Lett.}\ }\textbf {\bibinfo {volume} {B245}},\
  \bibinfo {pages} {561} (\bibinfo {year} {1990})}\BibitemShut {NoStop}%
\bibitem [{\citenamefont {Cohen}\ \emph
  {et~al.}(1991{\natexlab{a}})\citenamefont {Cohen}, \citenamefont {Kaplan},\
  and\ \citenamefont {Nelson}}]{Cohen:1990it}%
  \BibitemOpen
  \bibfield  {author} {\bibinfo {author} {\bibfnamefont {A.~G.}\ \bibnamefont
  {Cohen}}, \bibinfo {author} {\bibfnamefont {D.~B.}\ \bibnamefont {Kaplan}}, \
  and\ \bibinfo {author} {\bibfnamefont {A.~E.}\ \bibnamefont {Nelson}},\
  }\href {\doibase 10.1016/0550-3213(91)90395-E} {\bibfield  {journal}
  {\bibinfo  {journal} {Nucl.Phys.}\ }\textbf {\bibinfo {volume} {B349}},\
  \bibinfo {pages} {727} (\bibinfo {year} {1991}{\natexlab{a}})}\BibitemShut
  {NoStop}%
\bibitem [{\citenamefont {Cohen}\ \emph
  {et~al.}(1991{\natexlab{b}})\citenamefont {Cohen}, \citenamefont {Kaplan},\
  and\ \citenamefont {Nelson}}]{Cohen:1991iu}%
  \BibitemOpen
  \bibfield  {author} {\bibinfo {author} {\bibfnamefont {A.~G.}\ \bibnamefont
  {Cohen}}, \bibinfo {author} {\bibfnamefont {D.~B.}\ \bibnamefont {Kaplan}}, \
  and\ \bibinfo {author} {\bibfnamefont {A.~E.}\ \bibnamefont {Nelson}},\
  }\href {\doibase 10.1016/0370-2693(91)91711-4} {\bibfield  {journal}
  {\bibinfo  {journal} {Phys. Lett.}\ }\textbf {\bibinfo {volume} {B263}},\
  \bibinfo {pages} {86} (\bibinfo {year} {1991}{\natexlab{b}})}\BibitemShut
  {NoStop}%
\bibitem [{\citenamefont {Cohen}\ \emph {et~al.}(1993)\citenamefont {Cohen},
  \citenamefont {Kaplan},\ and\ \citenamefont {Nelson}}]{Cohen:1993nk}%
  \BibitemOpen
  \bibfield  {author} {\bibinfo {author} {\bibfnamefont {A.~G.}\ \bibnamefont
  {Cohen}}, \bibinfo {author} {\bibfnamefont {D.}~\bibnamefont {Kaplan}}, \
  and\ \bibinfo {author} {\bibfnamefont {A.}~\bibnamefont {Nelson}},\
  }\href@noop {} {\bibfield  {journal} {\bibinfo  {journal}
  {Ann.Rev.Nucl.Part.Sci.}\ }\textbf {\bibinfo {volume} {43}},\ \bibinfo
  {pages} {27} (\bibinfo {year} {1993})},\ \Eprint
  {http://arxiv.org/abs/hep-ph/9302210} {arXiv:hep-ph/9302210 [hep-ph]}
  \BibitemShut {NoStop}%
\bibitem [{\citenamefont {Kuzmin}\ \emph {et~al.}(1985)\citenamefont {Kuzmin},
  \citenamefont {Rubakov},\ and\ \citenamefont {Shaposhnikov}}]{Kuzmin:1985mm}%
  \BibitemOpen
  \bibfield  {author} {\bibinfo {author} {\bibfnamefont {V.~A.}\ \bibnamefont
  {Kuzmin}}, \bibinfo {author} {\bibfnamefont {V.~A.}\ \bibnamefont {Rubakov}},
  \ and\ \bibinfo {author} {\bibfnamefont {M.~E.}\ \bibnamefont
  {Shaposhnikov}},\ }\href {\doibase 10.1016/0370-2693(85)91028-7} {\bibfield
  {journal} {\bibinfo  {journal} {Phys. Lett.}\ }\textbf {\bibinfo {volume}
  {B155}},\ \bibinfo {pages} {36} (\bibinfo {year} {1985})}\BibitemShut
  {NoStop}%
\bibitem [{\citenamefont {D'Onofrio}\ and\ \citenamefont
  {Rummukainen}(2016)}]{DOnofrio:2015mpa}%
  \BibitemOpen
  \bibfield  {author} {\bibinfo {author} {\bibfnamefont {M.}~\bibnamefont
  {D'Onofrio}}\ and\ \bibinfo {author} {\bibfnamefont {K.}~\bibnamefont
  {Rummukainen}},\ }\href {\doibase 10.1103/PhysRevD.93.025003} {\bibfield
  {journal} {\bibinfo  {journal} {Phys. Rev.}\ }\textbf {\bibinfo {volume}
  {D93}},\ \bibinfo {pages} {025003} (\bibinfo {year} {2016})},\ \Eprint
  {http://arxiv.org/abs/1508.07161} {arXiv:1508.07161 [hep-ph]} \BibitemShut
  {NoStop}%
\bibitem [{\citenamefont {Ramsey-Musolf}(2020)}]{Ramsey-Musolf:2019lsf}%
  \BibitemOpen
  \bibfield  {author} {\bibinfo {author} {\bibfnamefont {M.~J.}\ \bibnamefont
  {Ramsey-Musolf}},\ }\href {\doibase 10.1007/JHEP09(2020)179} {\bibfield
  {journal} {\bibinfo  {journal} {JHEP}\ }\textbf {\bibinfo {volume} {09}},\
  \bibinfo {pages} {179} (\bibinfo {year} {2020})},\ \Eprint
  {http://arxiv.org/abs/1912.07189} {arXiv:1912.07189 [hep-ph]} \BibitemShut
  {NoStop}%
\bibitem [{\citenamefont {McDonald}(1994)}]{McDonald:1993ey}%
  \BibitemOpen
  \bibfield  {author} {\bibinfo {author} {\bibfnamefont {J.}~\bibnamefont
  {McDonald}},\ }\href {\doibase 10.1016/0370-2693(94)91229-7} {\bibfield
  {journal} {\bibinfo  {journal} {Phys. Lett.}\ }\textbf {\bibinfo {volume}
  {B323}},\ \bibinfo {pages} {339} (\bibinfo {year} {1994})}\BibitemShut
  {NoStop}%
\bibitem [{\citenamefont {Carena}\ \emph {et~al.}(1996)\citenamefont {Carena},
  \citenamefont {Quiros},\ and\ \citenamefont {Wagner}}]{Carena:1996wj}%
  \BibitemOpen
  \bibfield  {author} {\bibinfo {author} {\bibfnamefont {M.}~\bibnamefont
  {Carena}}, \bibinfo {author} {\bibfnamefont {M.}~\bibnamefont {Quiros}}, \
  and\ \bibinfo {author} {\bibfnamefont {C.~E.~M.}\ \bibnamefont {Wagner}},\
  }\href {\doibase 10.1016/0370-2693(96)00475-3} {\bibfield  {journal}
  {\bibinfo  {journal} {Phys. Lett. B}\ }\textbf {\bibinfo {volume} {380}},\
  \bibinfo {pages} {81} (\bibinfo {year} {1996})},\ \Eprint
  {http://arxiv.org/abs/hep-ph/9603420} {arXiv:hep-ph/9603420} \BibitemShut
  {NoStop}%
\bibitem [{\citenamefont {Davies}\ \emph {et~al.}(1996)\citenamefont {Davies},
  \citenamefont {Froggatt},\ and\ \citenamefont {Moorhouse}}]{Davies:1996qn}%
  \BibitemOpen
  \bibfield  {author} {\bibinfo {author} {\bibfnamefont {A.~T.}\ \bibnamefont
  {Davies}}, \bibinfo {author} {\bibfnamefont {C.~D.}\ \bibnamefont
  {Froggatt}}, \ and\ \bibinfo {author} {\bibfnamefont {R.~G.}\ \bibnamefont
  {Moorhouse}},\ }\href {\doibase 10.1016/0370-2693(96)00076-7} {\bibfield
  {journal} {\bibinfo  {journal} {Phys. Lett.}\ }\textbf {\bibinfo {volume}
  {B372}},\ \bibinfo {pages} {88} (\bibinfo {year} {1996})},\ \Eprint
  {http://arxiv.org/abs/hep-ph/9603388} {arXiv:hep-ph/9603388} \BibitemShut
  {NoStop}%
\bibitem [{\citenamefont {Huber}\ and\ \citenamefont
  {Schmidt}(2001)}]{Huber:2000mg}%
  \BibitemOpen
  \bibfield  {author} {\bibinfo {author} {\bibfnamefont {S.~J.}\ \bibnamefont
  {Huber}}\ and\ \bibinfo {author} {\bibfnamefont {M.~G.}\ \bibnamefont
  {Schmidt}},\ }\href {\doibase 10.1016/S0550-3213(01)00250-4} {\bibfield
  {journal} {\bibinfo  {journal} {Nucl. Phys.}\ }\textbf {\bibinfo {volume}
  {B606}},\ \bibinfo {pages} {183} (\bibinfo {year} {2001})},\ \Eprint
  {http://arxiv.org/abs/hep-ph/0003122} {arXiv:hep-ph/0003122} \BibitemShut
  {NoStop}%
\bibitem [{\citenamefont {Menon}\ \emph {et~al.}(2004)\citenamefont {Menon},
  \citenamefont {Morrissey},\ and\ \citenamefont {Wagner}}]{Menon:2004wv}%
  \BibitemOpen
  \bibfield  {author} {\bibinfo {author} {\bibfnamefont {A.}~\bibnamefont
  {Menon}}, \bibinfo {author} {\bibfnamefont {D.~E.}\ \bibnamefont
  {Morrissey}}, \ and\ \bibinfo {author} {\bibfnamefont {C.~E.~M.}\
  \bibnamefont {Wagner}},\ }\href {\doibase 10.1103/PhysRevD.70.035005}
  {\bibfield  {journal} {\bibinfo  {journal} {Phys. Rev.}\ }\textbf {\bibinfo
  {volume} {D70}},\ \bibinfo {pages} {035005} (\bibinfo {year} {2004})},\
  \Eprint {http://arxiv.org/abs/hep-ph/0404184} {arXiv:hep-ph/0404184}
  \BibitemShut {NoStop}%
\bibitem [{\citenamefont {Chung}\ \emph {et~al.}(2010)\citenamefont {Chung},
  \citenamefont {Garbrecht}, \citenamefont {Ramsey-Musolf},\ and\ \citenamefont
  {Tulin}}]{Chung:2009cb}%
  \BibitemOpen
  \bibfield  {author} {\bibinfo {author} {\bibfnamefont {D.~J.}\ \bibnamefont
  {Chung}}, \bibinfo {author} {\bibfnamefont {B.}~\bibnamefont {Garbrecht}},
  \bibinfo {author} {\bibfnamefont {M.~J.}\ \bibnamefont {Ramsey-Musolf}}, \
  and\ \bibinfo {author} {\bibfnamefont {S.}~\bibnamefont {Tulin}},\ }\href
  {\doibase 10.1103/PhysRevD.81.063506} {\bibfield  {journal} {\bibinfo
  {journal} {Phys.Rev.}\ }\textbf {\bibinfo {volume} {D81}},\ \bibinfo {pages}
  {063506} (\bibinfo {year} {2010})},\ \Eprint {http://arxiv.org/abs/0905.4509}
  {arXiv:0905.4509 [hep-ph]} \BibitemShut {NoStop}%
\bibitem [{\citenamefont {Carena}\ \emph {et~al.}(2012)\citenamefont {Carena},
  \citenamefont {Shah},\ and\ \citenamefont {Wagner}}]{Carena:2011jy}%
  \BibitemOpen
  \bibfield  {author} {\bibinfo {author} {\bibfnamefont {M.}~\bibnamefont
  {Carena}}, \bibinfo {author} {\bibfnamefont {N.~R.}\ \bibnamefont {Shah}}, \
  and\ \bibinfo {author} {\bibfnamefont {C.~E.~M.}\ \bibnamefont {Wagner}},\
  }\href {\doibase 10.1103/PhysRevD.85.036003} {\bibfield  {journal} {\bibinfo
  {journal} {Phys. Rev. D}\ }\textbf {\bibinfo {volume} {85}},\ \bibinfo
  {pages} {036003} (\bibinfo {year} {2012})},\ \Eprint
  {http://arxiv.org/abs/1110.4378} {arXiv:1110.4378 [hep-ph]} \BibitemShut
  {NoStop}%
\bibitem [{\citenamefont {Cohen}\ \emph {et~al.}(2012)\citenamefont {Cohen},
  \citenamefont {Morrissey},\ and\ \citenamefont {Pierce}}]{Cohen:2012zza}%
  \BibitemOpen
  \bibfield  {author} {\bibinfo {author} {\bibfnamefont {T.}~\bibnamefont
  {Cohen}}, \bibinfo {author} {\bibfnamefont {D.~E.}\ \bibnamefont
  {Morrissey}}, \ and\ \bibinfo {author} {\bibfnamefont {A.}~\bibnamefont
  {Pierce}},\ }\href {\doibase 10.1103/PhysRevD.86.013009} {\bibfield
  {journal} {\bibinfo  {journal} {Phys. Rev. D}\ }\textbf {\bibinfo {volume}
  {86}},\ \bibinfo {pages} {013009} (\bibinfo {year} {2012})},\ \Eprint
  {http://arxiv.org/abs/1203.2924} {arXiv:1203.2924 [hep-ph]} \BibitemShut
  {NoStop}%
\bibitem [{\citenamefont {Curtin}\ \emph {et~al.}(2012)\citenamefont {Curtin},
  \citenamefont {Jaiswal},\ and\ \citenamefont {Meade}}]{Curtin:2012aa}%
  \BibitemOpen
  \bibfield  {author} {\bibinfo {author} {\bibfnamefont {D.}~\bibnamefont
  {Curtin}}, \bibinfo {author} {\bibfnamefont {P.}~\bibnamefont {Jaiswal}}, \
  and\ \bibinfo {author} {\bibfnamefont {P.}~\bibnamefont {Meade}},\ }\href
  {\doibase 10.1007/JHEP08(2012)005} {\bibfield  {journal} {\bibinfo  {journal}
  {JHEP}\ }\textbf {\bibinfo {volume} {08}},\ \bibinfo {pages} {005} (\bibinfo
  {year} {2012})},\ \Eprint {http://arxiv.org/abs/1203.2932} {arXiv:1203.2932
  [hep-ph]} \BibitemShut {NoStop}%
\bibitem [{\citenamefont {Dong}\ \emph {et~al.}(2018)\citenamefont {Dong} \emph
  {et~al.}}]{CEPCStudyGroup:2018ghi}%
  \BibitemOpen
  \bibfield  {author} {\bibinfo {author} {\bibfnamefont {M.}~\bibnamefont
  {Dong}} \emph {et~al.} (\bibinfo {collaboration} {CEPC Study Group}),\
  }\href@noop {} {\  (\bibinfo {year} {2018})},\ \Eprint
  {http://arxiv.org/abs/1811.10545} {arXiv:1811.10545 [hep-ex]} \BibitemShut
  {NoStop}%
\bibitem [{\citenamefont {Abada}\ \emph
  {et~al.}(2019{\natexlab{a}})\citenamefont {Abada} \emph
  {et~al.}}]{FCC:2018byv}%
  \BibitemOpen
  \bibfield  {author} {\bibinfo {author} {\bibfnamefont {A.}~\bibnamefont
  {Abada}} \emph {et~al.} (\bibinfo {collaboration} {FCC}),\ }\href {\doibase
  10.1140/epjc/s10052-019-6904-3} {\bibfield  {journal} {\bibinfo  {journal}
  {Eur. Phys. J. C}\ }\textbf {\bibinfo {volume} {79}},\ \bibinfo {pages} {474}
  (\bibinfo {year} {2019}{\natexlab{a}})}\BibitemShut {NoStop}%
\bibitem [{\citenamefont {Abada}\ \emph
  {et~al.}(2019{\natexlab{b}})\citenamefont {Abada} \emph
  {et~al.}}]{FCC:2018evy}%
  \BibitemOpen
  \bibfield  {author} {\bibinfo {author} {\bibfnamefont {A.}~\bibnamefont
  {Abada}} \emph {et~al.} (\bibinfo {collaboration} {FCC}),\ }\href {\doibase
  10.1140/epjst/e2019-900045-4} {\bibfield  {journal} {\bibinfo  {journal}
  {Eur. Phys. J. ST}\ }\textbf {\bibinfo {volume} {228}},\ \bibinfo {pages}
  {261} (\bibinfo {year} {2019}{\natexlab{b}})}\BibitemShut {NoStop}%
\bibitem [{\citenamefont {Abada}\ \emph
  {et~al.}(2019{\natexlab{c}})\citenamefont {Abada} \emph
  {et~al.}}]{FCC:2018vvp}%
  \BibitemOpen
  \bibfield  {author} {\bibinfo {author} {\bibfnamefont {A.}~\bibnamefont
  {Abada}} \emph {et~al.} (\bibinfo {collaboration} {FCC}),\ }\href {\doibase
  10.1140/epjst/e2019-900087-0} {\bibfield  {journal} {\bibinfo  {journal}
  {Eur. Phys. J. ST}\ }\textbf {\bibinfo {volume} {228}},\ \bibinfo {pages}
  {755} (\bibinfo {year} {2019}{\natexlab{c}})}\BibitemShut {NoStop}%
\bibitem [{\citenamefont {Cepeda}\ \emph {et~al.}(2019)\citenamefont {Cepeda}
  \emph {et~al.}}]{Cepeda:2019klc}%
  \BibitemOpen
  \bibfield  {author} {\bibinfo {author} {\bibfnamefont {M.}~\bibnamefont
  {Cepeda}} \emph {et~al.},\ }\href {\doibase 10.23731/CYRM-2019-007.221}
  {\bibfield  {journal} {\bibinfo  {journal} {CERN Yellow Rep. Monogr.}\
  }\textbf {\bibinfo {volume} {7}},\ \bibinfo {pages} {221} (\bibinfo {year}
  {2019})},\ \Eprint {http://arxiv.org/abs/1902.00134} {arXiv:1902.00134
  [hep-ph]} \BibitemShut {NoStop}%
\bibitem [{\citenamefont {Curtin}\ \emph {et~al.}(2014)\citenamefont {Curtin},
  \citenamefont {Meade},\ and\ \citenamefont {Yu}}]{Curtin:2014jma}%
  \BibitemOpen
  \bibfield  {author} {\bibinfo {author} {\bibfnamefont {D.}~\bibnamefont
  {Curtin}}, \bibinfo {author} {\bibfnamefont {P.}~\bibnamefont {Meade}}, \
  and\ \bibinfo {author} {\bibfnamefont {C.-T.}\ \bibnamefont {Yu}},\ }\href
  {\doibase 10.1007/JHEP11(2014)127} {\bibfield  {journal} {\bibinfo  {journal}
  {JHEP}\ }\textbf {\bibinfo {volume} {11}},\ \bibinfo {pages} {127} (\bibinfo
  {year} {2014})},\ \Eprint {http://arxiv.org/abs/1409.0005} {arXiv:1409.0005
  [hep-ph]} \BibitemShut {NoStop}%
\bibitem [{\citenamefont {Huang}\ \emph
  {et~al.}(2016{\natexlab{a}})\citenamefont {Huang}, \citenamefont {Joglekar},
  \citenamefont {Li},\ and\ \citenamefont {Wagner}}]{Huang:2015tdv}%
  \BibitemOpen
  \bibfield  {author} {\bibinfo {author} {\bibfnamefont {P.}~\bibnamefont
  {Huang}}, \bibinfo {author} {\bibfnamefont {A.}~\bibnamefont {Joglekar}},
  \bibinfo {author} {\bibfnamefont {B.}~\bibnamefont {Li}}, \ and\ \bibinfo
  {author} {\bibfnamefont {C.~E.~M.}\ \bibnamefont {Wagner}},\ }\href {\doibase
  10.1103/PhysRevD.93.055049} {\bibfield  {journal} {\bibinfo  {journal} {Phys.
  Rev. D}\ }\textbf {\bibinfo {volume} {93}},\ \bibinfo {pages} {055049}
  (\bibinfo {year} {2016}{\natexlab{a}})},\ \Eprint
  {http://arxiv.org/abs/1512.00068} {arXiv:1512.00068 [hep-ph]} \BibitemShut
  {NoStop}%
\bibitem [{\citenamefont {Huang}\ \emph
  {et~al.}(2016{\natexlab{b}})\citenamefont {Huang}, \citenamefont {Long},\
  and\ \citenamefont {Wang}}]{Huang:2016cjm}%
  \BibitemOpen
  \bibfield  {author} {\bibinfo {author} {\bibfnamefont {P.}~\bibnamefont
  {Huang}}, \bibinfo {author} {\bibfnamefont {A.~J.}\ \bibnamefont {Long}}, \
  and\ \bibinfo {author} {\bibfnamefont {L.-T.}\ \bibnamefont {Wang}},\ }\href
  {\doibase 10.1103/PhysRevD.94.075008} {\bibfield  {journal} {\bibinfo
  {journal} {Phys. Rev. D}\ }\textbf {\bibinfo {volume} {94}},\ \bibinfo
  {pages} {075008} (\bibinfo {year} {2016}{\natexlab{b}})},\ \Eprint
  {http://arxiv.org/abs/1608.06619} {arXiv:1608.06619 [hep-ph]} \BibitemShut
  {NoStop}%
\bibitem [{\citenamefont {Chen}\ \emph {et~al.}(2017)\citenamefont {Chen},
  \citenamefont {Kozaczuk},\ and\ \citenamefont {Lewis}}]{Chen:2017qcz}%
  \BibitemOpen
  \bibfield  {author} {\bibinfo {author} {\bibfnamefont {C.-Y.}\ \bibnamefont
  {Chen}}, \bibinfo {author} {\bibfnamefont {J.}~\bibnamefont {Kozaczuk}}, \
  and\ \bibinfo {author} {\bibfnamefont {I.~M.}\ \bibnamefont {Lewis}},\ }\href
  {\doibase 10.1007/JHEP08(2017)096} {\bibfield  {journal} {\bibinfo  {journal}
  {JHEP}\ }\textbf {\bibinfo {volume} {08}},\ \bibinfo {pages} {096} (\bibinfo
  {year} {2017})},\ \Eprint {http://arxiv.org/abs/1704.05844} {arXiv:1704.05844
  [hep-ph]} \BibitemShut {NoStop}%
\bibitem [{\citenamefont {Reichert}\ \emph {et~al.}(2018)\citenamefont
  {Reichert}, \citenamefont {Eichhorn}, \citenamefont {Gies}, \citenamefont
  {Pawlowski}, \citenamefont {Plehn},\ and\ \citenamefont
  {Scherer}}]{Reichert:2017puo}%
  \BibitemOpen
  \bibfield  {author} {\bibinfo {author} {\bibfnamefont {M.}~\bibnamefont
  {Reichert}}, \bibinfo {author} {\bibfnamefont {A.}~\bibnamefont {Eichhorn}},
  \bibinfo {author} {\bibfnamefont {H.}~\bibnamefont {Gies}}, \bibinfo {author}
  {\bibfnamefont {J.~M.}\ \bibnamefont {Pawlowski}}, \bibinfo {author}
  {\bibfnamefont {T.}~\bibnamefont {Plehn}}, \ and\ \bibinfo {author}
  {\bibfnamefont {M.~M.}\ \bibnamefont {Scherer}},\ }\href {\doibase
  10.1103/PhysRevD.97.075008} {\bibfield  {journal} {\bibinfo  {journal} {Phys.
  Rev. D}\ }\textbf {\bibinfo {volume} {97}},\ \bibinfo {pages} {075008}
  (\bibinfo {year} {2018})},\ \Eprint {http://arxiv.org/abs/1711.00019}
  {arXiv:1711.00019 [hep-ph]} \BibitemShut {NoStop}%
\bibitem [{\citenamefont {Carena}\ \emph {et~al.}(2018)\citenamefont {Carena},
  \citenamefont {Liu},\ and\ \citenamefont {Riembau}}]{Carena:2018vpt}%
  \BibitemOpen
  \bibfield  {author} {\bibinfo {author} {\bibfnamefont {M.}~\bibnamefont
  {Carena}}, \bibinfo {author} {\bibfnamefont {Z.}~\bibnamefont {Liu}}, \ and\
  \bibinfo {author} {\bibfnamefont {M.}~\bibnamefont {Riembau}},\ }\href
  {\doibase 10.1103/PhysRevD.97.095032} {\bibfield  {journal} {\bibinfo
  {journal} {Phys. Rev. D}\ }\textbf {\bibinfo {volume} {97}},\ \bibinfo
  {pages} {095032} (\bibinfo {year} {2018})},\ \Eprint
  {http://arxiv.org/abs/1801.00794} {arXiv:1801.00794 [hep-ph]} \BibitemShut
  {NoStop}%
\bibitem [{\citenamefont {Gon\c{c}alves}\ \emph {et~al.}(2018)\citenamefont
  {Gon\c{c}alves}, \citenamefont {Han}, \citenamefont {Kling}, \citenamefont
  {Plehn},\ and\ \citenamefont {Takeuchi}}]{Goncalves:2018qas}%
  \BibitemOpen
  \bibfield  {author} {\bibinfo {author} {\bibfnamefont {D.}~\bibnamefont
  {Gon\c{c}alves}}, \bibinfo {author} {\bibfnamefont {T.}~\bibnamefont {Han}},
  \bibinfo {author} {\bibfnamefont {F.}~\bibnamefont {Kling}}, \bibinfo
  {author} {\bibfnamefont {T.}~\bibnamefont {Plehn}}, \ and\ \bibinfo {author}
  {\bibfnamefont {M.}~\bibnamefont {Takeuchi}},\ }\href {\doibase
  10.1103/PhysRevD.97.113004} {\bibfield  {journal} {\bibinfo  {journal} {Phys.
  Rev. D}\ }\textbf {\bibinfo {volume} {97}},\ \bibinfo {pages} {113004}
  (\bibinfo {year} {2018})},\ \Eprint {http://arxiv.org/abs/1802.04319}
  {arXiv:1802.04319 [hep-ph]} \BibitemShut {NoStop}%
\bibitem [{\citenamefont {Carena}\ \emph {et~al.}(2022)\citenamefont {Carena},
  \citenamefont {Kozaczuk}, \citenamefont {Liu}, \citenamefont {Ou},
  \citenamefont {Ramsey-Musolf}, \citenamefont {Shelton}, \citenamefont
  {Wang},\ and\ \citenamefont {Xie}}]{Carena:2022yvx}%
  \BibitemOpen
  \bibfield  {author} {\bibinfo {author} {\bibfnamefont {M.}~\bibnamefont
  {Carena}}, \bibinfo {author} {\bibfnamefont {J.}~\bibnamefont {Kozaczuk}},
  \bibinfo {author} {\bibfnamefont {Z.}~\bibnamefont {Liu}}, \bibinfo {author}
  {\bibfnamefont {T.}~\bibnamefont {Ou}}, \bibinfo {author} {\bibfnamefont
  {M.~J.}\ \bibnamefont {Ramsey-Musolf}}, \bibinfo {author} {\bibfnamefont
  {J.}~\bibnamefont {Shelton}}, \bibinfo {author} {\bibfnamefont
  {Y.}~\bibnamefont {Wang}}, \ and\ \bibinfo {author} {\bibfnamefont {K.-P.}\
  \bibnamefont {Xie}},\ }in\ \href@noop {} {\emph {\bibinfo {booktitle} {{2022
  Snowmass Summer Study}}}}\ (\bibinfo {year} {2022})\ \Eprint
  {http://arxiv.org/abs/2203.08206} {arXiv:2203.08206 [hep-ph]} \BibitemShut
  {NoStop}%
\bibitem [{\citenamefont {Andreev}\ \emph {et~al.}(2018)\citenamefont {Andreev}
  \emph {et~al.}}]{ACME:2018yjb}%
  \BibitemOpen
  \bibfield  {author} {\bibinfo {author} {\bibfnamefont {V.}~\bibnamefont
  {Andreev}} \emph {et~al.} (\bibinfo {collaboration} {ACME}),\ }\href
  {\doibase 10.1038/s41586-018-0599-8} {\bibfield  {journal} {\bibinfo
  {journal} {Nature}\ }\textbf {\bibinfo {volume} {562}},\ \bibinfo {pages}
  {355} (\bibinfo {year} {2018})}\BibitemShut {NoStop}%
\bibitem [{\citenamefont {Blum}\ \emph {et~al.}(2010)\citenamefont {Blum},
  \citenamefont {Delaunay}, \citenamefont {Losada}, \citenamefont {Nir},\ and\
  \citenamefont {Tulin}}]{Blum:2010by}%
  \BibitemOpen
  \bibfield  {author} {\bibinfo {author} {\bibfnamefont {K.}~\bibnamefont
  {Blum}}, \bibinfo {author} {\bibfnamefont {C.}~\bibnamefont {Delaunay}},
  \bibinfo {author} {\bibfnamefont {M.}~\bibnamefont {Losada}}, \bibinfo
  {author} {\bibfnamefont {Y.}~\bibnamefont {Nir}}, \ and\ \bibinfo {author}
  {\bibfnamefont {S.}~\bibnamefont {Tulin}},\ }\href {\doibase
  10.1007/JHEP05(2010)101} {\bibfield  {journal} {\bibinfo  {journal} {JHEP}\
  }\textbf {\bibinfo {volume} {05}},\ \bibinfo {pages} {101} (\bibinfo {year}
  {2010})},\ \Eprint {http://arxiv.org/abs/1003.2447} {arXiv:1003.2447
  [hep-ph]} \BibitemShut {NoStop}%
\bibitem [{\citenamefont {Cesarotti}\ \emph {et~al.}(2019)\citenamefont
  {Cesarotti}, \citenamefont {Lu}, \citenamefont {Nakai}, \citenamefont
  {Parikh},\ and\ \citenamefont {Reece}}]{Cesarotti:2018huy}%
  \BibitemOpen
  \bibfield  {author} {\bibinfo {author} {\bibfnamefont {C.}~\bibnamefont
  {Cesarotti}}, \bibinfo {author} {\bibfnamefont {Q.}~\bibnamefont {Lu}},
  \bibinfo {author} {\bibfnamefont {Y.}~\bibnamefont {Nakai}}, \bibinfo
  {author} {\bibfnamefont {A.}~\bibnamefont {Parikh}}, \ and\ \bibinfo {author}
  {\bibfnamefont {M.}~\bibnamefont {Reece}},\ }\href {\doibase
  10.1007/JHEP05(2019)059} {\bibfield  {journal} {\bibinfo  {journal} {JHEP}\
  }\textbf {\bibinfo {volume} {05}},\ \bibinfo {pages} {059} (\bibinfo {year}
  {2019})},\ \Eprint {http://arxiv.org/abs/1810.07736} {arXiv:1810.07736
  [hep-ph]} \BibitemShut {NoStop}%
\bibitem [{\citenamefont {Caprini}\ and\ \citenamefont
  {No}(2012)}]{Caprini:2011uz}%
  \BibitemOpen
  \bibfield  {author} {\bibinfo {author} {\bibfnamefont {C.}~\bibnamefont
  {Caprini}}\ and\ \bibinfo {author} {\bibfnamefont {J.~M.}\ \bibnamefont
  {No}},\ }\href {\doibase 10.1088/1475-7516/2012/01/031} {\bibfield  {journal}
  {\bibinfo  {journal} {JCAP}\ }\textbf {\bibinfo {volume} {1201}},\ \bibinfo
  {pages} {031} (\bibinfo {year} {2012})},\ \Eprint
  {http://arxiv.org/abs/1111.1726} {arXiv:1111.1726 [hep-ph]} \BibitemShut
  {NoStop}%
\bibitem [{\citenamefont {Baldes}\ \emph {et~al.}(2018)\citenamefont {Baldes},
  \citenamefont {Konstandin},\ and\ \citenamefont {Servant}}]{Baldes:2016rqn}%
  \BibitemOpen
  \bibfield  {author} {\bibinfo {author} {\bibfnamefont {I.}~\bibnamefont
  {Baldes}}, \bibinfo {author} {\bibfnamefont {T.}~\bibnamefont {Konstandin}},
  \ and\ \bibinfo {author} {\bibfnamefont {G.}~\bibnamefont {Servant}},\ }\href
  {\doibase 10.1016/j.physletb.2018.10.015} {\bibfield  {journal} {\bibinfo
  {journal} {Phys. Lett. B}\ }\textbf {\bibinfo {volume} {786}},\ \bibinfo
  {pages} {373} (\bibinfo {year} {2018})},\ \Eprint
  {http://arxiv.org/abs/1604.04526} {arXiv:1604.04526 [hep-ph]} \BibitemShut
  {NoStop}%
\bibitem [{\citenamefont {Baldes}\ \emph {et~al.}(2016)\citenamefont {Baldes},
  \citenamefont {Konstandin},\ and\ \citenamefont {Servant}}]{Baldes:2016gaf}%
  \BibitemOpen
  \bibfield  {author} {\bibinfo {author} {\bibfnamefont {I.}~\bibnamefont
  {Baldes}}, \bibinfo {author} {\bibfnamefont {T.}~\bibnamefont {Konstandin}},
  \ and\ \bibinfo {author} {\bibfnamefont {G.}~\bibnamefont {Servant}},\ }\href
  {\doibase 10.1007/JHEP12(2016)073} {\bibfield  {journal} {\bibinfo  {journal}
  {JHEP}\ }\textbf {\bibinfo {volume} {12}},\ \bibinfo {pages} {073} (\bibinfo
  {year} {2016})},\ \Eprint {http://arxiv.org/abs/1608.03254} {arXiv:1608.03254
  [hep-ph]} \BibitemShut {NoStop}%
\bibitem [{\citenamefont {Cline}\ \emph {et~al.}(2017)\citenamefont {Cline},
  \citenamefont {Kainulainen},\ and\ \citenamefont
  {Tucker-Smith}}]{Cline:2017qpe}%
  \BibitemOpen
  \bibfield  {author} {\bibinfo {author} {\bibfnamefont {J.~M.}\ \bibnamefont
  {Cline}}, \bibinfo {author} {\bibfnamefont {K.}~\bibnamefont {Kainulainen}},
  \ and\ \bibinfo {author} {\bibfnamefont {D.}~\bibnamefont {Tucker-Smith}},\
  }\href {\doibase 10.1103/PhysRevD.95.115006} {\bibfield  {journal} {\bibinfo
  {journal} {Phys. Rev. D}\ }\textbf {\bibinfo {volume} {95}},\ \bibinfo
  {pages} {115006} (\bibinfo {year} {2017})},\ \Eprint
  {http://arxiv.org/abs/1702.08909} {arXiv:1702.08909 [hep-ph]} \BibitemShut
  {NoStop}%
\bibitem [{\citenamefont {Long}\ \emph {et~al.}(2017)\citenamefont {Long},
  \citenamefont {Tesi},\ and\ \citenamefont {Wang}}]{Long:2017rdo}%
  \BibitemOpen
  \bibfield  {author} {\bibinfo {author} {\bibfnamefont {A.~J.}\ \bibnamefont
  {Long}}, \bibinfo {author} {\bibfnamefont {A.}~\bibnamefont {Tesi}}, \ and\
  \bibinfo {author} {\bibfnamefont {L.-T.}\ \bibnamefont {Wang}},\ }\href
  {\doibase 10.1007/JHEP10(2017)095} {\bibfield  {journal} {\bibinfo  {journal}
  {JHEP}\ }\textbf {\bibinfo {volume} {10}},\ \bibinfo {pages} {095} (\bibinfo
  {year} {2017})},\ \Eprint {http://arxiv.org/abs/1703.04902} {arXiv:1703.04902
  [hep-ph]} \BibitemShut {NoStop}%
\bibitem [{\citenamefont {Carena}\ \emph {et~al.}(2019)\citenamefont {Carena},
  \citenamefont {Quir\'os},\ and\ \citenamefont {Zhang}}]{Carena:2018cjh}%
  \BibitemOpen
  \bibfield  {author} {\bibinfo {author} {\bibfnamefont {M.}~\bibnamefont
  {Carena}}, \bibinfo {author} {\bibfnamefont {M.}~\bibnamefont {Quir\'os}}, \
  and\ \bibinfo {author} {\bibfnamefont {Y.}~\bibnamefont {Zhang}},\ }\href
  {\doibase 10.1103/PhysRevLett.122.201802} {\bibfield  {journal} {\bibinfo
  {journal} {Phys. Rev. Lett.}\ }\textbf {\bibinfo {volume} {122}},\ \bibinfo
  {pages} {201802} (\bibinfo {year} {2019})},\ \Eprint
  {http://arxiv.org/abs/1811.09719} {arXiv:1811.09719 [hep-ph]} \BibitemShut
  {NoStop}%
\bibitem [{\citenamefont {Glioti}\ \emph {et~al.}(2019)\citenamefont {Glioti},
  \citenamefont {Rattazzi},\ and\ \citenamefont {Vecchi}}]{Glioti:2018roy}%
  \BibitemOpen
  \bibfield  {author} {\bibinfo {author} {\bibfnamefont {A.}~\bibnamefont
  {Glioti}}, \bibinfo {author} {\bibfnamefont {R.}~\bibnamefont {Rattazzi}}, \
  and\ \bibinfo {author} {\bibfnamefont {L.}~\bibnamefont {Vecchi}},\ }\href
  {\doibase 10.1007/JHEP04(2019)027} {\bibfield  {journal} {\bibinfo  {journal}
  {JHEP}\ }\textbf {\bibinfo {volume} {04}},\ \bibinfo {pages} {027} (\bibinfo
  {year} {2019})},\ \Eprint {http://arxiv.org/abs/1811.11740} {arXiv:1811.11740
  [hep-ph]} \BibitemShut {NoStop}%
\bibitem [{\citenamefont {Ellis}\ \emph
  {et~al.}(2019{\natexlab{b}})\citenamefont {Ellis}, \citenamefont {Ipek},\
  and\ \citenamefont {White}}]{Ellis:2019flb}%
  \BibitemOpen
  \bibfield  {author} {\bibinfo {author} {\bibfnamefont {S.~A.~R.}\
  \bibnamefont {Ellis}}, \bibinfo {author} {\bibfnamefont {S.}~\bibnamefont
  {Ipek}}, \ and\ \bibinfo {author} {\bibfnamefont {G.}~\bibnamefont {White}},\
  }\href {\doibase 10.1007/JHEP08(2019)002} {\bibfield  {journal} {\bibinfo
  {journal} {JHEP}\ }\textbf {\bibinfo {volume} {08}},\ \bibinfo {pages} {002}
  (\bibinfo {year} {2019}{\natexlab{b}})},\ \Eprint
  {http://arxiv.org/abs/1905.11994} {arXiv:1905.11994 [hep-ph]} \BibitemShut
  {NoStop}%
\bibitem [{\citenamefont {Hall}\ \emph {et~al.}(2020)\citenamefont {Hall},
  \citenamefont {Konstandin}, \citenamefont {McGehee}, \citenamefont
  {Murayama},\ and\ \citenamefont {Servant}}]{Hall:2019ank}%
  \BibitemOpen
  \bibfield  {author} {\bibinfo {author} {\bibfnamefont {E.}~\bibnamefont
  {Hall}}, \bibinfo {author} {\bibfnamefont {T.}~\bibnamefont {Konstandin}},
  \bibinfo {author} {\bibfnamefont {R.}~\bibnamefont {McGehee}}, \bibinfo
  {author} {\bibfnamefont {H.}~\bibnamefont {Murayama}}, \ and\ \bibinfo
  {author} {\bibfnamefont {G.}~\bibnamefont {Servant}},\ }\href {\doibase
  10.1007/JHEP04(2020)042} {\bibfield  {journal} {\bibinfo  {journal} {JHEP}\
  }\textbf {\bibinfo {volume} {04}},\ \bibinfo {pages} {042} (\bibinfo {year}
  {2020})},\ \Eprint {http://arxiv.org/abs/1910.08068} {arXiv:1910.08068
  [hep-ph]} \BibitemShut {NoStop}%
\bibitem [{\citenamefont {Hall}\ \emph {et~al.}(2019)\citenamefont {Hall},
  \citenamefont {Konstandin}, \citenamefont {McGehee},\ and\ \citenamefont
  {Murayama}}]{Hall:2019rld}%
  \BibitemOpen
  \bibfield  {author} {\bibinfo {author} {\bibfnamefont {E.}~\bibnamefont
  {Hall}}, \bibinfo {author} {\bibfnamefont {T.}~\bibnamefont {Konstandin}},
  \bibinfo {author} {\bibfnamefont {R.}~\bibnamefont {McGehee}}, \ and\
  \bibinfo {author} {\bibfnamefont {H.}~\bibnamefont {Murayama}},\ }\href@noop
  {} {\  (\bibinfo {year} {2019})},\ \Eprint {http://arxiv.org/abs/1911.12342}
  {arXiv:1911.12342 [hep-ph]} \BibitemShut {NoStop}%
\bibitem [{\citenamefont {Buchmuller}\ \emph {et~al.}(2005)\citenamefont
  {Buchmuller}, \citenamefont {Di~Bari},\ and\ \citenamefont
  {Plumacher}}]{Buchmuller:2004nz}%
  \BibitemOpen
  \bibfield  {author} {\bibinfo {author} {\bibfnamefont {W.}~\bibnamefont
  {Buchmuller}}, \bibinfo {author} {\bibfnamefont {P.}~\bibnamefont {Di~Bari}},
  \ and\ \bibinfo {author} {\bibfnamefont {M.}~\bibnamefont {Plumacher}},\
  }\href {\doibase 10.1016/j.aop.2004.02.003} {\bibfield  {journal} {\bibinfo
  {journal} {Annals Phys.}\ }\textbf {\bibinfo {volume} {315}},\ \bibinfo
  {pages} {305} (\bibinfo {year} {2005})},\ \Eprint
  {http://arxiv.org/abs/hep-ph/0401240} {arXiv:hep-ph/0401240 [hep-ph]}
  \BibitemShut {NoStop}%
\bibitem [{\citenamefont {Fukugita}\ and\ \citenamefont
  {Yanagida}(1986)}]{Fukugita:1986hr}%
  \BibitemOpen
  \bibfield  {author} {\bibinfo {author} {\bibfnamefont {M.}~\bibnamefont
  {Fukugita}}\ and\ \bibinfo {author} {\bibfnamefont {T.}~\bibnamefont
  {Yanagida}},\ }\href {\doibase 10.1016/0370-2693(86)91126-3} {\bibfield
  {journal} {\bibinfo  {journal} {Phys. Lett.}\ }\textbf {\bibinfo {volume}
  {B174}},\ \bibinfo {pages} {45} (\bibinfo {year} {1986})}\BibitemShut
  {NoStop}%
\bibitem [{\citenamefont {Yanagida}(1980)}]{Yanagida:1980xy}%
  \BibitemOpen
  \bibfield  {author} {\bibinfo {author} {\bibfnamefont {T.}~\bibnamefont
  {Yanagida}},\ }\href@noop {} {\bibfield  {journal} {\bibinfo  {journal}
  {Prog.Theor.Phys.}\ }\textbf {\bibinfo {volume} {64}},\ \bibinfo {pages}
  {1103} (\bibinfo {year} {1980})}\BibitemShut {NoStop}%
\bibitem [{\citenamefont {Branco}\ \emph {et~al.}(2001)\citenamefont {Branco},
  \citenamefont {Morozumi}, \citenamefont {Nobre},\ and\ \citenamefont
  {Rebelo}}]{Branco:2001pq}%
  \BibitemOpen
  \bibfield  {author} {\bibinfo {author} {\bibfnamefont {G.~C.}\ \bibnamefont
  {Branco}}, \bibinfo {author} {\bibfnamefont {T.}~\bibnamefont {Morozumi}},
  \bibinfo {author} {\bibfnamefont {B.~M.}\ \bibnamefont {Nobre}}, \ and\
  \bibinfo {author} {\bibfnamefont {M.~N.}\ \bibnamefont {Rebelo}},\ }\href
  {\doibase 10.1016/S0550-3213(01)00425-4} {\bibfield  {journal} {\bibinfo
  {journal} {Nucl. Phys. B}\ }\textbf {\bibinfo {volume} {617}},\ \bibinfo
  {pages} {475} (\bibinfo {year} {2001})},\ \Eprint
  {http://arxiv.org/abs/hep-ph/0107164} {arXiv:hep-ph/0107164} \BibitemShut
  {NoStop}%
\bibitem [{\citenamefont {Branco}\ \emph {et~al.}(2003)\citenamefont {Branco},
  \citenamefont {Gonzalez~Felipe}, \citenamefont {Joaquim}, \citenamefont
  {Masina}, \citenamefont {Rebelo},\ and\ \citenamefont
  {Savoy}}]{Branco:2002xf}%
  \BibitemOpen
  \bibfield  {author} {\bibinfo {author} {\bibfnamefont {G.~C.}\ \bibnamefont
  {Branco}}, \bibinfo {author} {\bibfnamefont {R.}~\bibnamefont
  {Gonzalez~Felipe}}, \bibinfo {author} {\bibfnamefont {F.~R.}\ \bibnamefont
  {Joaquim}}, \bibinfo {author} {\bibfnamefont {I.}~\bibnamefont {Masina}},
  \bibinfo {author} {\bibfnamefont {M.~N.}\ \bibnamefont {Rebelo}}, \ and\
  \bibinfo {author} {\bibfnamefont {C.~A.}\ \bibnamefont {Savoy}},\ }\href
  {\doibase 10.1103/PhysRevD.67.073025} {\bibfield  {journal} {\bibinfo
  {journal} {Phys. Rev. D}\ }\textbf {\bibinfo {volume} {67}},\ \bibinfo
  {pages} {073025} (\bibinfo {year} {2003})},\ \Eprint
  {http://arxiv.org/abs/hep-ph/0211001} {arXiv:hep-ph/0211001} \BibitemShut
  {NoStop}%
\bibitem [{\citenamefont {Endoh}\ \emph {et~al.}(2002)\citenamefont {Endoh},
  \citenamefont {Kaneko}, \citenamefont {Kang}, \citenamefont {Morozumi},\ and\
  \citenamefont {Tanimoto}}]{Endoh:2002wm}%
  \BibitemOpen
  \bibfield  {author} {\bibinfo {author} {\bibfnamefont {T.}~\bibnamefont
  {Endoh}}, \bibinfo {author} {\bibfnamefont {S.}~\bibnamefont {Kaneko}},
  \bibinfo {author} {\bibfnamefont {S.~K.}\ \bibnamefont {Kang}}, \bibinfo
  {author} {\bibfnamefont {T.}~\bibnamefont {Morozumi}}, \ and\ \bibinfo
  {author} {\bibfnamefont {M.}~\bibnamefont {Tanimoto}},\ }\href {\doibase
  10.1103/PhysRevLett.89.231601} {\bibfield  {journal} {\bibinfo  {journal}
  {Phys. Rev. Lett.}\ }\textbf {\bibinfo {volume} {89}},\ \bibinfo {pages}
  {231601} (\bibinfo {year} {2002})},\ \Eprint
  {http://arxiv.org/abs/hep-ph/0209020} {arXiv:hep-ph/0209020} \BibitemShut
  {NoStop}%
\bibitem [{\citenamefont {Frampton}\ \emph {et~al.}(2002)\citenamefont
  {Frampton}, \citenamefont {Glashow},\ and\ \citenamefont
  {Yanagida}}]{Frampton:2002qc}%
  \BibitemOpen
  \bibfield  {author} {\bibinfo {author} {\bibfnamefont {P.~H.}\ \bibnamefont
  {Frampton}}, \bibinfo {author} {\bibfnamefont {S.~L.}\ \bibnamefont
  {Glashow}}, \ and\ \bibinfo {author} {\bibfnamefont {T.}~\bibnamefont
  {Yanagida}},\ }\href {\doibase 10.1016/S0370-2693(02)02853-8} {\bibfield
  {journal} {\bibinfo  {journal} {Phys. Lett. B}\ }\textbf {\bibinfo {volume}
  {548}},\ \bibinfo {pages} {119} (\bibinfo {year} {2002})},\ \Eprint
  {http://arxiv.org/abs/hep-ph/0208157} {arXiv:hep-ph/0208157} \BibitemShut
  {NoStop}%
\bibitem [{\citenamefont {Pascoli}\ \emph {et~al.}(2007)\citenamefont
  {Pascoli}, \citenamefont {Petcov},\ and\ \citenamefont
  {Riotto}}]{Pascoli:2006ci}%
  \BibitemOpen
  \bibfield  {author} {\bibinfo {author} {\bibfnamefont {S.}~\bibnamefont
  {Pascoli}}, \bibinfo {author} {\bibfnamefont {S.~T.}\ \bibnamefont {Petcov}},
  \ and\ \bibinfo {author} {\bibfnamefont {A.}~\bibnamefont {Riotto}},\ }\href
  {\doibase 10.1016/j.nuclphysb.2007.02.019} {\bibfield  {journal} {\bibinfo
  {journal} {Nucl. Phys. B}\ }\textbf {\bibinfo {volume} {774}},\ \bibinfo
  {pages} {1} (\bibinfo {year} {2007})},\ \Eprint
  {http://arxiv.org/abs/hep-ph/0611338} {arXiv:hep-ph/0611338} \BibitemShut
  {NoStop}%
\bibitem [{\citenamefont {Branco}\ \emph {et~al.}(2007)\citenamefont {Branco},
  \citenamefont {Gonzalez~Felipe},\ and\ \citenamefont
  {Joaquim}}]{Branco:2006ce}%
  \BibitemOpen
  \bibfield  {author} {\bibinfo {author} {\bibfnamefont {G.~C.}\ \bibnamefont
  {Branco}}, \bibinfo {author} {\bibfnamefont {R.}~\bibnamefont
  {Gonzalez~Felipe}}, \ and\ \bibinfo {author} {\bibfnamefont {F.~R.}\
  \bibnamefont {Joaquim}},\ }\href {\doibase 10.1016/j.physletb.2006.12.060}
  {\bibfield  {journal} {\bibinfo  {journal} {Phys. Lett. B}\ }\textbf
  {\bibinfo {volume} {645}},\ \bibinfo {pages} {432} (\bibinfo {year}
  {2007})},\ \Eprint {http://arxiv.org/abs/hep-ph/0609297}
  {arXiv:hep-ph/0609297} \BibitemShut {NoStop}%
\bibitem [{\citenamefont {Anisimov}\ \emph {et~al.}(2008)\citenamefont
  {Anisimov}, \citenamefont {Blanchet},\ and\ \citenamefont
  {Di~Bari}}]{Anisimov:2007mw}%
  \BibitemOpen
  \bibfield  {author} {\bibinfo {author} {\bibfnamefont {A.}~\bibnamefont
  {Anisimov}}, \bibinfo {author} {\bibfnamefont {S.}~\bibnamefont {Blanchet}},
  \ and\ \bibinfo {author} {\bibfnamefont {P.}~\bibnamefont {Di~Bari}},\ }\href
  {\doibase 10.1088/1475-7516/2008/04/033} {\bibfield  {journal} {\bibinfo
  {journal} {JCAP}\ }\textbf {\bibinfo {volume} {04}},\ \bibinfo {pages} {033}
  (\bibinfo {year} {2008})},\ \Eprint {http://arxiv.org/abs/0707.3024}
  {arXiv:0707.3024 [hep-ph]} \BibitemShut {NoStop}%
\bibitem [{\citenamefont {Molinaro}\ and\ \citenamefont
  {Petcov}(2009)}]{Molinaro:2009lud}%
  \BibitemOpen
  \bibfield  {author} {\bibinfo {author} {\bibfnamefont {E.}~\bibnamefont
  {Molinaro}}\ and\ \bibinfo {author} {\bibfnamefont {S.~T.}\ \bibnamefont
  {Petcov}},\ }\href {\doibase 10.1140/epjc/s10052-009-0985-3} {\bibfield
  {journal} {\bibinfo  {journal} {Eur. Phys. J. C}\ }\textbf {\bibinfo {volume}
  {61}},\ \bibinfo {pages} {93} (\bibinfo {year} {2009})},\ \Eprint
  {http://arxiv.org/abs/0803.4120} {arXiv:0803.4120 [hep-ph]} \BibitemShut
  {NoStop}%
\bibitem [{\citenamefont {Shimizu}\ \emph {et~al.}(2018)\citenamefont
  {Shimizu}, \citenamefont {Takagi},\ and\ \citenamefont
  {Tanimoto}}]{Shimizu:2017vwi}%
  \BibitemOpen
  \bibfield  {author} {\bibinfo {author} {\bibfnamefont {Y.}~\bibnamefont
  {Shimizu}}, \bibinfo {author} {\bibfnamefont {K.}~\bibnamefont {Takagi}}, \
  and\ \bibinfo {author} {\bibfnamefont {M.}~\bibnamefont {Tanimoto}},\ }\href
  {\doibase 10.1016/j.physletb.2017.12.065} {\bibfield  {journal} {\bibinfo
  {journal} {Phys. Lett. B}\ }\textbf {\bibinfo {volume} {778}},\ \bibinfo
  {pages} {6} (\bibinfo {year} {2018})},\ \Eprint
  {http://arxiv.org/abs/1711.03863} {arXiv:1711.03863 [hep-ph]} \BibitemShut
  {NoStop}%
\bibitem [{\citenamefont {Moffat}\ \emph {et~al.}(2019)\citenamefont {Moffat},
  \citenamefont {Pascoli}, \citenamefont {Petcov},\ and\ \citenamefont
  {Turner}}]{Moffat:2018smo}%
  \BibitemOpen
  \bibfield  {author} {\bibinfo {author} {\bibfnamefont {K.}~\bibnamefont
  {Moffat}}, \bibinfo {author} {\bibfnamefont {S.}~\bibnamefont {Pascoli}},
  \bibinfo {author} {\bibfnamefont {S.~T.}\ \bibnamefont {Petcov}}, \ and\
  \bibinfo {author} {\bibfnamefont {J.}~\bibnamefont {Turner}},\ }\href
  {\doibase 10.1007/JHEP03(2019)034} {\bibfield  {journal} {\bibinfo  {journal}
  {JHEP}\ }\textbf {\bibinfo {volume} {03}},\ \bibinfo {pages} {034} (\bibinfo
  {year} {2019})},\ \Eprint {http://arxiv.org/abs/1809.08251} {arXiv:1809.08251
  [hep-ph]} \BibitemShut {NoStop}%
\bibitem [{\citenamefont {Pilaftsis}\ and\ \citenamefont
  {Underwood}(2004)}]{Pilaftsis:2003gt}%
  \BibitemOpen
  \bibfield  {author} {\bibinfo {author} {\bibfnamefont {A.}~\bibnamefont
  {Pilaftsis}}\ and\ \bibinfo {author} {\bibfnamefont {T.~E.~J.}\ \bibnamefont
  {Underwood}},\ }\href {\doibase 10.1016/j.nuclphysb.2004.05.029} {\bibfield
  {journal} {\bibinfo  {journal} {Nucl. Phys. B}\ }\textbf {\bibinfo {volume}
  {692}},\ \bibinfo {pages} {303} (\bibinfo {year} {2004})},\ \Eprint
  {http://arxiv.org/abs/hep-ph/0309342} {arXiv:hep-ph/0309342} \BibitemShut
  {NoStop}%
\bibitem [{\citenamefont {Akhmedov}\ \emph {et~al.}(1998)\citenamefont
  {Akhmedov}, \citenamefont {Rubakov},\ and\ \citenamefont
  {Smirnov}}]{Akhmedov:1998qx}%
  \BibitemOpen
  \bibfield  {author} {\bibinfo {author} {\bibfnamefont {E.~K.}\ \bibnamefont
  {Akhmedov}}, \bibinfo {author} {\bibfnamefont {V.~A.}\ \bibnamefont
  {Rubakov}}, \ and\ \bibinfo {author} {\bibfnamefont {A.~{\relax Yu}.}\
  \bibnamefont {Smirnov}},\ }\href {\doibase 10.1103/PhysRevLett.81.1359}
  {\bibfield  {journal} {\bibinfo  {journal} {Phys. Rev. Lett.}\ }\textbf
  {\bibinfo {volume} {81}},\ \bibinfo {pages} {1359} (\bibinfo {year}
  {1998})},\ \Eprint {http://arxiv.org/abs/hep-ph/9803255}
  {arXiv:hep-ph/9803255 [hep-ph]} \BibitemShut {NoStop}%
\bibitem [{\citenamefont {Ahdida}\ \emph {et~al.}(2022)\citenamefont {Ahdida}
  \emph {et~al.}}]{SHIP:2021tpn}%
  \BibitemOpen
  \bibfield  {author} {\bibinfo {author} {\bibfnamefont {C.}~\bibnamefont
  {Ahdida}} \emph {et~al.} (\bibinfo {collaboration} {SHiP}),\ }\href {\doibase
  10.1140/epjc/s10052-022-10346-5} {\bibfield  {journal} {\bibinfo  {journal}
  {Eur. Phys. J. C}\ }\textbf {\bibinfo {volume} {82}},\ \bibinfo {pages} {486}
  (\bibinfo {year} {2022})},\ \Eprint {http://arxiv.org/abs/2112.01487}
  {arXiv:2112.01487 [physics.ins-det]} \BibitemShut {NoStop}%
\bibitem [{\citenamefont {Gorbunov}\ and\ \citenamefont
  {Shaposhnikov}(2007)}]{Gorbunov:2007ak}%
  \BibitemOpen
  \bibfield  {author} {\bibinfo {author} {\bibfnamefont {D.}~\bibnamefont
  {Gorbunov}}\ and\ \bibinfo {author} {\bibfnamefont {M.}~\bibnamefont
  {Shaposhnikov}},\ }\href {\doibase 10.1088/1126-6708/2007/10/015} {\bibfield
  {journal} {\bibinfo  {journal} {JHEP}\ }\textbf {\bibinfo {volume} {10}},\
  \bibinfo {pages} {015} (\bibinfo {year} {2007})},\ \bibinfo {note} {[Erratum:
  JHEP 11, 101 (2013)]},\ \Eprint {http://arxiv.org/abs/0705.1729}
  {arXiv:0705.1729 [hep-ph]} \BibitemShut {NoStop}%
\bibitem [{\citenamefont {Bondarenko}\ \emph {et~al.}(2018)\citenamefont
  {Bondarenko}, \citenamefont {Boyarsky}, \citenamefont {Gorbunov},\ and\
  \citenamefont {Ruchayskiy}}]{Bondarenko:2018ptm}%
  \BibitemOpen
  \bibfield  {author} {\bibinfo {author} {\bibfnamefont {K.}~\bibnamefont
  {Bondarenko}}, \bibinfo {author} {\bibfnamefont {A.}~\bibnamefont
  {Boyarsky}}, \bibinfo {author} {\bibfnamefont {D.}~\bibnamefont {Gorbunov}},
  \ and\ \bibinfo {author} {\bibfnamefont {O.}~\bibnamefont {Ruchayskiy}},\
  }\href {\doibase 10.1007/JHEP11(2018)032} {\bibfield  {journal} {\bibinfo
  {journal} {JHEP}\ }\textbf {\bibinfo {volume} {11}},\ \bibinfo {pages} {032}
  (\bibinfo {year} {2018})},\ \Eprint {http://arxiv.org/abs/1805.08567}
  {arXiv:1805.08567 [hep-ph]} \BibitemShut {NoStop}%
\bibitem [{\citenamefont {Drewes}\ \emph {et~al.}(2022)\citenamefont {Drewes},
  \citenamefont {Georis},\ and\ \citenamefont {Klari\'c}}]{Drewes:2021nqr}%
  \BibitemOpen
  \bibfield  {author} {\bibinfo {author} {\bibfnamefont {M.}~\bibnamefont
  {Drewes}}, \bibinfo {author} {\bibfnamefont {Y.}~\bibnamefont {Georis}}, \
  and\ \bibinfo {author} {\bibfnamefont {J.}~\bibnamefont {Klari\'c}},\ }\href
  {\doibase 10.1103/PhysRevLett.128.051801} {\bibfield  {journal} {\bibinfo
  {journal} {Phys. Rev. Lett.}\ }\textbf {\bibinfo {volume} {128}},\ \bibinfo
  {pages} {051801} (\bibinfo {year} {2022})},\ \Eprint
  {http://arxiv.org/abs/2106.16226} {arXiv:2106.16226 [hep-ph]} \BibitemShut
  {NoStop}%
\bibitem [{\citenamefont {Aitken}\ \emph {et~al.}(2017)\citenamefont {Aitken},
  \citenamefont {McKeen}, \citenamefont {Neder},\ and\ \citenamefont
  {Nelson}}]{Aitken:2017wie}%
  \BibitemOpen
  \bibfield  {author} {\bibinfo {author} {\bibfnamefont {K.}~\bibnamefont
  {Aitken}}, \bibinfo {author} {\bibfnamefont {D.}~\bibnamefont {McKeen}},
  \bibinfo {author} {\bibfnamefont {T.}~\bibnamefont {Neder}}, \ and\ \bibinfo
  {author} {\bibfnamefont {A.~E.}\ \bibnamefont {Nelson}},\ }\href {\doibase
  10.1103/PhysRevD.96.075009} {\bibfield  {journal} {\bibinfo  {journal} {Phys.
  Rev. D}\ }\textbf {\bibinfo {volume} {96}},\ \bibinfo {pages} {075009}
  (\bibinfo {year} {2017})},\ \Eprint {http://arxiv.org/abs/1708.01259}
  {arXiv:1708.01259 [hep-ph]} \BibitemShut {NoStop}%
\bibitem [{\citenamefont {Elor}\ \emph {et~al.}(2019)\citenamefont {Elor},
  \citenamefont {Escudero},\ and\ \citenamefont {Nelson}}]{Elor:2018twp}%
  \BibitemOpen
  \bibfield  {author} {\bibinfo {author} {\bibfnamefont {G.}~\bibnamefont
  {Elor}}, \bibinfo {author} {\bibfnamefont {M.}~\bibnamefont {Escudero}}, \
  and\ \bibinfo {author} {\bibfnamefont {A.}~\bibnamefont {Nelson}},\ }\href
  {\doibase 10.1103/PhysRevD.99.035031} {\bibfield  {journal} {\bibinfo
  {journal} {Phys. Rev. D}\ }\textbf {\bibinfo {volume} {99}},\ \bibinfo
  {pages} {035031} (\bibinfo {year} {2019})},\ \Eprint
  {http://arxiv.org/abs/1810.00880} {arXiv:1810.00880 [hep-ph]} \BibitemShut
  {NoStop}%
\bibitem [{\citenamefont {Elor}\ and\ \citenamefont
  {McGehee}(2021)}]{Elor:2020tkc}%
  \BibitemOpen
  \bibfield  {author} {\bibinfo {author} {\bibfnamefont {G.}~\bibnamefont
  {Elor}}\ and\ \bibinfo {author} {\bibfnamefont {R.}~\bibnamefont {McGehee}},\
  }\href {\doibase 10.1103/PhysRevD.103.035005} {\bibfield  {journal} {\bibinfo
   {journal} {Phys. Rev. D}\ }\textbf {\bibinfo {volume} {103}},\ \bibinfo
  {pages} {035005} (\bibinfo {year} {2021})},\ \Eprint
  {http://arxiv.org/abs/2011.06115} {arXiv:2011.06115 [hep-ph]} \BibitemShut
  {NoStop}%
\bibitem [{\citenamefont {Elahi}\ \emph {et~al.}(2022)\citenamefont {Elahi},
  \citenamefont {Elor},\ and\ \citenamefont {McGehee}}]{Elahi:2021jia}%
  \BibitemOpen
  \bibfield  {author} {\bibinfo {author} {\bibfnamefont {F.}~\bibnamefont
  {Elahi}}, \bibinfo {author} {\bibfnamefont {G.}~\bibnamefont {Elor}}, \ and\
  \bibinfo {author} {\bibfnamefont {R.}~\bibnamefont {McGehee}},\ }\href
  {\doibase 10.1103/PhysRevD.105.055024} {\bibfield  {journal} {\bibinfo
  {journal} {Phys. Rev. D}\ }\textbf {\bibinfo {volume} {105}},\ \bibinfo
  {pages} {055024} (\bibinfo {year} {2022})},\ \Eprint
  {http://arxiv.org/abs/2109.09751} {arXiv:2109.09751 [hep-ph]} \BibitemShut
  {NoStop}%
\bibitem [{\citenamefont {Alonso-\'Alvarez}\ \emph {et~al.}(2021)\citenamefont
  {Alonso-\'Alvarez}, \citenamefont {Elor},\ and\ \citenamefont
  {Escudero}}]{Alonso-Alvarez:2021qfd}%
  \BibitemOpen
  \bibfield  {author} {\bibinfo {author} {\bibfnamefont {G.}~\bibnamefont
  {Alonso-\'Alvarez}}, \bibinfo {author} {\bibfnamefont {G.}~\bibnamefont
  {Elor}}, \ and\ \bibinfo {author} {\bibfnamefont {M.}~\bibnamefont
  {Escudero}},\ }\href {\doibase 10.1103/PhysRevD.104.035028} {\bibfield
  {journal} {\bibinfo  {journal} {Phys. Rev. D}\ }\textbf {\bibinfo {volume}
  {104}},\ \bibinfo {pages} {035028} (\bibinfo {year} {2021})},\ \Eprint
  {http://arxiv.org/abs/2101.02706} {arXiv:2101.02706 [hep-ph]} \BibitemShut
  {NoStop}%
\bibitem [{\citenamefont {Hadjivasiliou}\ \emph {et~al.}(2022)\citenamefont
  {Hadjivasiliou} \emph {et~al.}}]{Belle:2021gmc}%
  \BibitemOpen
  \bibfield  {author} {\bibinfo {author} {\bibfnamefont {C.}~\bibnamefont
  {Hadjivasiliou}} \emph {et~al.} (\bibinfo {collaboration} {Belle}),\ }\href
  {\doibase 10.1103/PhysRevD.105.L051101} {\bibfield  {journal} {\bibinfo
  {journal} {Phys. Rev. D}\ }\textbf {\bibinfo {volume} {105}},\ \bibinfo
  {pages} {L051101} (\bibinfo {year} {2022})},\ \Eprint
  {http://arxiv.org/abs/2110.14086} {arXiv:2110.14086 [hep-ex]} \BibitemShut
  {NoStop}%
\bibitem [{\citenamefont {Rodr\'\i{}guez}\ \emph {et~al.}(2021)\citenamefont
  {Rodr\'\i{}guez}, \citenamefont {Chobanova}, \citenamefont {Cid~Vidal},
  \citenamefont {Soli\~no}, \citenamefont {Santos}, \citenamefont
  {Momb\"acher}, \citenamefont {Prouv\'e}, \citenamefont {Fern\'andez},\ and\
  \citenamefont {V\'azquez~Sierra}}]{Rodriguez:2021urv}%
  \BibitemOpen
  \bibfield  {author} {\bibinfo {author} {\bibfnamefont {A.~B.}\ \bibnamefont
  {Rodr\'\i{}guez}}, \bibinfo {author} {\bibfnamefont {V.}~\bibnamefont
  {Chobanova}}, \bibinfo {author} {\bibfnamefont {X.}~\bibnamefont
  {Cid~Vidal}}, \bibinfo {author} {\bibfnamefont {S.~L.}\ \bibnamefont
  {Soli\~no}}, \bibinfo {author} {\bibfnamefont {D.~M.}\ \bibnamefont
  {Santos}}, \bibinfo {author} {\bibfnamefont {T.}~\bibnamefont {Momb\"acher}},
  \bibinfo {author} {\bibfnamefont {C.}~\bibnamefont {Prouv\'e}}, \bibinfo
  {author} {\bibfnamefont {E.~X.~R.}\ \bibnamefont {Fern\'andez}}, \ and\
  \bibinfo {author} {\bibfnamefont {C.}~\bibnamefont {V\'azquez~Sierra}},\
  }\href {\doibase 10.1140/epjc/s10052-021-09762-w} {\bibfield  {journal}
  {\bibinfo  {journal} {Eur. Phys. J. C}\ }\textbf {\bibinfo {volume} {81}},\
  \bibinfo {pages} {964} (\bibinfo {year} {2021})},\ \Eprint
  {http://arxiv.org/abs/2106.12870} {arXiv:2106.12870 [hep-ph]} \BibitemShut
  {NoStop}%
\bibitem [{\citenamefont {Borsato}\ \emph {et~al.}(2022)\citenamefont {Borsato}
  \emph {et~al.}}]{Borsato:2021aum}%
  \BibitemOpen
  \bibfield  {author} {\bibinfo {author} {\bibfnamefont {M.}~\bibnamefont
  {Borsato}} \emph {et~al.},\ }\href {\doibase 10.1088/1361-6633/ac4649}
  {\bibfield  {journal} {\bibinfo  {journal} {Rept. Prog. Phys.}\ }\textbf
  {\bibinfo {volume} {85}},\ \bibinfo {pages} {024201} (\bibinfo {year}
  {2022})},\ \Eprint {http://arxiv.org/abs/2105.12668} {arXiv:2105.12668
  [hep-ph]} \BibitemShut {NoStop}%
\end{thebibliography}%

\end{document}